\documentclass[amsmath,amssymb,reprint,aps,prd,eqsecnum]{revtex4-2}
 
\usepackage{graphicx}
\usepackage{dcolumn}
\usepackage{bm}
\usepackage{mathrsfs}
\usepackage{color}
\usepackage[normalem]{ulem}

\allowdisplaybreaks

\begin{document}

\title{Quantization of a charged scalar field on a charged black hole background}

\author{Visakan Balakumar}

\email{V.Balakumar@sheffield.ac.uk}

\affiliation{Consortium for Fundamental Physics,
	School of Mathematics and Statistics,
	The University of Sheffield,
	Hicks Building,
	Hounsfield Road,
	Sheffield. S3 7RH United Kingdom}

\author{Rafael P.~Bernar}

\email{rbernar@ufpa.br}

\affiliation{Faculdade de F\'isica, 
Universidade Federal do Par\'a, 
66075-110, Bel\'em, Par\'a, Brazil}

\author{Elizabeth Winstanley}

\email{E.Winstanley@sheffield.ac.uk}

\affiliation{Consortium for Fundamental Physics,
School of Mathematics and Statistics,
The University of Sheffield,
Hicks Building,
Hounsfield Road,
Sheffield. S3 7RH United Kingdom}

\date{\today}

\begin{abstract}
We study the canonical quantization of a massless charged scalar field on a Reissner-Nordstr\"om black hole background.
Our aim is to construct analogues of the standard Boulware, Unruh and Hartle-Hawking quantum states which can be defined for a neutral scalar field, and to explore their physical properties by computing differences in expectation values of the scalar field condensate, current and stress-energy tensor operators between two quantum states. 
Each of these three states has a non-time-reversal-invariant ``past'' and ``future'' charged field generalization, whose properties are similar to those of the corresponding ``past'' and ``future'' states for a neutral scalar field on a Kerr black hole.  
In addition, we present some tentative, time-reversal-invariant, equilibrium states. The first is a ``Boulware''-like state which is as empty as possible at both future and past null infinity. 
Second, we posit a ``Hartle-Hawking''-like state which may correspond to a thermal distribution of particles. The construction of both these latter states relies on the use of nonstandard commutation relations for the creation and annihilation operators pertaining to superradiant modes. 
\end{abstract}

\maketitle

\section{Introduction}
\label{sec:intro}

In the absence of a definitive theory of quantum gravity, quantum field theory in curved space-time has proven to be a fruitful avenue for research.
In this approach, the space-time background is regarded as purely classical, and quantum fields propagating on a fixed background are studied.
Some of the earliest and deepest results arising from this set-up are pertinent to black hole physics, including the thermal Hawking radiation emitted by black holes formed by the gravitational collapse of a compact body \cite{Hawking:1974rv,Hawking:1974sw}.

The simplest black hole space-time is the Schwarzschild black hole, and quantum fields propagating on this background have been studied extensively.
The primary physical quantities of interest are expectation values of quantum operators in a particular quantum state. 
For example, the expectation value of the quantum stress-energy tensor (SET) operator ${\hat {{T}}}_{\mu \nu }$ governs the back-reaction of the quantum field on the space-time geometry via the semi-classical Einstein equations
\begin{equation}
G_{\mu \nu } = 8\pi \langle {\hat {{T}}}_{\mu \nu } \rangle ,
\label{eq:SCEE}
\end{equation}
where $G_{\mu \nu }$ is the Einstein tensor, $\langle \, \rangle $ denotes an expectation value, and we are using units in which $c=G=\hbar = k_{B}=1$, as we shall throughout this paper.
In order to compute the right-hand-side of (\ref{eq:SCEE}), one needs to first specify a quantum field, and then consider a particular quantum state. 
Defining quantum states on a general curved space-time background is nontrivial because the notion of particle is observer-dependent and, as a result, there may not be a unique or natural vacuum state.

On a Schwarzschild black hole, three standard quantum states have been studied in the literature \cite{Candelas:1980zt}:
\begin{description}
	\item[Boulware state \cite{Boulware:1974dm}]
	This is defined as the quantum state which is as empty as possible far from the black hole.
	However, this state diverges on the event horizon of the black hole and physically represents the vacuum state outside a star which does not have a horizon. 
	This state respects the symmetries of the underlying Schwarzschild space-time and, in particular, is time-reversal invariant.
	\item[Unruh state \cite{Unruh:1976db}]
	Unlike the Boulware state, the Unruh state is not time-reversal invariant. 
	It is the state pertinent to modelling a black hole formed by gravitational collapse. 
	While the state is empty at past null infinity, it contains an outwards flux of particles (the Hawking radiation) at future null infinity. 
	The Unruh state is regular across the future event horizon, but not the past event horizon of an eternal black hole.
	\item[Hartle-Hawking state \cite{Hartle:1976tp,Israel:1976ur}]
	This state represents a black hole surrounded by thermal radiation at the Hawking temperature. 
	As well as being time-reversal invariant, this state has attractive regularity properties, being regular across both the future and past event horizons. 
\end{description}
The properties of these three states for various quantum fields on Schwarzschild space-time have been extensively studied via computations of renormalized expectation values (for a sample of the literature, see 
\cite{Anderson:1989vg,Anderson:1993if,Anderson:1994hg,Candelas:1984pg,Carlson:2003ub,Elster:1984hu,Fawcett:1983dk,Howard:1984qp,Howard:1985yg,Jensen:1988rh,Jensen:1992mv,Jensen:1995qv,Levi:2015eea,Levi:2016quh,Levi:2016esr,Levi:2016paz}).

Prior to the discovery of Hawking radiation, it was already known that rotating  Kerr black holes emit quantum Unruh-Starobinskii radiation \cite{Starobinsky:1973aij,Unruh:1974bw}.
Classical bosonic fields propagating on a Kerr black hole space-time exhibit superradiance \cite{Chandrasekhar:1985kt,Brito:2015oca}, whereby low frequency modes incident from infinity are amplified upon reflection from the black hole. 
Unruh-Starobinskii radiation is the quantum analogue of classical superradiance, and occurs for fermionic as well as bosonic fields \cite{Unruh:1974bw}.

The presence of superradiant modes complicates the definition of quantum states on a Kerr black hole background, particularly for bosonic fields
\cite{Frolov:1989jh,Ottewill:2000qh,Casals:2012es,Casals:2005kr,Duffy:2005mz}.
For a quantum scalar field, it is no longer possible to define a Boulware-like state which is as empty as possible at both future and past null infinity \cite{Ottewill:2000qh,Casals:2005kr} (there is such a state for a quantum fermion field \cite{Casals:2012es}, but it diverges on the stationary limit surface).
Instead, the analogue of the Boulware state for a quantum scalar field is no longer time-reversal invariant. 
Although it is empty at past null infinity, it contains an outgoing flux of particles in the superradiant modes at future null infinity, corresponding to the Unruh-Starobinskii radiation \cite{Unruh:1974bw,Casals:2012es,Ottewill:2000qh,Matacz:1993hs}.
While the Unruh state is well-defined and has similar properties to that on Schwarzschild space-time \cite{Levi:2016exv}, this is not the case for the Hartle-Hawking state.
In particular, for a quantum scalar field on a Kerr black hole there does not exist a quantum state respecting all the symmetries of the space-time and which is regular across both the future and past event horizons \cite{Kay:1988mu,Kay:1992gr}.
Attempts to define analogues of the Hartle-Hawking state for either bosonic or fermionic fields on a Kerr black hole lead to states which are either divergent in at least part of the space-time exterior to the event horizon \cite{Frolov:1989jh,Ottewill:2000qh,Casals:2012es,Casals:2005kr,Duffy:2005mz} or which do not describe an equilibrium state \cite{Candelas:1981zv,Casals:2012es}.

The study of quantum field theory on a Kerr black hole is further complicated due to the fact that the space-time has fewer symmetries than a Schwarzschild black hole, being only axisymmetric rather than spherically symmetric.
Indeed, renormalized expectation values for the Unruh state have only been computed comparatively recently for the whole region exterior to the event horizon 
\cite{Levi:2016exv}.

One of the reasons why quantum field theory on Kerr black holes is so challenging is because there are two interlinked effects at play: superradiance and rotation.
Even in flat Minkowski space-time, defining rotating quantum states is nontrivial \cite{Letaw:1979wy,Vilenkin:1980zv,Davies:1996ks}, and rigidly-rotating thermal states do not exist for bosonic fields on the unbounded space-time \cite{Duffy:2002ss} (for fermionic fields, such states can be constructed \cite{Iyer:1982ah,Ambrus:2014uqa} but they are not regular everywhere \cite{Ambrus:2014uqa}).
The question then arises as to whether it is possible to disentangle these two effects.
As outlined above, it is possible to study the effects of rotation separately in flat space-time, but what about the consequences of superradiance?
There is a simpler black hole system which exhibits superradiance without rotation, and that is the focus of our work in this paper.

A classical charged scalar field propagating on a charged Reissner-Nordstr\"om (RN) black hole space-time exhibits the phenomenon of charge superradiance \cite{Bekenstein:1973mi} (see, for example, \cite{Benone:2015bst,DiMenza:2014vpa} for more recent work).
This is analogous to superradiance for bosonic fields on Kerr black holes, namely low-frequency modes can be amplified on scattering by the charged RN black hole, thereby extracting some of the charge of the black hole.
There is a quantum analogue of charge superradiance \cite{Gibbons:1975kk,Balakumar:2020gli}, and particles are spontaneously emitted by the black hole in those modes which are subject to charge superradiance. 
The interaction between the charge of the scalar field and the charge of the black hole also affects the Hawking radiation \cite{Carter:1974yx,Gibbons:1975kk,Page:1977um,Sampaio:2009tp,Sampaio:2009ra}
and hence also the evolution of an evaporating black hole (studied using an adiabatic approximation in \cite{Hiscock:1990ex,Ong:2019rnn,Ong:2019vnv,Xu:2019wak}).
Recently, the Unruh state for a charged scalar on an RN-de Sitter black hole has been constructed and its properties explored both inside and outside the event horizon \cite{Klein:2021ctt,Klein:2021les}.
However, the physical properties of the analogues of the other standard quantum states discussed above for charged fields on the RN space-time have been little studied to date.

Here we consider in detail the canonical quantization of a massless charged scalar field, minimally coupled to the space-time geometry, and propagating on an RN black hole. 
As well as fixing the classical space-time geometry, we regard the background electromagnetic field as fixed and classical.
We discuss in detail the construction of analogues of the Boulware, Unruh and Hartle-Hawking states on this background, paying particular attention to the consequences of charge superradiance. 
The physical properties of these states are then explored by considering differences in expectation values between two quantum states, which do not require renormalization.
In addition to the quantum stress-energy tensor operator, we also examine the expectation value of the scalar field current operator ${\hat {{J}}}^{\mu }$ (considered in \cite{Klein:2021ctt,Klein:2021les} on an RN-de Sitter black hole), which acts as a source for the semiclassical Maxwell equations
 \begin{equation}
 \nabla _{\mu }F^{\mu \nu }=4\pi \langle  {\hat {{J}}}^{\nu } \rangle 
 \label{eq:SCME}
 \end{equation}
 (in Gaussian units), where $F_{\mu \nu }$ is the electromagnetic gauge field strength.
 The semiclassical Maxwell equations (\ref{eq:SCME}) govern the backreaction of the quantum field on the electromagnetic field.
 We also consider the simplest nontrivial expectation value, the scalar field condensate.
 
The outline of this paper is as follows.
In Sec.~\ref{sec:classical} we briefly review the RN geometry and describe the classical charged scalar field modes which will be used extensively throughout the paper.
The canonical quantization of the charged scalar field is the focus of Sec.~\ref{sec:quantum}.
We construct in detail a wide range of quantum states, inspired by the standard Boulware, Unruh and Hartle-Hawking states on the Schwarzschild black hole.
Differences in expectation values of observables between two quantum states are studied in Sec.~\ref{sec:expvalues}, first by considering the asymptotic behaviour of the states near the horizon and infinity, which can be derived analytically for at least some states, and second by full numerical computations valid everywhere outside the event horizon. 
From these computations we examine some key properties of the states we have defined, including symmetry with respect to time-reversal, regularity (including on the event horizons) and the presence of fluxes.
Our conclusions are presented in Sec.~\ref{sec:conc}.

\section{Classical charged scalar field on a Reissner-Nordstr\"om black hole}
\label{sec:classical}

In this section we review the key properties of the RN black hole geometry, and define the classical charged scalar field modes on this background.

\subsection{Reissner-Nordstr\"om black hole geometry}
\label{sec:geometry}

\begin{figure}[h]
	\centering
	\includegraphics[width=0.5\textwidth]{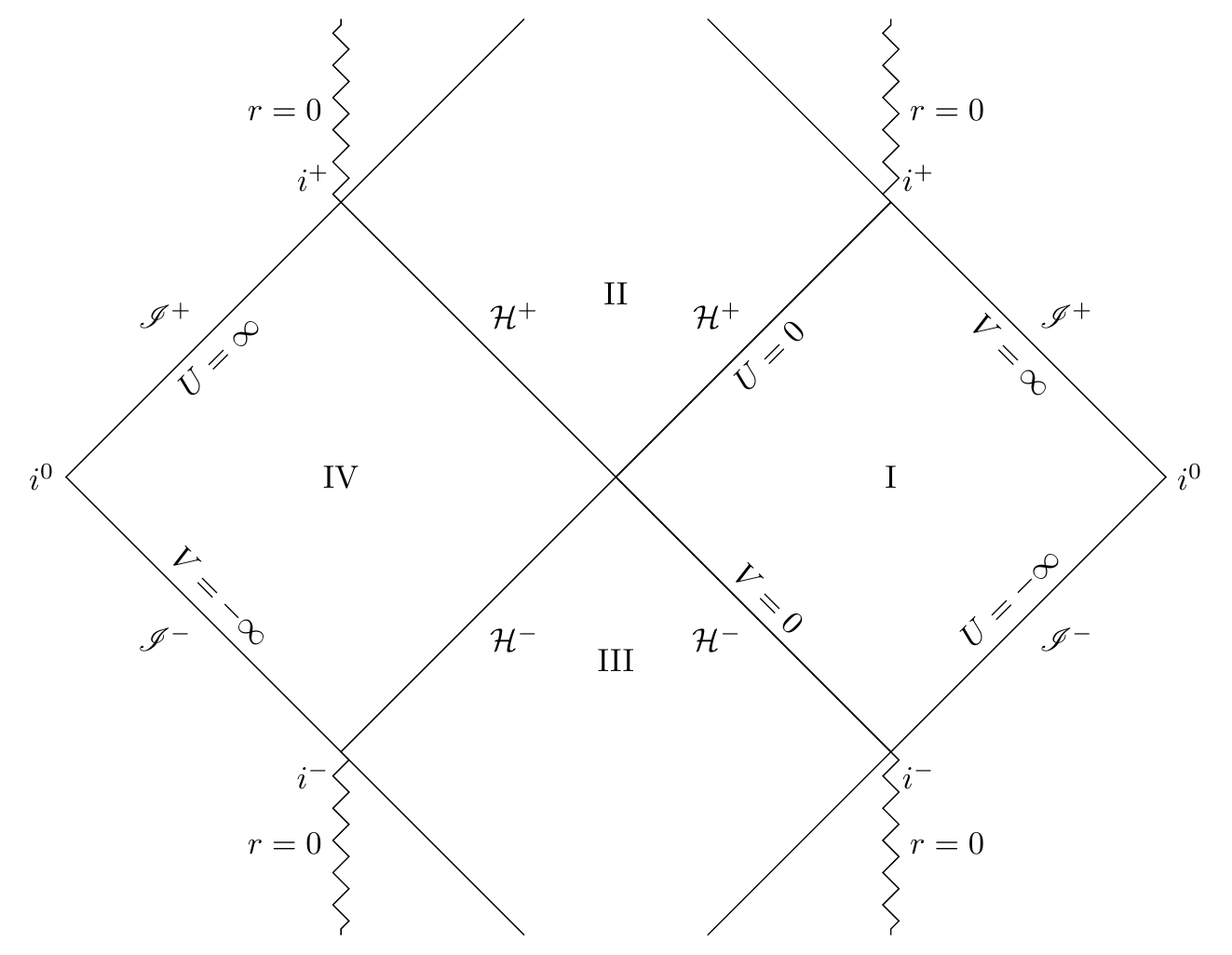}
	\caption{Penrose diagram of nonextremal RN space-time. The future and past event horizons are denoted ${\mathcal {H}}^{\pm }$, while ${\mathscr{I}}^{\pm }$ are future and past null infinity. Future and past timelike infinity are labelled $i^{\pm}$, and $i^{0}$ is space-like infinity. There is a space-time singularity at $r=0$. The diagram also shows regions I, II, III and IV, which will be required in our constructions.}
	\label{fig:RN}
\end{figure}

The background space-time is a four-dimensional, spherically symmetric RN black hole with metric
\begin{equation}
    ds^{2} = - f(r) \, dt^{2} + f(r)  ^{-1} dr^{2}+ r^{2} d\theta ^{2} + r^{2}\sin ^{2} \theta \, d\varphi ^{2} ,
\label{eq:RNmetric}
\end{equation}
where the metric function $f(r)$ is given by 
\begin{equation}
    f(r) = 1 - \frac{2M}{r} + \frac{Q^{2}}{r^{2}} ,
    \label{eq:fr}
\end{equation}
with $M$ the mass and $Q$ the electric charge of the black hole.
If $M^{2}>Q^{2}$, the metric function $f(r)$ has two zeros, at $r=r_{\pm }$, where
\begin{equation}
    r_{\pm } = M \pm {\sqrt {M^{2}-Q^{2}}}.
    \label{eq:rpm}
\end{equation}
In this case, $r_{+}$ is the location of the black hole event horizon and $r_{-}$ is the location of the inner horizon.
When $M^{2}=Q^{2}$, the two horizons coincide and the black hole is extremal. 
For $M^{2}<Q^{2}$, there is a naked singularity.
In this paper we restrict our attention to the case $M^{2}>Q^{2}$.
Part of the Penrose diagram for the nonextremal black hole is depicted in Fig.~\ref{fig:RN}.

Our primary interest in this paper is in defining states in the region exterior to the event horizon, region I in Fig.~\ref{fig:RN}.  
However, in order to do so we will need to employ scalar field modes which are defined in the other regions shown in Fig.~\ref{fig:RN}.
It is therefore useful to define Kruskal coordinates $U$, $V$ which are regular in all four of regions I--IV.
In region I, ingoing and outgoing null coordinates $u$, $v$ are given respectively by
\begin{equation}
u = t-r_{*}, \qquad v=t+r_{*}, 
\label{eq:uv}
\end{equation}
in terms of the usual ``tortoise'' coordinate $r_{*}$, defined by 
\begin{equation}
\frac{dr_{*}}{dr} = \frac{1}{f(r)},
\label{eq:rstar}
\end{equation}
where the metric function $f(r)$ is given by (\ref{eq:fr}).
In region I, the tortoise coordinate has the range $-\infty < r_{*}<\infty $.
Kruskal coordinates can then be written in terms of $u$ and $v$ in region I as follows:
\begin{equation}
U = -\frac{1}{\kappa }e^{-\kappa u}, \qquad  V= \frac{1}{\kappa }e^{\kappa v},
\label{eq:Kruskal}
\end{equation}
where 
\begin{equation}
\kappa = \frac{1}{2}f'(r_{+}) = \frac{1}{r_{+}^{2}} \left( r_{+}-M  \right)
\label{eq:kappa}
\end{equation}
is the surface gravity of the event horizon.
In region I, the Kruskal coordinate $U=0$ on the future event horizon ${\mathcal {H}}^{+}$, and tends to $-\infty $ at past null infinity ${\mathscr{I}}^{-}$.
Similarly, in region I, the Kruskal coordinate $V=0$ on the past event horizon ${\mathcal {H}}^{-}$ and tends to $\infty $ at future null infinity ${\mathscr{I}}^{+}$.
Values of $U$ and $V$ on some other key surfaces in the space-time are shown in Fig.~\ref{fig:RN}.

The RN black hole is a solution of Einstein's equations with an electromagnetic field.
The background electromagnetic potential has components $A_{\mu } = (A_{0}, 0 , 0 ,0 )$ where
\begin{equation}
    A_{0} = -\frac{Q}{r} ,
    \label{eq:gaugepot}
\end{equation}
and we have chosen a constant of integration so that the gauge field potential vanishes far from the black hole.
As observed in \cite{Klein:2021ctt,Klein:2021les}, by means of a gauge transformation it is possible to set the gauge field potential to zero at any fixed chosen value of $r$. 
In this paper we fix the gauge so that (\ref{eq:gaugepot}) holds throughout.
The electromagnetic potential $A_{\mu }$  (\ref{eq:gaugepot}) satisfies the Lorenz gauge condition $\nabla ^{\mu }A_{\mu } = 0$.

\subsection{Classical charged scalar field}
\label{sec:chargedscalar}

The focus of this paper is a massless, charged, complex scalar field $\Phi $ with charge $q$, minimally coupled to the space-time geometry, and satisfying the equation
\begin{equation}
D_{\mu} D^{\mu } \Phi =0,
\label{eq:KG}
\end{equation}
where $D_{\mu } = \nabla _{\mu } - iqA_{\mu }$ is the covariant derivative, with $A_{\mu } $ the electromagnetic potential (\ref{eq:gaugepot}).
We consider mode solutions of the scalar field equation (\ref{eq:KG}) of the form
\begin{equation}
    \phi _{\omega \ell m }(t,r,\theta ,\varphi ) 
    = \frac{e^{-i\omega t}}{r} {\mathcal {N}}_{\omega }X_{\omega \ell }(r)Y_{\ell m }(\theta ,\varphi ),
    \label{eq:modes}
\end{equation}
where we emphasize that the frequency $\omega $ may take any positive or negative value. 
In (\ref{eq:modes}), the integer $\ell = 0,1,2,\ldots $ is the total angular momentum quantum number, $m=-\ell, -\ell + 1, \ldots , \ell -1 , \ell $ is the azimuthal angular momentum quantum number, $\omega $ the frequency of the mode, ${\mathcal {N}}_{\omega }$ is a normalization constant   and $Y_{\ell m}(\theta ,\varphi )$ is a spherical harmonic. 
The spherical harmonics are given by 
\begin{equation}
Y_{\ell m}(\theta, \varphi ) =
{\sqrt {\frac{(2\ell + 1)}{4\pi }\frac{(\ell - m)!}{(\ell + m)!}}}
 P_{\ell }^{m}(\cos \theta )e^{im\varphi } ,
 \label{eq:harmonics}
\end{equation}
where $P_{\ell }^{m}$ is a real Legendre function and
we have fixed the normalization such that 
\begin{equation}
\int Y_{\ell m }(\theta, \varphi )Y_{\ell 'm'}^{*}(\theta, \varphi ) \, \sin \theta \, d\theta \, d\varphi = \delta _{\ell \ell'}\delta _{mm'},
\end{equation}
where ${}^{*}$ denotes complex conjugation.
In terms of the ``tortoise'' coordinate $r_{*}$, defined by (\ref{eq:rstar}), 
 the radial equation for $X_{\omega \ell }(r)$ takes the form
\begin{equation}
\left[    -\frac{d^{2}}{dr_{*}^{2}} + V_{\rm {eff}}(r) \right] X_{\omega \ell }(r) = 0, 
\label{eq:radial}
\end{equation}
where the effective potential $V_{\rm {eff}}(r)$ is
\begin{equation}
    V_{\rm {eff}}(r) = 
    \frac{f(r)}{r^{2}} \left[ \ell \left( \ell + 1 \right) +rf'(r) \right] - \left( \omega - \frac{qQ}{r} \right) ^{2}.
    \label{eq:Veff}
\end{equation}
Near the black hole event horizon, as $r\rightarrow r_{+}$ and $r_{*}\rightarrow -\infty $, and at infinity, as $r,r_{*}\rightarrow \infty $, the effective potential $V_{\rm {eff}}$ (\ref{eq:Veff}) has the asymptotic values
\begin{equation}
    V_{\rm {eff}}(r) \sim 
    \begin{cases}
    -{\widetilde {\omega }}^{2} =-\left( \omega - \frac{qQ}{r_{+}} \right) ^{2}, & r_{*} \rightarrow -\infty , \\
    -\omega ^{2}, & r_{*} \rightarrow \infty ,
    \end{cases}
    \label{eq:Vasympt}
\end{equation}
where we have defined the quantity
\begin{equation}
    {\widetilde {\omega }} = \omega - \frac{qQ}{r_{+}}.
    \label{eq:tomega}
\end{equation}
The charges of both the black hole and of the scalar field do not appear in the effective potential far from the black hole since we have chosen a gauge in which the electromagnetic potential vanishes there. 
Under a gauge transformation of the form
\begin{equation}
	A_{\mu} \rightarrow A_{\mu } + \partial _{\mu } \Upsilon, \qquad 
	\Phi \rightarrow e^{iq\Upsilon }\Phi , \qquad \Upsilon = \frac{Qt}{r_{0}},
	\label{eq:gaugetrans}
\end{equation}
for a constant $r_{0}$, the gauge potential $A_{\mu }$ transforms to  $({\underline{A}}_{0},0,0,0)$ with
\begin{equation}
	{\underline {A}}_{0} = -\frac{Q}{r} + \frac{Q}{r_{0}}.
\end{equation} 
We have chosen a gauge with $r_{0}=\infty $, but we could equally well have chosen $r_{0}=r_{+}$. In this case the gauge potential ${\underline {A}}_{0}$ would vanish at the event horizon rather than at infinity, and the effective potential $V_{\rm {eff}}$ (\ref{eq:Veff}) at the horizon would be independent of the charge.

Under a gauge transformation (\ref{eq:gaugetrans}), 
the frequency $\omega $ of a scalar field mode  (\ref{eq:modes}) is transformed to
\begin{equation}
	{\underline {\omega } } = \omega -\frac{qQ}{r_{0}}.
	\label{eq:freqgauge}
\end{equation}
Therefore the frequency of a scalar field mode is not a gauge-invariant quantity.
A constant shift in the frequency corresponds to a gauge transformation (\ref{eq:gaugetrans}), which will affect the final term in the effective potential $V_{\rm {eff}}$ (\ref{eq:Veff}), and hence the form of the scalar field modes near the horizon and at infinity.  
Our choice of gauge means that the quantity $\omega $ in (\ref{eq:modes}) has a natural physical interpretation; it is the frequency of a mode as measured by a static observer far from the black hole.

With our choice of gauge, we see from (\ref{eq:Vasympt}) that the charge does affect the form of the effective potential close to the horizon. This turns out to have important consequences for both the form of the scalar field modes, and, in Sec.~\ref{sec:quantum}, for the canonical quantization of the scalar field.
A further important feature of the radial equation (\ref{eq:radial}) is that it is {\it {not}} invariant under the transformation $\omega \rightarrow -\omega $.
This means that, while $X_{\omega \ell }^{*}(r)$ satisfies the same radial equation (\ref{eq:radial}) as $X_{\omega \ell }(r)$, the function $X_{\omega \ell }^{*}(r)$ is not the same as $X_{-\omega \ell}(r)$.
This subtlety will be important in Sec.~\ref{sec:quantum} when we quantize the field.

In region I, a basis of solutions to the radial equation (\ref{eq:radial}) consists of the usual ``in'' and ``up'' scalar field modes, which have the asymptotic forms
\begin{subequations}
\label{eq:inup}
\begin{equation}
    X_{\omega \ell }^{\rm {in}}(r) = 
    \begin{cases}
    B_{\omega \ell}^{\rm {in}} e^{-i{\widetilde {\omega }}r_{*}}, & r_{*} \rightarrow -\infty ,\\
    e^{-i\omega r_{*}} + A_{\omega \ell }^{\rm {in}} e^{i\omega r_{*}}, & r_{*}\rightarrow \infty ,
    \end{cases}
    \label{eq:in}
\end{equation}
and
\begin{equation}
   X_{\omega \ell }^{\rm {up}}(r) = 
    \begin{cases}
    e^{i{\widetilde {\omega }}r_{*}} +
    A_{\omega \ell}^{\rm {up}} e^{-i {\widetilde {\omega }}r_{*}}, & r_{*} \rightarrow -\infty ,\\
     B_{\omega \ell }^{\rm {up}} e^{i\omega r_{*}}, & r_{*}\rightarrow \infty ,
     \end{cases}
     \label{eq:up}
\end{equation}
\end{subequations}
respectively, where $A_{\omega \ell}^{\rm {in/up}}$ and $B_{\omega \ell}^{\rm {in/up}}$ are complex constants. 
The ``in'' modes represent scalar waves incoming from ${\mathscr {I}}^{-}$, which are partly reflected back to ${\mathscr {I}}^{+}$ and partly transmitted down the future horizon ${\mathcal {H}}^{+}$.
The ``up'' modes represent scalar waves which are outgoing near the past horizon ${\mathcal {H}}^{-}$, partly reflected back down the future horizon ${\mathcal {H}}^{+}$ and partly transmitted to ${\mathscr{I}}^{+}$.
Our ``in'' and ``up'' modes are the same as those constructed in \cite{Klein:2021ctt,Klein:2021les}, although our different choice of gauge means that the asymptotic forms (\ref{eq:in}, \ref{eq:up}) are not identical. 

\begin{figure}[h]
	\centering
	\begin{tabular}{cc}
		\includegraphics[width=0.25\textwidth]{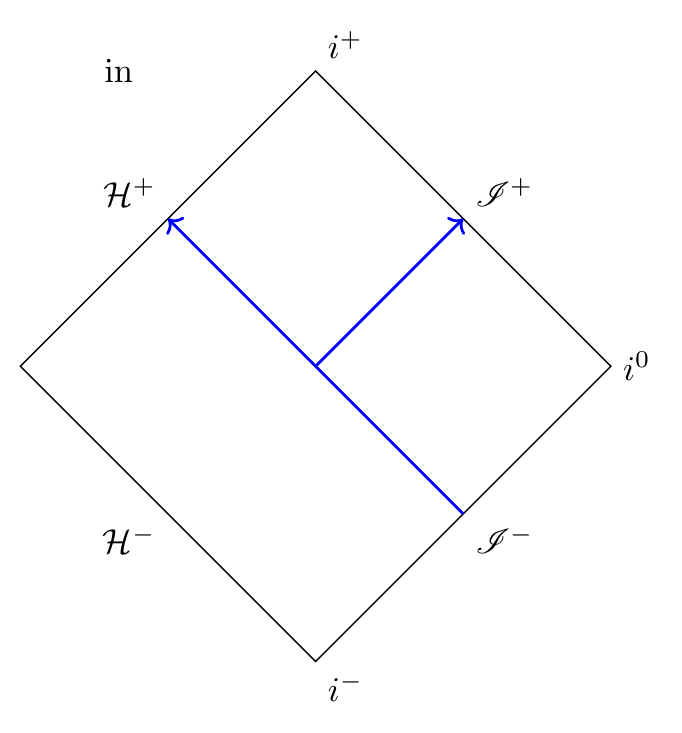}
		& \includegraphics[width=0.25\textwidth]{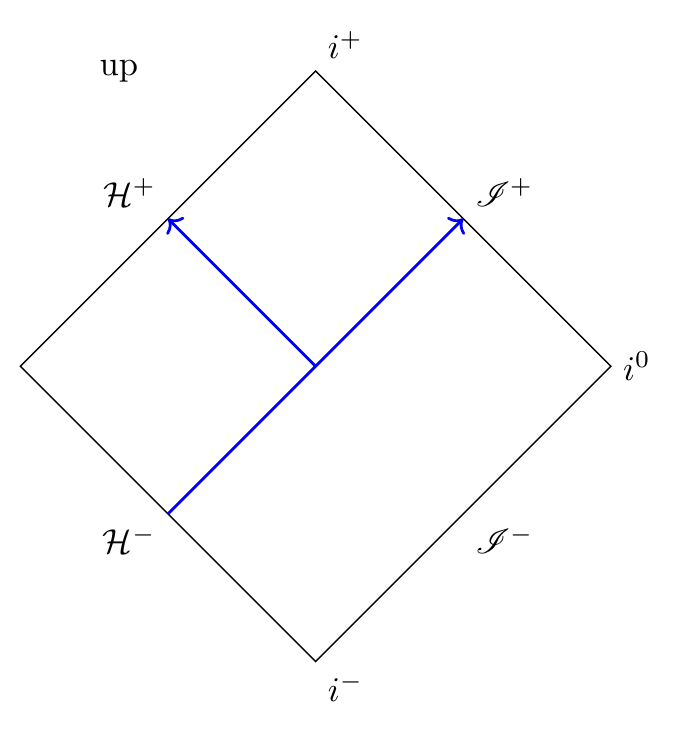} \\
		\includegraphics[width=0.25\textwidth]{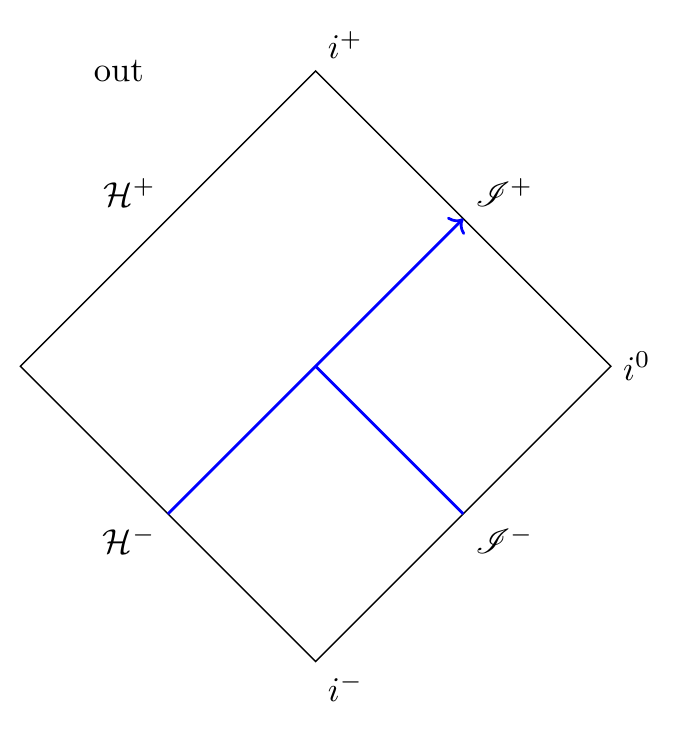}
		& \includegraphics[width=0.25\textwidth]{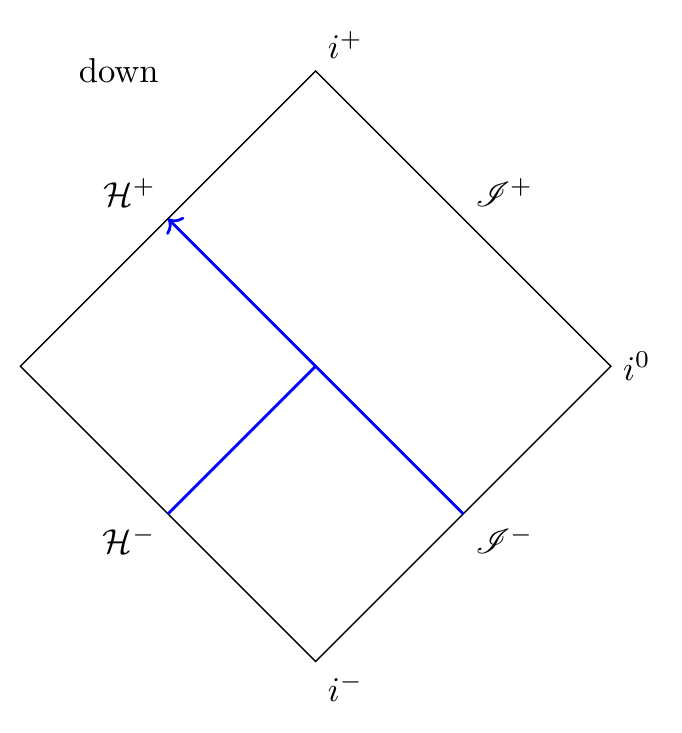}
	\end{tabular}
	\caption{``In'', ``up'', ``out'' and ``down'' modes depicted in region I of the RN space-time.}
	\label{fig:modes}
\end{figure}

We will also make use of an alternative basis in region I, given by the following ``out'' and ``down'' modes, whose radial functions are defined by:
\begin{subequations}
\label{eq:outdown}
\begin{equation}
	 X_{\omega \ell }^{\rm {out}}(r) = X_{\omega \ell }^{{\rm {in}}*}(r), \qquad 
	  X_{\omega \ell }^{\rm {down}}(r) = X_{\omega \ell }^{{\rm {up}}*}(r) ,
\end{equation}
and whose asymptotic forms are therefore
\begin{equation}
    X_{\omega \ell }^{\rm {out}}(r) =
    \begin{cases}
    B_{\omega \ell}^{{\rm {in}}*} e^{i {\widetilde {\omega }}r_{*}}, & r_{*} \rightarrow -\infty ,\\
    e^{i\omega r_{*}} + A_{\omega \ell }^{{\rm {in}}*} e^{-i\omega r_{*}}, & r_{*}\rightarrow \infty ,
    \end{cases}
    \label{eq:out}
\end{equation}
and
\begin{equation}
   X_{\omega \ell }^{\rm {down}}(r) =
    \begin{cases}
    e^{-i{\widetilde {\omega }}r_{*}} +
    A_{\omega \ell}^{{\rm {up}}*} e^{i {\widetilde {\omega }}r_{*}}, & r_{*} \rightarrow -\infty ,\\
     B_{\omega \ell }^{{\rm {up}}*} e^{-i\omega r_{*}}, & r_{*}\rightarrow \infty , 
     \end{cases}
     \label{eq:down}
\end{equation}
\end{subequations}
respectively.
The ``out'' and ``down'' radial mode functions can be written as linear combinations of the ``in'' and ``up'' radial mode functions as follows:
\begin{align}
    X_{\omega \ell }^{\rm {out}} (r) = & ~
    A_{\omega \ell }^{{\rm {in}}*}X_{\omega \ell}^{\rm {in}} (r)
    + B_{\omega \ell }^{{\rm {in}}*}X_{\omega \ell }^{\rm {up}} (r) , 
    \nonumber \\
    X_{\omega \ell }^{\rm {down}} (r) = & ~
    A_{\omega \ell }^{{\rm {up}}*}X_{\omega \ell }^{\rm {up}} (r)
    +B_{\omega \ell }^{{\rm {up}}*}X_{\omega \ell }^{\rm {in}} (r).
    \label{eq:inoutupdown}
\end{align}
The ``out'' modes correspond to a combination of ``in'' and ``up'' modes such that there is no flux going down the event horizon, while the ``down'' modes have no outwards flux at infinity.
The ``in'', ``up'', ``out'' and ``down'' modes are depicted in Fig.~\ref{fig:modes}.

Since the effective potential (\ref{eq:Veff}) in the radial equation (\ref{eq:radial}) is real, for any two solutions $X_{1}$, $X_{2}$ of the radial equation the Wronskians 
\begin{equation}
    X_{1}\frac{dX_{2}}{dr_{*}}- X_{2}\frac{dX_{1}}{dr_{*}}, \qquad
    X_{1}^{*}\frac{dX_{2}}{dr_{*}} - X_{2}\frac{dX_{1}^{*}}{dr_{*}}
\end{equation}
are independent of $r_{*}$.
Using the asymptotic forms (\ref{eq:inup}), we obtain the following Wronskian relations, valid for any value of the frequency $\omega $: 
\begin{align}
    \omega \left[ 1- \left| A^{\rm {in}}_{\omega \ell }\right| ^{2} \right] = & ~{\widetilde {\omega }} \left| B_{\omega \ell }^{\rm {in}} \right| ^{2} ,
    \nonumber \\
    {\widetilde {\omega }} \left[ 1- \left|   A^{\rm {up}}_{\omega \ell }\right| ^{2} \right] = & ~ \omega  \left| B_{\omega \ell }^{\rm {up}} \right| ^{2} ,
    \nonumber \\ 
    {\widetilde {\omega }} B^{\rm {in}}_{\omega \ell } =  & ~ \omega B^{\rm {up}}_{\omega \ell} ,
    \nonumber \\
    {\widetilde {\omega }} A^{{\rm {up}}*}_{\omega \ell } B^{\rm {in}}_{\omega \ell } = & ~ - \omega A^{\rm {in}}_{\omega \ell} B^{{\rm {up}}*}_{\omega \ell },
    \label{eq:wronskians}
\end{align}
where it should be stressed that both $\omega $ and ${\widetilde {\omega }}$ can take any real value.
For scalar field modes with $\omega {\widetilde {\omega }}<0$, the reflection coefficient $\left| A_{\omega \ell }\right| ^{2}>1$.
This is the classical phenomenon of charge superradiance \cite{Bekenstein:1973mi}.
An ``in'' mode with $\omega {\widetilde {\omega }}<0$ will be reflected back to ${\mathscr {I}}^{+}$ with an amplitude greater than it had coming in from ${\mathscr {I}}^{-}$, and, similarly, an ``up'' mode with $\omega {\widetilde {\omega }}<0$ will be reflected back down ${\mathcal {H}}^{+}$ with an amplitude greater than it had coming out from ${\mathcal {H}}^{-}$.

\begin{figure}[h]
	\includegraphics[width=0.45\textwidth]{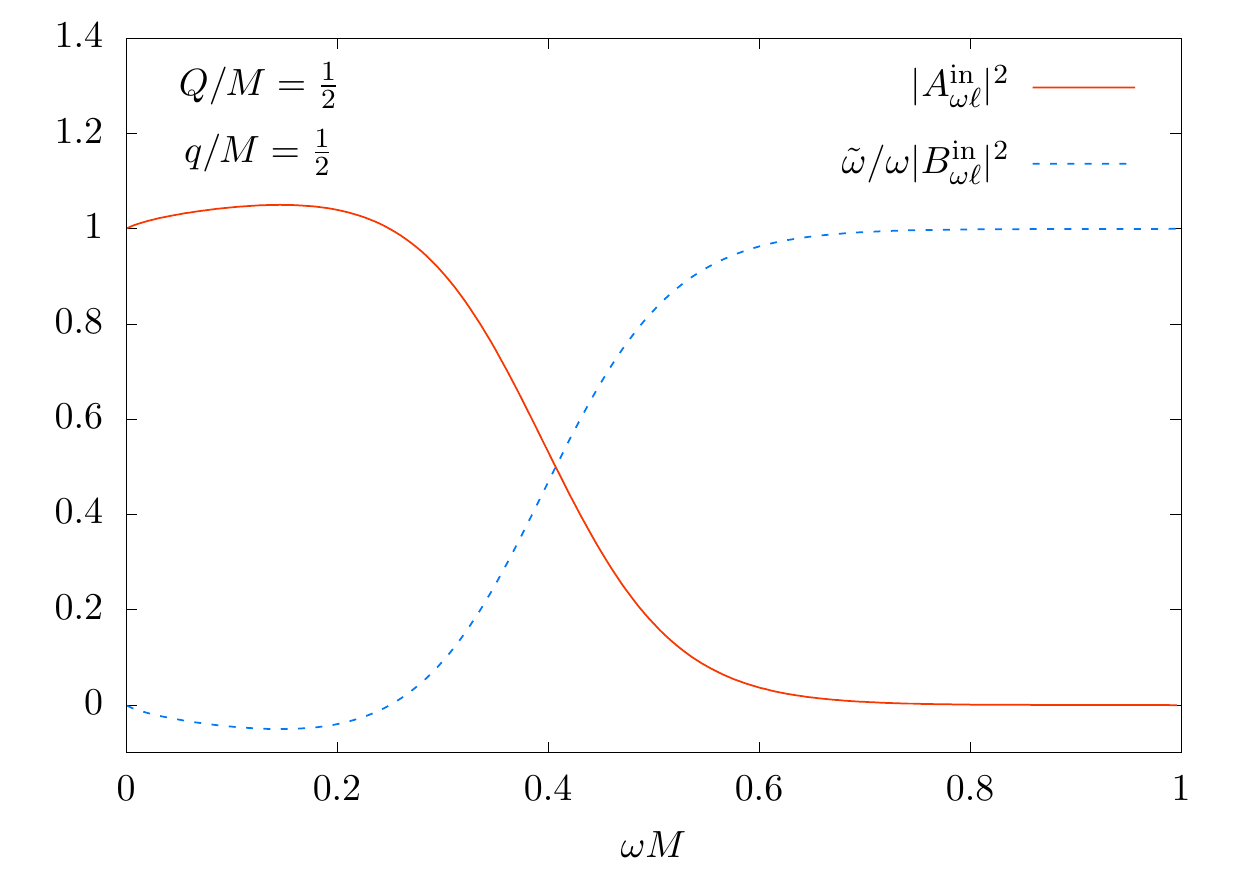}
	\caption{Reflection $\left| A^{\rm {in}}_{\omega \ell }\right| ^{2}$ and transmission ${\widetilde {\omega }} \left| B_{\omega \ell }^{\rm {in}} \right| ^{2}/\omega $   coefficients for ``in'' modes with $\ell =0$ as a function of frequency $\omega $, for a particular choice of scalar field charge $q=M/2$ and black hole charge $Q=M/{2}$. Superradiance occurs when the reflection coefficient is greater than unity.}
	\label{fig:superradiance}
\end{figure}

In Fig.~\ref{fig:superradiance} we show the reflection $\left| A^{\rm {in}}_{\omega \ell }\right| ^{2}$ and transmission ${\widetilde {\omega }} \left| B_{\omega \ell }^{\rm {in}} \right| ^{2}/\omega $   coefficients for ``in'' modes with $\ell =0$, and fixed scalar field and black hole charges. We find similar qualitative behaviour for other values of these parameters. 
It can be seen that for small positive frequency $\omega $, we have $\left| A^{\rm {in}}_{\omega \ell }\right| ^{2}>1$ and hence superradiance. 
In this frequency range, we have ${\widetilde{\omega }}<0$ and hence the transmission coefficient is negative.
One notable feature of charge superradiance is that the amplification of low-frequency waves is much greater than the corresponding effect on Kerr black hole backgrounds \cite{Brito:2015oca} (cf. Fig.~16 in \cite{Macedo:2012zz}).

The inner product $\langle \Phi _{1}, \Phi _{2}\rangle $ between any two solutions $\Phi _{1}$, $\Phi _{2}$ of the scalar field equation (\ref{eq:KG}) is defined by 
\begin{align}
  \langle \Phi _{1}, \Phi _{2}\rangle 
  = & ~ i \int _{\Sigma } \left[ 
  \left( D_{\mu }\Phi _{1} \right) ^{*} \Phi _{2} - \Phi _{1}^{*}D_{\mu }\Phi _{2} 
  \right] {\sqrt {-g}} \, d\Sigma ^{\mu }
  \nonumber \\ = & ~
  i\int _{\Sigma }
  \left[ 
  \left( \nabla _{\mu }\Phi _{1}^{*} \right) \Phi _{2} - \Phi _{1}^{*} \nabla _{\mu }\Phi _{2} 
  \right. \nonumber \\ & \left. \, \, 
  +2iqA_{\mu }\Phi _{1}^{*}\Phi _{2} 
  \right] {\sqrt {-g}} \, d\Sigma ^{\mu },
  \label{eq:innerprod}
\end{align}
where $\Sigma $ is a Cauchy surface.
The inner product (\ref{eq:innerprod}) depends on the electromagnetic potential $A_{\mu }$, and this will have an effect on the normalization of the scalar field modes.
We compute the inner product of two ``in'' or ``up'' scalar field modes (\ref{eq:inup}) on a Cauchy surface close to ${\mathcal {H}}^{-}\cup {\mathscr {I}}^{-}$.
The ``in'' modes vanish close to ${\mathcal {H}}^{-}$, and hence we find
\begin{subequations}
\label{eq:innerprodmodes}
\begin{align}
    \langle \phi ^{\rm {in}}_{\omega \ell m },\phi ^{\rm {in}}_{\omega ' \ell ' m ' } \rangle 
    = & ~  4\pi \omega \, {\mathcal {N}}_{\omega }^{{\rm {in}}*}{\mathcal {N}}^{\rm {in}}_{\omega ' }
    \delta \left( \omega - \omega '\right) \delta _{\ell \ell '}\delta _{mm'}.
    \nonumber \\  
    \label{eq:innerprodmodesin}
\end{align}
Similarly, the ``up'' modes vanish close to ${\mathscr {I}}^{-}$ and we obtain
\begin{align}
    \langle \phi ^{\rm {up}}_{\omega \ell m },\phi ^{\rm {up}}_{\omega ' \ell ' m ' } \rangle 
    = & ~ 4\pi {\widetilde {\omega }} \, {\mathcal {N}}_{\omega }^{{\rm {up}}*}{\mathcal {N}}^{\rm {up}}_{\omega ' }
    \delta \left( \omega - \omega '\right) \delta _{\ell \ell '}\delta _{mm'}.
    \nonumber \\
    \label{eq:innerprodmodesup}
\end{align}
For the ``out'' and ``down'' modes (\ref{eq:outdown}), it is most convenient to perform the integration over a Cauchy surface close to ${\mathcal {H}}^{+} \cup {\mathscr{I}}^{+}$.
The ``out'' modes vanish close to ${\mathcal {H}}^{+}$, giving
\begin{align}
    \langle \phi ^{\rm {out}}_{\omega \ell m },\phi ^{\rm {out}}_{\omega ' \ell ' m ' } \rangle 
    = & ~4\pi \omega \, {\mathcal {N}}_{\omega }^{{\rm {out}}*}{\mathcal {N}}^{\rm {out}}_{\omega ' }
    \delta \left( \omega - \omega '\right) \delta _{\ell \ell '}\delta _{mm'},
    \nonumber \\
\end{align}
while the ``down'' modes vanish close to ${\mathscr{I}}^{+}$ and we have
\begin{align}
    \langle \phi ^{\rm {down}}_{\omega \ell m },\phi ^{\rm {down}}_{\omega ' \ell ' m ' } \rangle 
    = & ~ 4\pi {\widetilde {\omega }} \, {\mathcal {N}}_{\omega }^{{\rm {down}}*}{\mathcal {N}}^{\rm {down}}_{\omega ' }
    \delta \left( \omega - \omega '\right) \delta _{\ell \ell '}\delta _{mm'}.
    \nonumber \\
\end{align}
\end{subequations}
In all cases, modes with different values of the frequency $\omega $ and quantum numbers $\ell $ and $m$ are orthogonal. It is also straightforward to see that any ``in'' mode is orthogonal to any ``up'' mode and any ``out'' mode is orthogonal to any ``down'' mode. 
From (\ref{eq:innerprodmodes}), the modes are normalized by taking
\begin{equation}
    {\mathcal {N}}_{\omega }^{\rm {in/out}} = \frac{1}{{\sqrt {4\pi \! \left| \omega \right| }}},
    \qquad
    {\mathcal {N}}_{\omega }^{\rm {up/down}} = \frac{1}{{\sqrt {4\pi \! \left| {\widetilde {\omega }} \right| }}}.
    \label{eq:normconsts}
\end{equation}
The ``in'' and ``out'' modes then have positive ``norm'' when $\omega >0$, while the ``up'' and ``down'' modes have positive ``norm'' when ${\widetilde {\omega }}>0$.
This will turn out to be crucial when we perform the canonical quantization of the scalar field in the next section.
Using the normalization constants (\ref{eq:normconsts}) and the relationships (\ref{eq:inoutupdown}) between the radial mode functions for the ``in'', ``up'', ``out'' and ``down'' modes, we find the following equations  connecting the ``in'', ``up'', ``out'' and ``down'' modes:
\begin{align}
	\phi ^{{\rm {out}}}_{\omega \ell m} = & ~
	A^{{\rm {in}}*}_{\omega \ell } \phi ^{\rm {in}}_{\omega \ell m}
	+ \left| \frac{{\widetilde {\omega }}}{\omega } \right| ^{\frac{1}{2}} B^{{\rm {in}}*}_{\omega \ell } \phi ^{\rm {up}}_{\omega \ell m} ,
	\nonumber \\
	\phi ^{{\rm {down}}}_{\omega \ell m} = & ~
	A^{{\rm {up}}*}_{\omega \ell } \phi ^{\rm {up}}_{\omega \ell m}
	+ \left| \frac{\omega }{{\widetilde {\omega }}} \right| ^{\frac{1}{2}} B^{{\rm {up}}*}_{\omega \ell } \phi ^{\rm {in}}_{\omega \ell m} .
	\label{eq:inoutupdownmodes}
	\end{align}
We will make use of these results in the quantization of the scalar field in the next section.

\section{Canonical quantization of the charged scalar field}
\label{sec:quantum}

In this section we schematically review the method of canonical quantization for defining states of a charged quantum scalar field on a general static curved space-time, before applying this method to the charged scalar field on the RN space-time.
We will see that the presence of superradiant modes complicates the canonical quantization of a charged scalar field compared with the neutral case. 
Similar challenges occur due to superradiance on a Kerr black hole \cite{Frolov:1989jh,Ottewill:2000qh,Casals:2012es,Casals:2005kr,Duffy:2005mz}.

\subsection{General approach}
\label{sec:general}

We begin with an orthonormal basis of classical mode solutions $\phi _{j}$ of the charged scalar field equation (\ref{eq:KG}), labelled by an index $j$. 
The modes are normalized using the inner product (\ref{eq:innerprod}), so that
\begin{equation}
    \langle \phi _{j}, \phi _{j'} \rangle = \eta _{j} \delta _{j j'},
    \label{eq:orthonorm}
\end{equation}
where $\delta _{jj'}$ is either the Kronecker delta or the Dirac delta function, depending on whether the label $j$ is discrete or continuous.
The product (\ref{eq:innerprod}) is not, strictly speaking, an inner product because the ``norm'' of any mode $\phi _{j}$ is not necessarily positive.
We have therefore defined a quantity $\eta _{j}$ given by 
\begin{equation}
    \eta _{j} = \begin{cases}
    1, & {\mbox {if $\phi _{j}$ has positive ``norm''}},\\
    -1, & {\mbox {if $\phi _{j}$ has negative ``norm''}}.
    \end{cases}
    \label{eq:etadef}
\end{equation}
The basis modes are then split into two sets, corresponding to positive and negative frequency modes. 
Consider a scalar field mode $\phi _{j}$ having harmonic dependence on a particular time-like coordinate $T$, so that 
\begin{equation}
\frac{\partial }{\partial T} \phi _{j} = -i \varpi \phi _{j}, 
\label{eq:posfreq}
\end{equation}
where $\varpi \in {\mathbb{R}}$ is the frequency of the mode. Such a mode is positive frequency if $\varpi>0$. From (\ref{eq:freqgauge}), this definition depends on the choice of gauge.  
More generally, a mode is positive frequency with respect to the coordinate $T$ if, when Fourier decomposed with respect to the time coordinate $T$ it only contains positive frequency components, which means that the mode, considered as a function of $T$, is analytic in the lower-half of the complex plane. 
The way in which the basis of field modes is split into positive and negative frequency components therefore depends on a choice of time coordinate $T$.
Denoting the positive frequency modes by $\phi _{j}^{+}$ and the negative frequency modes by $\phi _{j}^{-}$, any classical solution $\Phi $ of the scalar field equation (\ref{eq:KG}) can therefore be written schematically as
\begin{equation}
    \Phi = \sum _{j} \left( a_{j}\phi _{j}^{+} + b_{j}^{\dagger }\phi _{j}^{-} \right) ,
\end{equation}
where $a_{j}$, $b_{j}^{\dagger}$ are complex constants and the sum is taken over the basis of modes.

The scalar field $\Phi  $ is quantized by promoting the expansion coefficients $a_{j}$, $b_{j}^{\dagger }$ to operators:
\begin{equation}
    {\hat {\Phi }} = \sum _{j} \left( {\hat {a}}_{j}\phi _{j}^{+} + {\hat {b}}_{j}^{\dagger }\phi _{j}^{-} \right)  .
    \label{eq:modeexpgen}
\end{equation}
Since we are considering a charged, complex scalar field, we have distinct operators ${\hat {a}}_{j}$ for particles and ${\hat {b}}_{j}$ for antiparticles.

With a choice of time coordinate $T$, the canonical momentum conjugate to the field operator ${\hat {\Phi }}$ is, since we are assuming that the space-time is static,
\begin{equation}
{\hat {\Pi }} =\frac{1}{2} g^{T\mu } \left( D_{\mu  } {\hat {\Phi }} \right)^{*} =\frac{1}{2}g^{T\mu } \left( \partial _{\mu }+iqA_{\mu } \right) {\hat {\Phi}}^{*} .    
\end{equation}
The quantum scalar field ${\hat {\Phi }}$ and its conjugate momentum ${\hat {\Pi }}$ then satisfy the equal-time canonical commutation relations
\begin{align}
    \left[ {\hat {\Phi }}(T,{\bm {x}} ), {\hat {\Pi }}(T,{\bm {x}}') \right]
    =  & ~i\delta ^{3}({\bm {x}}, {\bm {x}}' ),
    \nonumber \\
    \left[ {\hat {\Phi }}(T,{\bm {x}} ), {\hat {\Phi }}(T,{\bm {x}}') \right]
    = & ~ 0 = 
    \left[ {\hat {\Pi }}(T,{\bm {x}} ), {\hat {\Pi }}(T,{\bm {x}}') \right] .
\end{align}
Using the orthonormality relations (\ref{eq:orthonorm}), the operators ${\hat {a}}_{j}$, ${\hat {b}}_{j}$ are found to satisfy the commutation relations
\begin{align}
\left[ {\hat {a}}_{j}, {\hat {a}}^{\dagger}_{j'} \right] = \eta _{j}^{+} \delta _{j j'} , \qquad & \left[ {\hat {b}}_{j}, {\hat {b}}^{\dagger}_{j'} \right] =  - \eta _{j}^{-} \delta _{j j'} ,
\nonumber \\
\left[ {\hat {a}}_{j}, {\hat {a}}_{j'} \right] = 0 = 
\left[ {\hat {a}}_{j}^{\dagger }, {\hat {a}}^{\dagger}_{j'} \right] ,
\qquad & 
\left[ {\hat {b}}_{j}, {\hat {b}}_{j'} \right] =  0 = 
\left[ {\hat {b}}_{j}^{\dagger }, {\hat {b}}^{\dagger}_{j'} \right] ,
\label{eq:canonical}
\end{align}
where $\eta _{j}^{+}$ and $\eta _{j}^{-}$ are the quantities in (\ref{eq:orthonorm}) for the positive and negative frequency modes respectively.

In the standard approach to canonical quantization, the positive frequency modes $\phi _{j}^{+}$ are such that $\eta _{j}^{+}=1$ for all $j$, while the negative frequency modes $\phi _{j}^{-}$ are such that $\eta _{j}^{-}=-1$ for all $j$.
In this case the nonzero commutation relations (\ref{eq:canonical}) take the usual form
\begin{equation}
\left[ {\hat {a}}_{j}, {\hat {a}}^{\dagger}_{j'} \right] =  \delta _{j j'} , \qquad \left[ {\hat {b}}_{j}, {\hat {b}}^{\dagger}_{j'} \right] =  \delta _{j j'} ,
\label{eq:ccstandard}
\end{equation}
leading to the interpretation of the operators ${\hat {a}}_{j}$, ${\hat {b}}_{j}$ as annihilation operators and ${\hat {a}}_{j}^{\dagger }$, ${\hat {b}}_{j}^{\dagger }$ as creation operators.

On the other hand, suppose there exist positive frequency modes $\phi _{j}^{+}$ with negative ``norm'' for which $\eta _{j}^{+}=-1$ and/or negative frequency modes $\phi _{j}^{-}$ with positive ``norm'' so that $\eta _{j}^{-}=1$.
In this situation one could consider that the modes have effectively been ``mislabelled''.
In this section, we will be defining positive and negative frequency modes using various physical choices of the time coordinate $T$, and we will see that in some cases this leads to effectively ``mislabelled'' modes.
A similar situation arises in the quantization of a neutral scalar field on a Kerr black hole space-time (see \cite{Frolov:1989jh}, where details of the ``$\eta $-formalism'' developed to deal with such ``mislabelled'' modes can be found).

Nonetheless, in this situation, as in standard quantum field theory, once we have an expansion of the scalar field of the form (\ref{eq:modeexpgen}), a natural ``vacuum'' state  $|0\rangle $ can be defined as that state which is annihilated by the operators ${\hat {a}}_{j}$ and ${\hat {b}}_{j}$:
\begin{equation}
	{\hat {a}}_{j}|0\rangle  =0 , \qquad  {\hat {b}}_{j}|0\rangle  =0.
\end{equation}
When the operators ${\hat {a}}_{j}$ and ${\hat {b}}_{j}$ satisfy the conventional commutation relations (\ref{eq:ccstandard}), the state $ |0\rangle$ contains zero quanta in both the  $\phi _{j}^{+}$ and $\phi _{j}^{-}$ modes, as measured by the standard number operators $ {\hat {a}}_{j}^{\dagger}{\hat {a}}_{j}$ and ${\hat {b}}_{j}^{\dagger}{\hat {b}}_{j}$.
When there are ``mislabelled'' modes, the definition of the number operators ${\hat {n}}_{aj}$, ${\hat {n}}_{bj}$ for particles and anti-particles respectively are modified to be \cite{Frolov:1989jh}:
\begin{equation}
	{\hat {n}}_{aj}  = \eta _{j}^{+} {\hat {a}}_{j}^{\dagger}{\hat {a}}_{j},
	\qquad 
	{\hat {n}}_{bj}  = \eta _{j}^{-} {\hat {b}}_{j}^{\dagger}{\hat {b}}_{j} .
	\label{eq:numberop}
\end{equation} 
Therefore the ``vacuum'' state $|0\rangle $ still contains zero quanta, since $	{\hat {n}}_{aj}|0\rangle  =0$ and ${\hat {n}}_{bj}|0\rangle  =0$. 
Furthermore, applying an operator ${\hat {a}}_{j}^{\dagger}$ or ${\hat {b}}_{j}^{\dagger}$ to the state $|0\rangle $ results in a state which contains one quantum, as measured by the  relevant number operator (\ref{eq:numberop}), so that the operators  ${\hat {a}}_{j}^{\dagger}$ and ${\hat {b}}_{j}^{\dagger}$ have their usual interpretation as creation operators (and ${\hat {a}}_{j}$, ${\hat {b}}_{j}$ are annihilation operators).

\subsection{``Past'' quantum states}
\label{sec:past}

We first consider the construction of ``past'' quantum states, defined with respect to a Cauchy surface close to ${\mathcal{H}}^{-}\cup {\mathscr{I}}^{-}$.
The ``past'' Boulware state $|{\mathrm {B}}^{-}\rangle$ considered in Sec.~\ref{sec:pastB} was constructed in \cite{Balakumar:2020gli} (where it was referred to as the ``in'' vacuum), while the ``past'' Unruh state $|{\mathrm {U}}^{-}\rangle$ (Sec.~\ref{sec:pastU}) was first studied by Gibbons \cite{Gibbons:1975kk}, and more recently in \cite{Klein:2021ctt,Klein:2021les}. 

\subsubsection{``Past'' Boulware state}
\label{sec:pastB}

Near past null infinity ${\mathscr {I}}^{-}$, it is natural to use the Schwarzschild-like coordinate $t$ as the time coordinate. 
This corresponds to the proper time of a static observer far from the black hole.
Restricting attention to region I, a suitable set of ``in'' modes having positive frequency with respect to $t$ near ${\mathscr {I}}^{-}$ is then
\begin{subequations}
\label{eq:pastin+-}
\begin{equation}
\phi ^{{\rm {in}}+}_{\omega \ell m} =   \frac{e^{-i\omega t}}{{\sqrt {4 \pi \! \left|\omega \right| }} r} X_{\omega \ell }^{\rm {in}}(r)Y_{\ell m}(\theta ,\varphi )   , \qquad \omega >0 , 
\label{eq:pastin+}
\end{equation}
where $X_{\omega \ell }^{\rm {in}}(r)$ is given by (\ref{eq:in}). 
From (\ref{eq:innerprodmodesin}), these modes have positive norm.
Similarly, in region I, a suitable set of ``in'' modes having negative frequency with respect to $t$ near ${\mathscr {I}}^{-}$ is
\begin{equation}
\phi ^{{\rm {in}}-}_{\omega \ell m} =   \frac{e^{-i\omega t}}{{\sqrt {4 \pi \! \left| \omega \right| }} r} X_{\omega \ell }^{\rm {in}}(r)Y_{\ell m}(\theta ,\varphi )   , \qquad \omega <0 . 
\label{eq:pastin-}
\end{equation} 
\end{subequations}
A key subtlety here is that $\phi ^{{\rm {in}}-}_{\omega \ell m} \neq \phi ^{{\rm {in}}+*}_{\omega \ell m}$ because the radial equation (\ref{eq:radial}) is not invariant under the transformation $\omega \rightarrow -\omega $ and hence $X_{-\omega \ell }(r) \neq X^{*}_{\omega \ell }(r)$.

Near the past event horizon ${\mathcal {H}}^{-}$, the natural time coordinate for a static (and hence accelerating) observer is still the Schwarzschild-like coordinate $t$.
However, from (\ref{eq:innerprodmodesup}), the ``up'' modes $\phi ^{\rm {up}}_{\omega \ell m}$ have positive norm only if ${\widetilde {\omega }}>0$, where ${\widetilde {\omega }}$ is given by (\ref{eq:tomega}). 
Working in region I only, we therefore consider the set of positive norm ``up'' modes given by 
\begin{subequations}
	\label{eq:pastup+-}
\begin{equation}
    \phi ^{{\rm {up}}+}_{\omega \ell m} =  \frac{e^{-i\omega t}}{{\sqrt {4 \pi \! \left| {\widetilde {\omega }} \right| }}r} X_{\omega \ell }^{\rm {up}}(r)Y_{\ell m}(\theta ,\varphi ) , \qquad {\widetilde {\omega }}>0 .
    \label{eq:pastup+}
\end{equation}
If $qQ>0$, these modes all have $\omega >0$ and hence have positive frequency with respect to $t$, but if $qQ<0$, then some of the positive norm ``up'' modes (\ref{eq:pastup+}) will have $\omega <0$ and hence be considered to have negative frequency as measured by a static observer.
Similarly, in region I, the set of negative norm ``up'' modes is given by
\begin{equation}
\phi ^{{\rm {up}}-}_{\omega \ell m} =  \frac{e^{-i\omega t}}{{\sqrt {4 \pi \! \left|{\widetilde {\omega }}\right| }}r} X_{\omega \ell }^{\rm {up}}(r)Y_{\ell m}(\theta ,\varphi ) , \qquad {\widetilde {\omega }}<0 .
\label{eq:pastup-}
\end{equation}
\end{subequations}
If $qQ<0$, then all the negative norm ``up'' modes have $\omega <0$ and are negative frequency with respect to $t$, but if $qQ>0$, some negative norm ``up'' modes will have $\omega >0$ and hence a static observer will regard them as having positive frequency.

The quantum scalar field is then expanded in terms of these ``in'' and ``up'' modes as follows:
\begin{align}
    {\hat {\Phi }} = & ~ 
    \sum _{\ell =0}^{\infty }\sum _{m=-\ell }^{\ell } \left\{ 
    \int _{0}^{\infty } d\omega \,
    {\hat {a}}_{\omega \ell m}^{\rm {in}} \phi _{\omega \ell m}^{{\rm {in}}+} +
    \int _{-\infty }^{0} d\omega \,{\hat {b}}_{\omega \ell m }^{{\rm {in}}\dagger } \phi _{\omega \ell m}^{{\rm {in}}-} 
    \right. \nonumber \\ & \left. 
    + \int _{0}^{\infty }d{\widetilde {\omega }}  \,
    {\hat {a}}_{\omega \ell m}^{\rm {up}} \phi _{\omega \ell m}^{{\rm {up}}+} +
    \int _{-\infty }^{0}d{\widetilde {\omega }} \, {\hat {b}}_{\omega \ell m }^{{\rm {up}}\dagger } \phi _{\omega \ell m}^{{\rm {up}}-}  
    \right\} .
\end{align}
Note that the ``in'' operators ${\hat {a}}_{\omega \ell m}^{\rm {in}}$ are defined for $\omega >0$ and ${\hat {b}}_{\omega \ell m}^{\rm {in}}$ are defined for $\omega <0$, while the ``up'' operators ${\hat {a}}_{\omega \ell m}^{\rm {up}}$ have ${\widetilde {\omega }}>0$, and the ${\hat {b}}_{\omega \ell m}^{\rm {up}}$ have ${\widetilde {\omega }}<0$.
Since the $\phi ^{{\rm {in/up}}+}$ modes all have positive norm, and the $\phi ^{{\rm {in/up}}-}$ modes all have negative norm, the operators ${\hat {a}}$, ${\hat {b}}$ satisfy standard commutation relations:
\begin{align}
    \left[ {\hat {a}}_{\omega \ell m}^{\rm {in}}, {\hat {a}}_{\omega '\ell 'm'}^{{\rm {in}}\dagger } \right] =  & ~\delta _{\ell \ell' }\delta _{mm'} \delta (\omega - \omega ') , \qquad \omega >0,
    \nonumber \\ 
    \left[ {\hat {b}}_{\omega \ell m}^{\rm {in}}, {\hat {b}}_{\omega '\ell 'm'}^{{\rm {in}}\dagger } \right] =  & ~\delta _{\ell \ell' }\delta _{mm'} \delta (\omega - \omega ') , \qquad \omega <0,
    \nonumber \\
  \left[ {\hat {a}}_{\omega \ell m}^{\rm {up}}, {\hat {a}}_{\omega '\ell 'm'}^{{\rm {up}}\dagger } \right] =  & ~\delta _{\ell \ell' }\delta _{mm'} \delta (\omega - \omega ') , \qquad {\widetilde {\omega }}>0,
  \nonumber \\ 
  \left[ {\hat {b}}_{\omega \ell m}^{\rm {up}}, {\hat {b}}_{\omega '\ell 'm'}^{{\rm {up}}\dagger } \right] =  & ~\delta _{\ell \ell' }\delta _{mm'} \delta (\omega - \omega ') , \qquad {\widetilde {\omega }}<0.
\end{align}
All commutators not given explicitly above vanish.
The ``past'' Boulware state $| {\rm {B}}^{-} \rangle $ is then defined as the vacuum state which is annihilated by the following ${\hat {a}}$ and ${\hat {b}}$ operators:
\begin{align}
    {\hat {a}}^{{\rm {in}}}_{\omega \ell m} | {\rm {B}}^{-} \rangle  = & ~0 ,
    \qquad \omega >0,
    \nonumber \\ 
    {\hat {b}}^{{\rm {in}}}_{\omega \ell m} | {\rm {B}}^{-} \rangle = & ~0,
    \qquad \omega <0,
    \nonumber \\
     {\hat {a}}^{{\rm {up}}}_{\omega \ell m} | {\rm {B}}^{-} \rangle  = & ~0 ,
     \qquad {\widetilde {\omega }} >0,
     \nonumber \\ 
    {\hat {b}}^{{\rm {up}}}_{\omega \ell m} | {\rm {B}}^{-} \rangle  = & ~0 ,
    \qquad {\widetilde {\omega }}<0 .
\end{align}
In \cite{Balakumar:2020gli} we referred to this state as the ``in'' vacuum. 
The ``past'' Boulware state has no particles or antiparticles incoming from past null infinity ${\mathscr {I}}^{-}$ nor emanating from the past horizon ${\mathcal {H}}^{-}$. 
It is therefore that state which is as empty as possible as seen by a static observer at past null infinity.
However, this state is not empty as seen by a static observer at future null infinity ${\mathscr {I}}^{+}$, where it contains an outgoing flux of particles in the superradiant regime \cite{Balakumar:2020gli}. 

\subsubsection{``Past'' Unruh state}
\label{sec:pastU}

To define the ``past'' Unruh state $| {\rm {U}}^{-} \rangle $, we consider ``in'' modes having positive frequency with respect to Schwarzschild time $t$ near past null infinity ${\mathscr{I}}^{-}$. 
Therefore, in region I, the positive frequency ``in'' modes are given by (\ref{eq:pastin+}) and the negative frequency ``in'' modes are given by (\ref{eq:pastin-}).

Near the past horizon ${\mathcal {H}}^{-}$, the natural choice of time coordinate is Kruskal retarded time $U$, which is an affine parameter along the null generators of the past horizon.  
We now describe in some detail how to construct a set of ``up'' modes which have positive frequency with respect to $U$ near the past horizon ${\mathcal {H}}^{-}$,
since this differs in some respects from the corresponding derivation for a neutral scalar field (see, for example, \cite{Novikov:1989sz,Unruh:1976db}).
Our construction is analogous to that presented in \cite{Klein:2021ctt,Klein:2021les}, although we use a different gauge for the electromagnetic potential, which affects the detailed form of the scalar field modes. 

\begin{widetext}
The ``up'' modes are defined in region I by their radial function $ X^{\rm {up}}_{\omega \ell} (r) $ (\ref{eq:up}), and this definition can be extended across the past horizon ${\mathcal {H}}^{-}$ into region III.
In terms of the Kruskal coordinates $U$, $V$, near the past horizon ${\mathcal {H}}^{-}$ an ``up'' mode $\phi _{\omega \ell m}^{\rm {up}}$ takes the form
\begin{equation}
\phi _{\omega \ell m}^{\rm {up}} = 
\frac{1}{{\sqrt {4 \pi \! \left| {\widetilde {\omega }} \right| }} r}
e^{\frac{i \left( \omega + {\widetilde {\omega }} \right) }{2\kappa } \ln \left( - \kappa U \right) } 
e^{-\frac{i\left( \omega - {\widetilde {\omega }} \right) }{2\kappa} \ln \left( \kappa V \right) }Y_{\ell m }(\theta, \varphi ) \Theta (-U) ,
\label{eq:upH-}
\end{equation}
where the surface gravity $\kappa $ is given by (\ref{eq:kappa}), and $\Theta $ the Heaviside step function
\begin{equation}
    \Theta (x) = \begin{cases}
    1, & x \ge 0 ,  \\
    0, & {\mbox {otherwise}}.
    \end{cases}
\end{equation}
By definition, the ``up'' modes vanish in region IV of the space-time.
Next we define a set of modes denoted by $\psi _{\omega \ell m}^{\rm {down}}$ in regions II and IV by  making the transformation $U\rightarrow -U$, $V\rightarrow -V$ in the  ``up'' modes $\phi _{\omega \ell m}^{\rm {up}}$.
It should be emphasized that the modes $\psi _{\omega \ell m}^{\rm {down}}$, as defined, are nonzero in region IV of the space-time diagram in Fig.~\ref{fig:RN} and therefore are not the same as the ``down'' modes whose radial functions are given by (\ref{eq:down}), and which vanish in region IV.

Near the surface $V=0$, the modes $\psi _{\omega \ell m}^{\rm {down}}$ take the form
\begin{equation}
\psi _{\omega \ell m }^{\rm {down}} =  
\frac{1}{{\sqrt {4 \pi \! \left| {\widetilde {\omega }} \right| }} r}
e^{\frac{i \left( \omega + {\widetilde {\omega }} \right) }{2\kappa } \ln \left(  \kappa U \right) } 
e^{-\frac{i\left( \omega - {\widetilde {\omega }} \right) }{2\kappa} \ln \left( -\kappa V \right) }Y_{\ell m }(\theta, \varphi ) \Theta (U) .
\label{eq:downH}
\end{equation}
Like the ``up'' modes $\phi ^{\rm {up}}_{\omega \ell m}$, the norm of the modes $\psi _{\omega \ell m}^{\rm {down}}$ depends on the sign of ${\widetilde {\omega }}$. 
However, unlike the ``up'' modes $\phi ^{\rm {up}}_{\omega \ell m}$, we find that the 
modes $\psi _{\omega \ell m}^{\rm {down}}$ have negative norm when ${\widetilde {\omega }}>0$ and positive norm when ${\widetilde {\omega }}<0$. 
This difference is crucial in the construction below.

To define a set of modes having positive frequency with respect to the Kruskal coordinate $U$, we make use of the Lemma in Appendix H of \cite{Novikov:1989sz}, which states that, for positive real ${\mathfrak{p}}$ and arbitrary real ${\mathfrak {q}}$
\begin{equation}
    \int _{-\infty }^{\infty }dX \, e^{-i{\mathfrak {p}}X} \left[
    e^{-i{\mathfrak {q}}\ln X } \Theta (X) + e^{-\pi {\mathfrak {q}}}e^{-i{\mathfrak {q}}\ln \left( -X \right) } \Theta (-X) 
    \right] =0 .
    \label{eq:FNlemma}
\end{equation}
We wish to apply (\ref{eq:FNlemma}) to a linear combination of the ``up'' modes (\ref{eq:upH-}) and the modes $\psi _{\omega \ell m}^{\rm {down}}$ (\ref{eq:downH}), integrating over a surface close to $V=0$ for which $V>0$.
Comparing (\ref{eq:upH-}) and (\ref{eq:FNlemma}), we take $X=U$ and ${\mathfrak {q}}=-\left( \omega + {\widetilde {\omega }}\right) /2\kappa $.
Before we can apply (\ref{eq:FNlemma}), we need to simplify the terms involving $\ln (-\kappa V)$ in (\ref{eq:downH}).
Positive frequency modes are analytic in the lower half plane and therefore we need to use a branch of the logarithm which is also analytic in the lower half plane. 
Making an appropriate branch cut (for example along the positive imaginary axis), we have $\ln (-1) = -i\pi $ and hence, bearing in mind that $\kappa V >0$ for the surface over which we want to integrate,
\begin{equation}
\exp \left[ -\frac{i\left( \omega - {\widetilde{\omega }} \right)} {2\kappa } \ln ( -\kappa V) \right] 
= \exp \left[ -\frac{\pi \left( \omega - {\widetilde{\omega }} \right)} {2\kappa }\right]  \exp \left[ -\frac{i\left( \omega - {\widetilde{\omega }} \right)} {2\kappa } \ln ( \kappa V) \right].
\end{equation}
Applying (\ref{eq:FNlemma}) then gives 
\begin{equation}
 \int _{-\infty }^{\infty }dU \, e^{-i{\mathfrak {p}}U} \left[ 
 e^{\frac{\pi {\widetilde {\omega }}}{2\kappa }}\phi _{\omega \ell m}^{\rm {up}}  
 + e^{-\frac{\pi {\widetilde{\omega }} } {2\kappa }} \psi _{\omega \ell m}^{{\rm {down}}} 
 \right]
 =0 , \qquad {\mathfrak{p}}>0.
 \label{eq:Ustep1}
\end{equation}
From this we deduce that the modes $e^{\frac{\pi {\widetilde {\omega }}}{2\kappa }}\phi _{\omega \ell m}^{\rm {up}} + e^{-\frac{\pi {\widetilde {\omega }}}{2\kappa }}\psi _{\omega \ell m}^{\rm {down}}$ have positive frequency with respect to Kruskal time $U$ for {\it {any}} value of ${\widetilde {\omega }}$.
Therefore a set of normalized modes having positive frequency with respect to $U$ near the past horizon ${\mathcal {H}}^{-}$ is, for all values of ${\widetilde {\omega }}$,
\begin{subequations}
	\label{eq:chiup+-}
\begin{equation}
\chi _{\omega \ell m}^{{\rm {up}}+} =
\frac{1}{{\sqrt {2 \left| \sinh \left( \frac{\pi {\widetilde{\omega }} }{\kappa }\right)  \right| }}}\left( e^{\frac{\pi {\widetilde {\omega }}}{2\kappa }}\phi _{\omega \ell m}^{\rm {up}} + e^{-\frac{\pi {\widetilde {\omega }}}{2\kappa }} \psi _{\omega \ell m}^{{\rm {down}}} \right)  , \qquad
{\mbox {all ${\widetilde {\omega }}$}}.
\label{eq:chiup+}
\end{equation}
Similarly, a set of normalized modes having negative frequency with respect to $U$ near the past horizon is, for all values of ${\widetilde {\omega }}$,
\begin{equation}
\chi _{\omega \ell m}^{{\rm {up}}-} =\frac{1}{{\sqrt {2 \left| \sinh \left( \frac{\pi {\widetilde{\omega }} }{\kappa }\right)  \right| }}}\left( e^{-\frac{\pi {\widetilde {\omega }}}{2\kappa }}\phi _{\omega \ell m}^{\rm {up}} + e^{\frac{\pi {\widetilde {\omega }}}{2\kappa }} \psi _{\omega \ell m}^{{\rm {down}}} \right)  , \qquad {\mbox {all ${\widetilde {\omega }}$}}.
\label{eq:chiup-}
\end{equation}
\end{subequations}
The modes $\chi ^{{\rm {up}}\pm}_{\omega \ell m}$ are defined throughout regions I--IV.
It is straightforward to check that the positive frequency modes $\chi ^{{\rm {up}}+}_{\omega \ell m}$ have positive norm for all ${\widetilde {\omega }}$, while the negative frequency modes $\chi ^{{\rm {up}}-}_{\omega \ell m}$ have negative norm for all ${\widetilde {\omega }}$.

We now expand the quantum scalar field in terms of the modes 
$\phi _{\omega \ell m}^{\rm {in}}$ and $\chi _{\omega \ell m}^{\rm {up}}$. 
Working only in region I, the modes $\psi _{\omega \ell m}^{\rm {down}}$ vanish and we are left with
\begin{equation}
    {\hat {\Phi }}
    = \sum _{\ell =0}^{\infty } \sum _{m=-\ell }^{\ell }
    \left\{ 
    \int _{0}^{\infty } d\omega \,
    {\hat {c}}_{\omega \ell m}^{\rm {in}} \phi _{\omega \ell m}^{{\rm {in}}+} +
    \int _{-\infty }^{0} d\omega \,
    {\hat {d}}_{\omega \ell m}^{{\rm {in}}\dagger } \phi _{\omega \ell m}^{{\rm {in}}-} 
    + \int _{-\infty }^{\infty } d{\widetilde {\omega } }
    \frac{1}{{\sqrt {2 \left| \sinh \left( \frac{\pi {\widetilde{\omega }} }{\kappa }\right)  \right| }}}
    \phi _{\omega \ell m}^{\rm {up}} 
    \left[ 
    e^{\frac{\pi {\widetilde{\omega }}}{2\kappa }}
    {\hat {c}}_{\omega \ell m}^{\rm {up}}
    + e^{-\frac{\pi {\widetilde{\omega }}}{2\kappa }}
    {\hat {d}}_{\omega \ell m}^{{\rm {up}}\dagger }
    \right] 
    \right\} .
\end{equation}
\end{widetext}
The commutation relations satisfied by the ${\hat {c}}$ and ${\hat {d}}$ operators take the standard form (commutators not given explicitly below vanish)
\begin{align}
\left[ {\hat {c}}_{\omega \ell m}^{\rm {in}}, {\hat {c}}_{\omega '\ell 'm'}^{{\rm {in}}\dagger } \right] =  & ~\delta _{\ell \ell' }\delta _{mm'} \delta (\omega - \omega ') , \qquad \omega >0,
\nonumber \\ 
\left[ {\hat {d}}_{\omega \ell m}^{\rm {in}}, {\hat {d}}_{\omega '\ell 'm'}^{{\rm {in}}\dagger } \right] =  & ~\delta _{\ell \ell' }\delta _{mm'} \delta (\omega - \omega ') , \qquad \omega <0,
\nonumber \\
\left[ {\hat {c}}_{\omega \ell m}^{\rm {up}}, {\hat {c}}_{\omega '\ell 'm'}^{{\rm {up}}\dagger } \right] =  & ~\delta _{\ell \ell' }\delta _{mm'} \delta (\omega - \omega ') , \qquad {\mbox {all ${\widetilde {\omega }}$}},
\nonumber \\ 
\left[ {\hat {d}}_{\omega \ell m}^{\rm {up}}, {\hat {d}}_{\omega '\ell 'm'}^{{\rm {up}}\dagger } \right] =  & ~\delta _{\ell \ell' }\delta _{mm'} \delta (\omega - \omega ') , \qquad {\mbox {all ${\widetilde {\omega }}$}},
\end{align}
since the modes $\phi _{\omega \ell m}^{{\rm {in}}+}$, $\chi _{\omega \ell m}^{{\rm {in}}+}$ that we have designated to be positive frequency have positive norm, and the negative frequency modes have negative norm.
The ``past'' Unruh state $| {\rm {U}}^{-} \rangle $ is then defined as that state which is annihilated by the following ${\hat {c}}$ and ${\hat {d}}$ operators:
\begin{align}
    {\hat {c}}^{{\rm {in}}}_{\omega \ell m} | {\rm {U}}^{-} \rangle  =0
    , \qquad & \omega >0,
    \nonumber \\  
     {\hat {d}}^{{\rm {in}}}_{\omega \ell m} | {\rm {U}}^{-} \rangle =0 , 
     \qquad & \omega <0,
    \nonumber \\
     {\hat {c}}^{{\rm {up}}}_{\omega \ell m} | {\rm {U}}^{-} \rangle  =0 ,
     \qquad & {\mbox {all ${\widetilde {\omega }}$}},
     \nonumber \\
    {\hat {d}}^{{\rm {up}}}_{\omega \ell m} | {\rm {U}}^{-} \rangle =0,
    \qquad  & {\mbox {all ${\widetilde {\omega }}$}} .
\end{align}
Like the ``past'' Boulware state $|{\rm {B}}^{-}\rangle $, the ``past'' Unruh state $|{\rm {U}}^{-}\rangle $ contains no particles or antiparticles as seen by a static observer at past null infinity ${\mathscr{I}}^{-}$.
There is however an outgoing thermal flux of particles/antiparticles as seen by a static observer at future null infinity ${\mathscr{I}}^{+}$, corresponding to the Hawking radiation at all frequencies, in agreement with Ref.~\cite{Gibbons:1975kk}.

\subsubsection{``Past'' CCH state}
\label{sec:pastCCH}

We next define a further ``past'' quantum state, denoted by $|{\rm {CCH}}^{-}\rangle$ (where ``CCH'' stands for Candelas, Chrzanowski and Howard) \cite{Candelas:1981zv}.  
As we shall see in Sec.~\ref{sec:pastUexp}, the ``past'' Unruh state $|{\rm {U}}^{-} \rangle$ contains a thermal distribution of particles in the ``up'' modes and no particles in the ``in'' modes.
The ``past'' CCH state $|{\rm {CCH}}^{-}\rangle$ will also contain a thermal distribution of particles in the ``in'' modes, although, as we shall find in Sec.~\ref{sec:pastCCHexp}, the thermal factors in the ``in'' and ``up'' modes are not the same.

We construct this state by employing a suitable orthonormal basis of field modes.
Our basis is formed of the $\chi ^{{\rm {up}}\pm }_{\omega \ell m}$ modes (\ref{eq:chiup+-}),
together with a set of modes $\chi ^{{\rm {in}}\pm }_{\omega \ell m}$, which are constructed from the ``in'' modes $\phi _{\omega \ell m}^{\rm {in}}$ using a method similar to that for the $\chi ^{{\rm {up}}\pm }_{\omega \ell m}$ modes.

In region I, the ``in'' modes are defined by their radial function $ X^{\rm {in}}_{\omega \ell} (r) $ (\ref{eq:in}), and this definition can be extended across the future horizon ${\mathcal {H}}^{+}$ into region II. By definition, the ``in'' modes vanish in regions III and IV.
We define a set of modes $\psi _{\omega \ell m}^{{\rm {out}}}$ in regions III and IV (and vanishing in regions I and II) by taking the ``in'' modes $\phi _{\omega \ell m}^{\rm {in}}$ and performing the mapping $U\rightarrow - U$, $V\rightarrow -V$.
From this definition, the modes $\psi _{\omega \ell m}^{{\rm {out}}}$ have negative norm for $\omega >0$ and positive norm for $\omega <0$.

Using the ``in'' modes $\phi _{\omega \ell m}^{\rm {in}}$ and the  $\psi _{\omega \ell m}^{{\rm {out}}}$ modes, we then define, for all $\omega $, and throughout regions I--IV,
\begin{subequations}
\label{eq:chiin+-}
\begin{equation}
\chi _{\omega \ell m}^{{\rm {in}}+} =
\frac{1}{{\sqrt {2 \left| \sinh \left( \frac{\pi {\omega } }{\kappa }\right)  \right| }}}\left( e^{\frac{\pi {\omega }}{2\kappa }}\phi _{\omega \ell m}^{\rm {in}} + e^{-\frac{\pi {\omega }}{2\kappa }} \psi _{\omega \ell m}^{{\rm {out}}} \right)  , 
\label{eq:chiin+}
\end{equation}
and (again for all $\omega $  and throughout regions I--IV)
\begin{equation}
\chi _{\omega \ell m}^{{\rm {in}}-} =
\frac{1}{{\sqrt {2 \left| \sinh \left( \frac{\pi {\omega } }{\kappa }\right)  \right| }}}\left( e^{-\frac{\pi {\omega }}{2\kappa }}\phi _{\omega \ell m}^{\rm {in}} + e^{\frac{\pi {\omega }}{2\kappa }} \psi _{\omega \ell m}^{{\rm {out}}} \right) .
\label{eq:chiin-}
\end{equation}
\end{subequations}
The modes  $\chi _{\omega \ell m}^{{\rm {in}}+}$ have positive norm for all $\omega$, and the modes $\chi _{\omega \ell m}^{{\rm {in}}-}$ have negative norm for all $\omega $.
Therefore we may use (\ref{eq:chiin+-}), together with the modes (\ref{eq:chiup+-}) constructed in the previous subsection, to form an orthonormal basis of field modes.

\begin{widetext}
The question is then whether the modes (\ref{eq:chiin+-}) have a natural interpretation in terms of being positive or negative frequency with respect to a particular coordinate.
Near the surface $U=0$ (${\mathcal{H}}^{+}$ in region I), the ``in'' modes $\phi _{\omega \ell m}^{\rm {in}}$ take the form
\begin{equation}
\phi _{\omega \ell m}^{\rm {in}}=\frac{B^{\rm {in}}_{\omega \ell}}{{\sqrt {4 \pi \! \left|\omega \right| }}r}
e^{\frac{i\left(\omega -{\widetilde {\omega }}\right)}{2\kappa }\ln (-\kappa U)}
e^{-\frac{i\left(\omega +{\widetilde {\omega }}\right)}{2\kappa }\ln (\kappa V)} Y_{\ell m }(\theta, \varphi ) \Theta (V) .
\label{eq:inI-}
\end{equation}
Similarly, near the surface $U=0$ the modes $\psi _{\omega \ell m}^{{\rm {out}}}$
are given by 
\begin{equation}
\psi _{\omega \ell m}^{{\rm {out}}} = \frac{B^{\rm {in}}_{\omega \ell}}{{\sqrt {4 \pi \! \left| \omega \right|}}r}
e^{\frac{i\left(\omega -{\widetilde {\omega }}\right)}{2\kappa }\ln (\kappa U)}
e^{-\frac{i\left(\omega +{\widetilde {\omega }}\right)}{2\kappa }\ln (-\kappa V)} Y_{\ell m }(\theta, \varphi ) \Theta (-V) .
\label{eq:outH}
\end{equation}
We now seek to combine (\ref{eq:inI-}, \ref{eq:outH}) using Lemma (\ref{eq:FNlemma}).
Considering modes having positive frequency with respect to the Kruskal coordinate $V$ (an affine parameter along the null generators of the future event horizon ${\mathcal {H}}^{+}$) we set, as before, $\ln (-1)=-i\pi $. Integrating over a surface for which $U=-\epsilon <0$, we have
\begin{equation}
\exp \left[ \frac{i\left( \omega - {\widetilde{\omega }} \right)} {2\kappa } \ln ( \kappa U) \right] 
= \exp \left[ \frac{\pi \left( \omega - {\widetilde{\omega }} \right)} {2\kappa }\right]  \exp \left[ \frac{i\left( \omega - {\widetilde{\omega }} \right)} {2\kappa } \ln ( -\kappa U) \right].	
\end{equation}
Applying (\ref{eq:FNlemma}) with $X=V$ and ${\mathfrak{q}}=\left( \omega + {\widetilde{\omega }}\right)/2\kappa $ then gives
\begin{equation}
	\int _{-\infty}^{\infty } dV \, e^{-i{\mathfrak{p}}V} \left[  \phi _{\omega \ell m}^{\rm {in}} +  e^{-\frac{\pi \omega }{\kappa }} \psi _{\omega \ell m}^{{\rm {out}}}  \right] =0,  \qquad {\mathfrak{p}}>0,
\end{equation}
 with the result that the modes (\ref{eq:chiin+}) have positive frequency with respect to the Kruskal coordinate $V$, along the surface $U=-\epsilon <0$, part of which lies close to the future horizon ${\mathcal {H}}^{+}$ in region I of the space-time.
 Similarly, the modes (\ref{eq:chiin-}) have negative frequency with respect to $V$.

Restricting attention to region I of the space-time, the modes $\psi _{\omega \ell m}^{\rm {out}}$ vanish, and  we therefore expand the quantum scalar field as
\begin{align}
{\hat {\Phi }}
& ~ =  \sum _{\ell =0}^{\infty } \sum _{m=-\ell }^{\ell }
\left\{ 
\int _{-\infty }^{\infty } d\omega \,
\frac{1}{{\sqrt {2 \left| \sinh \left( \frac{\pi {{\omega }} }{\kappa }\right)  \right| }}}
\phi _{\omega \ell m}^{\rm {in}} 
\left[ 
e^{\frac{\pi {{\omega }}}{2\kappa }}
{\hat {f}}_{\omega \ell m}^{\rm {in}}
+ e^{-\frac{\pi {{\omega }}}{2\kappa }}
{\hat {g}}_{\omega \ell m}^{{\rm {in}}\dagger }
\right]  \right. 
\nonumber \\ & \left. \qquad \qquad 
+ \int _{-\infty }^{\infty } d{\widetilde {\omega } }
\frac{1}{{\sqrt {2 \left| \sinh \left( \frac{\pi {\widetilde{\omega }} }{\kappa }\right)  \right| }}}
\phi _{\omega \ell m}^{\rm {up}} 
\left[ 
e^{\frac{\pi {\widetilde{\omega }}}{2\kappa }}
{\hat {f}}_{\omega \ell m}^{\rm {up}}
+ e^{-\frac{\pi {\widetilde{\omega }}}{2\kappa }}
{\hat {g}}_{\omega \ell m}^{{\rm {up}}\dagger }
\right] 
\right\} , 
\label{eq:CCH-exp}
\end{align}
\end{widetext}
and define the ``past'' CCH state $|{\rm {CCH}}^{-}\rangle$ as that state annihilated by the following ${\hat {f}}$ and ${\hat {g}}$ operators:
\begin{align}
{\hat {f}}^{{\rm {in}}}_{\omega \ell m} | {\rm {CCH}}^{-} \rangle  =0
, \qquad & {\mbox {all ${{\omega }}$}} ,
\nonumber \\  
{\hat {g}}^{{\rm {in}}}_{\omega \ell m} | {\rm {CCH}}^{-} \rangle =0 , 
\qquad &  {\mbox {all ${{\omega }}$}} ,
\nonumber \\
{\hat {f}}^{{\rm {up}}}_{\omega \ell m} | {\rm {CCH}}^{-} \rangle  =0 ,
\qquad & {\mbox {all ${\widetilde {\omega }}$}},
\nonumber \\
{\hat {g}}^{{\rm {up}}}_{\omega \ell m} | {\rm {CCH}}^{-} \rangle =0,
\qquad  & {\mbox {all ${\widetilde {\omega }}$}} .
\end{align}
Since the $\chi _{\omega \ell m}^{{\rm {in/up}}+} $ modes all have positive norm, while the $\chi _{\omega \ell m}^{{\rm {in/up}}-} $ modes all have negative norm, the operators ${\hat {f}}_{\omega \ell m}^{{\rm {in/up}}}$ and ${\hat {g}}_{\omega \ell m}^{{\rm {in/up}}}$ satisfy the standard commutation relations (all other commutators vanish)
\begin{align}
\left[ {\hat {f}}_{\omega \ell m}^{\rm {in}}, {\hat {f}}_{\omega '\ell 'm'}^{{\rm {in}}\dagger } \right] =  & ~\delta _{\ell \ell' }\delta _{mm'} \delta (\omega - \omega ') , \qquad {\mbox {all ${{\omega }}$}},
\nonumber \\ 
\left[ {\hat {g}}_{\omega \ell m}^{\rm {in}}, {\hat {g}}_{\omega '\ell 'm'}^{{\rm {in}}\dagger } \right] =  & ~\delta _{\ell \ell' }\delta _{mm'} \delta (\omega - \omega ') , \qquad {\mbox {all ${{\omega }}$}},
\nonumber \\
\left[ {\hat {f}}_{\omega \ell m}^{\rm {up}}, {\hat {f}}_{\omega '\ell 'm'}^{{\rm {up}}\dagger } \right] =  & ~\delta _{\ell \ell' }\delta _{mm'} \delta (\omega - \omega ') , \qquad {\mbox {all ${\widetilde {\omega }}$}},
\nonumber \\ 
\left[ {\hat {g}}_{\omega \ell m}^{\rm {up}}, {\hat {g}}_{\omega '\ell 'm'}^{{\rm {up}}\dagger } \right] =  & ~\delta _{\ell \ell' }\delta _{mm'} \delta (\omega - \omega ') , \qquad {\mbox {all ${\widetilde {\omega }}$}}.
\end{align}
The properties of the state  $|{\rm {CCH}}^{-}\rangle$ will be investigated further in Sec.~\ref{sec:expvalues}, along with the other ``past'' quantum states defined in this section.
From the expansion (\ref{eq:CCH-exp}), we anticipate a thermal distribution of particles/antiparticles in both the ``in'' and ``up'' modes, but with the frequency $\omega $ in the thermal factor for the ``in'' modes, while the thermal factor for the ``up'' modes contains the quantity ${\widetilde {\omega }}$ (\ref{eq:tomega}).

\subsection{``Future'' quantum states}
\label{sec:future}

Following \cite{Ottewill:2000qh,Casals:2012es}, we next define ``future'' Boulware, Unruh and CCH states which are the time-reverse of the ``past'' Boulware, Unruh and CCH states constructed in the previous subsection.
The ``future'' states are defined by using the ``out'' and ``down'' basis modes rather than the ``in'' and ``up'' modes as considered for the ``past'' quantum states.
The ``future'' Boulware state $|{\mathrm {B}}^{+}\rangle$ which we construct in Sec.~\ref{sec:futureB} was previously considered in Ref.~\cite{Balakumar:2020gli}, where it was called the ``out'' vacuum state.  

\subsubsection{``Future'' Boulware state}
\label{sec:futureB}

Near future null infinity ${\mathscr{I}}^{+}$, we consider a set of ``out'' modes which have positive frequency with respect to Schwarzschild time $t$, which is the natural frequency for a static observer in this region:
\begin{subequations}
	\label{eq:futureout+-}
\begin{equation}
    \phi ^{\rm {out}+}_{\omega \ell m} = \frac{e^{-i\omega t}}{{\sqrt {4 \pi \! \left| \omega \right|}} r} X^{\rm {out}}_{\omega \ell }(r) Y_{\ell m}(\theta , \varphi ), \qquad \omega >0 ,
    \label{eq:pfout}
\end{equation}
where $X^{\rm {out}}_{\omega \ell } (r) $ is given by (\ref{eq:out}).
These modes have positive norm.
The corresponding ``out'' modes with negative frequency and negative norm are
\begin{equation}
\phi ^{\rm {out}-}_{\omega \ell m} = \frac{e^{-i\omega t}}{{\sqrt {4 \pi \! \left| \omega \right|}} r} X^{\rm {out}}_{\omega \ell }(r) Y_{\ell m}(\theta , \varphi ), \qquad \omega <0 .
\label{eq:nfout}
\end{equation}
\end{subequations}
Near the future horizon ${\mathcal {H}}^{+}$, we consider positive norm ``down'' modes 
\begin{subequations}
	\label{eq:futuredown+-}
\begin{equation}
    \phi ^{\rm {down}+}_{\omega \ell m} = \frac{e^{-i\omega t}}{{\sqrt {4 \pi \! \left|{\widetilde {\omega }} \right| }} r} X^{\rm {down}}_{\omega \ell }(r) Y_{\ell m}(\theta , \varphi ), \qquad {\widetilde {\omega }}>0 ,
    \label{eq:pfdown}
\end{equation}
where $X^{\rm {down}}_{\omega \ell} (r) $ is given by (\ref{eq:down}).
The restriction ${\widetilde {\omega }}>0$ is required for these modes to have positive norm.
Similarly, a suitable set of ``down'' modes having negative frequency (and negative norm) is
\begin{equation}
\phi ^{\rm {down}-}_{\omega \ell m} = \frac{e^{-i\omega t}}{{\sqrt {4 \pi \! \left|{\widetilde {\omega }} \right| }} r} X^{\rm {down}}_{\omega \ell }(r) Y_{\ell m}(\theta , \varphi ), \qquad {\widetilde {\omega }}<0 .
\label{eq:nfdown}
\end{equation}
\end{subequations}
The modes (\ref{eq:futureout+-}, \ref{eq:futuredown+-}) form an orthonormal basis in region I, hence,
expanding the scalar field in terms of these modes, we find
\begin{align}
{\hat {\Phi }} = & ~ 
\sum _{\ell =0}^{\infty }\sum _{m=-\ell }^{\ell } \left\{ 
\int _{0}^{\infty } d\omega \,
{\hat {a}}_{\omega \ell m}^{\rm {out}} \phi _{\omega \ell m}^{{\rm {out}}+} +
\int _{-\infty }^{0} d\omega \,{\hat {b}}_{\omega \ell m }^{{\rm {out}}\dagger } \phi _{\omega \ell m}^{{\rm {out}}-} 
\right. \nonumber \\ & \left. 
+ \int _{0}^{\infty }d{\widetilde {\omega }}  \,
{\hat {a}}_{\omega \ell m}^{\rm {down}} \phi _{\omega \ell m}^{{\rm {down}}+} +
\int _{-\infty }^{0}d{\widetilde {\omega }} \, {\hat {b}}_{\omega \ell m }^{{\rm {down}}\dagger } \phi _{\omega \ell m}^{{\rm {down}}-}  
\right\} ,
\label{eq:expBplus}
\end{align}   
where the operators $ \hat{a} $ and $ \hat{b} $ satisfy the standard nonzero commutation relations
\begin{align}
\left[ {\hat {a}}_{\omega \ell m}^{\rm {out}}, {\hat {a}}_{\omega '\ell 'm'}^{{\rm {out}}\dagger } \right] =  & ~\delta _{\ell \ell' }\delta _{mm'} \delta (\omega - \omega ') , \qquad \omega >0,
\nonumber \\ 
\left[ {\hat {b}}_{\omega \ell m}^{\rm {out}}, {\hat {b}}_{\omega '\ell 'm'}^{{\rm {out}}\dagger } \right] =  & ~\delta _{\ell \ell' }\delta _{mm'} \delta (\omega - \omega ') , \qquad \omega <0,
\nonumber \\
\left[ {\hat {a}}_{\omega \ell m}^{\rm {down}}, {\hat {a}}_{\omega '\ell 'm'}^{{\rm {down}}\dagger } \right] =  & ~\delta _{\ell \ell' }\delta _{mm'} \delta (\omega - \omega ') , \qquad {\widetilde {\omega }}>0,
\nonumber \\ 
\left[ {\hat {b}}_{\omega \ell m}^{\rm {down}}, {\hat {b}}_{\omega '\ell 'm'}^{{\rm {down}}\dagger } \right] =  & ~\delta _{\ell \ell' }\delta _{mm'} \delta (\omega - \omega ') , \qquad {\widetilde {\omega }}<0,
\end{align}
and all commutators not given above vanish.
The ``future'' Boulware state $| {\rm {B}}^{+} \rangle $ is then defined as the state which is annihilated by the following ${\hat {a}}$ and ${\hat {b}}$ operators:
\begin{align}
{\hat {a}}^{{\rm {out}}}_{\omega \ell m} | {\rm {B}}^{+} \rangle  = & ~0 ,
\qquad \omega >0,
\nonumber \\ 
{\hat {b}}^{{\rm {out}}}_{\omega \ell m} | {\rm {B}}^{+} \rangle = & ~0,
\qquad \omega <0,
\nonumber \\
{\hat {a}}^{{\rm {down}}}_{\omega \ell m} | {\rm {B}}^{+} \rangle  = & ~0 ,
\qquad {\widetilde {\omega }} >0,
\nonumber \\
{\hat {b}}^{{\rm {down}}}_{\omega \ell m} | {\rm {B}}^{+} \rangle  = & ~0 ,
\qquad {\widetilde {\omega }}<0 .
\end{align}
The ``future'' Boulware state $| {\rm {B}}^{+} \rangle $  was referred to as the ``out'' vacuum in our previous work \cite{Balakumar:2020gli}. It corresponds to an absence of  outgoing particles/antiparticles as seen by a static observer at future null infinity \cite{Balakumar:2020gli}.

\begin{widetext}

\subsubsection{``Future'' Unruh state}
\label{sec:futureU}

In a similar fashion, we next define the ``future'' Unruh state $|{\rm {U}}^{+}\rangle$.
The ``out'' modes take the form (\ref{eq:futureout+-}) in region I (which can be extended into region III) and have positive/negative frequency with respect to Schwarzschild time $t$ near future null infinity $ \mathscr{I}^+ $. 
We consider ``down'' modes having positive/negative frequency with respect to Kruskal time $V$ near the future horizon ${\mathcal {H}}^{+}$.
The derivation of these modes follows that for the ``up'' modes in Sec.~\ref{sec:pastU}.

The ``down'' modes are defined in region I by their radial function $ X^{\rm {down}}_{\omega \ell} (r) $ (\ref{eq:down}), and this definition can be extended across the future horizon ${\mathcal {H}}^{+}$ into region II.  Near ${\mathcal {H}}^{+}$ we have: 
\begin{equation}
    \phi _{\omega \ell m}^{\rm {down}} 
    = \frac{1}{{\sqrt {4 \pi \! \left| {\widetilde {\omega }} \right| }} r}
    e^{\frac{i\left( \omega - {\widetilde{\omega }} \right) }{2\kappa } \ln \left( -\kappa U\right) } e^{-\frac{i\left( \omega + {\widetilde {\omega }} \right) }{2\kappa } \ln \left( \kappa V \right) } Y_{\ell m}(\theta , \varphi ) \Theta (V).
    \label{eq:downH+}
\end{equation}
We then define a set of modes denoted by $\psi _{\omega \ell m}^{\rm {up}}$, which are obtained by taking the ``down'' modes $\phi _{\omega \ell m}^{\rm {down}}$ and making the coordinate transformation $U\rightarrow -U$, $V\rightarrow -V$.
These new modes are nonvanishing in regions III and IV and have negative norm when ${\widetilde {\omega }}>0$ and positive norm when ${\widetilde {\omega }}<0$.
Near the surface $U=0$ they take the form 
\begin{equation}
    \psi _{\omega \ell m}^{\rm {up}} 
    = \frac{1}{{\sqrt {4 \pi \! \left| {\widetilde {\omega }}\right| }} r}
     e^{\frac{i\left( \omega - {\widetilde{\omega }} \right) }{2\kappa } \ln \left( \kappa U\right) } e^{-\frac{i\left( \omega + {\widetilde {\omega }} \right) }{2\kappa } \ln \left( -\kappa V \right) } Y_{\ell m}(\theta , \varphi ) \Theta (-V).
     \label{eq:upH}
\end{equation}
In a similar fashion to the approach of Sec.~\ref{sec:pastU}, we seek to apply the Lemma (\ref{eq:FNlemma}) to a suitable combination of the modes (\ref{eq:downH+}, \ref{eq:upH}) with $X=V$ and ${\mathfrak {q}}=\left( \omega + \widetilde{\omega} \right) /2\kappa $, integrating over a surface near $U=0$ with $U=\epsilon >0$ (part of this surface lies close to the future event horizon ${\mathcal {H}}^{+}$ in region II of the space-time).
Again we use a branch of the logarithm which is analytic in the lower half plane to give, for $\kappa U >0$,
\begin{equation}
\exp \left[ \frac{i\left( \omega - {\widetilde{\omega }} \right)} {2\kappa } \ln ( -\kappa U) \right] 
= \exp \left[ \frac{\pi \left( \omega - {\widetilde{\omega }} \right)} {2\kappa }\right] 
\exp \left[ \frac{i\left( \omega - {\widetilde{\omega }} \right)} {2\kappa } \ln ( \kappa U) \right] .
\label{eq:Uepsilon>0}
\end{equation}
We can now apply (\ref{eq:FNlemma}) to obtain
\begin{equation}
    \int _{-\infty }^{\infty } dV \, e^{-i{\mathfrak {p}}V} \left[ 
    e^{-\frac{\pi \left( \omega - {\widetilde{\omega }} \right)}{2\kappa }}\phi _{\omega \ell m}^{\rm {down}} + e^{-\frac{\pi \left( \omega +{\widetilde{\omega }}\right) }{2\kappa } }\psi _{\omega \ell m}^{\rm {up}} \right] =0, \qquad {\mathfrak {p}}>0,
\label{eq:FNlemma1}
\end{equation}
from which we deduce that the modes in square brackets in (\ref{eq:FNlemma1}) have positive frequency with respect to the Kruskal coordinate $V$ for any value of ${\widetilde {\omega }}$. 
Therefore a suitable orthonormal basis of field modes defined throughout regions I--IV and having positive frequency with respect to $V$ near ${\mathcal {H}}^{+}$ is
\begin{subequations}
	\label{eq:chidown+-}
\begin{equation}
    \chi _{\omega \ell m}^{{\rm {down}}+} 
    =\frac{1}{{\sqrt {2\left| \sinh \left( \frac{\pi {\widetilde{\omega }}}{\kappa } \right) \right| }}} \left( 
    e^{\frac{\pi {\widetilde{\omega }}}{2\kappa }} \phi _{\omega \ell m}^{\rm {down}} + e^{-\frac{\pi {\widetilde{\omega }}}{2\kappa }} \psi _{\omega \ell m}^{{\rm {up}}} \right) , \qquad {\mbox {all ${\widetilde {\omega }}$}},
    \label{eq:chidown+}
\end{equation}
while an orthonormal set of modes having negative frequency with respect to $V$ near ${\mathcal {H}}^{+}$ is
\begin{equation}
    \chi _{\omega \ell m}^{{\rm {down}}-} 
    =\frac{1}{{\sqrt {2\left| \sinh \left( \frac{\pi {\widetilde{\omega }}}{\kappa } \right) \right| }}} \left( 
    e^{-\frac{\pi {\widetilde{\omega }}}{2\kappa }} \phi _{\omega \ell m}^{\rm {down}} + e^{\frac{\pi {\widetilde{\omega }}}{2\kappa }} \psi _{\omega \ell m}^{{\rm {up}}} \right) , \qquad {\mbox {all ${\widetilde {\omega }}$}}.
    \label{eq:chidown-}
\end{equation}
\end{subequations}
All our positive frequency modes $\chi _{\omega \ell m}^{{\rm {down}}+}$ (\ref{eq:chidown+})  have positive norm, and all negative frequency modes  $\chi _{\omega \ell m}^{{\rm {down}}-}$ (\ref{eq:chidown-}) have negative norm.
In region I (where the $\psi _{\omega \ell m}^{\rm {up}}$ modes vanish), the expansion of the field in terms of the modes $\phi _{\omega \ell m}^{\rm {out}}$ and $\chi ^{\rm {down}}_{\omega \ell m}$ takes the form
\begin{equation}
    {\hat {\Phi }}
    = \sum _{\ell =0}^{\infty } \sum _{m=-\ell }^{\ell }
    \left\{ 
    \int _{0}^{\infty } d\omega \,
    {\hat {c}}_{\omega \ell m}^{\rm {out}} \phi _{\omega \ell m}^{{\rm {out}}+} +
    \int _{-\infty }^{0} d\omega \,
    {\hat {d}}_{\omega \ell m}^{{\rm {out}}\dagger } \phi _{\omega \ell m}^{{\rm {out}}-} 
    + \int _{-\infty }^{\infty } d{\widetilde {\omega } }
    \frac{1}{{\sqrt {2 \left| \sinh \left( \frac{\pi {\widetilde{\omega }} }{\kappa }\right)  \right| }}}
    \phi _{\omega \ell m}^{\rm {down}} 
    \left[ 
    e^{\frac{\pi {\widetilde{\omega }}}{2\kappa }}
    {\hat {c}}_{\omega \ell m}^{\rm {down}}
    + e^{-\frac{\pi {\widetilde{\omega }}}{2\kappa }}
    {\hat {d}}_{\omega \ell m}^{{\rm {down}}\dagger }
    \right] 
    \right\} .
\end{equation}
\end{widetext}
The commutation relations satisfied by the ${\hat {c}}$ and ${\hat {d}}$ operators take the standard form (with all commutators not given below vanishing)
\begin{align}
\left[ {\hat {c}}_{\omega \ell m}^{\rm {out}}, {\hat {c}}_{\omega '\ell 'm'}^{{\rm {out}}\dagger } \right] =  & ~\delta _{\ell \ell' }\delta _{mm'} \delta (\omega - \omega ') , \qquad \omega >0,
\nonumber \\ 
\left[ {\hat {d}}_{\omega \ell m}^{\rm {out}}, {\hat {d}}_{\omega '\ell 'm'}^{{\rm {out}}\dagger } \right] =  & ~\delta _{\ell \ell' }\delta _{mm'} \delta (\omega - \omega ') , \qquad \omega <0,
\nonumber \\
\left[ {\hat {c}}_{\omega \ell m}^{\rm {down}}, {\hat {c}}_{\omega '\ell 'm'}^{{\rm {down}}\dagger } \right] =  & ~\delta _{\ell \ell' }\delta _{mm'} \delta (\omega - \omega ') , \qquad {\mbox {all ${\widetilde {\omega }}$}},
\nonumber \\ 
\left[ {\hat {d}}_{\omega \ell m}^{\rm {down}}, {\hat {d}}_{\omega '\ell 'm'}^{{\rm {down}}\dagger } \right] =  & ~\delta _{\ell \ell' }\delta _{mm'} \delta (\omega - \omega ') , \qquad {\mbox {all ${\widetilde {\omega }}$}}.
\end{align}
The ``future'' Unruh state $| {\rm {U}}^{+} \rangle $ is then defined as that state which is annihilated by the following ${\hat {c}}$ and ${\hat {d}}$ operators:
\begin{align}
{\hat {c}}^{{\rm {out}}}_{\omega \ell m} | {\rm {U}}^{+} \rangle  =0
, \qquad & \omega >0,
\nonumber \\  
{\hat {d}}^{{\rm {out}}}_{\omega \ell m} | {\rm {U}}^{+} \rangle =0 , 
\qquad & \omega <0,
\nonumber \\
{\hat {c}}^{{\rm {down}}}_{\omega \ell m} | {\rm {U}}^{+} \rangle  =0 ,
\qquad & {\mbox {all ${\widetilde {\omega }}$}},
\nonumber \\
{\hat {d}}^{{\rm {down}}}_{\omega \ell m} | {\rm {U}}^{+} \rangle =0,
\qquad  & {\mbox {all ${\widetilde {\omega }}$}} .
\end{align}
In this state there are no outgoing particles/antiparticles as seen by a static observer at ${\mathscr{I}}^{+}$, but the ``down'' modes are thermally populated.

\subsubsection{``Future'' CCH state}
\label{sec:futureCCH}

Our final ``future'' state is 
the ``future'' CCH state $|{\rm {CCH}}^{+} \rangle $, in which the ``down'' modes are thermalized with respect to the quantity ${\widetilde {\omega }} $ near the future event horizon ${\mathcal {H}}^{+}$, as in the ``future'' Unruh state $|{\rm {U}}^{+}\rangle $.
In addition, the ``out'' modes are also thermalized, but with respect to $\omega $, which is the natural frequency for a static observer far from the black hole.

Our orthonormal basis of modes consists of $\chi_{\omega \ell m}^{\rm {down}\pm}$, (\ref{eq:chidown+-}) and new modes $\chi_{\omega \ell m}^{\rm {out}\pm}$, which we construct using the same approach as in Sec.~\ref{sec:pastCCH} and which are also defined throughout regions I--IV.
We define modes $\psi ^{\rm {in}}_{\omega \ell m}$ by taking $\phi _{\omega \ell m}^{\rm {out}}$ (defined in region I by their radial function $ X^{\rm {out}}_{\omega \ell} (r) $ (\ref{eq:out}), and extended across the past horizon ${\mathcal {H}}^{-}$ into region III) and performing the substitution $U\rightarrow -U$, $V\rightarrow -V$.
The resulting modes $\psi ^{\rm {in}}_{\omega \ell m}$ (which are nonvanishing in regions II and IV) have negative norm for $\omega >0$ and positive norm for $\omega <0$.
We then define the modes $\chi_{\omega \ell m}^{\rm {out}\pm}$, for all $\omega $, as follows:
\begin{subequations}
	\label{eq:chiout+-}
\begin{align}
\chi _{\omega \ell m}^{{\rm {out}}+} & =
\frac{1}{{\sqrt {2 \left| \sinh \left( \frac{\pi {\omega } }{\kappa }\right)  \right| }}}\left( e^{\frac{\pi {\omega }}{2\kappa }}\phi _{\omega \ell m}^{\rm {out}} + e^{-\frac{\pi {\omega }}{2\kappa }} \psi _{\omega \ell m}^{{\rm {in}}} \right)  , 
\label{eq:chiout+}
	\\
\chi _{\omega \ell m}^{{\rm {out}}-} & =
\frac{1}{{\sqrt {2 \left| \sinh \left( \frac{\pi {\omega } }{\kappa }\right)  \right| }}}\left( e^{-\frac{\pi {\omega }}{2\kappa }}\phi _{\omega \ell m}^{\rm {out}} + e^{\frac{\pi {\omega }}{2\kappa }} \psi _{\omega \ell m}^{{\rm {in}}} \right) .
\label{eq:chiout-}
\end{align}
\end{subequations}

\begin{widetext}
\noindent It is straightforward to show that the $\chi_{\omega \ell m}^{\rm {out}+}$ modes have positive norm for all $\omega $, while the $\chi_{\omega \ell m}^{\rm {out}-}$ modes have negative norm for all $\omega $.

To investigate whether the modes $\chi_{\omega \ell m}^{\rm {out}\pm}$ have an interpretation as being positive/negative frequency with respect to a particular coordinate, we consider the form of the ``out'' modes $\phi _{\omega \ell m}^{\rm {out}}$ and the modes $\psi _{\omega \ell m}^{{\rm {in}}}$ near the surface $V=0$.
We have 
\begin{equation}
\phi _{\omega \ell m}^{\rm {out}}=\frac{B^{{\rm {in}}*}_{\omega \ell}}{{\sqrt {4 \pi \! \left| \omega \right|}}r}
e^{\frac{i\left(\omega +{\widetilde {\omega }}\right) }{2\kappa }\ln (-\kappa U)}
e^{-\frac{i\left(\omega -{\widetilde {\omega }}  \right) }{2\kappa }\ln (\kappa V)}
 Y_{\ell m }(\theta, \varphi ) \Theta (-U) ,
\label{eq:outI+}
\end{equation}
and
\begin{equation}
\psi _{\omega \ell m}^{{\rm {in}}} = 
\frac{B^{{\rm {in}}*}_{\omega \ell}}{{\sqrt {4 \pi \! \left| \omega \right|}}r}
	e^{\frac{i\left(\omega +{\widetilde {\omega }} \right) }{2\kappa }\ln (\kappa U)}
	e^{-\frac{i\left(\omega -{\widetilde {\omega }} \right) }{2\kappa }\ln (-\kappa V)}
Y_{\ell m }(\theta, \varphi ) \Theta (U) . 
\label{eq:inH}
\end{equation}
For positive frequency modes, we take $\ln (-1)=-i\pi $ as previously, then, integrating over the surface $V=-\epsilon<0$ we have 
\begin{equation}
\exp \left[ -\frac{i\left( \omega - {\widetilde{\omega }} \right)} {2\kappa } \ln ( \kappa V) \right] 
= \exp \left[ -\frac{\pi \left( \omega - {\widetilde{\omega }} \right)} {2\kappa }\right]  \exp \left[ -\frac{i\left( \omega - {\widetilde{\omega }} \right)} {2\kappa } \ln ( -\kappa V) \right].	
\end{equation}
Applying (\ref{eq:FNlemma}) with $X=U$ and ${\mathfrak {q}}=-\left(\omega + {\widetilde {\omega  }} \right)/2\kappa $ gives 
\begin{equation}
	 \int _{-\infty }^{\infty } dU \, e^{-i{\mathfrak {p}}U} \left[ \psi _{\omega \ell m}^{\rm {in}} 
	 + e^{\frac{\pi \omega  } {\kappa }}
	 \phi _{\omega \ell m}^{\rm {out}}\right] =0, \qquad {\mathfrak {p}}>0,
\end{equation}
from which we deduce that the modes (\ref{eq:chiout+}) have positive frequency with respect to the Kruskal coordinate $U$ along the surface $V=-\epsilon <0$, part of which lies close to the past event horizon ${\mathcal {H}}^{-}$ in region II of the space-time.
By a similar argument, the modes (\ref{eq:chiout-}) have negative frequency with respect to $U$ along the same surface. 

As previously, we now consider the quantum field on  region I of the space-time, where the modes $\psi _{\omega \ell m}^{\rm {in}}$ vanish.
Therefore the expansion of the quantum scalar field is
	\begin{align}
	{\hat {\Phi }}
	& ~ =  \sum _{\ell =0}^{\infty } \sum _{m=-\ell }^{\ell }
	\left\{ 
	\int _{-\infty }^{\infty } d\omega \,
	\frac{1}{{\sqrt {2 \left| \sinh \left( \frac{\pi {{\omega }} }{\kappa }\right)  \right| }}}
	\phi _{\omega \ell m}^{\rm {out}} 
	\left[ 
	e^{\frac{\pi {{\omega }}}{2\kappa }}
	{\hat {f}}_{\omega \ell m}^{\rm {out}}
	+ e^{-\frac{\pi {{\omega }}}{2\kappa }}
	{\hat {g}}_{\omega \ell m}^{{\rm {out}}\dagger }
	\right]  \right. 
	\nonumber \\ & \left. \qquad \qquad 
	+ \int _{-\infty }^{\infty } d{\widetilde {\omega } }
	\frac{1}{{\sqrt {2 \left| \sinh \left( \frac{\pi {\widetilde{\omega }} }{\kappa }\right)  \right| }}}
	\phi _{\omega \ell m}^{\rm {down}} 
	\left[ 
	e^{\frac{\pi {\widetilde{\omega }}}{2\kappa }}
	{\hat {f}}_{\omega \ell m}^{\rm {down}}
	+ e^{-\frac{\pi {\widetilde{\omega }}}{2\kappa }}
	{\hat {g}}_{\omega \ell m}^{{\rm {down}}\dagger }
	\right] 
	\right\} . 
	\end{align}
\end{widetext}
The operators ${\hat {f}}_{\omega \ell m}^{{\rm {in/up}}}$ and ${\hat {g}}_{\omega \ell m}^{{\rm {in/up}}}$ satisfy the standard nonzero commutation relations, since the $\chi _{\omega \ell m}^{{\rm {out/down}}+} $ modes all have positive norm, while the $\chi _{\omega \ell m}^{{\rm {out/down}}-} $ modes all have negative norm, 
\begin{align}
\left[ {\hat {f}}_{\omega \ell m}^{\rm {out}}, {\hat {f}}_{\omega '\ell 'm'}^{{\rm {out}}\dagger } \right] =  & ~\delta _{\ell \ell' }\delta _{mm'} \delta (\omega - \omega ') , \qquad {\mbox {all ${{\omega }}$}},
\nonumber \\ 
\left[ {\hat {g}}_{\omega \ell m}^{\rm {out}}, {\hat {g}}_{\omega '\ell 'm'}^{{\rm {out}}\dagger } \right] =  & ~\delta _{\ell \ell' }\delta _{mm'} \delta (\omega - \omega ') , \qquad {\mbox {all ${{\omega }}$}},
\nonumber \\
\left[ {\hat {f}}_{\omega \ell m}^{\rm {down}}, {\hat {f}}_{\omega '\ell 'm'}^{{\rm {down}}\dagger } \right] =  & ~\delta _{\ell \ell' }\delta _{mm'} \delta (\omega - \omega ') , \qquad {\mbox {all ${\widetilde {\omega }}$}},
\nonumber \\ 
\left[ {\hat {g}}_{\omega \ell m}^{\rm {down}}, {\hat {g}}_{\omega '\ell 'm'}^{{\rm {down}}\dagger } \right] =  & ~\delta _{\ell \ell' }\delta _{mm'} \delta (\omega - \omega ') , \qquad {\mbox {all ${\widetilde {\omega }}$}},
\end{align}
with all commutators not given explicitly above vanishing.
Finally we define the ``future'' CCH state $|{\rm {CCH}}^{+}\rangle$ as that state annihilated by the following ${\hat {f}}$ and ${\hat {g}}$ operators:
\begin{align}
{\hat {f}}^{{\rm {out}}}_{\omega \ell m} | {\rm {CCH}}^{+} \rangle  =0
, \qquad & {\mbox {all ${{\omega }}$}} ,
\nonumber \\  
{\hat {g}}^{{\rm {out}}}_{\omega \ell m} | {\rm {CCH}}^{+} \rangle =0 , 
\qquad &  {\mbox {all ${{\omega }}$}} ,
\nonumber \\
{\hat {f}}^{{\rm {down}}}_{\omega \ell m} | {\rm {CCH}}^{+} \rangle  =0 ,
\qquad & {\mbox {all ${\widetilde {\omega }}$}},
\nonumber \\
{\hat {g}}^{{\rm {down}}}_{\omega \ell m} | {\rm {CCH}}^{+} \rangle =0,
\qquad  & {\mbox {all ${\widetilde {\omega }}$}} .
\end{align}
We have now defined six states: three ``past'' and three ``future'' states. 
The properties of these states will be studied in detail in Sec.~\ref{sec:expvalues}.

\subsection{``Boulware''-like state} 
\label{sec:Boulware}

In Secs.~\ref{sec:pastB}, \ref{sec:futureB} we defined the ``past'' and ``future'' Boulware states (dubbed the ``in'' and ``out'' vacua in \cite{Balakumar:2020gli}). 
The ``past'' Boulware state $|{\rm {B}}^{-}\rangle$ is empty of incoming particles as seen by a static observer at ${\mathscr{I}}^{-}$ but, as we shall find in Sec.~\ref{sec:expvalues}, a static observer at ${\mathscr{I}}^{+}$ sees an outgoing flux of particles \cite{Balakumar:2020gli}. 
Similarly, the ``future'' Boulware state $|{\rm {B}}^{+}\rangle $ has no outgoing flux as seen by a static observer at ${\mathscr{I}}^{+}$, but is not empty at ${\mathscr{I}}^{-}$.
The question then arises as to whether it is possible to define a quantum state which is as empty as possible as seen by static observers at {\it {both}} past null infinity ${\mathscr{I}}^{-}$ and  future null infinity ${\mathscr{I}}^{+}$.
To define a state which is as empty as possible at ${\mathscr{I}}^{\pm }$, we seek to expand the classical scalar field in terms of the positive and negative frequency ``in'' (\ref{eq:pastin+-}) and ``out'' modes (\ref{eq:futureout+-}).
Working in region I, the ``in'' and ``up'' modes form a basis and therefore the classical scalar field can be expanded in terms of these modes. Using (\ref{eq:inoutupdownmodes}), each ``up'' mode can be written in terms of an ``in'' and an ``out'' mode, leading to an expansion of the classical scalar field in terms of ``in'' and ``out'' modes, as follows:
\begin{widetext}
\begin{align}
   \Phi  =  &~ \sum _{\ell =0}^{\infty }\sum _{m=-\ell }^{\ell } \left\{ 
    \int _{0}^{\infty } d\omega \left[ 
    {\widetilde {h}}_{\omega \ell m}^{\rm {in}}\phi _{\omega \ell m}^{\rm {in}+}
    + {\widetilde {h}}_{\omega \ell m}^{\rm {out}}\phi _{\omega \ell m}^{\rm {out}+}
    \right] 
    +\int _{-\infty }^{0} d\omega \left[ 
     {\widetilde {k}}_{\omega \ell m}^{{\rm {in}}\dagger} \phi _{\omega \ell m}^{{\rm {in}}-}
    + {\widetilde {k}}_{\omega \ell m}^{{\rm {out}}\dagger} \phi _{\omega \ell m}^{{\rm {out}}-} 
    \right] \right\} .
    \label{eq:inoutsum}
\end{align}
Both the ``in'' and ``out'' modes have positive norm for $\omega >0$, so those modes in (\ref{eq:inoutsum}) which we have identified as having positive frequency also have positive norm, while the negative frequency modes have negative norm.

However, there is a complication. 
The ``in'' and ``out'' modes are not orthogonal, and therefore we cannot directly quantize the field using the expansion (\ref{eq:inoutsum}).  
We therefore write the ``out'' modes in terms of the ``in'' and ``up'' modes (which are orthogonal) using (\ref{eq:inoutupdownmodes}), and hence obtain the classical field expansion
\begin{align}
	\Phi  =  &~ \sum _{\ell =0}^{\infty }\sum _{m=-\ell }^{\ell } \left\{ 
	\int _{0}^{\infty } d\omega  \,
	h_{\omega \ell m}^{\rm {in}}\phi _{\omega \ell m}^{\rm {in}+}
	+ \int _{-\infty }^{0} d\omega \,  k_{\omega \ell m}^{{\rm {in}}\dagger} \phi _{\omega \ell m}^{{\rm {in}}-}
	\right. \nonumber \\ & \left. 
+ \int _{\max \{\frac{qQ}{r_{+}}, 0 \}}^{\infty } d\omega \,
	h_{\omega \ell m}^{\rm {up}}\phi _{\omega \ell m}^{\rm {up}+}
	+ \int _{0}^{\max \{\frac{qQ}{r_{+}}, 0 \}} d\omega \,
	h_{\omega \ell m}^{\rm {up}}\phi _{\omega \ell m}^{\rm {up}-}
	+\int _{\min \{\frac{qQ}{r_{+}}, 0 \}}^{0} d\omega \, 
k_{\omega \ell m}^{{\rm {up}}\dagger} \phi _{\omega \ell m}^{{\rm {up}}+} 
+\int _{-\infty }^{\min \{\frac{qQ}{r_{+}}, 0 \}} d\omega \, 
k_{\omega \ell m}^{{\rm {up}}\dagger} \phi _{\omega \ell m}^{{\rm {up}}-} 
\right\} ,
	\label{eq:inupsum}
\end{align}
\end{widetext}
where the coefficients in the expansion are given by 
\begin{align}
    h_{\omega \ell m}^{\rm {in}} = & ~
    {\widetilde {h}}_{\omega \ell m}^{\rm {in}} +
    A_{\omega \ell }^{{\rm {in}}*}{\widetilde {h}}_{\omega \ell m}^{\rm {out}},
\qquad    \omega >0 ,
    \nonumber \\
    k_{\omega \ell m}^{{\rm {in}}\dagger} = & ~
    {\widetilde {k}}_{\omega \ell m}^{{\rm {in}}\dagger } +
    A_{\omega \ell }^{{\rm {in}}*}{\widetilde {k}}_{\omega \ell m}^{{\rm {out}}\dagger }
    , \qquad \omega <0 ,
    \nonumber \\
    h_{\omega \ell m}^{\rm {up}} = & ~
    \left| \frac{{\widetilde {\omega }}}{\omega } \right| ^{\frac{1}{2}} B_{\omega \ell }^{{\rm {in}}*} {\widetilde {h}}_{\omega \ell m}^{\rm {out}}
    , \qquad \omega >0 ,
    \nonumber \\
    k_{\omega \ell m}^{{\rm {up}}\dagger} = & ~
     \left| \frac{{\widetilde {\omega }}}{\omega } \right| ^{\frac{1}{2}} B_{\omega \ell }^{{\rm {in}}*} {\widetilde {k}}_{\omega \ell m}^{{\rm {out}}\dagger},
    \qquad \omega <0.
\end{align}
In (\ref{eq:inupsum}) we have now expanded the classical scalar field in terms of an orthonormal basis of field modes, and therefore we can proceed to quantize the field by promoting the expansion coefficients $h$ and $k$ to operators.
At this point a subtlety arises.
 As can be seen in (\ref{eq:inupsum}),  depending on the sign of $qQ/r_{+}$, we either have some positive norm ``up'' modes $\phi _{\omega \ell m}^{{\rm {up}}+}$ which are multiplied by operators ${\hat {k}}^{\dagger}$ which we would like to interpret as creation operators, or else there are negative norm ``up'' modes $\phi _{\omega \ell m}^{{\rm {up}}-}$ which are multiplied by operators ${\hat {h}}$ which we would like to interpret as annihilation operators. 

As discussed in Sec.~\ref{sec:general}, we therefore find that the operators $ \hat{h}^{\mathrm{in}} $ and $ \hat{k}^{\mathrm{in}} $ satisfy the usual commutation relations (with those commutators not given explicitly below vanishing)
\begin{align}
    \left[ {\hat {h}}_{\omega \ell m}^{{\rm {in}}}, {\hat {h}}_{\omega ' \ell 'm '}^{{\rm {in}}\dagger} \right]
    = & ~ \delta _{\ell \ell' }\delta _{mm'}\delta (\omega - \omega ' ) ,
    \qquad \omega >0,
    \nonumber \\
    \left[ {\hat {k}}_{\omega \ell m}^{{\rm {in}}}, {\hat {k}}_{\omega ' \ell 'm '}^{{\rm {in}}\dagger} \right]
    = & ~ \delta _{\ell \ell' }\delta _{mm'}\delta (\omega - \omega ' ) ,
    \qquad \omega <0,
\end{align}
but that the operators $ \hat{h}^{\mathrm{up}} $ and $ \hat{k}^{\mathrm{up}} $ satisfy modified commutation relations (the remaining commutators vanish as usual):
\begin{align}
    \left[ {\hat {h}}_{\omega \ell m}^{{\rm {up}}}, {\hat {h}}_{\omega ' \ell 'm '}^{{\rm {up}}\dagger} \right]
    = & ~ \eta _{\omega {\widetilde {\omega }}}\delta _{\ell \ell' }\delta _{mm'}\delta (\omega - \omega ' ) ,
    \qquad \omega >0,
    \nonumber \\
    \left[ {\hat {k}}_{\omega \ell m}^{{\rm {up}}}, {\hat {k}}_{\omega ' \ell 'm '}^{{\rm {up}}\dagger} \right]
    = & ~ \eta _{\omega {\widetilde {\omega }}}\delta _{\ell \ell' }\delta _{mm'}\delta (\omega - \omega ' ) ,
    \qquad \omega <0,
    \label{eq:Bccrs} 
\end{align}
where we have defined
\begin{equation}
    \eta _{\omega {\widetilde{\omega }}} = \begin{cases}
    1, & {\mbox {if $\omega {\widetilde{\omega }}>0$}}, \\
    -1, & {\mbox {if $\omega {\widetilde{\omega }}<0$}}.
    \end{cases}
    \label{eq:etadef1}
\end{equation}
Essentially what is happening is that the ``up'' modes with $\omega >0$ but ${\widetilde {\omega }}<0$ have been ``mislabelled'' as positive frequency modes in the expansion (\ref{eq:inupsum}), despite the fact that they have negative ``norm''.
Similarly, ``up'' modes with $\omega <0$ but ${\widetilde {\omega }}>0$ have been ``mislabelled'' as negative frequency modes in the expansion,  since they have positive ``norm''.

Despite the unconventional commutation relations (\ref{eq:Bccrs}), following the discussion in Sec.~\ref{sec:general} and Ref.~\cite{Frolov:1989jh}, we posit a ``Boulware''-like state $|{\rm {B}} \rangle $ as that state which is annihilated by the ${\hat {h}}$ and ${\hat {k}}$ operators as follows:
\begin{align}
	{\hat {h}}^{{\rm {in}}}_{\omega \ell m} | {\rm {B}} \rangle  = & ~0 ,
	\qquad \omega >0,
	\nonumber \\ 
	{\hat {k}}^{{\rm {in}}}_{\omega \ell m} | {\rm {B}} \rangle = & ~0,
	\qquad \omega <0,
	\nonumber \\
	{\hat {h}}^{{\rm {up}}}_{\omega \ell m} | {\rm {B}} \rangle  = & ~0 ,
	\qquad \omega  >0,
	\nonumber \\ 
	{\hat {k}}^{{\rm {up}}}_{\omega \ell m} | {\rm {B}} \rangle  = & ~0 ,
	\qquad \omega <0 .
\end{align}
This state corresponds to an absence of quanta in the ``in'' modes, as is the case for the ``past'' Boulware state $| {\rm {B}}^{-} \rangle $.
As in $| {\rm {B}}^{-} \rangle $, there are also no particles in the nonsuperradiant ``up'' modes. 
Our new state $| {\rm {B}} \rangle $ differs from $| {\rm {B}}^{-} \rangle $ in its quanta content in the superradiant ``up'' modes.
This will be evident when we study expectation values of observables in this state in Sec.~\ref{sec:Boulwareexp}.

\medskip

\subsection{``Hartle-Hawking''-like states}
\label{sec:HH}

The Hartle-Hawking state $| {\rm {H}} \rangle $ \cite{Hartle:1976tp} on Schwarzschild space-time is constructed by considering both the ``in'' and ``up'' modes to be thermalized.
This is equivalent to considering a set of ``up'' modes which are positive frequency with respect to the Kruskal coordinate $U$ on the past horizon ${\mathcal {H}}^{-}$ and a set of ``down'' modes which are positive frequency with respect to the Kruskal coordinate $V$ on the future horizon ${\mathcal {H}}^{+}$. 
We now examine whether it is possible to define a corresponding ``Hartle-Hawking''-like state for the charged scalar field on RN.

In Sec.~\ref{sec:pastU} we have already constructed sets of ``up'' modes $\chi _{\omega \ell m}^{{\rm {up}}\pm  }$ (\ref{eq:chiup+-}) which have positive/negative  frequency with respect to $U$ near ${\mathcal {H}}^{-}$, and in Sec.~\ref{sec:futureU} we have a similar set of ``down'' modes  $\chi _{\omega \ell m}^{{\rm {down}}\pm  }$ (\ref{eq:chidown+-}) having positive/negative frequency with respect to $V$ near ${\mathcal {H}}^{+}$.
To expand the classical scalar field in terms of these modes, first note that the modes $\chi _{\omega \ell m}^{{\rm {up}}\pm  }$ (\ref{eq:chiup+-}) and $\chi _{\omega \ell m}^{{\rm {in}}\pm  }$ (\ref{eq:chiin+-}) form a basis of modes in regions I--IV. Using (\ref{eq:inoutupdownmodes}) and the corresponding relationships between the modes ${\psi }^{\rm {in}}_{\omega \ell m}$,  ${\psi }^{\rm {up}}_{\omega \ell m}$,  ${\psi }^{\rm {out}}_{\omega \ell m}$,  ${\psi }^{\rm {down}}_{\omega \ell m}$, each $\chi _{\omega \ell m}^{{\rm {in}}\pm  }$ mode can be written as a linear combination of the modes $\chi _{\omega \ell m}^{{\rm {up}}\pm  }$ (\ref{eq:chiup+-}) and $\chi _{\omega \ell m}^{{\rm {down}}\pm  }$ (\ref{eq:chidown+-}). Therefore we can write the classical scalar field in terms of the $\chi _{\omega \ell m}^{{\rm {up}}\pm  }$ and $\chi _{\omega \ell m}^{{\rm {down}}\pm  }$ modes  as follows:
\begin{widetext}
	\begin{equation}
	 \Phi 
	= \sum _{\ell =0}^{\infty } \sum _{m=-\ell }^{\ell }
	\int_{-\infty }^{\infty } d{\widetilde {\omega }} \,
	\left\{ 
	{\widetilde {p}}^{\rm {up}}_{\omega \ell m}\chi ^{{\rm {up}}+}_{\omega \ell m}
	+ {\widetilde {s}}^{{\rm {up}}\dagger}_{\omega \ell m}\chi ^{{\rm {up}}-}_{\omega \ell m}
	+	{\widetilde {p}}^{\rm {down}}_{\omega \ell m}\chi ^{{\rm {down}}+}_{\omega \ell m}
	+ {\widetilde {s}}^{{\rm {down}}\dagger}_{\omega \ell m}\chi ^{{\rm {down}}-}_{\omega \ell m}
	\right\} ,
	\end{equation}
	which, in region I, equals
\begin{equation}
     \Phi 
    = \sum _{\ell =0}^{\infty } \sum _{m=-\ell }^{\ell }
    \int_{-\infty }^{\infty } d{\widetilde {\omega }} \,
    \frac{1}{{\sqrt {2 \left| \sinh \left( \frac{\pi {\widetilde{\omega }} }{\kappa }\right)  \right| }}}
    \left\{ 
    \phi _{\omega \ell m}^{\rm {up}} 
    \left[ 
    e^{\frac{\pi {\widetilde{\omega }}}{2\kappa }}
    {\widetilde {p}}_{\omega \ell m}^{\rm {up}}
    + e^{-\frac{\pi {\widetilde{\omega }}}{2\kappa }}
   {\widetilde {s}}_{\omega \ell m}^{{\rm {up}}\dagger }
    \right] 
    +
    \phi _{\omega \ell m}^{\rm {down}} 
    \left[ 
    e^{\frac{\pi {\widetilde{\omega }}}{2\kappa }}
   {\widetilde {p}}_{\omega \ell m}^{\rm {down}}
    + e^{-\frac{\pi {\widetilde{\omega }}}{2\kappa }}
    {\widetilde {s}}_{\omega \ell m}^{{\rm {down}}\dagger }
    \right] 
    \right\} .
\end{equation}
\end{widetext}
We now have a problem similar to that in Sec.~\ref{sec:Boulware}, in that the ``up'' modes $\chi _{\omega \ell m}^{{\rm {up}}\pm  }$ and ``down'' modes $\chi _{\omega \ell m}^{{\rm {down}}\pm  }$ are not orthogonal.
The  ``up'' modes $\chi _{\omega \ell m}^{{\rm {up}}\pm  }$ are orthogonal to the ``in'' modes $\chi _{\omega \ell m}^{{\rm {in}}\pm  }$ (\ref{eq:chiin+-}), but the ``in'' modes $\chi _{\omega \ell m}^{{\rm {in}}\pm  }$ lead to thermal factors depending on the frequency $\omega $ (as in the state $|{\rm {CCH}}^{-} \rangle $) rather than ${\widetilde {\omega }}$ as in the modes $\chi _{\omega \ell m}^{{\rm {up}}\pm  }$ and $\chi _{\omega \ell m}^{{\rm {down}}\pm  }$.
The ``in'' modes $\chi _{\omega \ell m}^{{\rm {in}}\pm  }$ constructed in Sec.~\ref{sec:pastCCH} have positive frequency with respect to the Kruskal coordinate $V$ on the surface $U=-\epsilon <0$, part of which lies close to the future horizon ${\mathcal {H}}^{+}$ in region I of the space-time. 
Here we take an alternative approach and instead construct an alternative set of ``in'' modes ${\widetilde {\chi }}_{\omega \ell m}^{{\rm {in}}\pm  }$ which are positive frequency on the surface $U=\epsilon >0$.
Using the asymptotic forms (\ref{eq:inI-}, \ref{eq:outH}), and the result (\ref{eq:Uepsilon>0}), as in the construction of the ``future'' Unruh state $| {\rm {U}}^{-} \rangle $ we can apply the Lemma (\ref{eq:FNlemma}) with  $X=V$ and ${\mathfrak{q}}=\left(\omega + {\widetilde {\omega }} \right)/2\kappa $ to give, for ${\mathfrak {p}}>0$,
\begin{equation}
	\int _{-\infty }^{\infty } dV \, e^{-i{\mathfrak {p}}V} \left[ 
	e^{-\frac{\pi \left(\omega - {\widetilde {\omega }}\right)}{2\kappa }}\phi ^{\rm {in}}_{\omega \ell m } + 
	e^{-\frac{\pi \left(\omega + {\widetilde {\omega }}\right)}{2\kappa }} \psi ^{\rm {out}}_{\omega \ell m}\right] =0.
\end{equation}
From this (and a similar argument) we deduce that the modes 
\begin{subequations}
\begin{align}
{\widetilde {\chi }}_{\omega \ell m}^{{\rm {in}}+} = &~
\frac{1}{{\sqrt {2 \left| \sinh \left( \frac{\pi {\widetilde {\omega }} }{\kappa }\right)  \right| }}}\left( e^{\frac{\pi {\widetilde {\omega }}}{2\kappa }}\phi _{\omega \ell m}^{\rm {in}} + e^{-\frac{\pi {\widetilde {\omega }}}{2\kappa }} \psi _{\omega \ell m}^{{\rm {out}}} \right)  , 
 \\
{\widetilde {\chi }}_{\omega \ell m}^{{\rm {in}}-} = &~
\frac{1}{{\sqrt {2 \left| \sinh \left( \frac{\pi {\widetilde {\omega }} }{\kappa }\right)  \right| }}}\left( e^{-\frac{\pi {\widetilde {\omega }}}{2\kappa }}\phi _{\omega \ell m}^{\rm {in}} + e^{\frac{\pi {\widetilde {\omega }}}{2\kappa }} \psi _{\omega \ell m}^{{\rm {out}}} \right)  ,
\end{align}
\end{subequations}
(defined throughout regions I--IV) are positive and negative frequency with respect to the Kruskal coordinate $V$ on the surface $U=\epsilon >0$.
Furthermore, these new ``in'' modes ${\widetilde {\chi }}_{\omega \ell m}^{{\rm {in}}\pm}$ are orthogonal to the ``up'' modes ${{\chi }}_{\omega \ell m}^{{\rm {up}}\pm}$, as desired.
However, the positive frequency modes ${\widetilde {\chi }}_{\omega \ell m}^{{\rm {in}}+}$ have positive ``norm'' only if $\omega {\widetilde {\omega }}>0$ and have negative ``norm'' if $\omega {\widetilde {\omega }}<0$. Similarly, the negative frequency modes ${\widetilde {\chi }}_{\omega \ell m}^{{\rm {in}}-}$ have negative ``norm'' if $\omega {\widetilde {\omega }}>0$ but positive ``norm'' if $\omega {\widetilde {\omega }}<0$.
The situation is therefore similar to that encountered in Sec.~\ref{sec:Boulware}, in that modes ${\widetilde {\chi }}_{\omega \ell m}^{{\rm {in}}\pm}$ for which 
$\omega {\widetilde {\omega }}<0$ will be ``mislabelled'' according to their frequency, rather than their ``norm''.

We therefore write the ``down'' $\chi _{\omega \ell m}^{{\rm {down}}\pm  }$ modes in terms of the  new ``in'' modes ${\widetilde {\chi }}_{\omega \ell m}^{{\rm {in}}\pm  }$ and ``up'' modes $\chi _{\omega \ell m}^{{\rm {up}}\pm  }$. 
To do this, we first use the relationship (\ref{eq:inoutupdownmodes}) between the modes in region I, to give, in region IV, 
\begin{equation}
\psi ^{\rm {up}}_{\omega \ell m} = A^{{\rm {up}}*}_{\omega \ell m}\psi ^{\rm {down}}_{\omega \ell m}+ \left| \frac{\omega }{{\widetilde {\omega }}} \right| ^{\frac{1}{2}} B^{{\rm {up}}*} _{\omega \ell m} \psi ^{\rm {out}}_{\omega \ell m} ,
\end{equation}
for all $\omega $, ${\widetilde {\omega }}$,
where the modes $\psi ^{\rm {up}}_{\omega \ell m}$ are defined in Sec.~\ref{sec:futureU}, the modes $\psi ^{\rm {down}}_{\omega \ell m}$ in Sec.~\ref{sec:pastU} and the modes $\psi ^{\rm {out}}_{\omega \ell m}$ are defined in Sec.~\ref{sec:pastCCH}.
Therefore, using the definitions (\ref{eq:chiup+-}, \ref{eq:chiin+-}, \ref{eq:chidown+-}), we find
\begin{equation}
\chi _{\omega \ell m}^{{\rm {down}}\pm}
= A^{{\rm {up}}*}_{\omega \ell m}\chi _{\omega \ell m}^{{\rm {up}}\pm }
+ \left| \frac{\omega }{{\widetilde {\omega }}} \right|^{\frac{1}{2}} 
B^{{\rm {up}}*}_{\omega \ell m}
{\widetilde {\chi }}_{\omega \ell m}^{{\rm {in}}\pm } .
\end{equation}
We thus write the expansion of the classical field as
\begin{align}
\Phi 
= & ~\sum _{\ell =0}^{\infty } \sum _{m=-\ell }^{\ell }
\left\{ 
\int_{-\infty }^{\infty } d{\widetilde {\omega }} \,
\left[  
p^{\rm {up}}_{\omega \ell m}\chi ^{{\rm {up}}+}_{\omega \ell m}
+ s^{{\rm {up}}\dagger}_{\omega \ell m}\chi ^{{\rm {up}}-}_{\omega \ell m}
\right]
\right. \nonumber \\ & ~ \left.
+ \int _{-\infty }^{\infty } d\omega \, \left[ 
	p^{\rm {in}}_{\omega \ell m}{\widetilde {\chi }}^{{\rm {in}}+}_{\omega \ell m}
+ s^{{\rm {in}}\dagger}_{\omega \ell m}{\widetilde {\chi }}^{{\rm {in}}-}_{\omega \ell m}
\right] 
\right\} ,
\label{eq:newexpanH}
\end{align}
where we have rewritten the integral over the ``in'' modes in terms of $\omega $ rather than ${\widetilde{\omega }}$ and the coefficients in the expansion are
\begin{align}
p_{\omega \ell m}^{\rm {up}} =  & ~
{\widetilde {p}}_{\omega \ell m}^{\rm {up}} + A_{\omega \ell}^{{\rm {up}}*}{\widetilde {p}}_{\omega \ell m}^{\rm {down}},
\nonumber \\
s_{\omega \ell m}^{{\rm {up}}\dagger } = &~
{\widetilde {s}}_{\omega \ell m}^{{\rm {up}} \dagger} + A_{\omega \ell}^{{\rm {up}}*}{\widetilde {s}}_{\omega \ell m}^{{\rm {down}}\dagger },
\nonumber \\ 
p_{\omega \ell m}^{\rm {in}} = & ~\left| \frac{\omega }{{\widetilde {\omega }}} \right|^{\frac{1}{2}} 
 B_{\omega \ell}^{{\rm {up}}*} {\widetilde {p}}_{\omega \ell m}^{\rm {down}},
\nonumber \\  
s_{\omega \ell m}^{{\rm {in}} \dagger} = &~ \left| \frac{\omega }{{\widetilde {\omega }}} \right|^{\frac{1}{2}} 
 B_{\omega \ell}^{{\rm {up}}*} {\widetilde {s}}_{\omega \ell m}^{{\rm {down}}\dagger} .
\end{align}

In region I, the expansion (\ref{eq:newexpanH}) takes the form
\begin{widetext}
\begin{equation}
    \Phi 
    = \sum _{\ell =0}^{\infty } \sum _{m=-\ell }^{\ell }
    \int_{-\infty }^{\infty } d{\widetilde {\omega }} \,
    \frac{1}{{\sqrt {2 \left| \sinh \left( \frac{\pi {\widetilde{\omega }} }{\kappa }\right)  \right| }}}
    \left\{ 
    \phi _{\omega \ell m}^{\rm {up}} 
    \left[ 
    e^{\frac{\pi {\widetilde{\omega }}}{2\kappa }}
   p_{\omega \ell m}^{\rm {up}}
    + e^{-\frac{\pi {\widetilde{\omega }}}{2\kappa }}
   s_{\omega \ell m}^{{\rm {up}}\dagger }
    \right] 
    +
    \phi _{\omega \ell m}^{\rm {in}} 
    \left[ 
    e^{\frac{\pi {\widetilde{\omega }}}{2\kappa }}
   p_{\omega \ell m}^{\rm {in}}
    + e^{-\frac{\pi {\widetilde{\omega }}}{2\kappa }}
   s_{\omega \ell m}^{{\rm {in}}\dagger }
    \right] 
    \right\} ,
    \label{eq:inupHsum}
\end{equation}
\end{widetext}
which will lead to thermal factors depending on the frequency ${\widetilde {\omega}}$ for all modes.

As we now have an expansion (\ref{eq:newexpanH}) of the scalar field in terms of an orthonormal basis of field modes, we can promote the expansion coefficients $p$ and $s$ to operators. 
As in Sec.~\ref{sec:Boulware}, there is a subtlety due to the ``norm'' of the new ``in'' modes ${\widetilde {\chi }}_{\omega \ell m}^{{\rm {in}}\pm  }$.
The operators $ \hat{p}^{\mathrm{up}} $ and $ \hat{s}^{\mathrm{up}} $ satisfy the usual commutation relations (commutators not given explicitly vanish)
\begin{align}
    \left[ {\hat {p}}_{\omega \ell m}^{{\rm {up}}}, {\hat {p}}_{\omega ' \ell 'm '}^{{\rm {up}}\dagger} \right]
    = & ~ \delta _{\ell \ell' }\delta _{mm'}\delta (\omega - \omega ' ) ,
     \qquad {\mbox {all ${\widetilde {\omega }}$}},
    \nonumber \\
    \left[ {\hat {s}}_{\omega \ell m}^{{\rm {up}}}, {\hat {s}}_{\omega ' \ell 'm '}^{{\rm {up}}\dagger} \right]
    = & ~ \delta _{\ell \ell' }\delta _{mm'}\delta (\omega - \omega ' ) ,
    \qquad {\mbox {all ${\widetilde {\omega }}$}},
\end{align}
but the operators $ \hat{p}^{\mathrm{in}} $ and $ \hat{s}^{\mathrm{in}} $ satisfy modified commutation relations:
\begin{align}
    \left[ {\hat {p}}_{\omega \ell m}^{{\rm {in}}}, {\hat {p}}_{\omega ' \ell 'm '}^{{\rm {in}}\dagger} \right]
    = & ~ \eta _{\omega {\widetilde {\omega }}}\delta _{\ell \ell' }\delta _{mm'}\delta (\omega - \omega ' ) ,
    \qquad {\mbox {all ${\widetilde {\omega }}$}},
    \nonumber \\
    \left[ {\hat {s}}_{\omega \ell m}^{{\rm {in}}}, {\hat {s}}_{\omega ' \ell 'm '}^{{\rm {in}}\dagger} \right]
    = & ~ \eta _{\omega {\widetilde {\omega }}}\delta _{\ell \ell' }\delta _{mm'}\delta (\omega - \omega ' ) ,
    \qquad {\mbox {all ${\widetilde {\omega }}$}},
    \label{eq:Hccrs}
\end{align}
where $\eta _{\omega {\widetilde {\omega }}}$ is given in (\ref{eq:etadef1}) and all other commutators vanish.
When, in the next section, we calculate expectation values of observables in this state, we will need to take into account the modified commutation relations (\ref{eq:Hccrs}). 

Nonetheless, we proceed by defining a tentative ``Hartle-Hawking''-like state $|{\rm {H}}\rangle$ as that state annihilated by the ${\hat {p}}$ and ${\hat {s}}$ operators:
\begin{align}
	{\hat {p}}^{{\rm {in}}}_{\omega \ell m} | {\rm {H}} \rangle  = & ~0,
	\qquad {\mbox {all ${\widetilde {\omega }}$}},
	\nonumber \\ 
	{\hat {s}}^{{\rm {in}}}_{\omega \ell m} | {\rm {H}} \rangle = & ~0,
	\qquad {\mbox {all ${\widetilde {\omega }}$}},
	\nonumber \\
	{\hat {p}}^{{\rm {up}}}_{\omega \ell m} | {\rm {H}} \rangle  = & ~0,
	\qquad {\mbox {all ${\widetilde {\omega }}$}},
	\nonumber \\ 
	{\hat {s}}^{{\rm {up}}}_{\omega \ell m} | {\rm {H}} \rangle  = & ~0 ,
	\qquad {\mbox {all ${\widetilde {\omega }}$}}.
\end{align}
This state has no quanta in the $\chi _{\omega \ell m}^{{\rm {up}}}$ modes, as in both the ``past'' Unruh $|{\rm {U}}^{-}\rangle $ and CCH $| {\rm {CCH}}^{-} \rangle $ states, and hence a thermal distribution of particles/antiparticles in the ``up'' modes $\phi ^{\rm {up}}_{\omega \ell m}$.
However, the distribution of quanta in the ``in'' modes will be different from both the states $|{\rm {U}}^{-}\rangle $ and $| {\rm {CCH}}^{-} \rangle $. 

In defining the state $| {\rm {H}}\rangle $, we encountered ``mislabelled'' superradiant ``in'' modes. Suppose instead that we expand the classical field in terms of the $\chi ^{{\rm {up}}\pm }_{\omega \ell m}$ and ${\widetilde {\chi }}^{{\rm {in}}\pm }_{\omega \ell m}$ modes, but with the expansion coefficients denoted as annihilation/creation operators according to the ``norm'' of the modes.
The resulting expansion is
\begin{widetext}
\begin{align}
\Phi = & ~ \sum _{\ell =0}^{\infty } \sum _{m=-\ell }^{\ell }
\left\{ 
\int_{-\infty }^{\infty } d{\widetilde {\omega }} \,
\left[  
u^{\rm {up}}_{\omega \ell m}\chi ^{{\rm {up}}+}_{\omega \ell m}
+ v^{{\rm {up}}\dagger}_{\omega \ell m}\chi ^{{\rm {up}}-}_{\omega \ell m}
\right]
+ \int _{-\infty }^{\min \{\frac{qQ}{r_{+}},0\} } d\omega \, \left[ 
u^{\rm {in}}_{\omega \ell m}{\widetilde {\chi }}^{{\rm {in}}+}_{\omega \ell m}
+ v^{{\rm {in}}\dagger}_{\omega \ell m}{\widetilde {\chi }}^{{\rm {in}}-}_{\omega \ell m}
\right]
\right. \nonumber \\ & ~ \left.  
+ \int _{\min \{\frac{qQ}{r_{+}},0\} }^{\max \{\frac{qQ}{r_{+}},0\} } d\omega \, \left[ 
u^{\rm {in}}_{\omega \ell m}{\widetilde {\chi }}^{{\rm {in}}-}_{\omega \ell m}
+ v^{{\rm {in}}\dagger}_{\omega \ell m}{\widetilde {\chi }}^{{\rm {in}}+}_{\omega \ell m}
\right] 
+ \int _{\max \{\frac{qQ}{r_{+}},0\} }^{\infty } d\omega \, \left[ 
u^{\rm {in}}_{\omega \ell m}{\widetilde {\chi }}^{{\rm {in}}+}_{\omega \ell m}
+ v^{{\rm {in}}\dagger}_{\omega \ell m}{\widetilde {\chi }}^{{\rm {in}}-}_{\omega \ell m}
\right] 
\right\} ,
\label{eq:FTexp}
\end{align}
which reduces, in region I, to
\begin{align}
	\Phi = & ~ \sum _{\ell =0}^{\infty } \sum _{m=-\ell }^{\ell }
	\left\{ 
	\int_{-\infty }^{\infty } d{\widetilde {\omega }} \,
	\frac{1}{{\sqrt {2 \left| \sinh \left( \frac{\pi {\widetilde{\omega }} }{\kappa }\right)  \right| }}}
	\phi ^{\rm {up}}_{\omega \ell m}\left[  
	e^{\frac{\pi {\widetilde{\omega }}}{2\kappa }}u^{\rm {up}}_{\omega \ell m}
	+ e^{-\frac{\pi {\widetilde{\omega }}}{2\kappa }}v^{{\rm {up}}\dagger}_{\omega \ell m}
	\right]
	\right. \nonumber \\ & \left. 
	+ \int _{-\infty }^{\min \{\frac{qQ}{r_{+}},0\} } d\omega \, 
	\frac{1}{{\sqrt {2 \left| \sinh \left( \frac{\pi {\widetilde{\omega }} }{\kappa }\right)  \right| }}}
	\phi ^{\rm {in}}_{\omega \ell m} \left[ 
	e^{\frac{\pi {\widetilde{\omega }}}{2\kappa }}u^{\rm {in}}_{\omega \ell m}
	+ e^{-\frac{\pi {\widetilde{\omega }}}{2\kappa }}v^{{\rm {in}}\dagger}_{\omega \ell m}
	\right]
	\right. \nonumber \\ & ~ \left.  
	+ \int _{\min \{\frac{qQ}{r_{+}},0\} }^{\max \{\frac{qQ}{r_{+}},0\} } d\omega \, 
	\frac{1}{{\sqrt {2 \left| \sinh \left( \frac{\pi {\widetilde{\omega }} }{\kappa }\right)  \right| }}}
	\phi ^{\rm {in}}_{\omega \ell m} \left[ 
	e^{-\frac{\pi {\widetilde{\omega }}}{2\kappa }}u^{\rm {in}}_{\omega \ell m}
	+ e^{\frac{\pi {\widetilde{\omega }}}{2\kappa }}v^{{\rm {in}}\dagger}_{\omega \ell m}
	\right] 
	\right. \nonumber \\ & \left. 
	+ \int _{\max \{\frac{qQ}{r_{+}},0\} }^{\infty } d\omega \, 
	\frac{1}{{\sqrt {2 \left| \sinh \left( \frac{\pi {\widetilde{\omega }} }{\kappa }\right)  \right| }}}
	\phi ^{\rm {in}}_{\omega \ell m} \left[ 
	e^{\frac{\pi {\widetilde{\omega }}}{2\kappa }}u^{\rm {in}}_{\omega \ell m}
	+ e^{-\frac{\pi {\widetilde{\omega }}}{2\kappa }}v^{{\rm {in}}\dagger}_{\omega \ell m}
	\right] 
	\right\} .
	\label{eq:FTexpI}
\end{align}
\end{widetext}
When $\min \{\frac{qQ}{r_{+}},0\} < \omega <\max \{\frac{qQ}{r_{+}},0\} $, we have $\omega {\widetilde {\omega }}<0$ and hence we have relabelled the ${\widetilde {\chi }}^{{\rm {in}}+}_{\omega \ell m}$ modes as negative frequency (as they have negative ``norm'') and the ${\widetilde {\chi }}^{{\rm {in}}-}_{\omega \ell m}$ modes as positive frequency (since they have positive ``norm'').
Promoting the expansion coefficients to operators, they now satisfy the standard commutation relations
\begin{align}
	\left[ {\hat {u}}_{\omega \ell m}^{\rm {in}}, {\hat {u}}_{\omega '\ell 'm'}^{{\rm {in}}\dagger } \right] =  & ~\delta _{\ell \ell' }\delta _{mm'} \delta (\omega - \omega ') , \qquad {\mbox {all ${{\omega }}$}},
	\nonumber \\ 
	\left[ {\hat {v}}_{\omega \ell m}^{\rm {in}}, {\hat {v}}_{\omega '\ell 'm'}^{{\rm {in}}\dagger } \right] =  & ~\delta _{\ell \ell' }\delta _{mm'} \delta (\omega - \omega ') , \qquad {\mbox {all ${{\omega }}$}},
	\nonumber \\
	\left[ {\hat {u}}_{\omega \ell m}^{\rm {up}}, {\hat {u}}_{\omega '\ell 'm'}^{{\rm {up}}\dagger } \right] =  & ~\delta _{\ell \ell' }\delta _{mm'} \delta (\omega - \omega ') , \qquad {\mbox {all ${\widetilde {\omega }}$}},
	\nonumber \\ 
	\left[ {\hat {v}}_{\omega \ell m}^{\rm {up}}, {\hat {v}}_{\omega '\ell 'm'}^{{\rm {up}}\dagger } \right] =  & ~\delta _{\ell \ell' }\delta _{mm'} \delta (\omega - \omega ') , \qquad {\mbox {all ${\widetilde {\omega }}$}}.
\end{align} 
The expansion (\ref{eq:FTexp}) corresponds to the expansion used in \cite{Frolov:1989jh} to define a ``Hartle-Hawking''-like state for a neutral scalar field on a Kerr space-time. 
We therefore use the notation $|{\rm {FT}} \rangle $ (where FT stands for Frolov and Thorne) for the state annihilated by the ${\hat {u}}$ and ${\hat {v}}$ operators, as follows:
\begin{align}
	{\hat {u}}^{{\rm {in}}}_{\omega \ell m} | {\rm {FT}} \rangle  = & ~0 ,
	\qquad {\mbox {all $\omega $}},
	\nonumber \\ 
	{\hat {v}}^{{\rm {in}}}_{\omega \ell m} | {\rm {FT}} \rangle = & ~0,
	\qquad {\mbox {all $\omega $}},
	\nonumber \\
	{\hat {u}}^{{\rm {up}}}_{\omega \ell m} | {\rm {FT}} \rangle  = & ~0 ,
	\qquad {\mbox {all ${\widetilde {\omega }}$}},
	\nonumber \\ 
	{\hat {v}}^{{\rm {up}}}_{\omega \ell m} | {\rm {FT}} \rangle  = & ~0 ,
	\qquad {\mbox {all ${\widetilde {\omega }}$}}.
\end{align}
From the expansion (\ref{eq:FTexpI}), we anticipate that this state will have both the ``in'' and ``up'' modes thermally populated, with the frequency ${\widetilde {\omega}}$ in all thermal factors, although the superradiant ``in'' modes will require careful treatment.

\section{Expectation values of observables}
\label{sec:expvalues}

In the previous section we were able to define a plethora of quantum states for a charged scalar field on a charged black hole background. 
In this section we consider the expectation values of observables in these various quantum states.
The observables we study are the scalar field condensate, the current and the stress-energy tensor.

\subsection{Observables}
\label{sec:observables}

The simplest nontrivial observable for a charged scalar field is the scalar field condensate (or vacuum polarization) 
\begin{equation}
{\widehat {{{SC}}}} = \frac{1}{2}\left[ {\hat {\Phi}} {\hat {\Phi }}^{\dagger }+{\hat {\Phi}}^{\dagger} {\hat {\Phi }} \right] ,
\label{eq:CalC}
\end{equation}
which classically is simply the square of the magnitude of the scalar field:
\begin{equation}
SC = \left|\Phi \right|^{2}.
\end{equation}
Since we are considering static states on a static and spherically symmetric black hole, the expectation values $\langle {\widehat {{{SC}}}} \rangle $ will be functions of the radial coordinate $r$ only.

The next simplest nontrivial observable is the charged scalar field current $J^{\mu }$, which is given classically by
\begin{equation}
    J^{\mu } = -\frac{q}{4\pi } \Im \left[ 
    \Phi ^{*}D^{\mu }\Phi 
    \right] .
    \label{eq:current}
\end{equation}
We therefore define the current operator to be
\begin{align}
{\hat {{{J}}}}^{\mu } = & ~ \frac{iq}{16\pi } \left[ {\hat {\Phi }} ^{\dagger} \left( D^{\mu } {\hat {\Phi }} \right) + \left( D^{\mu } {\hat {\Phi }} \right){\hat {\Phi}}^{\dagger}
\right. \nonumber \\ &  ~ \left. 
 - {\hat {\Phi }} \left( D^{\mu} {\hat {\Phi  }} \right) ^{\dagger } - \left( D^{\mu} {\hat {\Phi  }} \right) ^{\dagger }{\hat {\Phi }} \right] .
\label{eq:CalJ}
\end{align}
The classical current $J^{\mu }$ is conserved, as are the expectation values of the current operator \cite{Balakumar:2019djw}:
\begin{equation}
\nabla _{\mu }\langle {\hat {{{J}}}}^{\mu } \rangle =0.
\end{equation}
Since we are considering only static states which do not evolve with time, the above equation governing the conservation of the current reduces to 
\begin{equation}
\partial _{r} \langle {\hat {{{J}}}}^{r} \rangle + \frac{2}{r} \langle {\hat {{{J}}}}^{r} \rangle =0.
\end{equation}
This is readily integrated to give, for any quantum state,
\begin{equation}
\langle {\hat {{{J}}}}^{r} \rangle = -\frac{{\mathcal{K}}}{r^{2}},
\label{eq:Jr}
\end{equation}
where ${\mathcal {K}}$ is a constant whose value depends on the quantum state under consideration.
Physically, ${\mathcal {K}}$ is the flux of charge emitted by the black hole in that particular quantum state.  
The black hole is losing charge if ${\mathcal {K}}$ has the same sign as the black hole charge $Q$. 
In App.~\ref{sec:nonren} we prove that this component of the current does not require renormalization. 
It is shown in \cite{Klein:2021les} that while the component $\langle {\hat {{{J}}}}^{t} \rangle$ requires renormalization, for a suitable choice of point-splitting on an RN-de Sitter black hole background, the renormalization counterterms are finite. While we expect that result to hold also for an RN black hole, given the remaining numerical difficulties in computing the renormalized expectation value of that component, we shall restrict our attention in this paper to differences in expectation values between two quantum states.

Later in this section we shall wish to study the regularity of our quantum states at the event horizon, for which it is helpful to have the nonzero components of the current in terms of Kruskal coordinates (\ref{eq:Kruskal}):
\begin{align}
	J^{U} =  & ~ \kappa U \left[ - J^{t} + f(r)^{-1} J^{r} \right] , 
	\nonumber \\
	J^{V} = & ~ \kappa V \left[ J^{t} + f(r)^{-1} J^{r} \right] .
	\label{eq:JKruskal} 
	\end{align} 

Our final observable is the stress-energy tensor $T_{\mu \nu }$, which has the following classical expression for a massless, minimally coupled, charged complex scalar field
\begin{equation}
T_{\mu \nu } =    \Re \left\{
 \left( D_{\mu }\Phi \right) ^{*} D_{\nu } \Phi  
-\frac{1}{2}
g_{\mu \nu } g^{\rho \sigma  } \left( D_{\rho }\Phi \right) ^{*} D_{\sigma }\Phi  \right\} ,
    \label{eq:stressenergy}
\end{equation}
and for which the corresponding quantum operator is thus
\begin{align}
	{\hat {{{T}}}}_{\mu \nu }= & ~
	\frac{1}{4} \left\{  \left( D_{\mu }{\hat {\Phi }} \right)^{\dagger} D_{\nu }{\hat {\Phi }} 
	+ D_{\nu }{\hat {\Phi }} \left( D_{\mu }{\hat {\Phi }} \right)^{\dagger}
	\right. \nonumber  \\ & ~ \left.
	+\left( D_{\nu }{\hat {\Phi }} \right)^{\dagger} D_{\mu }{\hat {\Phi }} 
	+ D_{\mu }{\hat {\Phi }} \left( D_{\nu }{\hat {\Phi }} \right)^{\dagger}
	\right. \nonumber  \\ & ~ \left.
-	\frac{1}{2}g_{\mu \nu }g^{\rho \sigma} \left[  \left(D_{\rho }{\hat {\Phi }}\right)^{\dagger}  D_{\sigma } {\hat {\Phi }}  + 
D_{\sigma } {\hat {\Phi }} \left(D_{\rho }{\hat {\Phi }}\right)^{\dagger} 
\right. \right. \nonumber \\ & ~ \left. \left. 
+\left(D_{\sigma }{\hat {\Phi }}\right)^{\dagger}  D_{\rho } {\hat {\Phi }}  + 
D_{\rho } {\hat {\Phi }} \left(D_{\sigma }{\hat {\Phi }}\right)^{\dagger}
\right]
	 \right\} .
	 \label{eq:CalT}
	\end{align}
The expectation value of the stress-energy tensor operator ${\hat {{T}}}_{\mu \nu }$ is not conserved \cite{Balakumar:2019djw}, due to the coupling between the scalar field and the electromagnetic field strength. 
Expectation values of the stress-energy tensor operator should instead satisfy \cite{Balakumar:2019djw}:
\begin{equation}
\nabla ^{\mu }\langle {\hat {{{T}}}}_{\mu \nu } \rangle = 4\pi F_{\mu \nu }
\langle {\hat {{{J}}}}^{\mu } \rangle ,
\label{eq:Tmunuconserve}
\end{equation}
where $F_{\mu \nu }= \nabla _{\mu }A_{\nu }- \nabla _{\nu }A_{\mu }$ is the background electromagnetic field strength and we are using Gaussian units. 
For static states on a spherically symmetric black hole, the stress-energy tensor expectation value will have the nonzero components 
\begin{equation}
\langle {\hat {{{T}}}}^{\mu }_{\nu } \rangle = \left(
\begin{array}{cccc}
{\mathcal {A}}(r)  & -{\mathcal {P}}(r)f(r)^{-1} & 0 & 0 \\
{\mathcal {P}}(r) f(r) & {\mathsf {T}}(r) - {\mathcal {A}}(r) - 2{\mathcal {Q}}(r) & 0 & 0 \\
0 & 0 & {\mathcal {Q}}(r) & 0 \\
0 & 0 & 0 & {\mathcal {Q}}(r)
\end{array}
\right) ,
\label{eq:Tmunuform}
\end{equation}
where $f(r)$ is the metric function (\ref{eq:fr}), ${\mathcal {A}}(r)$, ${\mathcal {P}}(r)$ and ${\mathcal {Q}}(r)$ are (presently unknown) functions of the radial coordinate $r$ only, and ${\mathsf{T}}(r)$ is the trace of the stress-energy tensor.
For a charged scalar field minimally coupled to the space-time curvature, the trace when the space-time background has vanishing Ricci scalar is given by \cite{Balakumar:2019djw}
\begin{align}
{\mathsf{T}} (r) = & ~ \frac{1}{2880\pi ^{2}}R_{\alpha \beta \gamma \delta }R^{\alpha \beta \gamma \delta } 
- \frac{1}{2880 \pi ^{2}} R_{\alpha \beta }R^{\alpha \beta }
\nonumber \\ & ~
- \frac{q^{2}}{192 \pi ^{2}} F^{\alpha \beta }F_{\alpha \beta }
-\frac{1}{2}\Box  \langle {\widehat {{{SC}}}} \rangle .
\label{eq:trace}
\end{align}
The final, state-dependent term arises because of the minimal coupling of the scalar field to the space-time geometry (it would be absent if the field were conformally coupled).
For the RN metric (\ref{eq:RNmetric}) and gauge potential (\ref{eq:gaugepot}), the trace has the expression
\begin{equation}
{\mathsf{T}} (r) = \frac{13Q^{2}-24MQ^{2}r+12M^{2}r^{2}}{720 \pi ^{2}r^{8}}
- \frac{q^{2}Q^{2}}{96 \pi ^{2}r^{4}} -\frac{1}{2}\Box  \langle {\widehat {{{SC}}}} \rangle .
\end{equation}
The $t$-component of the conservation equations (\ref{eq:Tmunuconserve}) for a stress-energy tensor expectation value having the form (\ref{eq:Tmunuform}) can be readily solved to give
\begin{equation}
\langle {\hat {{{T}}}}^{r}_{t} \rangle  = -\frac{{\mathcal {L}}}{r^{2}} +\frac{4\pi Q {\mathcal {K}}}{r^{3}},
\label{eq:Ttr} 
\end{equation}
where ${\mathcal {L}}$ is another constant depending on the particular quantum state under consideration.
Physically, ${\mathcal {L}}$ gives the flux of energy emitted by the black hole, and if ${\mathcal {L}}>0$ then the black hole is losing energy. 
In App.~\ref{sec:nonren} we prove that this component of the stress-energy tensor also does not require renormalization. 

The nonzero components of the stress-energy tensor in Kruskal coordinates (\ref{eq:Kruskal}) are
\begin{align}
	T_{UU} = & ~ \frac{1}{4} \kappa ^{-2}U^{-2} \left[ T_{tt} - 2f(r)T_{tr} + f(r)^{2}T_{rr} \right] ,
	\nonumber \\ 
	T_{UV} = & ~ - \frac{1}{4} \kappa ^{-2}U^{-1}V^{-1} \left[ T_{tt}  -f(r)^{2}T_{rr} \right] ,
	\nonumber \\ 
	T_{VV} = & ~ \frac{1}{4} \kappa ^{-2}V^{-2} \left[ T_{tt} + 2f(r)T_{tr} + f(r)^{2}T_{rr} \right]  ,
	\label{eq:TKruskal}		
\end{align}
with $T_{\theta \theta }$ and $T_{\varphi \varphi }$ unchanged by the transformation to Kruskal coordinates. 

All three quantum operators (\ref{eq:CalC}, \ref{eq:CalJ}, \ref{eq:CalT}) involve products of field operators at the same space-time point and are therefore divergent. 
One would ideally like to compute renormalized expectation values for the  states
constructed in Sec.~\ref{sec:quantum}.
However, while the general formalism for the Hadamard renormalization of these expectation values has been developed \cite{Balakumar:2019djw}, implementing this into a practical procedure for the computation of renormalized expectation values on black hole spacetimes is in its infancy (see the recent work \cite{Klein:2021ctt,Klein:2021les}).
Therefore in this paper we consider the differences in expectation values in two quantum states.
Since, for a Hadamard state, the divergent parts of the Feynman Green's function for the charged scalar field are independent of the quantum state of the field \cite{Balakumar:2019djw}, renormalization can be performed for one chosen quantum state.
Renormalized expectation values for any other quantum state can then be constructed using the differences we study here.

To write the expectation values for the various states we consider in a comparatively compact form, let ${\hat {{{O}}}}$ denote one of the quantum observables  (\ref{eq:CalC}, \ref{eq:CalJ}, \ref{eq:CalT}), corresponding to a classical quantity $O$, and let $o_{\omega \ell m}^{{\rm {in/up/out/down}}}$ be the classical value of $O$ calculated for the ``in'', ``up'', ``out'' or ``down'' modes (\ref{eq:inup}, \ref{eq:outdown}).  
For the scalar field condensate $SC$, we have simply
\begin{equation}
sc_{\omega \ell m} = \left| \phi _{\omega \ell m}\right| ^{2}.
\label{eq:scmode}
\end{equation}
In App.~\ref{sec:JT} we derive the nonzero components of the current and stress-energy tensor. These simplify if we sum over the azimuthal quantum number $m$.
The nonzero components of the current are then:
\begin{widetext}
 	\begin{subequations}
 	\begin{align}
 		j^{t}_{\omega \ell} = \sum _{m=-\ell }^{\ell }j^{t}_{\omega \ell m} = & ~ 
 		-\frac{q\left(2\ell + 1\right)}{16\pi ^{2} r^{2} f(r)}\left|{\mathcal {N}}_{\omega }\right| ^{2} \left|X_{\omega \ell }(r) \right |^{2}\left( \omega - \frac{qQ}{r} \right)  ,
 		\\
 		j^{r}_{\omega \ell} = \sum _{m=-\ell }^{\ell }j^{r}_{\omega \ell m} = & ~
 		-\frac{qf(r)\left(2 \ell + 1 \right)}{16\pi  ^{2}} \left| {\mathcal {N}}_{\omega } \right| ^{2} \Im \left[ \frac{X^{*}_{\omega \ell }(r)}{r} \frac{d}{dr} \left( \frac{X_{\omega \ell }(r)}{r} \right)  
 		\right] ,
 	\end{align}
\end{subequations}
 and for the stress-energy tensor the nonzero components are 
\begin{subequations}
	\begin{align}
		t_{tt, \omega \ell} = & ~
		\frac{2\ell + 1}{8\pi }\left| {\mathcal {N}}_{\omega }\right|^{2}
		\left\{ 
		\left[ \frac{1}{r^{2}} \left(  \omega - \frac{qQ}{r} \right) ^{2} + \frac{\ell \left(\ell + 1 \right) f(r)}{r^{4}}  \right]  \left| X_{\omega \ell }(r) \right|^{2}
		+ f(r)^{2}
		\left| \frac{d}{dr} \left( \frac{X_{\omega \ell } (r)}{r}\right) \right| ^{2} 
		\right\} ,
		\\
		t_{tr, \omega \ell  } = & ~
		- \frac{2\ell + 1}{4\pi }\left( \omega - \frac{qQ}{r}\right)  \left| {\mathcal {N}}_{\omega } \right| ^{2}  \Im \left[ 
		\frac {X_{\omega \ell }^{*}(r)}{r} \frac{d}{dr}\left( \frac{X_{\omega \ell}(r)}{r} \right) 
		\right] ,
		\\
		t_{rr, \omega \ell } = & ~
		\frac{2\ell + 1}{8\pi }\left| {\mathcal {N}}_{\omega }\right|^{2}
		\left\{ 
		\left[ \frac{1}{f(r)^{2}r^{2}} \left(  \omega - \frac{qQ}{r} \right) ^{2} - \frac{\ell \left(\ell + 1 \right) }{r^{4}f(r)}  \right]  \left| X_{\omega \ell }(r) \right|^{2}
		+ 
		\left| \frac{d}{dr} \left( \frac{X_{\omega \ell } (r)}{r}\right) \right| ^{2} 
		\right\} ,
		\\
		t_{\theta \theta , \omega \ell } = & ~
		\frac{2\ell + 1}{8\pi }\left| {\mathcal {N}}_{\omega }\right|^{2}
		\left\{ 
		\frac{1}{f(r)} \left(  \omega - \frac{qQ}{r} \right) ^{2} \left| X_{\omega \ell }(r) \right|^{2}
		- f(r)r^{2} 
		\left| \frac{d}{dr} \left( \frac{X_{\omega \ell } (r)}{r}\right) \right| ^{2} 
		\right\} ,
		\\
		t_{\varphi \varphi ,\omega \ell } = & ~ t_{\theta \theta , \omega \ell }\sin ^{2} \theta  ,
	\end{align}
\end{subequations}
where we have defined
\begin{equation}
	t_{\mu \nu ,\omega \ell } = \sum _{m=-\ell}^{\ell } t_{\mu \nu , \omega \ell m},
\end{equation}
and the symbol $\Im $ denotes the imaginary part of a complex quantity.

The (unrenormalized) expectation values of ${\hat {{O}}}$ in each of the ``past'' quantum states defined in Sec.~\ref{sec:past} can be written as sums over the ``in'' and ``up'' modes:
\begin{subequations}
	\label{eq:expvaluesstates}
\begin{align}
    \langle {\rm {B}}^{-} |{\hat {{O}}} | {\rm {B}}^{-} \rangle = & ~
    \frac{1}{2}\sum _{\ell =0}^{\infty }\sum _{m=-\ell}^{\ell }\left[
    \int _{-\infty }^{\infty }d\omega \, o _{\omega \ell m}^{{\rm {in}}}  
     + \int _{-\infty }^{\infty }d{\widetilde{\omega }} \,
    o _{\omega \ell m}^{{\rm {up}}} 
    \right], 
\\
\langle {\rm {U}}^{-} | {\hat {{O}}} | {\rm {U}}^{-} \rangle = & ~
    \frac{1}{2}\sum _{\ell =0}^{\infty }\sum _{m=-\ell}^{\ell }\left[
    \int _{-\infty }^{\infty }d\omega \, o _{\omega \ell m}^{\rm {in}}  + \int _{-\infty }^{\infty }d{\widetilde{\omega }} \,
     o _{\omega \ell m}^{\rm {up}} 
    \coth \left| \frac{\pi {\widetilde{\omega }}}{\kappa } \right| 
    \right],
    \\
\langle {\rm {CCH}}^{-}| {\hat {{O}}} |  {\rm {CCH}}^{-} \rangle = & ~
    \frac{1}{2}\sum _{\ell =0}^{\infty }\sum _{m=-\ell}^{\ell }\left[
    \int _{-\infty }^{\infty }d\omega \, 
     o _{\omega \ell m}^{\rm {in}} 
    \coth \left| \frac{\pi {\omega }}{\kappa } \right| 
     + \int _{-\infty }^{\infty }d{\widetilde{\omega }} \,
     o _{\omega \ell m}^{\rm {up}} 
    \coth \left| \frac{\pi {\widetilde{\omega }}}{\kappa } \right| 
    \right],   
 \end{align}
while those for the ``future'' quantum states defined in Sec.~\ref{sec:future} have corresponding expressions in terms of the ``out'' and ``down'' modes:
\begin{align}
        \langle {\rm {B}}^{+} | {\hat {{O}}} |  {\rm {B}}^{+} \rangle = & ~
   \frac{1}{2} \sum _{\ell =0}^{\infty }\sum _{m=-\ell}^{\ell }\left[
    \int _{-\infty }^{\infty }d\omega \, o _{\omega \ell m}^{\rm {out}}  + \int _{-\infty }^{\infty }d{\widetilde{\omega }} \,
    o _{\omega \ell m}^{\rm {down}} 
    \right], 
\\
\langle {\rm {U}}^{+} | {\hat {{O}}} |  {\rm {U}}^{+} \rangle = & ~
    \frac{1}{2}\sum _{\ell =0 }^{\infty }\sum _{m=-\ell}^{\ell }\left[
    \int _{-\infty }^{\infty }d\omega \, o _{\omega \ell m}^{\rm {out}}  + \int _{-\infty }^{\infty }d{\widetilde{\omega }} \,
  o _{\omega \ell m}^{\rm {down}} 
    \coth \left| \frac{\pi {\widetilde{\omega }}}{\kappa } \right|
    \right],
    \\
\langle {\rm {CCH}}^{+} | {\hat {{O}}} |  {\rm {CCH}}^{+} \rangle = & ~
    \frac{1}{2}\sum _{\ell =0}^{\infty }\sum _{m=-\ell}^{\ell }\left[
    \int _{-\infty }^{\infty }d\omega \, 
    o _{\omega \ell m}^{\rm {out}} 
    \coth \left| \frac{\pi {\omega }}{\kappa } \right| 
     + \int _{-\infty }^{\infty }d{\widetilde{\omega }} \,
     o _{\omega \ell m}^{\rm {down}} 
    \coth \left| \frac{\pi {\widetilde{\omega }}}{\kappa } \right| 
    \right] .   
\end{align}
For our proposed ``Boulware''-like $| {\rm {B}}\rangle $ and ``Hartle-Hawking''-like ($| {\rm {H}}\rangle $  and $| {\rm {FT}}\rangle $) states, we have
\begin{align}
   \langle {\rm {B}} | {\hat {{O}}} |  {\rm {B}}\rangle = & ~
     \frac{1}{2} \sum _{\ell =0}^{\infty }\sum _{m=-\ell}^{\ell }
    \left\{ 
    \int _{-\infty }^{\infty } d\omega  \left[ 
     o _{\omega \ell m}^{\rm {in}}  +
    o _{\omega \ell m}^{\rm {up}} 
    \right] 
    - 2\int _{\min \{ 0, \frac{qQ}{r_{+}}\} } ^{\max \{ 0, \frac{qQ}{r_{+}} \} }
    d\omega  \,
    o _{\omega \ell m}^{\rm {up}}    \right\} ,
    \label{eq:Bexp}
    \\
    \langle {\rm {H}} |{\hat {{O}}}|  {\rm {H}} \rangle = & ~
     \frac{1}{2} \sum _{\ell =0}^{\infty }\sum _{m=-\ell}^{\ell }
    \left\{ \int _{-\infty }^{\infty }d{\widetilde{\omega }}  \,
    \left[
    o _{\omega \ell m}^{\rm {in}} +
    o _{\omega \ell m}^{\rm {up}} \right]
    \coth \left| \frac{\pi {\widetilde{\omega }}}{\kappa } \right| 
    - 2 \int _{\min \{ 0, \frac{qQ}{r_{+}}\} } ^{\max \{ 0, \frac{qQ}{r_{+}} \} }
    d\omega  \,
    o _{\omega \ell m}^{\rm {in}}  \coth \left| \frac{\pi {\widetilde{\omega }}}{\kappa } \right| 
     \right\} ,
    \label{eq:Hexp}
    \\ 
    \langle {\rm {FT}} |{\hat {{O}}}|  {\rm {FT}} \rangle = & ~
    \frac{1}{2} \sum _{\ell =0}^{\infty }\sum _{m=-\ell}^{\ell }
    \left\{ \int _{-\infty }^{\infty }d{\widetilde{\omega }}  \,
    \left[
    o _{\omega \ell m}^{\rm {in}} +
    o _{\omega \ell m}^{\rm {up}} \right]
    \coth \left| \frac{\pi {\widetilde{\omega }}}{\kappa } \right| 
    \right\} .
    \label{eq:FTexp2}
    \end{align}
\end{subequations}
\end{widetext}
The expectation values of the candidate ``Boulware''-like state $| {\rm {B}}\rangle $ (\ref{eq:Bexp}) and ``Hartle-Hawking''-like states $| {\rm {H}} \rangle $ (\ref{eq:Hexp}) take into account the fact that, in defining those states, we have ``mislabelled'' some modes according to their frequency rather than their ``norm''.
This leads to contributions to the expectation values from the superradiant modes which have the opposite sign to those for the nonsuperradiant modes.

In the following subsections, we examine differences in expectation values between two of the above states listed in  (\ref{eq:expvaluesstates}). 
We consider reference states which simplify these differences in an asymptotic region either close to the horizon or at infinity.
Useful expressions for the corresponding differences in expectation values can then be derived in the relevant asymptotic regions, and these will aid the physical interpretation of the states. 
However, in order to study the differences in expectation values everywhere outside the event horizon and not just in the asymptotic regions,  numerical computation is required. 

We find the ``in'' and ``up'' modes by numerically integrating the radial equation (\ref{eq:radial}).
For the ``in'' modes, we use the boundary conditions (\ref{eq:in}) for $r$ close to the horizon and integrate the radial equation (\ref{eq:radial}) outwards to find $X_{\omega \ell }^{\rm {in}}(r)/B^{\rm {in}}_{\omega \ell}$ since the constants $B^{\rm {in}}_{\omega \ell}$ are not known {\it {a priori}}.
The constants $A^{\rm {in}}_{\omega \ell}$ and $B^{\rm {in}}_{\omega \ell }$ are then determined by comparing the numerical solution with the boundary conditions (\ref{eq:in}) at infinity.
A similar process is used for the ``up'' modes, starting with the boundary conditions (\ref{eq:up}) for large $r$ and integrating the radial equation (\ref{eq:radial}) inwards to find $X_{\omega \ell }^{\rm {up}}/B_{\omega \ell }^{\rm {up}}$, then matching with the boundary conditions near the horizon.

For all the differences in expectation values that we consider, either the integrals over frequency converge very rapidly due to an exponential factor in the denominator or else they are taken over a finite interval of values of the frequency. The sum over $m$ is also straightforward. It remains then to find the sum over $\ell $.
We find that this is dominated by the low-$\ell$ modes and that summing over modes with values of $\ell $ up to $40$ gives results for the differences in expectation values which are accurate to at least three significant figures.

\begin{widetext}

\subsection{``Past'' and ``future'' quantum states}
\label{sec:pastfuture}

\subsubsection{``Past'' Unruh state}
\label{sec:pastUexp}

We begin by examining the ``past'' quantum states defined in Sec.~\ref{sec:past}.
Considering the difference in expectation values between the ``past'' Unruh $|{\rm {U}}^{-}\rangle $ and ``past'' Boulware $|{\rm {B}}^{-}\rangle$ states, we find 
\begin{equation}
\langle {\rm {U}}^{-} | {\hat {{O}}} | {\rm {U}}^{-} \rangle  
- \langle {\rm {B}}^{-} | {\hat {{O}}} | {\rm {B}}^{-} \rangle
=
\sum _{\ell =0}^{\infty }\sum _{m=-\ell}^{\ell }
\int _{-\infty }^{\infty }d{\widetilde{\omega }} \,
\frac{1}{\exp \left| \frac{2\pi {\widetilde{\omega }}}{\kappa } \right| -1 } 
o^{\rm {up}}_{\omega \ell m}. 
\label{eq:UminusBminusgen}
\end{equation}
As $r\rightarrow \infty $, the ``up'' modes take a particularly simple form (\ref{eq:up}) and we find the following leading order behaviour of the expectation values:
\begin{subequations}
	\label{eq:UminusBminus}
\begin{align}
\langle {\rm {U}}^{-} | {\widehat {{SC}}} | {\rm {U}}^{-} \rangle  
- \langle {\rm {B}}^{-} | {\widehat {{SC}}} | {\rm {B}}^{-} \rangle
\sim & \frac{1}{16 \pi  ^{2} r^{2}}  
\sum _{\ell =0}^{\infty }
\int _{-\infty }^{\infty }d{\widetilde{\omega }} \, \frac{2\ell + 1}{\left| {\widetilde {\omega }}\right| \left( \exp \left| \frac{2\pi {\widetilde{\omega }}}{\kappa } \right| -1 \right) }  \left| B^{\rm {up}}_{\omega \ell} \right| ^{2} ,
\\ 
\langle {\rm {U}}^{-} | {\hat {{J}}}^{\mu } | {\rm {U}}^{-} \rangle  
- \langle {\rm {B}}^{-} | {\hat {{J}}}^{\mu } | {\rm {B}}^{-} \rangle
\sim & - \frac{q}{64 \pi^3 r^2} \sum_{\ell = 0}^\infty  \int_{- \infty}^{\infty} d \widetilde{\omega} \frac{\omega \left( 2\ell + 1 \right) }{\left| \widetilde{\omega} \right| \left( \exp{ \left| \frac{2 \pi \widetilde{\omega}}{\kappa} \right|} - 1 \right)} {\left| B^{\mathrm{up}}_{\omega \ell} \right|}^2  {\left( 1, 1 , 0, 0 \right)}^\intercal ,
\\
\langle {\rm {U}}^{-} | {\hat {{T}}}^{\mu }_{\nu } | {\rm {U}}^{-} \rangle  
- \langle {\rm {B}}^{-} | {\hat {{T}}}^{\mu }_{\nu } | {\rm {B}}^{-} \rangle
\sim & \frac{1}{16 \pi ^{2} r^2} \sum_{\ell = 0}^\infty \int_{- \infty}^{\infty} d \widetilde{\omega} \frac{\omega ^{2} \left( 2\ell + 1 \right) }{\left| \widetilde{\omega} \right| \left( \exp{\left| \frac{2 \pi \widetilde{\omega}}{\kappa} \right|} - 1 \right)} {\left| B^{\mathrm{up}}_{\omega \ell} \right|}^2  \begin{pmatrix}
   - 1 & 1 & 0 & 0 \\
   - 1 & 1 & 0 & 0 \\
   0 & 0 & \mathcal{O} (r^{-2}) & 0 \\
   0 & 0 & 0 & \mathcal{O} (r^{-2})
  \end{pmatrix} .
\end{align}
\end{subequations}
By virtue of the Wronskian relations (\ref{eq:wronskians}), the integrands in (\ref{eq:UminusBminus}) are regular when ${\widetilde {\omega }}=0$.
As seen by a static observer far from the black hole, the ``past'' Unruh state $|{\rm {U}}^{-}\rangle $ contains a flux of particles at infinity relative to the ``past'' Boulware state $|{\rm {B}}^{-}\rangle $.
The ``past'' Boulware state is defined to be as empty as possible at past null infinity ${\mathscr{I}}^{-}$, and contains an outgoing flux of radiation in the superradiant modes \cite{Gibbons:1975kk,Balakumar:2020gli}, given by the following expectation values, as $r\rightarrow \infty $:
	\begin{subequations}
	\label{eq:expvaluesBminus}
	\begin{align}
	\langle {\rm {B}}^{-}| {\hat {{{J}}}}^{r}|{\rm {B}}^{-} \rangle & \sim  
	-\frac{q}{64\pi ^{3}r^{2}}\sum _{\ell =0}^{\infty } 
	\int _{\min \{\frac{qQ}{r_{+}},0\}}^{\max \{ \frac{qQ}{r^{+}},0\}} d\omega 
	\frac{\omega }{\left| {\widetilde {\omega }} \right| }\left( 2\ell + 1 \right) \left| B^{\rm {up}}_{\omega \ell } \right| ^{2}  
	,
	\label{eq:JrBminus}
	\\
	\langle {\rm {B}}^{-}|{\hat {{{T}}}}_{t}^{r}|{\rm {B}}^{-} \rangle & \sim  -\frac{1}{16 \pi ^{2} r^{2}} \sum _{\ell =0}^{\infty } 
	\int _{\min \{\frac{qQ}{r_{+}},0\}}^{\max \{ \frac{qQ}{r^{+}},0\}} d\omega 
	\frac{\omega ^{2}}{\left| {\widetilde {\omega }} \right| }\left( 2\ell + 1 \right) \left| B^{\rm {up}}_{\omega \ell } \right| ^{2}  ,
	\label{eq:TtrBminus}
	\end{align}
\end{subequations}
and the resulting fluxes of charge and energy are \cite{Balakumar:2020gli}:
\begin{subequations}
	\label{eq:fluxesBminus}
	\begin{align}
		{\mathcal {K}}_{{\mathrm {B}}^{-}} & =  
		\frac{q}{64\pi ^{3}}\sum _{\ell =0}^{\infty } 
		\int _{\min \{\frac{qQ}{r_{+}},0\}}^{\max \{ \frac{qQ}{r^{+}},0\}} d\omega 
		\frac{\omega }{\left| {\widetilde {\omega }} \right| }\left( 2\ell + 1 \right) \left| B^{\rm {up}}_{\omega \ell } \right| ^{2}  
		,
		\label{eq:chargeBminus}
		\\
			{\mathcal {L}}_{{\mathrm {B}}^{-}} & =    \frac{1}{16 \pi ^{2}} \sum _{\ell =0}^{\infty } 
		\int _{\min \{\frac{qQ}{r_{+}},0\}}^{\max \{ \frac{qQ}{r^{+}},0\}} d\omega 
		\frac{\omega ^{2}}{\left| {\widetilde {\omega }} \right| }\left( 2\ell + 1 \right) \left| B^{\rm {up}}_{\omega \ell } \right| ^{2}  .
		\label{eq:energyBminus}
		\end{align}
	\end{subequations}
As observed in \cite{Balakumar:2020gli}, in the ``past'' Boulware state the flux of charge ${\mathcal {K}}_{{\mathrm {B}}^{-}}$ always has the same sign as the black hole charge $Q$, so that in this state the black hole is losing charge by the emission of particles in the superradiant modes. 
Similarly, the flux of energy ${\mathcal {L}}_{{\mathrm {B}}^{-}}$ is always positive, so the black hole is losing energy.
  
\begin{figure}[p]
	\includegraphics[scale=0.63]{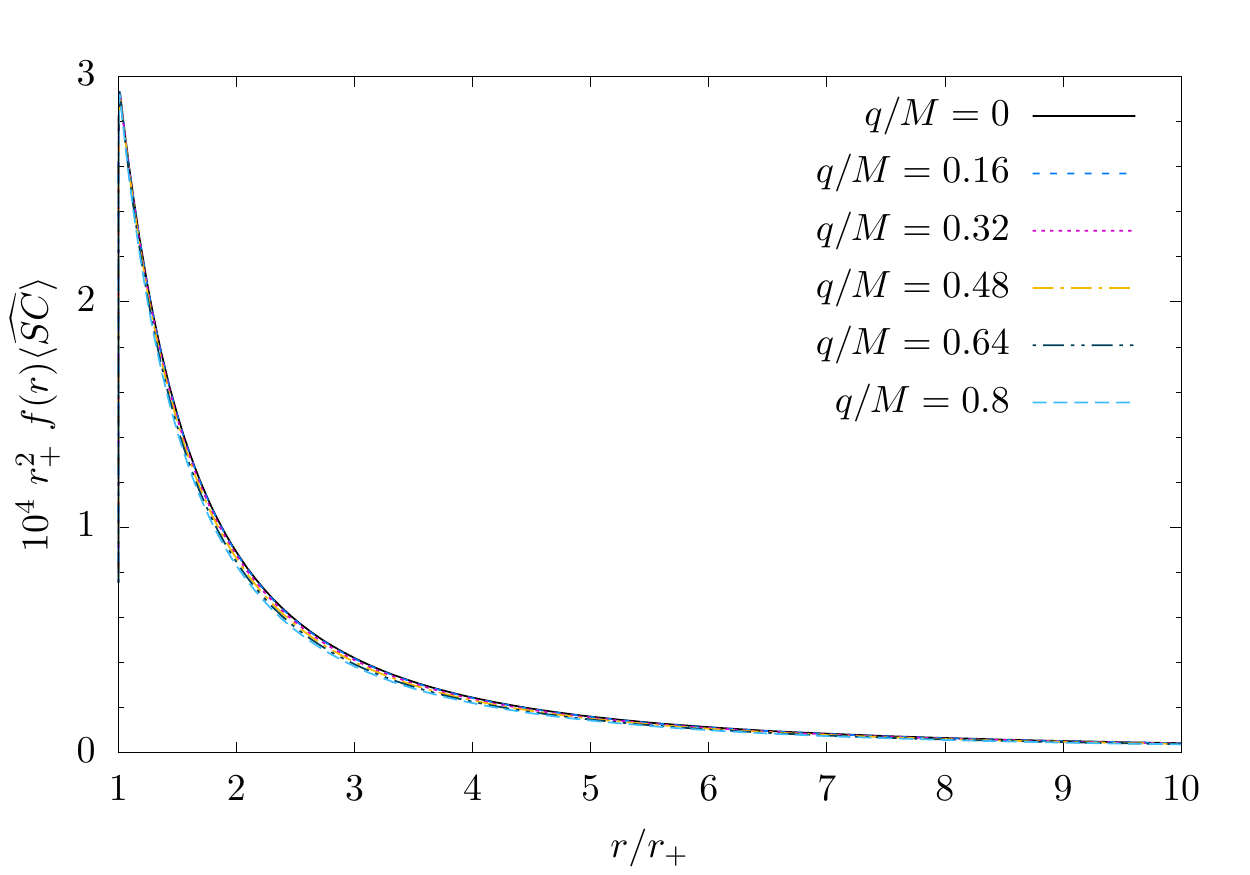}\\ 
	\includegraphics[scale=0.61]{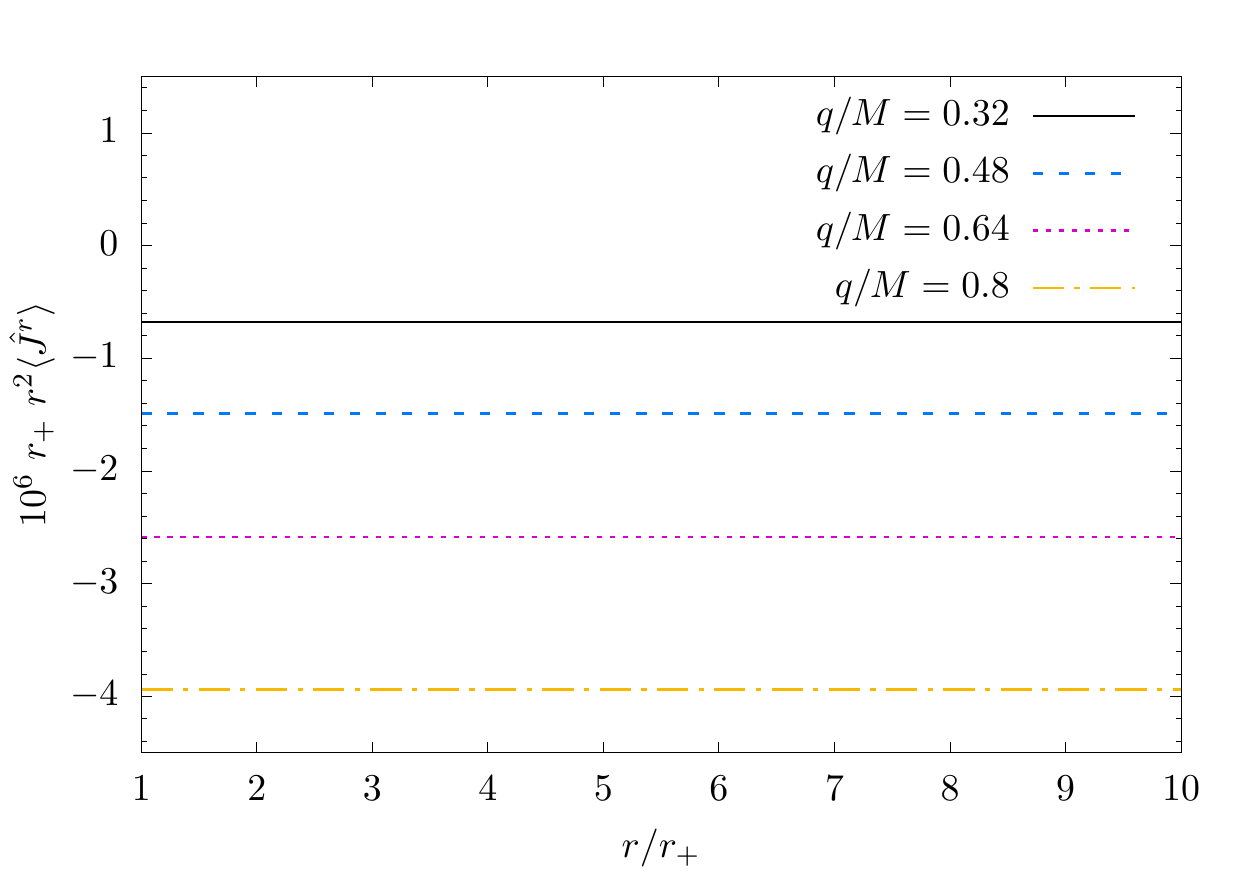}
	\includegraphics[scale=0.61]{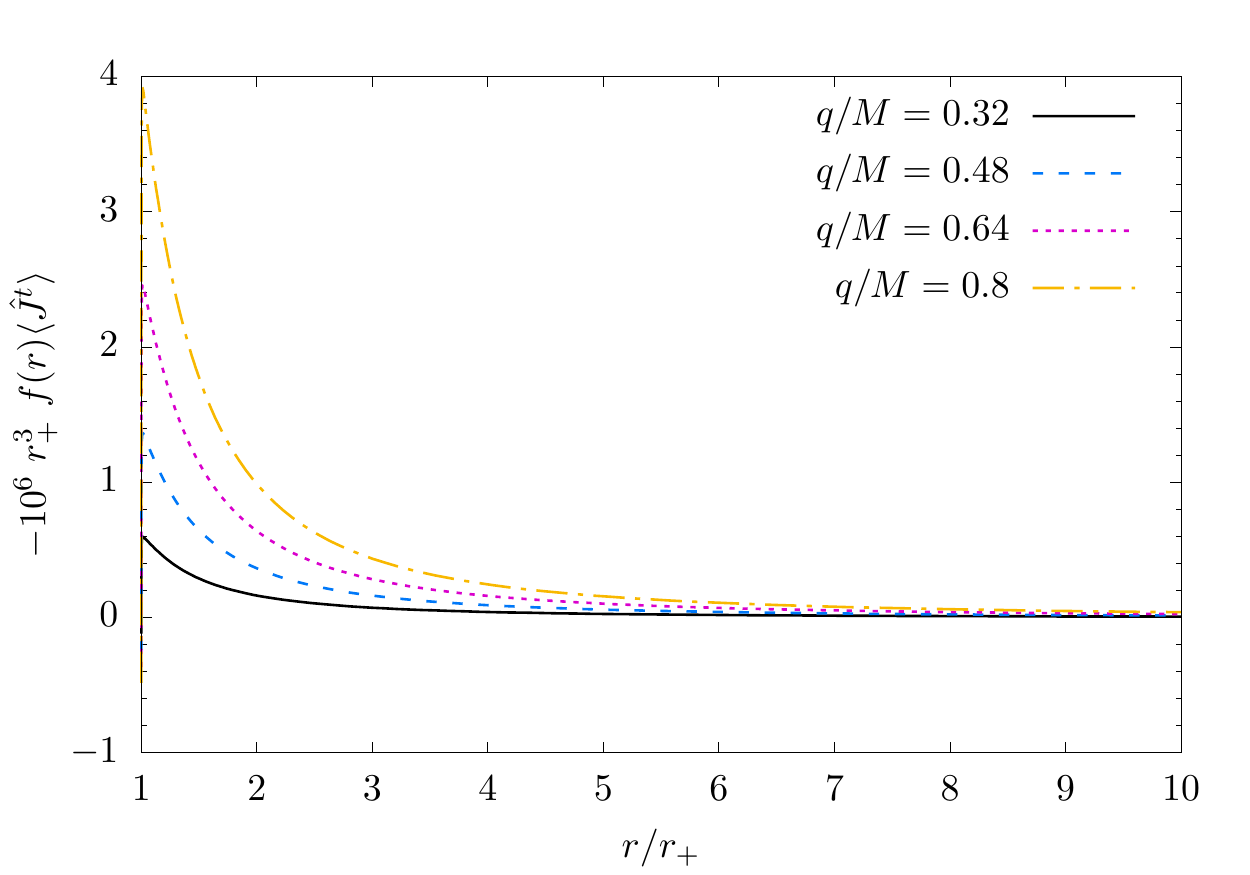}
	\\ \includegraphics[scale=0.61]{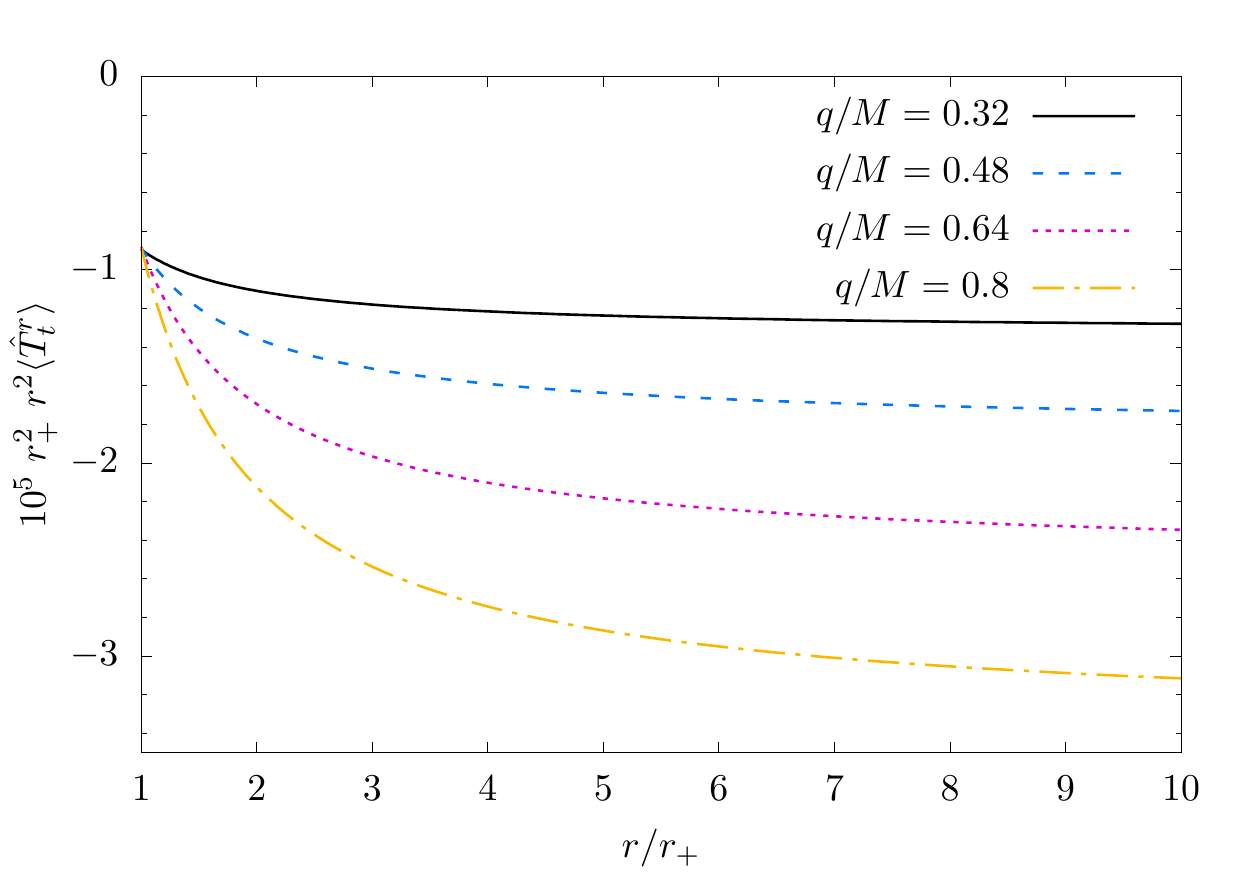}
	\includegraphics[scale=0.61]{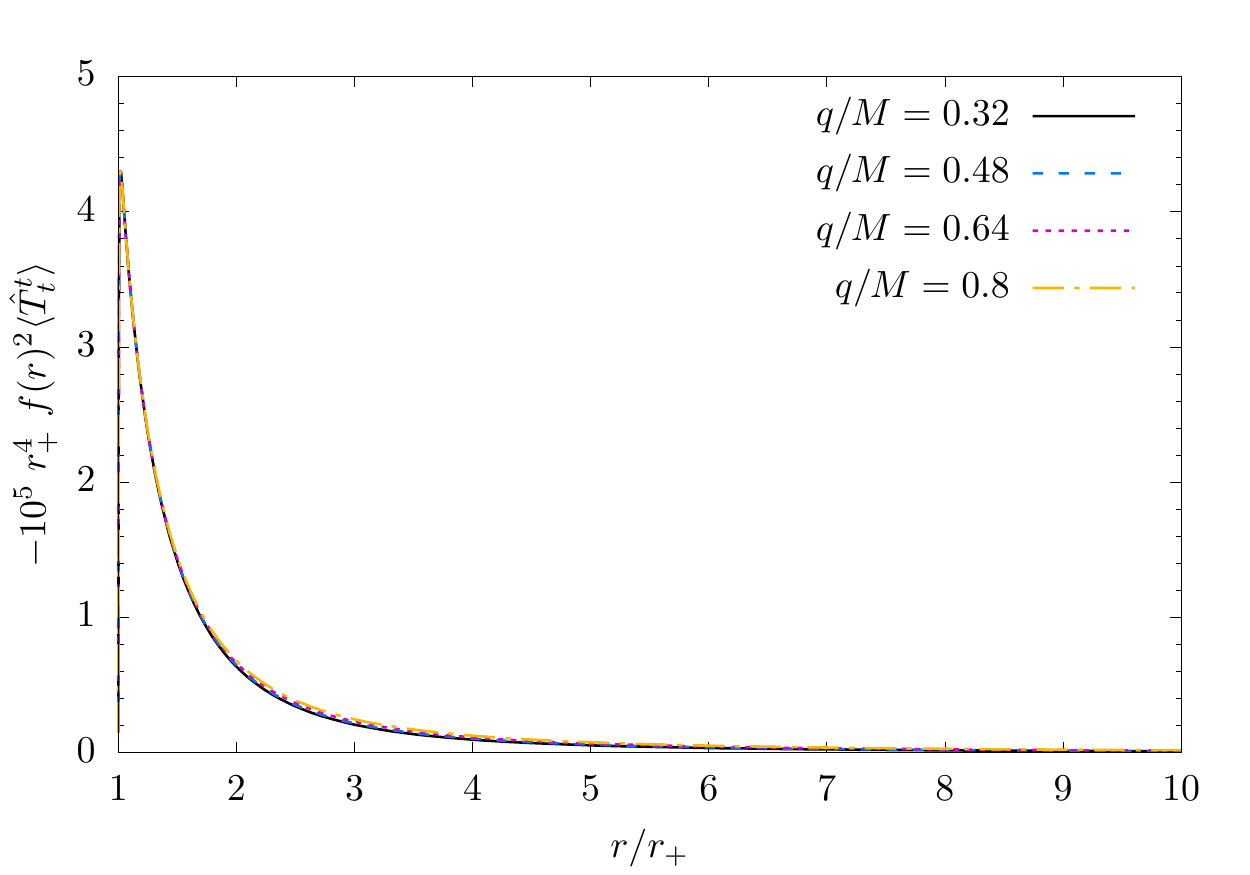} \\
	\includegraphics[scale=0.61]{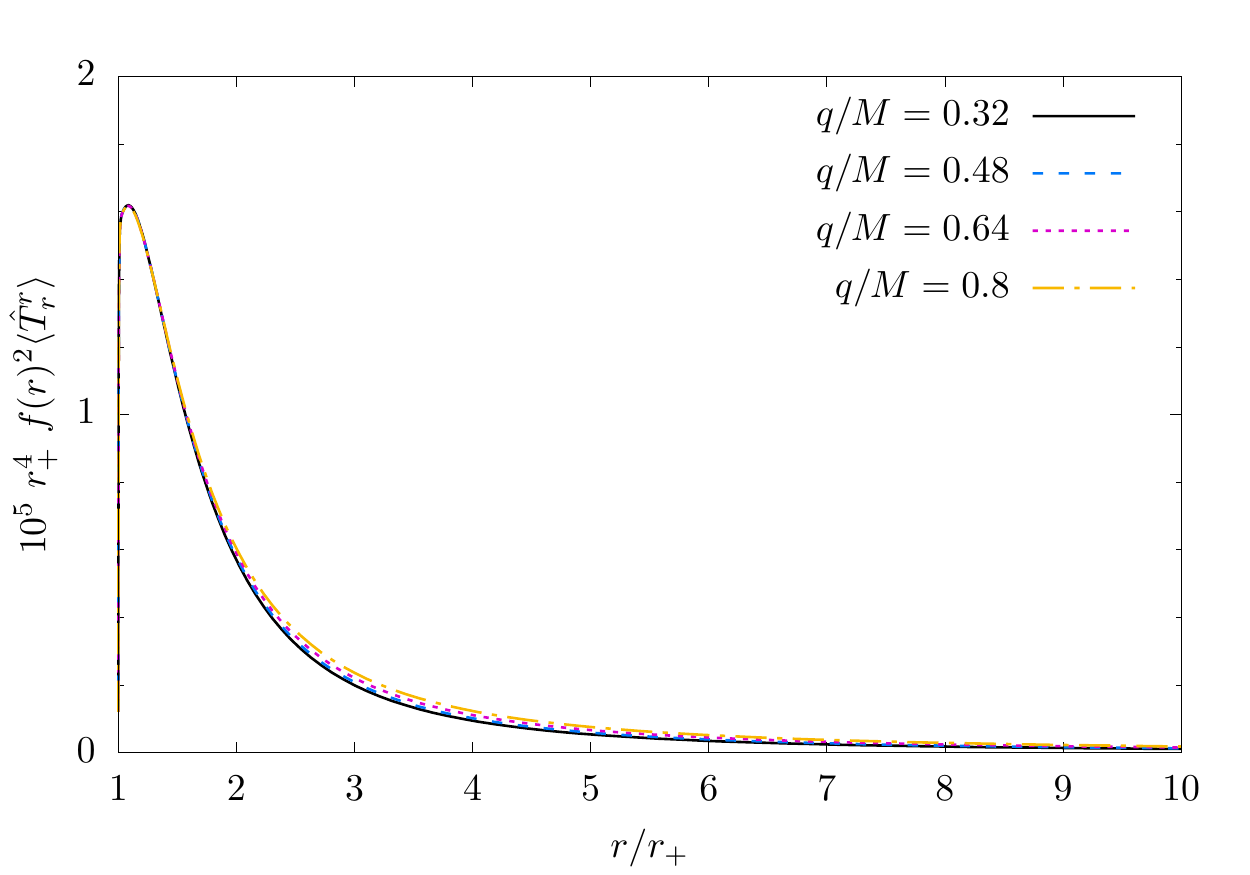}
	\includegraphics[scale=0.61]{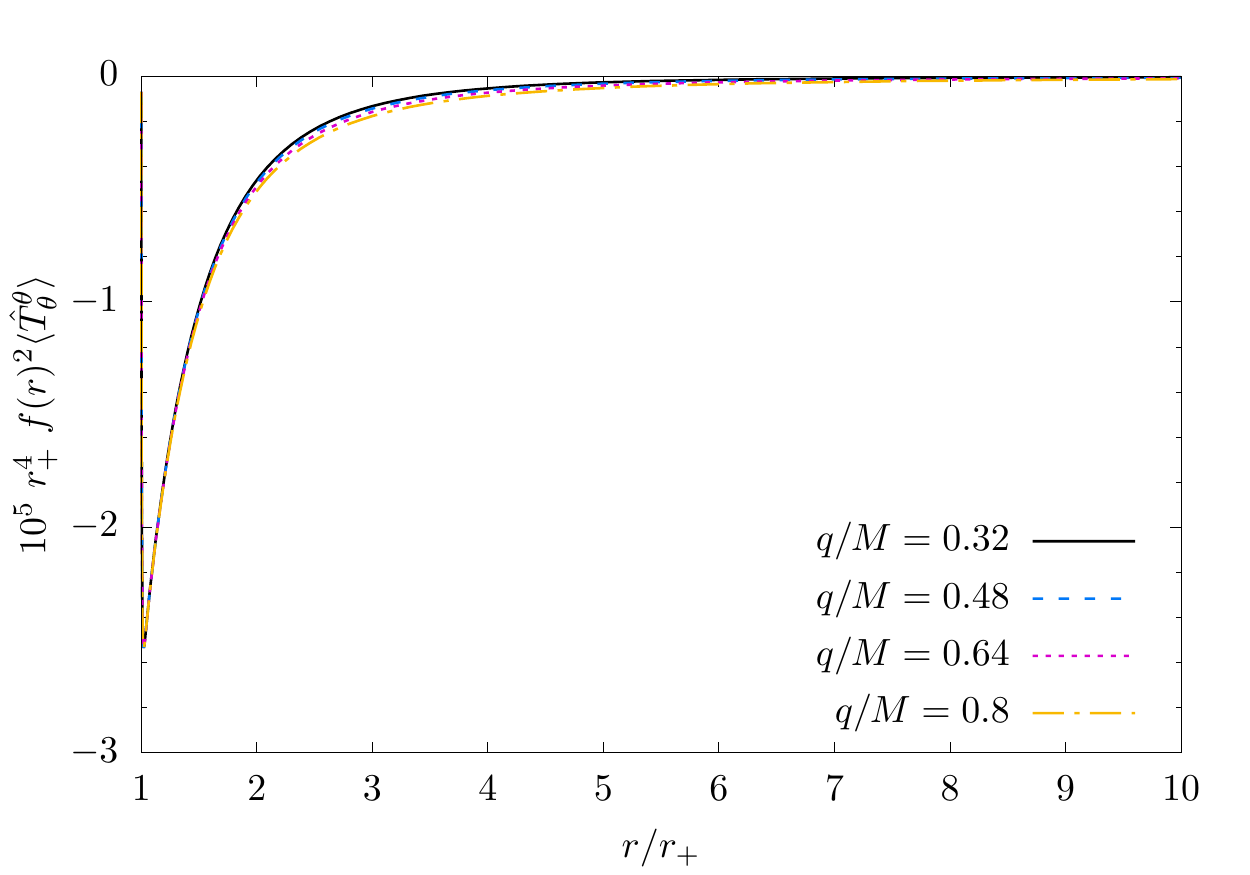} \\
	\vspace{.1cm}
	\caption{Difference in expectation values for the scalar condensate operator and components of the current and stress-energy tensor operators, between the ``past'' Unruh, $| {\rm{U}}^{-} \rangle$, and ``past'' Boulware state, $|{\rm{B}}^{-} \rangle$, in the spacetime of a RN black hole with $Q=0.8M$. All expectation values are multiplied by powers of $f(r)$ so that the resulting quantities are regular at $r=r_{+}$.}
	\label{fig:Uminus-Bminus}
\end{figure}

Since ${\hat {J}}^{r}$ and ${\hat {T}}^{r}_{t}$ do not require renormalization (see App.~\ref{sec:nonren}), adding the relevant components in  (\ref{eq:UminusBminus}, \ref{eq:expvaluesBminus}), we find
\begin{subequations}
	\label{eq:expvaluesUminus}
	\begin{align}
	\langle {\rm {U}}^{-}| {\hat {{{J}}}}^{r}|{\rm {U}}^{-} \rangle & \sim - \frac{q}{64 \pi^{3} r^2} \sum_{\ell = 0}^\infty  \int_0^\infty d \omega 
	\left(2 \ell + 1 \right)\omega 
	\left[ \frac{ {\left| {B}^\mathrm{up}_{\omega \ell} \right|}^2}{{\widetilde {\omega }}\left( \exp{ \left[ \frac{2\pi {\widetilde {\omega }}}{\kappa } \right]} - 1 \right)} - \frac{ {\left| {B}^\mathrm{up}_{-\omega \ell} \right|}^2}{{\overline {\omega }}\left( \exp{ \left[ \frac{2\pi {\overline {\omega }}}{\kappa }  \right]} - 1 \right)} \right]  ,
	\label{eq:JrUminus}
	\\
	\langle {\rm {U}}^{-}|{\hat {{{T}}}}_{t}^{r}|{\rm {U}}^{-} \rangle & \sim - \frac{1}{16 \pi^2 r^2}\sum_{\ell = 0}^\infty \int_0^\infty d \omega 
	\left(2 \ell + 1 \right)\omega ^{2}
	\left[ \frac{ {\left| {B}^\mathrm{up}_{\omega \ell} \right|}^2}{{\widetilde {\omega }}\left( \exp{ \left[ \frac{2\pi {\widetilde {\omega }}}{\kappa } \right]} - 1 \right)} + \frac{ {\left| {B}^\mathrm{up}_{-\omega \ell} \right|}^2}{{\overline {\omega }}\left( \exp{ \left[ \frac{2\pi {\overline {\omega }}}{\kappa }  \right]} - 1 \right)} \right] ,
	\label{eq:TtrUminus}
	\end{align}
	\end{subequations}
where ${\widetilde {\omega }}$ is given by (\ref{eq:tomega}) and ${\overline {\omega }}$ is
\begin{equation}
	{\overline {\omega }} = \omega + \frac{qQ}{r_{+}} .
	\label{eq:baromega}
\end{equation}
The flux of charge resulting from (\ref{eq:JrUminus}) agrees with that in Ref.~\cite{Gibbons:1975kk}:
\begin{subequations}
	\label{eq:fluxesUminus}
\begin{equation}
	{\mathcal {K}}_{{\mathrm {U}}^{-}} =  \frac{q}{64 \pi^{3}} \sum_{\ell = 0}^\infty  \int_0^\infty d \omega 
	\left(2 \ell + 1 \right)\omega 
	\left[ \frac{ {\left| {B}^\mathrm{up}_{\omega \ell} \right|}^2}{{\widetilde {\omega }}\left( \exp{ \left[ \frac{2\pi {\widetilde {\omega }}}{\kappa } \right]} - 1 \right)} - \frac{ {\left| {B}^\mathrm{up}_{-\omega \ell} \right|}^2}{{\overline {\omega }}\left( \exp{ \left[ \frac{2\pi {\overline {\omega }}}{\kappa }  \right]} - 1 \right)} \right]  ,
	\label{eq:chargeUminus}
\end{equation}
while the flux of energy in the ``past'' Unruh state is
\begin{equation}
	{\mathcal {L}}_{{\mathrm {U}}^{-}} =  \frac{1}{16 \pi^{2}} \sum_{\ell = 0}^\infty  \int_0^\infty d \omega 
	\left(2 \ell + 1 \right)\omega  ^{2}
	\left[ \frac{ {\left| {B}^\mathrm{up}_{\omega \ell} \right|}^2}{{\widetilde {\omega }}\left( \exp{ \left[ \frac{2\pi {\widetilde {\omega }}}{\kappa } \right]} - 1 \right)} + \frac{ {\left|{B}^\mathrm{up}_{-\omega \ell} \right|}^2}{{\overline {\omega }}\left( \exp{ \left[ \frac{2\pi {\overline {\omega }}}{\kappa }  \right]} - 1 \right)} \right]  .
	\label{eq:energyUminus}
\end{equation}
\end{subequations}
The integrals in (\ref{eq:fluxesUminus}) are taken over positive frequencies $\omega >0$. The first term in each case is the contribution of modes which have positive frequency $\omega $ as seen by a static observer far from the black hole, while the modes in the second term have negative frequency as seen by the same observer.
We see that there is thermal emission of particles with an effective chemical potential $qQ/r_{+}$ \cite{Hawking:1974sw,Gibbons:1975kk}. The chemical potential has the opposite sign for negative frequency particles as compared to positive frequency particles. 

Consider first the flux of charge (\ref{eq:chargeUminus}). 
Here the emission of positive frequency modes gives a contribution to ${\mathcal {K}}_{{\mathrm {U}}^{-}}$ which has the same sign as the scalar field charge $q$, while the emission of negative frequency modes gives a contribution to ${\mathcal {K}}_{{\mathrm {U}}^{-}}$ having the opposite sign to $q$.
On the other hand, both the positive and negative frequency modes give a positive contribution to the energy flux ${\mathcal {L}}_{{\mathrm {U}}^{-}}$ (\ref{eq:energyUminus}), so that the black hole is losing energy due to the emission of Hawking radiation.
As $\kappa \rightarrow 0$, the temperature vanishes and the fluxes  (\ref{eq:fluxesUminus}) reduce to the superradiant flux obtained in the ``past'' Boulware state (\ref{eq:fluxesBminus}).

We find that the expectation values (\ref{eq:UminusBminus}) diverge as $r\rightarrow r_{+}$ and the event horizon is approached. 
We anticipate that this is due to the divergence of the ``past'' Boulware state $|{\mathrm{B}}^{-}\rangle$ on the horizon, although a computation of renormalized expectation values would be required to confirm this conjecture (see \cite{Klein:2021ctt,Klein:2021les} for recent work for the Unruh state on an RN-de Sitter black hole).
It is expected that the ``past'' Unruh state $|{\mathrm{U}}^{-}\rangle$, in analogy with the Unruh state on a Schwarzschild black hole, is regular at the future horizon ${\mathcal {H}}^{+}$ but not the past horizon ${\mathcal {H}}^{-}$.

In Fig.~\ref{fig:Uminus-Bminus} we plot the differences in expectation values for the scalar condensate and the components of the current and stress-energy tensor between the ``past'' Unruh $| {\mathrm {U}}^{-} \rangle$ and Boulware $|{\mathrm{B}}^{-} \rangle$ states.
The charge of the black hole, $Q$, is fixed, and a selection of values of the scalar field charge $q$ are considered.   
All expectation values have been multiplied by an appropriate power of the metric function $f(r)$ (\ref{eq:fr}) to give quantities which are finite and nonzero on the horizon.

We see that the difference in expectation values of the scalar field condensate does not vary much with the scalar field charge $q$. Furthermore, it is positive, indicating that the  expectation value of the scalar field condensate in the ``past'' Unruh state $| {\rm {U}}^{-} \rangle$ is greater than in the ``past'' Boulware state  $| {\rm {B}}^{-} \rangle$, at least for a scalar field whose charge has the same sign as the black hole charge. 
Near the horizon, the scalar field condensate diverges like $f(r)^{-1}$, which we suspect is due to the ``past'' Boulware state $| {\rm {B}}^{-} \rangle$ rather than the ``past'' Unruh state $| {\rm {U}}^{-} \rangle$.

As expected from (\ref{eq:Jr}), the difference in expectation values of the radial component of the current is proportional to $-r^{-2}$, with the constant of proportionality ${\mathcal{K}}$ equal to zero when $q=0$ and increasing as the scalar field charge increases. 
Since ${\mathcal {K}}$ is positive, the black hole is losing charge as expected.
Therefore, while the charge flux in the ``past'' Boulware state $| {\rm {B}}^{-} \rangle$ increases as the scalar field charge increases \cite{Balakumar:2020gli}, the charge flux in the ``past'' Unruh state $| {\rm {U}}^{-} \rangle$ increases more rapidly with increasing scalar field charge.
The ``past'' Boulware state contains an outgoing flux of particles in the superradiant modes only \cite{Balakumar:2020gli}, while the ``past'' Unruh state contains a thermal distribution of particles emitted due to Hawking radiation.
We deduce that the loss of charge due to Hawking radiation increases more rapidly with scalar field charge than the loss of charge due to quantum superradiance. 

The magnitude of the difference in expectation values of the time component of the current also vanishes when $q=0$ and increases  significantly as the scalar field charge increases (similar behaviour is found for a massless, conformally coupled scalar field on an RN-de Sitter black hole in \cite{Klein:2021les}).
This also diverges like $f(r)^{-1}$ as the horizon is approached. 
As with the scalar condensate, we find that the components of the difference in expectation values of the current in Kruskal coordinates (\ref{eq:JKruskal}) diverge on the event horizon.  

The  differences in expectation values of the diagonal SET components generally do not change much as the scalar field charge increases and exhibit similar behaviour to the scalar field condensate.  All diverge like $f(r)^{-2}$ as the horizon is approached. 
The difference in energy density between the two states considered here is positive, with the energy density in the ``past'' Unruh state being greater than that in the ``past'' Boulware state. 
In contrast, the difference in the energy flux increases rapidly as the scalar field charge increases.  
The component $\langle {\hat {T}}_{t}^{r}\rangle $ is negative, indicating that it is dominated by the flux of energy ${\mathcal {L}}$ rather than the flux of charge ${\mathcal {K}}$.    
All differences in expectation values between the ``past'' Unruh and ``past'' Boulware states tend to zero like $r^{-2}$ far from the black hole. 
We anticipate that the expectation values in the ``past'' Boulware state $| {\rm {B}}^{-} \rangle$ will vanish at infinity, since this state is empty at infinity apart from the  outgoing flux of particles in the superradiant modes. Therefore we  conjecture that renormalized expectation values in the ``past'' Unruh state $| {\rm {U}}^{-} \rangle$ will also tend to zero far from the black hole. 

\subsubsection{``Past'' CCH state} 
\label{sec:pastCCHexp}

To examine the properties of the ``past'' CCH state, it is convenient to consider the differences
\begin{equation}
\langle {\rm {CCH}}^{-} | {\hat {{O}}} | {\rm {CCH}}^{-} \rangle  
- \langle {\rm {U}}^{-} | {\hat {{O}}} | {\rm {U}}^{-} \rangle
=
\sum _{\ell =0}^{\infty }\sum _{m=-\ell}^{\ell }
\int _{-\infty }^{\infty }d\omega  \,
\frac{1}{\exp \left| \frac{2\pi \omega }{\kappa } \right| -1 } 
o^{\rm {in}}_{\omega \ell m}. 
\end{equation}
As $r_* \rightarrow -\infty $, and $r\rightarrow r_{+}$, the ``in'' modes take a particularly simple form (\ref{eq:in}) and we find the following leading order behaviour of the expectation values:
\begin{subequations}
	\label{eq:CCHminusUminus}
	\begin{align}
		\langle {\rm {CCH}}^{-} | {\widehat {{SC}}} | {\rm {CCH}}^{-} \rangle  
		- \langle {\rm {U}}^{-} | {\widehat {{SC}}} | {\rm {U}}^{-} \rangle
		\sim & \frac{1}{16 \pi ^{2} r^{2}}  
		\sum _{\ell =0}^{\infty }
		\int _{-\infty }^{\infty }d\omega \, \frac{ 2\ell + 1 }{\left| \omega \right| \left( \exp \left| \frac{2\pi \omega }{\kappa } \right| -1 \right) }  \left| B^{\rm {in}}_{\omega \ell} \right| ^{2} ,
		\\ 
		\langle {\rm {CCH}}^{-} | {\hat {{J}}}^{\mu } | {\rm {CCH}}^{-} \rangle  
		- \langle {\rm {U}}^{-} | {\hat {{J}}}^{\mu } | {\rm {U}}^{-} \rangle
		\sim & \frac{q}{64 \pi^3 r^2} \sum_{\ell = 0}^\infty  \int_{- \infty}^{\infty} d \omega \frac{{\widetilde{\omega }}\left( 2\ell + 1 \right) }{\left| \omega \right| \left( \exp{\left| \frac{2 \pi \omega}{\kappa} \right|} - 1 \right)} {\left| B^{\mathrm{in}}_{\omega \ell} \right|}^2  {\left( - f( r)^{-1}, 1, 0, 0 \right)}^\intercal ,
		\\
		\langle {\rm {CCH}}^{-} | {\hat {{T}}}^{\mu }_{\nu } | {\rm {CCH}}^{-} \rangle  
		- \langle {\rm {U}}^{-} | {\hat {{T}}}^{\mu }_{\nu } | {\rm {U}}^{-} \rangle
		\sim & \frac{1}{16 \pi ^{2} r^2} \sum_{\ell = 0}^\infty  \int_{- \infty}^{\infty} d \omega \frac{{\widetilde{\omega}}^{2}\left( 2\ell + 1 \right) }{\left| \omega \right| \left( \exp{\left| \frac{2 \pi \omega}{\kappa} \right|} - 1 \right)} {\left| B^{\mathrm{in}}_{\omega \ell} \right|}^2 
		\nonumber \\ &  \qquad \qquad \times
  \begin{pmatrix}
   - f( r )^{-1} & - f( r)^{-2} & 0 & 0 \\
   1 & f(r)^{-1} & 0 & 0 \\
   0 & 0 & \mathcal{O} \left( 1 \right) & 0 \\
   0 & 0 & 0 & \mathcal{O} \left( 1 \right)
  \end{pmatrix} .
	\end{align}
\end{subequations}
From the Wronskian relations (\ref{eq:wronskians}), the integrands in (\ref{eq:CCHminusUminus}) are regular when $\omega =0$.
Transforming to Kruskal coordinates $U$, $V$ (\ref{eq:Kruskal}) and using (\ref{eq:JKruskal}, \ref{eq:TKruskal}), we find that the leading order divergences in the expectation values (\ref{eq:CCHminusUminus}) cancel on the future horizon ${\mathcal {H}}^{+}$. 
We anticipate that the ``past'' Unruh state is regular on ${\mathcal {H}}^{+}$, so this implies that the expectation value of the current in the ``past'' CCH state is also regular on the future horizon ${\mathcal {H}}^{+}$. 
However, we are not able at this stage to make a similar deduction about the expectation value of the SET in the ``past'' CCH state.  The asymptotic results (\ref{eq:CCHminusUminus}) only indicate that any divergence in the SET at the future horizon ${\mathcal {H}}^{+}$ is no more severe than $f(r)^{-1}$. We will explore this question in more detail below.

Using (\ref{eq:fluxesUminus}) and the relevant components of (\ref{eq:CCHminusUminus}), we can find the fluxes of charge and energy in the ``past'' CCH state as integrals over positive frequency modes with $\omega >0$ (where we have used the Wronskian relations (\ref{eq:wronskians})):
\begin{subequations}
	\label{eq:fluxesCCHminus}
	\begin{align}
	{\mathcal {K}}_{{\mathrm {CCH}}^{-}} & = {\mathcal {K}}_{{\mathrm {U}}^{-}}
	-\frac{q}{64 \pi^{3}} \sum_{\ell = 0}^\infty  \int_0^\infty d \omega 
	\left(2 \ell + 1 \right) 
		\frac{\omega }{\left( \exp \left[ \frac{2\pi \omega }{\kappa }  \right] -1\right) }
	\left[
\frac{1}{{\widetilde {\omega }}}\left| B_{\omega \ell }^{\rm {up}}\right| ^{2} 
	- \frac {1}{{\overline {\omega }}}\left| B_{-\omega \ell }^{\rm {up}}\right| ^{2}
	 \right]
	,
	\label{eq:chargeCCHminus}
	\\
	{\mathcal {L}}_{{\mathrm {CCH}}^{-}} & = {\mathcal {L}}_{{\mathrm {U}}^{-}}
	-\frac{1}{16 \pi^{2}} \sum_{\ell = 0}^\infty  \int_0^\infty d \omega 
	\left(2 \ell + 1 \right)
		\frac{\omega ^{2}}{\left( \exp \left[ \frac{2\pi \omega }{\kappa }  \right] -1\right) }
	\left[
\frac{1}{{\widetilde {\omega }}}\left| B_{\omega \ell }^{\rm {up}}\right| ^{2} 
+ \frac {1}{{\overline {\omega }}}\left| B_{-\omega \ell }^{\rm {up}}\right| ^{2}
\right] .
	\label{eq:energyCCHminus}	
			\end{align}
	\end{subequations}
It can be seen that the difference in fluxes between the ``past'' CCH and ``past'' Unruh states consists of a thermal spectrum of particles, but without a chemical potential. 
The nonsuperradiant modes reduce the flux of energy in the ``past'' CCH state compared to the ``past'' Unruh state, while the flux of energy in the superradiant modes (with either ${\widetilde {\omega }}<0$ or ${\overline {\omega }}<0$) is enhanced.
The flux of charge in the ``past'' CCH state compared with that in the ``past'' Unruh state has a complex form. Nonsuperradiant positive frequency modes give a charge flux having the opposite sign to the scalar field charge $q$, and superradiant positive frequency modes give  contribution to the charge flux which has the same sign as $q$. The opposite is true for modes with negative frequency $\omega$. 

We deduce that, while the ``past'' CCH state has attractive regularity properties, it does not represent an equilibrium state, since it has nonzero fluxes of charge and energy. 
This is to be expected since the ``in'' and ``up'' modes are thermalized with different thermal factors. 

\begin{figure}[p]
  \includegraphics[scale=0.63]{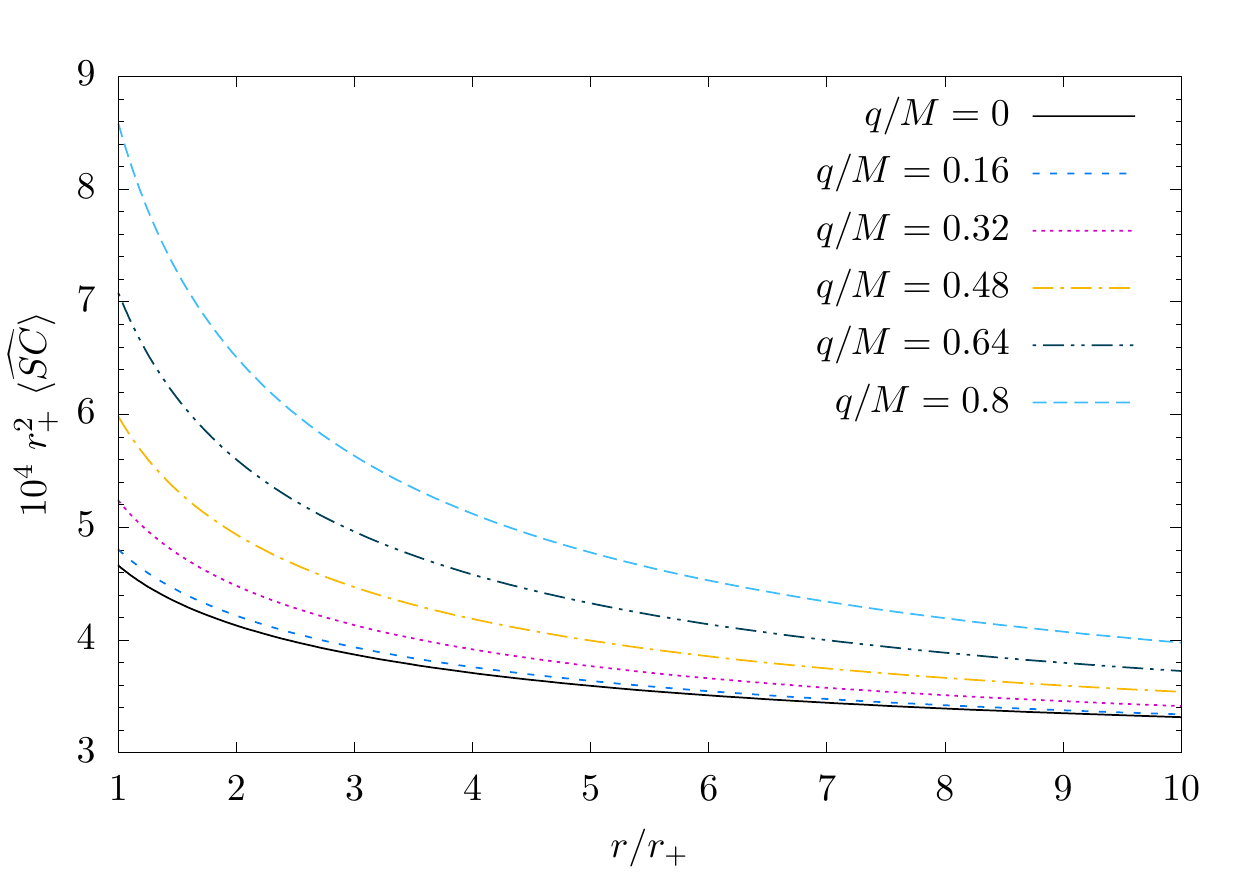}\\ 
  \includegraphics[scale=0.61]{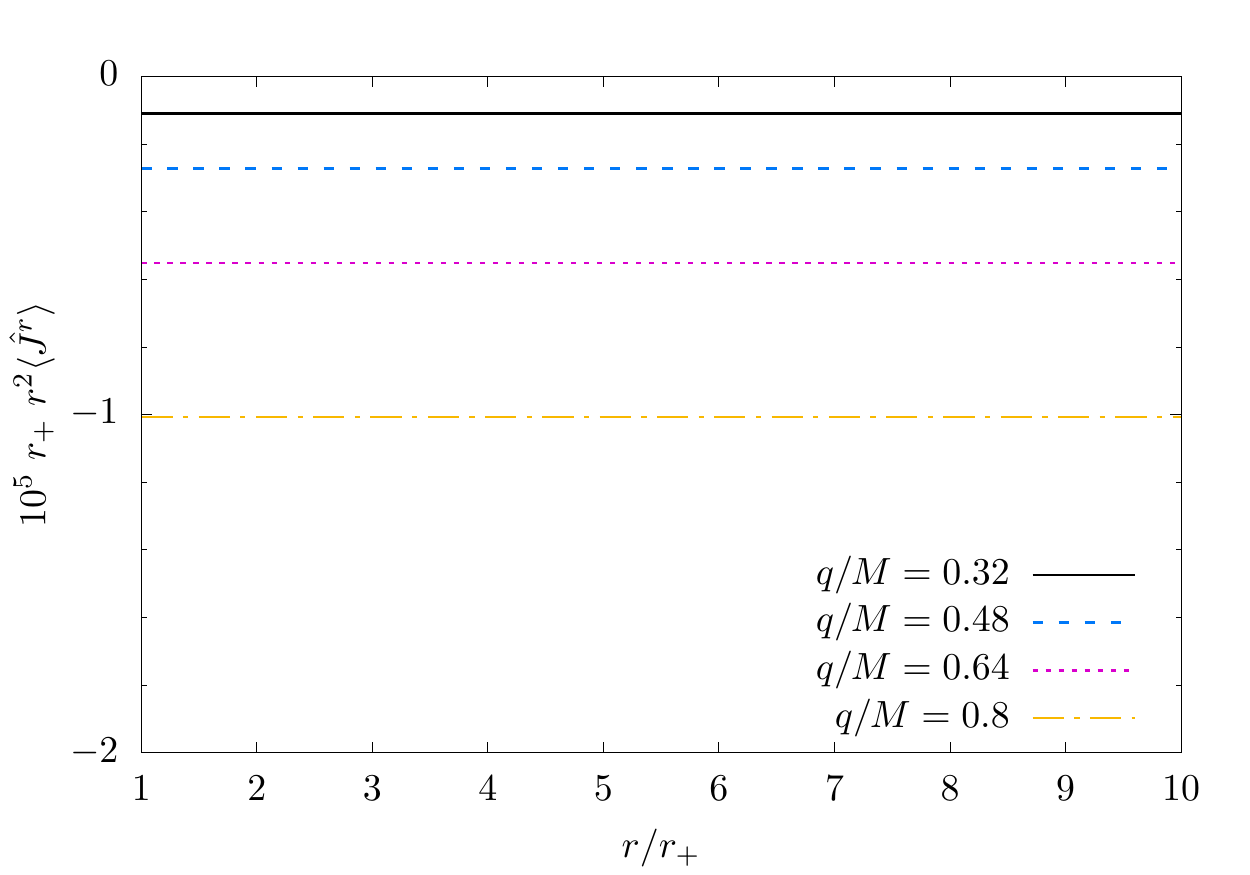}
  \includegraphics[scale=0.61]{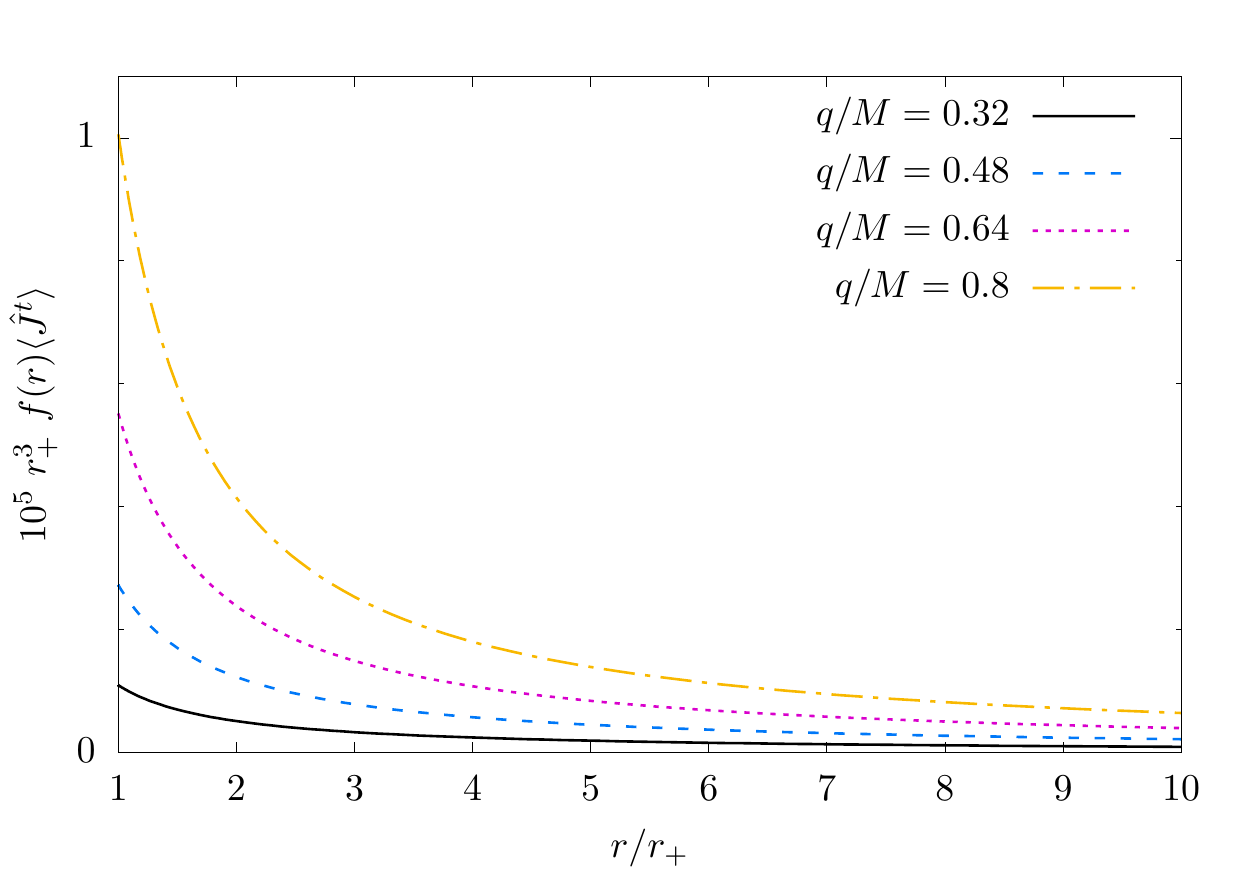}
   \\ \includegraphics[scale=0.61]{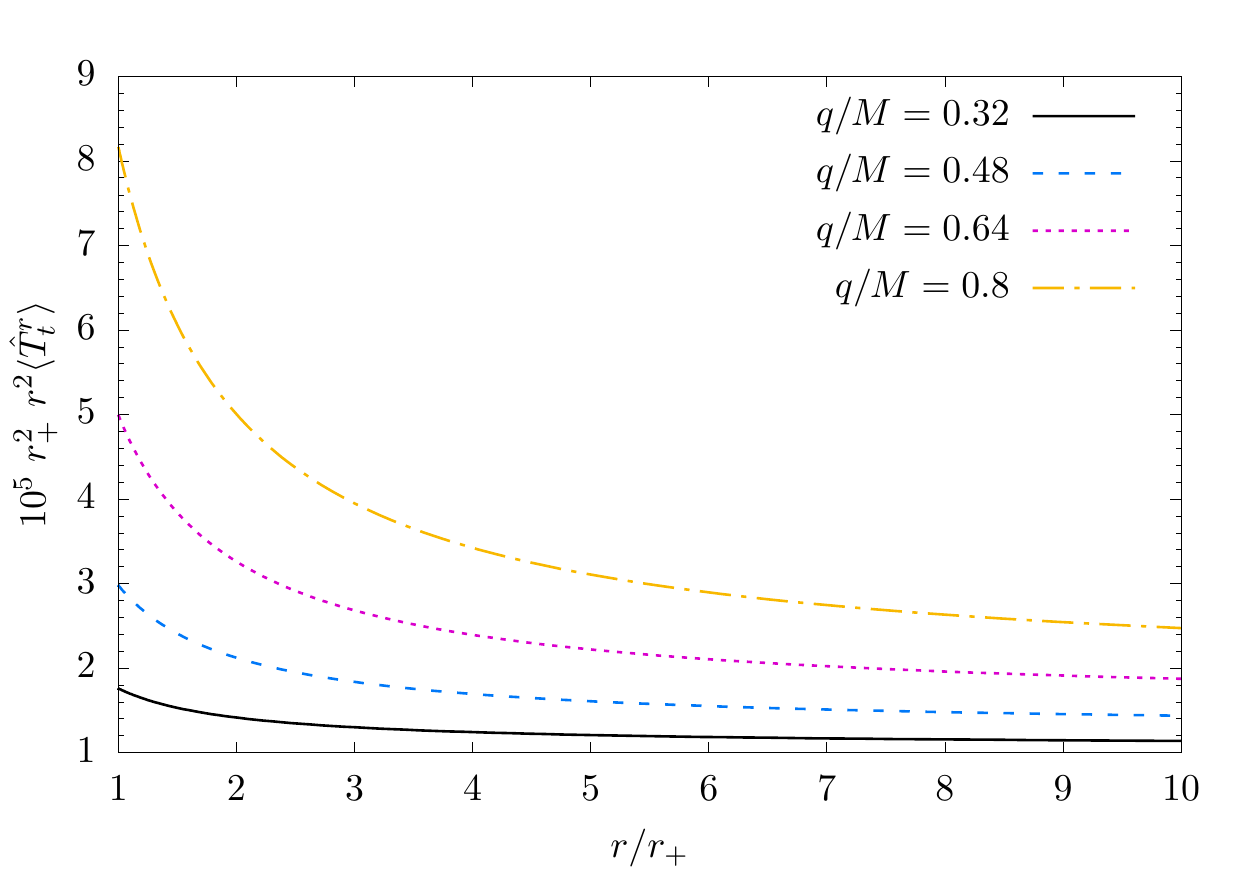}
  \includegraphics[scale=0.61]{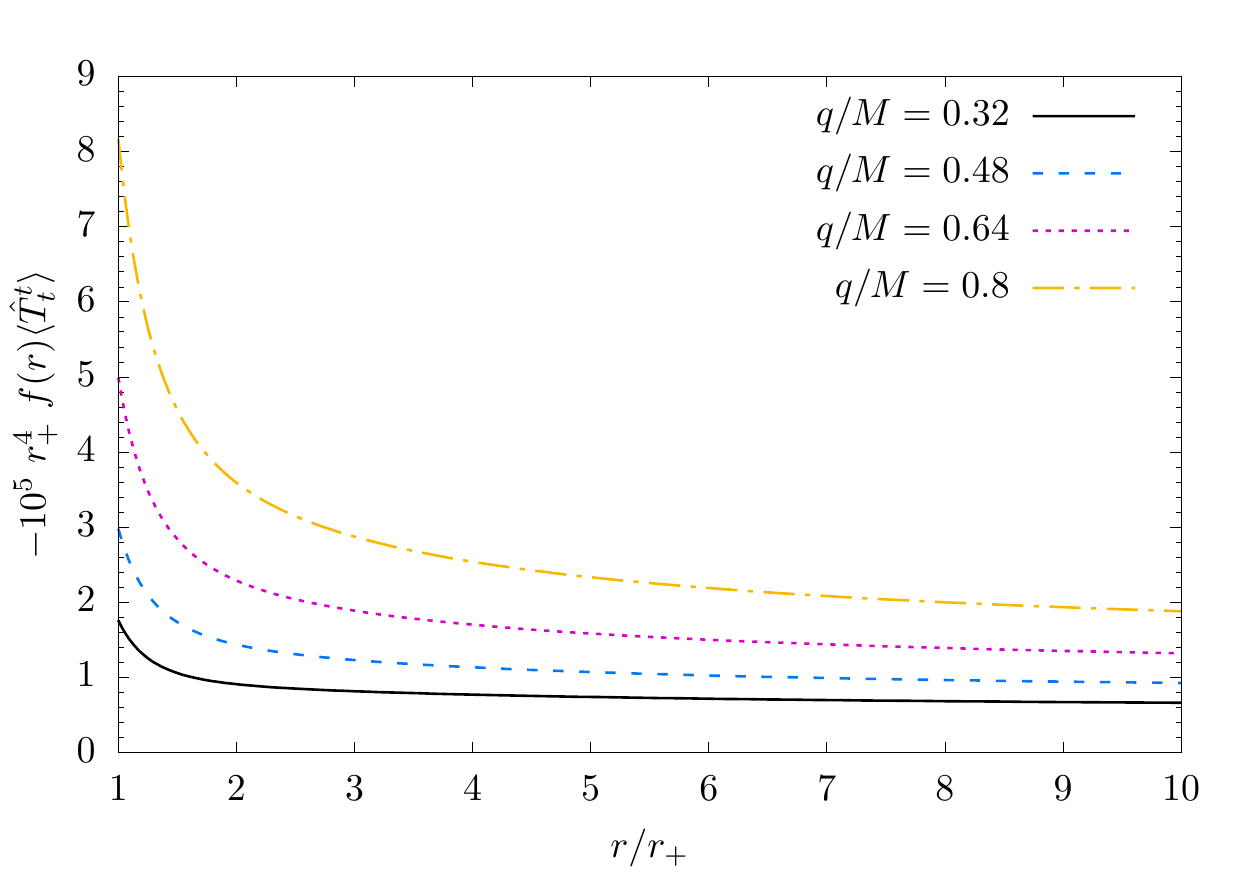} \\
  \includegraphics[scale=0.61]{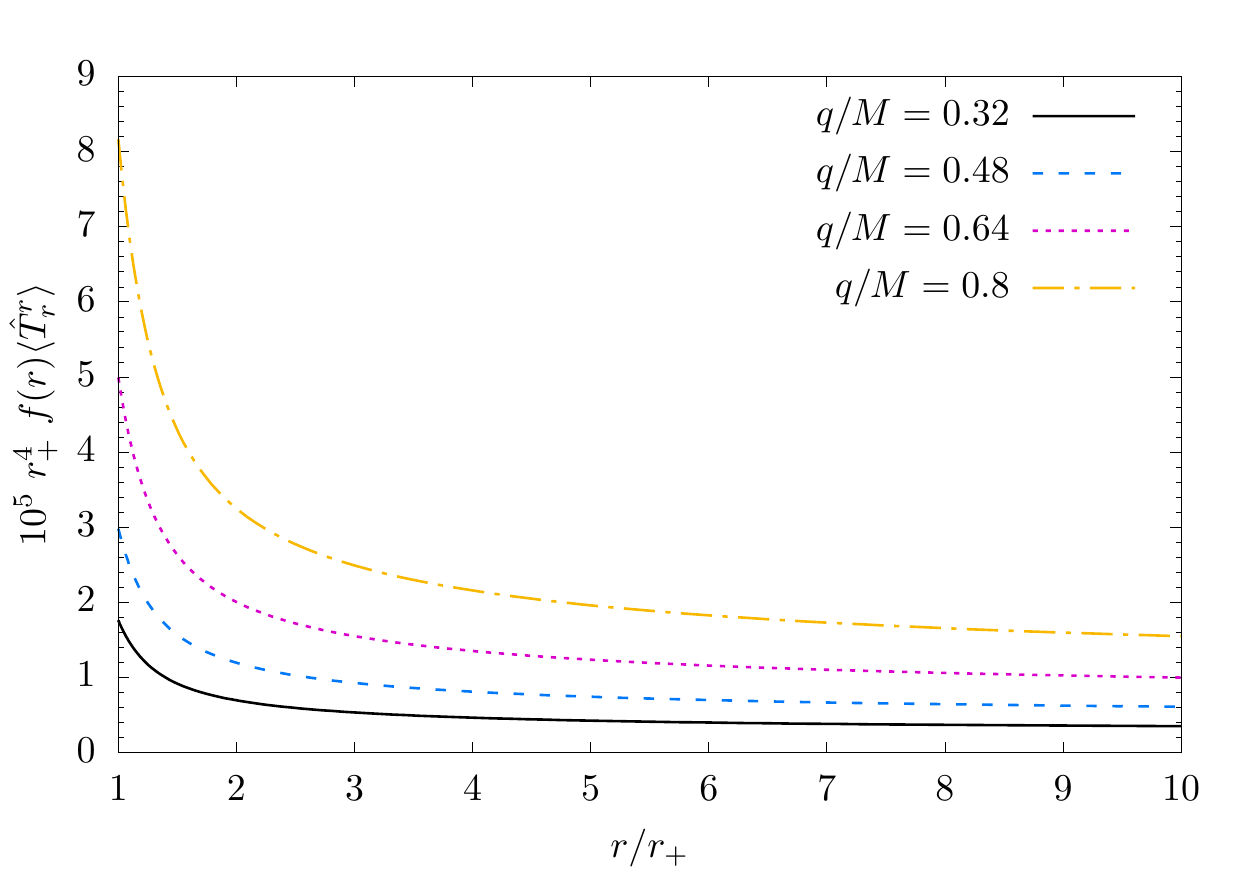}
  \includegraphics[scale=0.61]{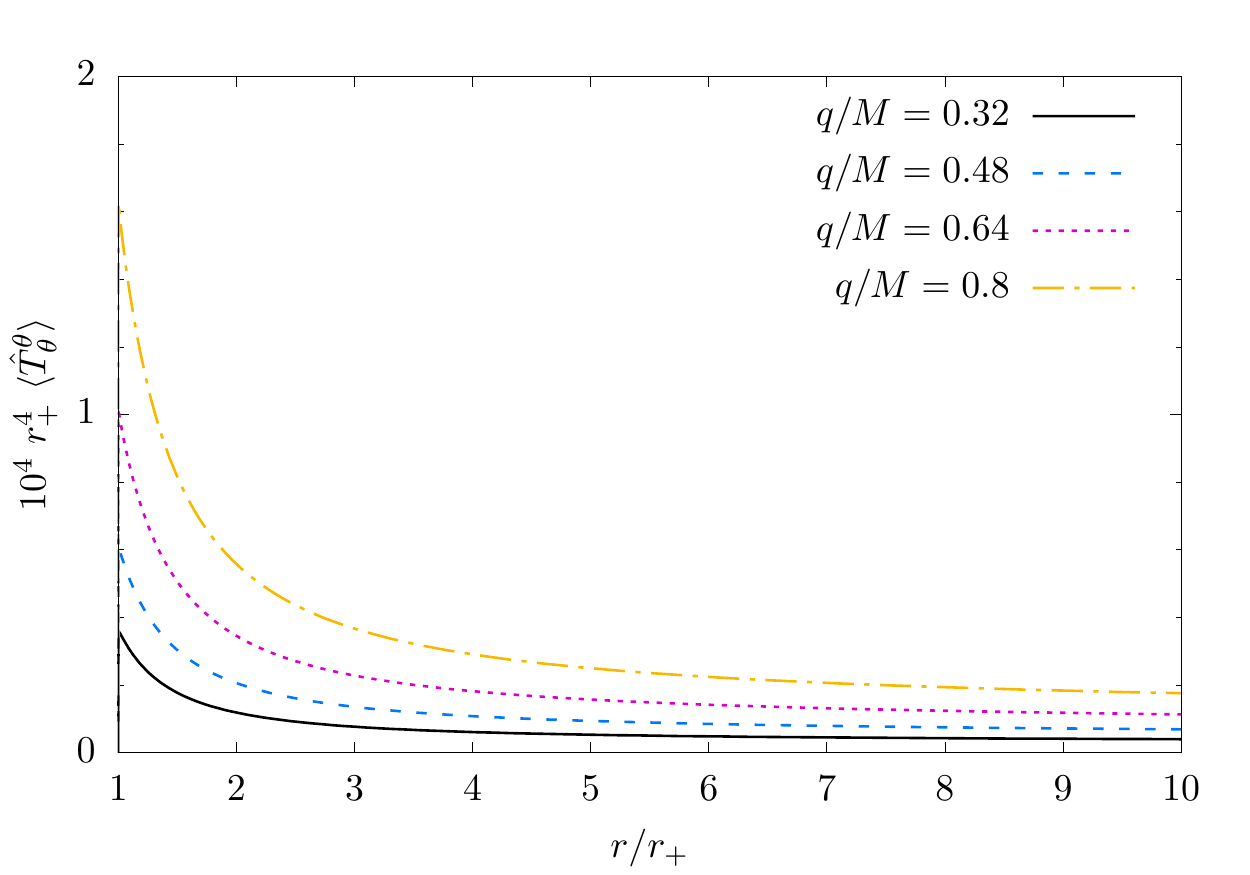} \\
  \vspace{.1cm}
  \caption{Difference in expectation values for the scalar condensate operator and components of the current and stress-energy tensor operators, between the ``past'' CCH, $| {\rm {CCH}}^{-} \rangle$, and ``past'' Unruh state, $| {\rm {U}}^{-} \rangle$, in the spacetime of a RN black hole with $Q=0.8M$. All expectation values are multiplied by powers of $f(r)$ so that the resulting quantities are regular at $r=r_{+}$.}
  \label{fig:CCHminus-Uminus}
\end{figure}

\begin{figure}
	\includegraphics[scale=0.61]{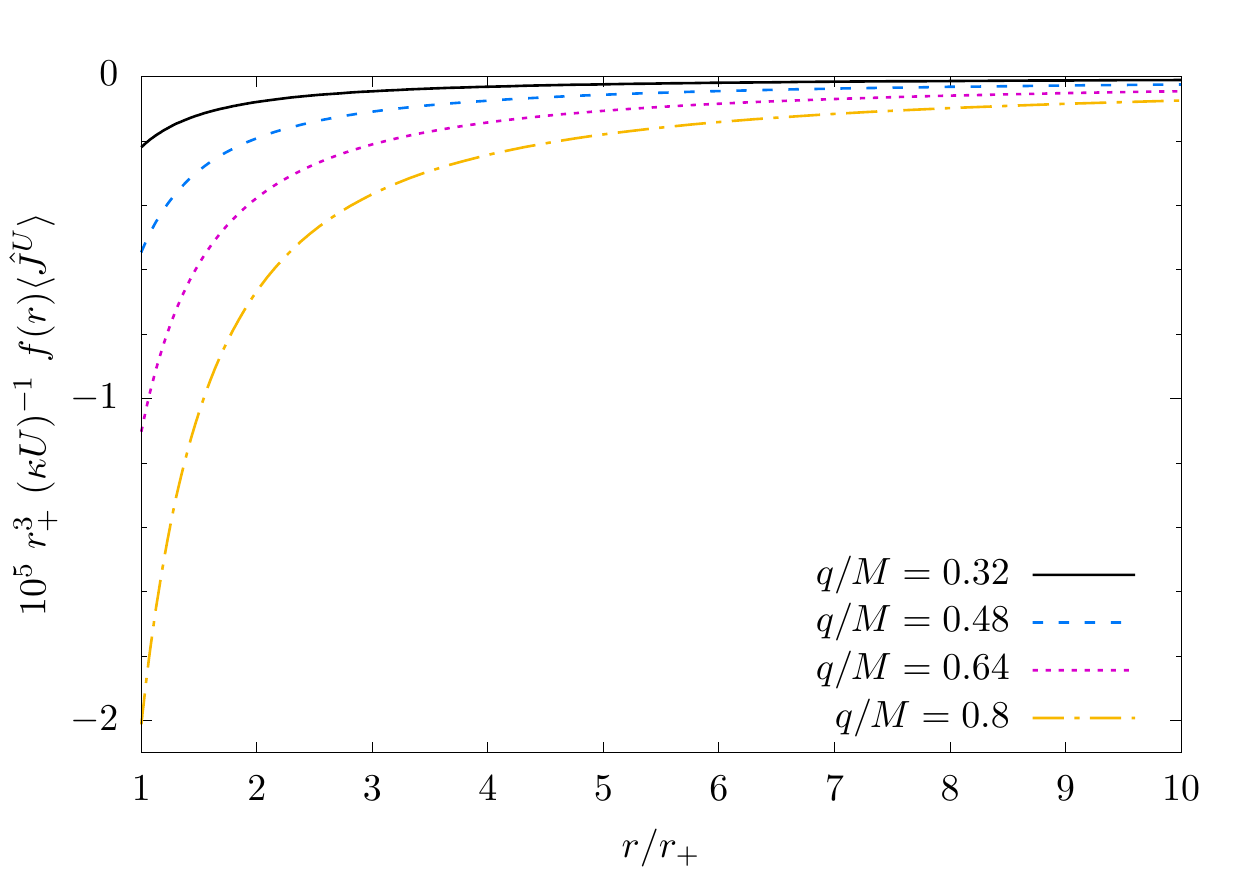}
	\includegraphics[scale=0.61]{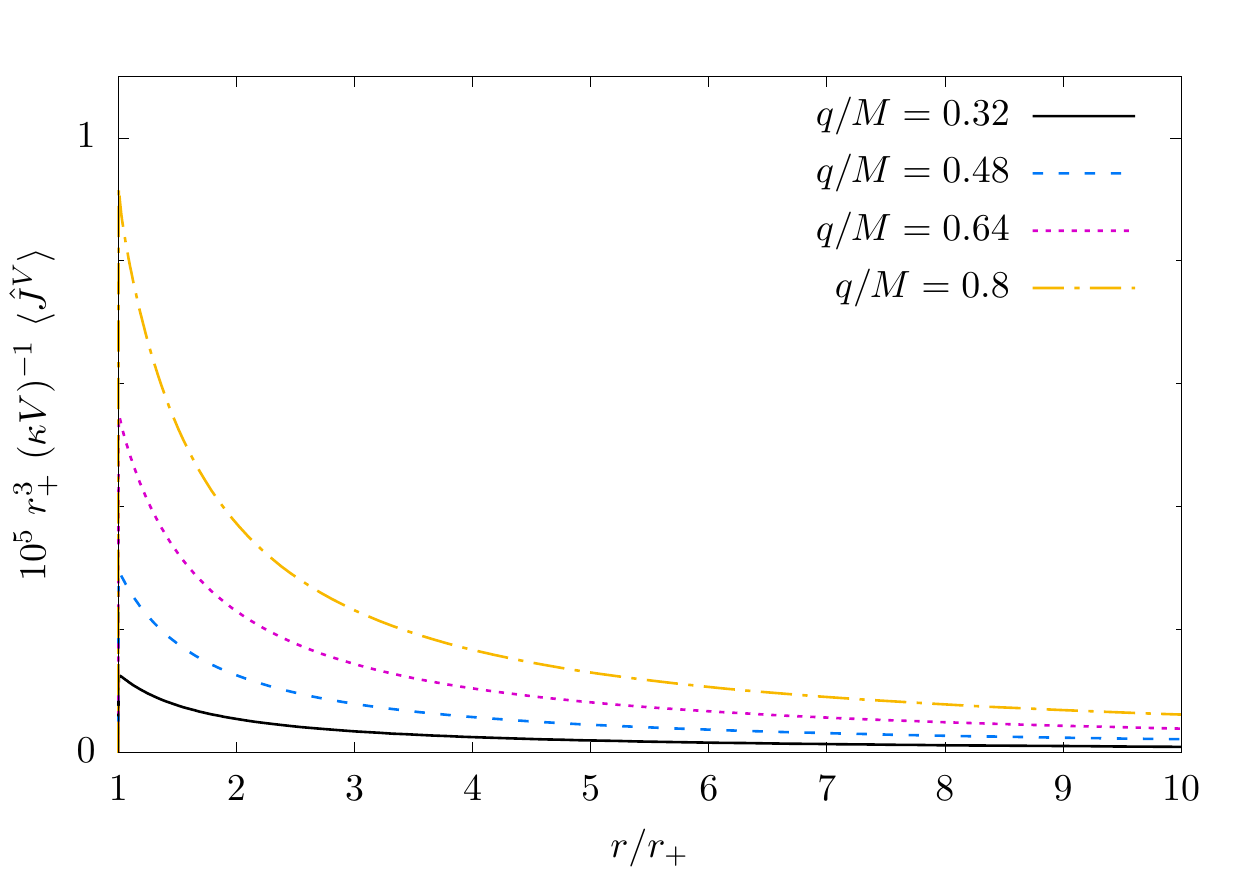}
	\\
	\includegraphics[scale=0.61]{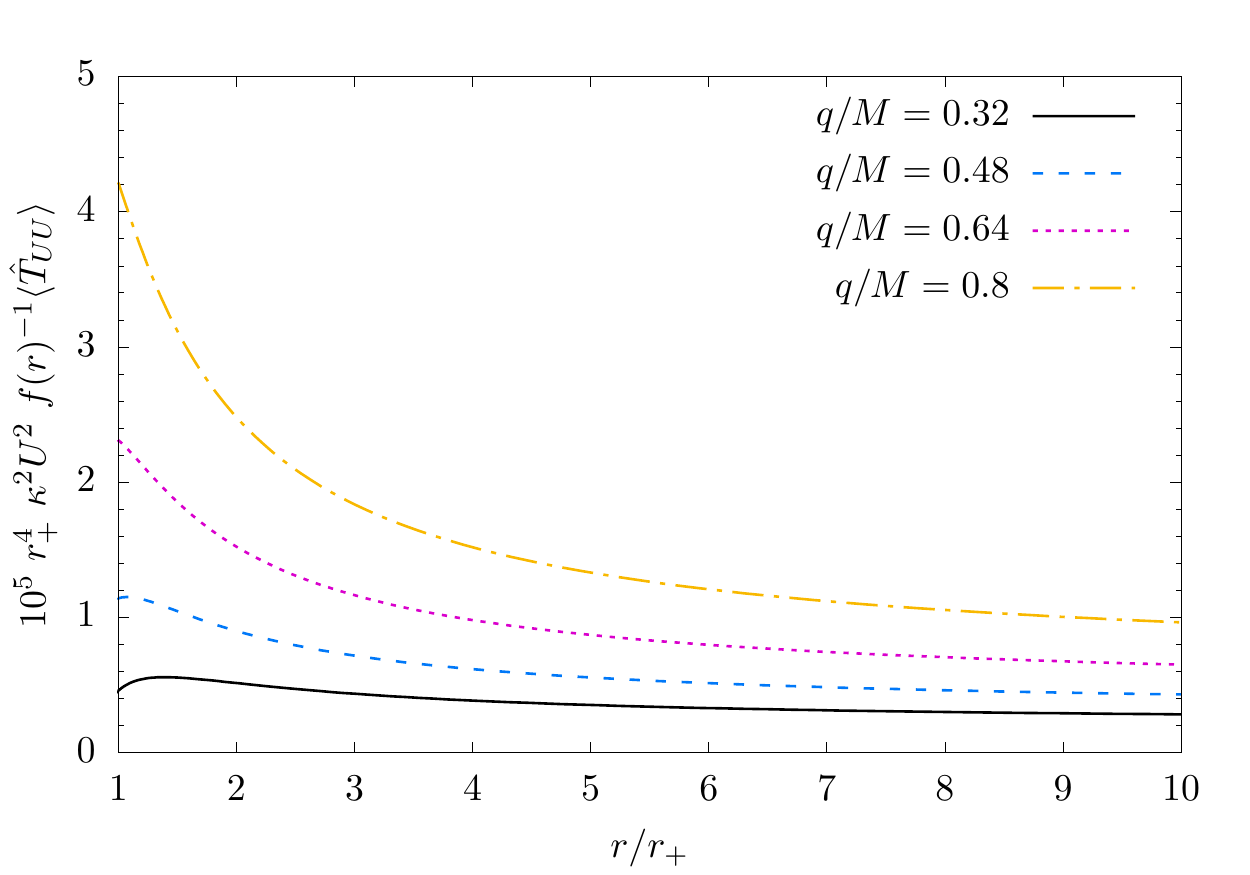}
	\includegraphics[scale=0.61]{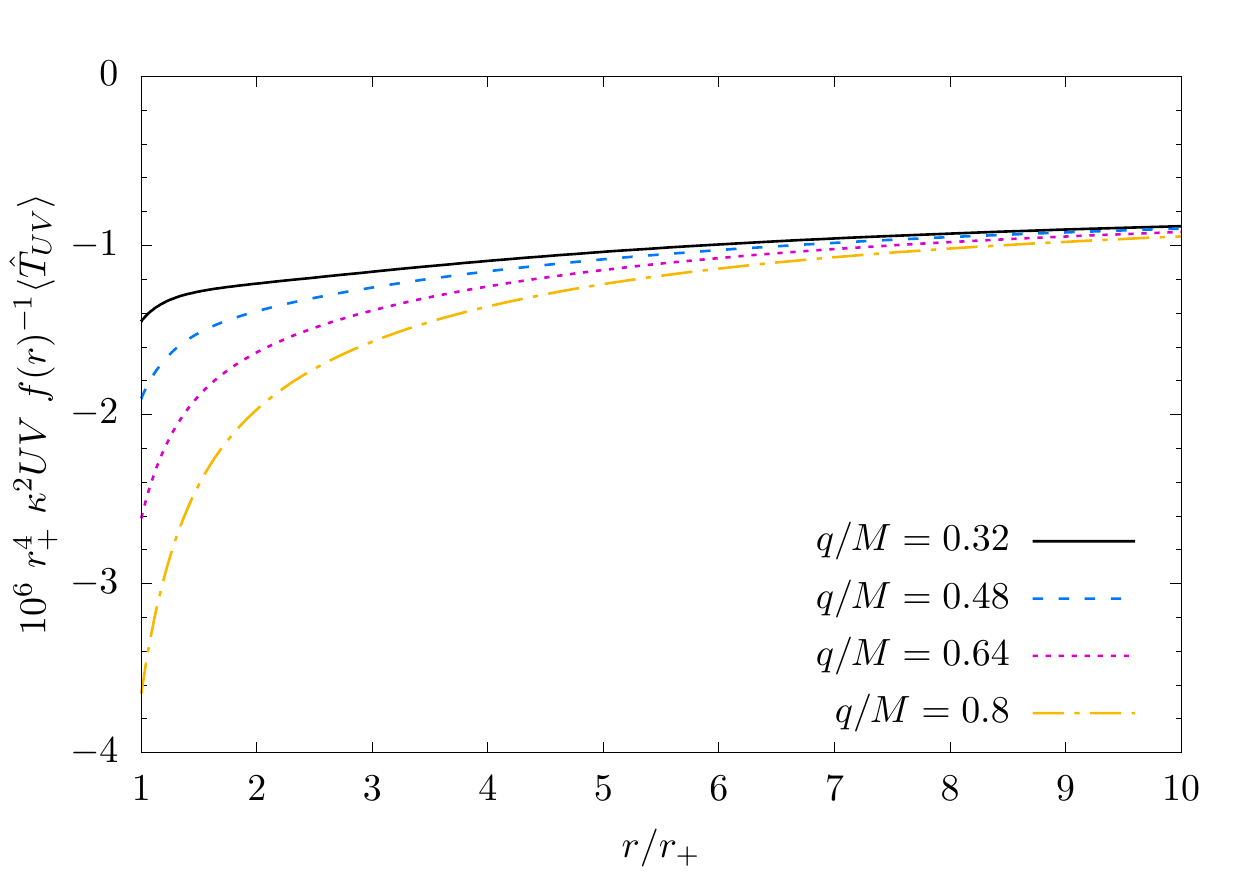}
	\\
	\includegraphics[scale=0.61]{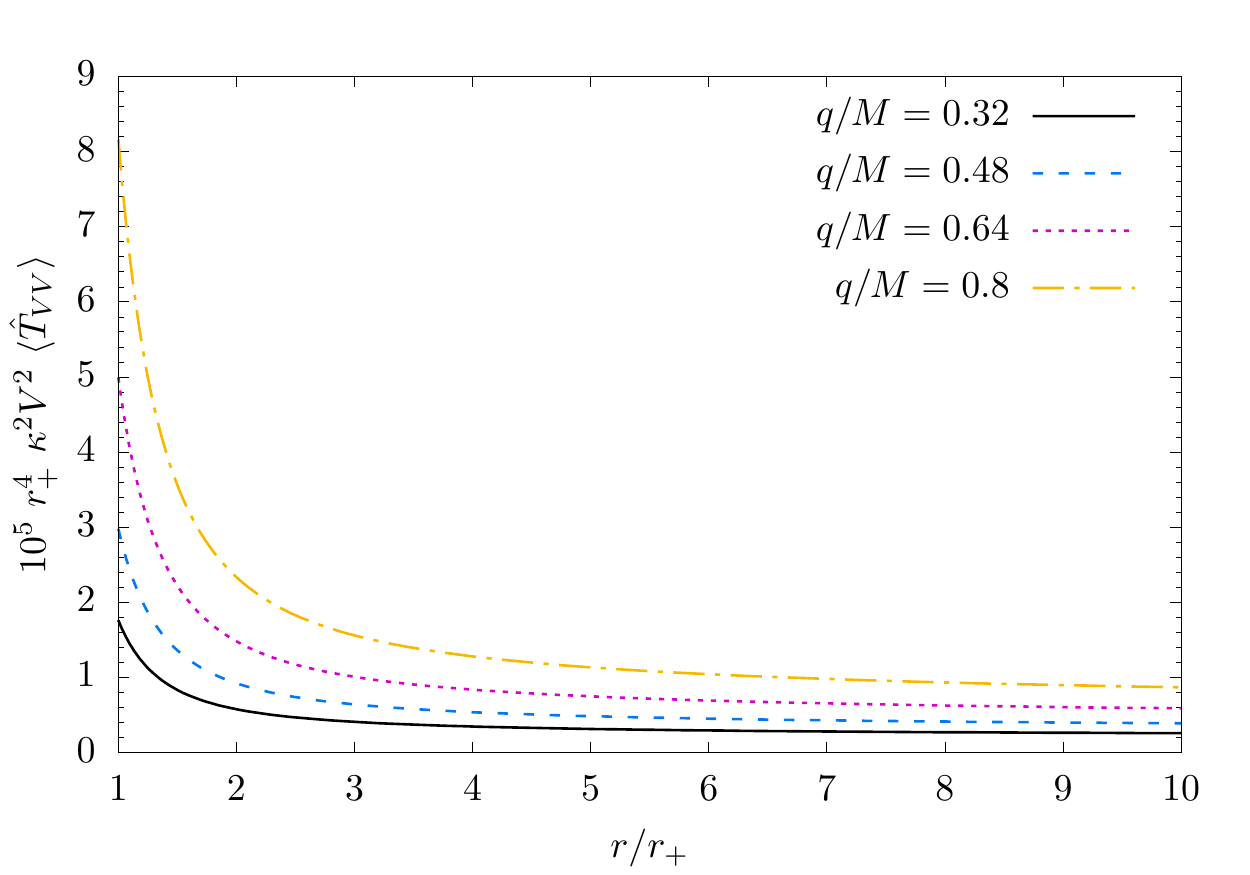}
	\vspace{.1cm}
	\caption{Difference in expectation values for the  components of the current and stress-energy tensor operators in Kruskal coordinates (\ref{eq:Kruskal}), between the ``past'' CCH, $| {\rm {CCH}}^{-} \rangle$, and ``past'' Unruh state, $| {\rm {U}}^{-} \rangle$, in the spacetime of a RN black hole with $Q=0.8M$. All expectation values are multiplied by powers of $f(r)$ so that the resulting quantities are regular at $r=r_{+}$.}
	\label{fig:CCHminus-Uminus_Kruskal}
\end{figure}

Differences in expectation values between the ``past'' CCH state $| {\rm {CCH}}^{-} \rangle$ and ``past'' Unruh state $|{\rm {U}}^{-} \rangle$ are shown in Fig.~\ref{fig:CCHminus-Uminus}.
The black hole charge and scalar field charge have the same values as in Fig.~\ref{fig:Uminus-Bminus}.
We immediately see a much greater variation in these differences as the scalar field charge varies, compared with the differences in expectation values shown in Fig.~\ref{fig:Uminus-Bminus}. 
We also note that the differences in expectation values no longer tend to zero far from the black hole. This indicates that the ``past'' CCH state $| {\rm {CCH}}^{-} \rangle$ is not empty at infinity. 

Examining our numerical results for the difference in expectation values of the scalar field condensate, we find that this quantity is regular at the event horizon. Therefore either both the ``past'' CCH and ``past'' Unruh states are regular on the horizon, or both diverge there. Since we expect that the ``past'' Unruh state is regular on the future event horizon but divergent on the past horizon, we conjecture that the same holds for the ``past'' CCH state.

From the difference in expectation values of the radial component of the current, 
the black hole is losing charge in the ``past'' CCH state $| {\rm {CCH}}^{-} \rangle$.
The magnitude of the difference in  charge flux between the ``past'' CCH and Unruh states is about two and a half times that between the ``past'' Unruh and Boulware states.
This indicates that the ``past'' CCH state has considerably more outgoing charge flux in the ``in'' modes than the ``past'' Unruh state has in the ``up'' modes, due to the different thermal factor for the ``in'' modes in the ``past'' CCH state.

The sign of  ${\mathcal {K}}_{{\mathrm {CCH}}^{-}}- {\mathcal {K}}_{{\mathrm {U}}^{-}}$ is not immediately constrained by (\ref{eq:chargeCCHminus}), but our numerical results show that this quantity is positive (at least for $qQ>0$). We deduce that the contribution to the charge flux of the superradiant positive frequency modes and nonsuperradiant negative frequency modes dominates that of the nonsuperradiant positive frequency modes and superradiant negative frequency modes. 

In contrast to the case for the difference in expectation values between the  ``past'' Unruh and ``past'' Boulware states, for the difference between the ``past'' CCH and ``past'' Unruh states the time component of the current is positive, and increases as the scalar field charge increases. 

The components of the current in Kruskal coordinates (\ref{eq:JKruskal}) are of particular interest for  the properties of the  ``past'' CCH state. We expect the ``past'' Unruh state to be regular on the future horizon where the Kruskal coordinate $U$ vanishes, but divergent on the past horizon (where  the Kruskal coordinate $V$ is zero). Examining the components $\langle {\hat {{J}}}^U \rangle $ and $\langle {\hat {{J}}}^V \rangle $ shown in Fig.~\ref{fig:CCHminus-Uminus_Kruskal}, we see that $V^{-1}\langle {\hat {{J}}}^V \rangle $ is regular as $r\rightarrow r_+$, but that $U^{-1}\langle {\hat {{J}}}^U \rangle $ diverges like $f(r)^{-1}$. This means that $\langle {\hat {{J}}}^V \rangle $ is regular on both the future and past horizons, but $\langle {\hat {{J}}}^U \rangle $ is regular only on the future horizon.
If our assumptions about the regularity of the ``past'' Unruh state are correct, we would deduce that the ``past'' CCH state is also regular on the future horizon but not the past horizon.

Turning now to the components of the stress-energy tensor, the diagonal components reveal greater variation as the scalar charge increases than was observed for the differences between the ``past'' Unruh and ``past'' Boulware states. Furthermore, these components do not decay at infinity, but instead appear to approach constant values, as might be expected for a thermal state.  

The flux $\langle {\hat {{T}}}_t^r\rangle$ for the difference between the ``past'' CCH and ``past'' Unruh states is positive for all values of $r$ examined, in contrast to the negative values seen for the difference between the ``past'' Unruh and ``past'' Boulware states. Since ${\mathcal {K}}_{{\rm {CCH}}^{-}}-{\mathcal {K}}_{{\rm {U}}^{-}}>0$, this implies that ${\mathcal {L}}_{{\rm {CCH}}^{-}}-{\mathcal {L}}_{{\rm {U}}^{-}}$ is negative. From (\ref{eq:energyCCHminus}), we deduce that for the difference between the ``past'' CCH and ``past'' Unruh states, the nonsuperradiant modes dominate the energy flux compared to the superradiant modes. 

Examining the components of the SET in Kruskal coordinates (\ref{eq:TKruskal}), our numerical results in Fig.~\ref{fig:CCHminus-Uminus_Kruskal} reveal that $U^2f(r)^{-1}\langle {\hat {{T}}}_{UU}\rangle$, $UVf(r)^{-1}\langle {\hat {{T}}}_{UV}\rangle$ and $V^2\langle {\hat {{T}}}_{VV}\rangle$ are all finite and nonzero on the horizon. 
Therefore $\langle {\hat {{T}}}_{VV}\rangle$ will diverge on the past horizon where $V=0$, but is regular on the future horizon where $V$ is finite and nonzero. 
As the future horizon is approached, $U\sim {\mathcal {O}}(f(r))$, and hence $\langle {\hat {{T}}}_{UV}\rangle$ is regular there. Similarly, $\langle {\hat {{T}}}_{UV}\rangle$ is also regular on the past horizon. 
However, $\langle {\hat {{T}}}_{UU}\rangle$ will vanish on the past horizon (where $U$ is finite and nonzero) but will diverge as $f(r)^{-1}$ as the future horizon is approached. 
Assuming that the ``past'' Unruh state is regular on the future horizon, we therefore find a mild divergence in the SET for the ``past'' CCH state on ${\mathcal {H}}^{+}$. 
Since we expect that the ``past'' Unruh state will be divergent on the past horizon, we are unable to make any deductions about the regularity of the ``past'' CCH state on the past horizon.

\subsubsection{Future states}
The ``future'' quantum states constructed in Sec.~\ref{sec:future} are the time-reverse of the ``past'' quantum states discussed above. To see this, we consider the differences
\begin{subequations}
\begin{align}
	\langle {\rm {U}}^{+} | {\hat {{O}}} | {\rm {U}}^{+} \rangle  
	- \langle {\rm {B}}^{+} | {\hat {{O}}} | {\rm {B}}^{+} \rangle
	= &~ 
	\sum _{\ell =0}^{\infty }\sum _{m=-\ell}^{\ell }
	\int _{-\infty }^{\infty }d{\widetilde{\omega }} \,
	\frac{1}{\exp \left| \frac{2\pi {\widetilde{\omega }}}{\kappa } \right| -1 } 
	o^{\rm {down}}_{\omega \ell m},
	\\ 
	\langle {\rm {CCH}}^{+} | {\hat {{O}}} | {\rm {CCH}}^{+} \rangle  
	- \langle {\rm {U}}^{+} | {\hat {{O}}} | {\rm {U}}^{+} \rangle
	= &~
	\sum _{\ell =0}^{\infty }\sum _{m=-\ell}^{\ell }
	\int _{-\infty }^{\infty }d\omega  \,
	\frac{1}{\exp \left| \frac{2\pi \omega }{\kappa } \right| -1 } 
	o^{\rm {out}}_{\omega \ell m}. 
\end{align}
\end{subequations}
As $r\rightarrow \infty $, we find
\begin{subequations}
	\label{eq:UplusBplus}
	\begin{align}
		\langle {\rm {U}}^{+} | {\widehat {{SC}}} | {\rm {U}}^{+} \rangle  
		- \langle {\rm {B}}^{+} | {\widehat {{SC}}} | {\rm {B}}^{+} \rangle
		\sim & \frac{1}{16 \pi ^{2} r^{2}}  
		\sum _{\ell =0}^{\infty }
		\int _{-\infty }^{\infty }d{\widetilde{\omega }} \, \frac{ 2\ell + 1  }{\left| {\widetilde {\omega }}\right| \left( \exp \left| \frac{2\pi {\widetilde{\omega }}}{\kappa } \right| -1 \right) }  \left| B^{\rm {up}}_{\omega \ell } \right| ^{2} ,
		\\ 
		\langle {\rm {U}}^{+} | {\hat {{J}}}^{\mu } | {\rm {U}}^{+} \rangle  
		- \langle {\rm {B}}^{+} | {\hat {{J}}}^{\mu } | {\rm {B}}^{+} \rangle
		\sim & \frac{q}{64 \pi^3 r^2} \sum_{\ell = 0}^\infty  \int_{- \infty}^{\infty} d \widetilde{\omega} \frac{\omega \left( 2\ell + 1 \right) }{\left| \widetilde{\omega} \right| \left( \exp{\left| \frac{2 \pi \widetilde{\omega}}{\kappa} \right|} - 1 \right)} {\left| B^{\mathrm{up}}_{\omega \ell} \right|}^2  {\left( - 1, 1 , 0, 0 \right)}^\intercal ,
		\\
		\langle {\rm {U}}^{+} | {\hat {{T}}}^{\mu }_{\nu } | {\rm {U}}^{+} \rangle  
		- \langle {\rm {B}}^{+} | {\hat {{T}}}^{\mu }_{\nu } | {\rm {B}}^{+} \rangle
		\sim & \frac{1}{16 \pi ^{2} r^2} \sum_{\ell = 0}^\infty \int_{- \infty}^{\infty} d \widetilde{\omega} \frac{\omega ^{2} \left( 2\ell + 1 \right) }{\left| \widetilde{\omega} \right| \left( \exp{\left| \frac{2 \pi \widetilde{\omega}}{\kappa} \right|} - 1 \right)} {\left| B^{\mathrm{up}}_{\omega \ell} \right|}^2
		\begin{pmatrix}
   - 1 & - 1 & 0 & 0 \\
   1& 1 & 0 & 0 \\
   0 & 0 & \mathcal{O} \left( r^{-2} \right) & 0 \\
   0 & 0 & 0 & \mathcal{O} \left( r^{-2} \right)
  \end{pmatrix} ,
	\end{align}
	\end{subequations}
while as $r\rightarrow r_{+}$, we have
\begin{subequations}
		\label{eq:CCHplusUplus}
	\begin{align}
		\langle {\rm {CCH}}^{+} | {\widehat {{SC}}} | {\rm {CCH}}^{+} \rangle  
		- \langle {\rm {U}}^{+} | {\widehat {{SC}}} | {\rm {U}}^{+} \rangle
		\sim & \frac{1}{16 \pi ^{2} r^{2}}  
		\sum _{\ell =0}^{\infty }
		\int _{-\infty }^{\infty }d\omega \, \frac{ 2\ell + 1 }{\left| \omega \right| \left( \exp \left| \frac{2\pi \omega }{\kappa } \right| -1 \right) }  \left| B^{\rm {in}}_{\omega \ell } \right| ^{2} ,
		\\ 
		\langle {\rm {CCH}}^{+} | {\hat {{J}}}^{\mu } | {\rm {CCH}}^{+} \rangle  
		- \langle {\rm {U}}^{+} | {\hat {{J}}}^{\mu } | {\rm {U}}^{+} \rangle
		\sim & - \frac{q}{64 \pi^3 r^2} \sum_{\ell = 0}^\infty \int_{- \infty}^{\infty} d \omega \frac{{\widetilde {\omega }}\left( 2\ell + 1 \right) }{\left| \omega \right| \left( \exp{\left| \frac{2 \pi \omega}{\kappa} \right|} - 1 \right)} {\left| B^{\mathrm{in}}_{\omega \ell} \right|}^2  {\left( f(r)^{-1}, 1, 0, 0 \right)}^\intercal ,
		\\
		\langle {\rm {CCH}}^{+} | {\hat {{T}}}^{\mu }_{\nu } | {\rm {CCH}}^{+} \rangle  
		- \langle {\rm {U}}^{+} | {\hat {{T}}}^{\mu }_{\nu } | {\rm {U}}^{+} \rangle
		\sim & \frac{1}{16 \pi ^{2} r^2} \sum_{\ell = 0}^\infty  \int_{- \infty}^{\infty} d \omega \frac{{\widetilde{\omega}}^{2}\left( 2\ell + 1 \right) }{\left| \omega \right| \left( \exp{\left| \frac{2 \pi \omega}{\kappa} \right|} - 1 \right)} {\left| B^{\mathrm{in}}_{\omega \ell} \right|}^2 
		\nonumber \\ 
  & \qquad \qquad \times \begin{pmatrix}
   - f(r)^{-1} & f(r)^{-2} & 0 & 0 \\
   - 1 & f(r)^{-1} & 0 & 0 \\
   0 & 0 & \mathcal{O} \left( 1 \right) & 0 \\
   0 & 0 & 0 & \mathcal{O} \left( 1 \right)
  \end{pmatrix} .
	\end{align}
	\end{subequations}
As expected, the scalar condensate does not distinguish between ``past'' and ``future'' states.
The expectation values of the current and stress-energy tensor in (\ref{eq:UplusBplus}, \ref{eq:CCHplusUplus}) are obtained from those in (\ref{eq:UminusBminus}, \ref{eq:CCHminusUminus}) by making the coordinate transformation $t\rightarrow -t$.
By virtue of the Wronskian relations (\ref{eq:wronskians}), the integrands in (\ref{eq:UplusBplus}) are regular at ${\widetilde {\omega }}=0$ and those in (\ref{eq:CCHplusUplus}) are regular at $\omega =0$.
Given that we have already explored the properties of the ``past'' Unruh and CCH states in some detail, we will not consider the ``future'' states further.

\subsection{``Boulware''-like state}
\label{sec:Boulwareexp}

Now we turn to the first of the new states defined in this paper, namely the tentative ``Boulware''-like state constructed in Sec.~\ref{sec:Boulware}. 
To examine the properties of the state $| {\rm {B}} \rangle $, since the properties of the ``past'' and ``future'' Boulware states are well-understood, we consider the differences
\begin{subequations}
	\label{eq:BminusBpmfirst}
\begin{align}
 \langle {\rm {B}} |{\hat {{O}}} | {\rm {B}}\rangle 
- \langle {\rm {B}}^{-}|{\hat {{O}}} |{\rm {B}}^{-} \rangle
 & ~ = 
- \sum _{\ell =0}^{\infty }\sum _{m=-\ell}^{\ell }
\int _{\min \left\{\frac{qQ}{r_{+}},0\right\} } ^{\max\left\{\frac{qQ}{r_{+}},0\right\} } d\omega  \,
o_{\omega \ell m}^{\rm {up}},
\\
 \langle {\rm {B}} |{\hat {{O}}} | {\rm {B}}\rangle 
- \langle {\rm {B}}^{+}|{\hat {{O}}} |{\rm {B}}^{+} \rangle
& ~ = 
- \sum _{\ell =0}^{\infty }\sum _{m=-\ell}^{\ell }
\int _{\min \left\{\frac{qQ}{r_{+}},0\right\} } ^{\max\left\{\frac{qQ}{r_{+}},0\right\} } d\omega  \,
o_{\omega \ell m}^{\rm {down}}. 
\end{align}
\end{subequations}
These involve only the superradiant ``up'' and ``down'' modes.
Since the ``down'' modes are the time-reverse of the ``up'' modes, the differences (\ref{eq:BminusBpmfirst}) are the time-reverse of each other, which suggests that the state $|{\mathrm {B}}\rangle $ is time-reversal invariant.

Since the scalar field condensate does not distinguish between ``past'' and ``future'' states, consider first the expectation values of the current and stress-energy tensor.
 As $r\rightarrow \infty $, we find
\begin{subequations}
	\label{eq:BminusBpm}
	\begin{align}
		\langle {\rm {B}} | {\hat {{J}}}^{\mu } | {\rm {B}} \rangle  
		- \langle {\rm {B}}^{-} | {\hat {{J}}}^{\mu } | {\rm {B}}^{-} \rangle
		\sim & \frac{q}{64 \pi^3 r^2} \sum_{\ell = 0}^\infty  \int_{\mathrm{min} \left\{ \frac{qQ}{r_+}, 0 \right\} }^{\mathrm{max} \left\{ \frac{qQ}{r_+}, 0 \right\} } d \omega \, \frac{\omega }{\left| \widetilde{\omega} \right|} \left( 2\ell + 1 \right)  {\left| B^{\mathrm{up}}_{\omega \ell} \right|}^2 {\left( 1, 1, 0, 0 \right)}^\intercal ,
		\\
		\langle {\rm {B}} | {\hat {{T}}}^{\mu }_{\nu } | {\rm {B}} \rangle  
		- \langle {\rm {B}}^{-} | {\hat {{T}}}^{\mu }_{\nu } | {\rm {B}}^{-} \rangle
		\sim & - \frac{1}{16 \pi^{2} r^2} \sum_{\ell = 0}^\infty  \int_{\mathrm{min} \left\{ \frac{qQ}{r_+}, 0 \right\} }^{\mathrm{max} \left\{ \frac{qQ}{r_+}, 0 \right\} } d \omega \, \frac{\omega ^{2} }{\left| \widetilde{\omega} \right|} \left( 2\ell + 1 \right) {\left| B^{\mathrm{up}}_{\omega \ell} \right|}^2 \begin{pmatrix}
   - 1 & 1 & 0 & 0 \\
   - 1 & 1 & 0 & 0 \\
   0 & 0 & \mathcal{O} \left( r^{-2} \right) & 0 \\
   0 & 0 & 0 & \mathcal{O} \left( r^{-2} \right)
  \end{pmatrix} ,
\\
\langle {\rm {B}} | {\hat {{J}}}^{\mu } | {\rm {B}} \rangle  
- \langle {\rm {B}}^{+} | {\hat {{J}}}^{\mu } | {\rm {B}}^{+} \rangle
\sim & \frac{q}{64 \pi^3 r^2} \sum_{\ell = 0}^\infty  \int_{\mathrm{min} \left\{ \frac{qQ}{r_+}, 0 \right\} }^{\mathrm{max} \left\{ \frac{qQ}{r_+}, 0 \right\} } d \omega \, \frac{\omega }{\left| \widetilde{\omega} \right|} \left( 2\ell + 1 \right) {\left| B^{\mathrm{up}}_{\omega \ell} \right|}^2  {\left( 1, -1, 0, 0 \right)}^\intercal ,
\\
\langle {\rm {B}} | {\hat {{T}}}^{\mu }_{\nu } | {\rm {B}} \rangle  
- \langle {\rm {B}}^{+} | {\hat {{T}}}^{\mu }_{\nu } | {\rm {B}}^{+} \rangle
\sim & - \frac{1}{16 \pi ^{2} r^2} \sum_{\ell = 0}^\infty  \int_{\mathrm{min} \left\{ \frac{qQ}{r_+}, 0 \right\} }^{\mathrm{max} \left\{ \frac{qQ}{r_+}, 0 \right\} } d \omega \, \frac{\omega ^{2}}{\left| \widetilde{\omega} \right|} \left( 2\ell + 1 \right) {\left| B^{\mathrm{up}}_{\omega \ell} \right|}^2 \begin{pmatrix}
	- 1 & -1 & 0 & 0 \\
	1 & 1 & 0 & 0 \\
	0 & 0 & \mathcal{O} \left( r^{-2} \right) & 0 \\
	0 & 0 & 0 & \mathcal{O} \left( r^{-2} \right)
\end{pmatrix} .
	\end{align}
\end{subequations}
The integrands in these expectation values are all regular at ${\widetilde {\omega }}=0$ due to the Wronskian relations (\ref{eq:wronskians}). 
Using (\ref{eq:fluxesBminus})  and the relevant components in (\ref{eq:BminusBpm}), we find that the fluxes of charge and energy in the state $|{\mathrm {B}}\rangle $ vanish:
\begin{equation}
	{\mathcal {K}}_{\mathrm {B}}=0, \qquad  {\mathcal {L}}_{\mathrm {B}}=0.
\label{eq:fluxesB}
\end{equation}
Therefore $|{\mathrm {B}}\rangle $ is an equilibrium state and is, indeed, time-reversal invariant.

\begin{figure}[p]
	\includegraphics[scale=0.63]{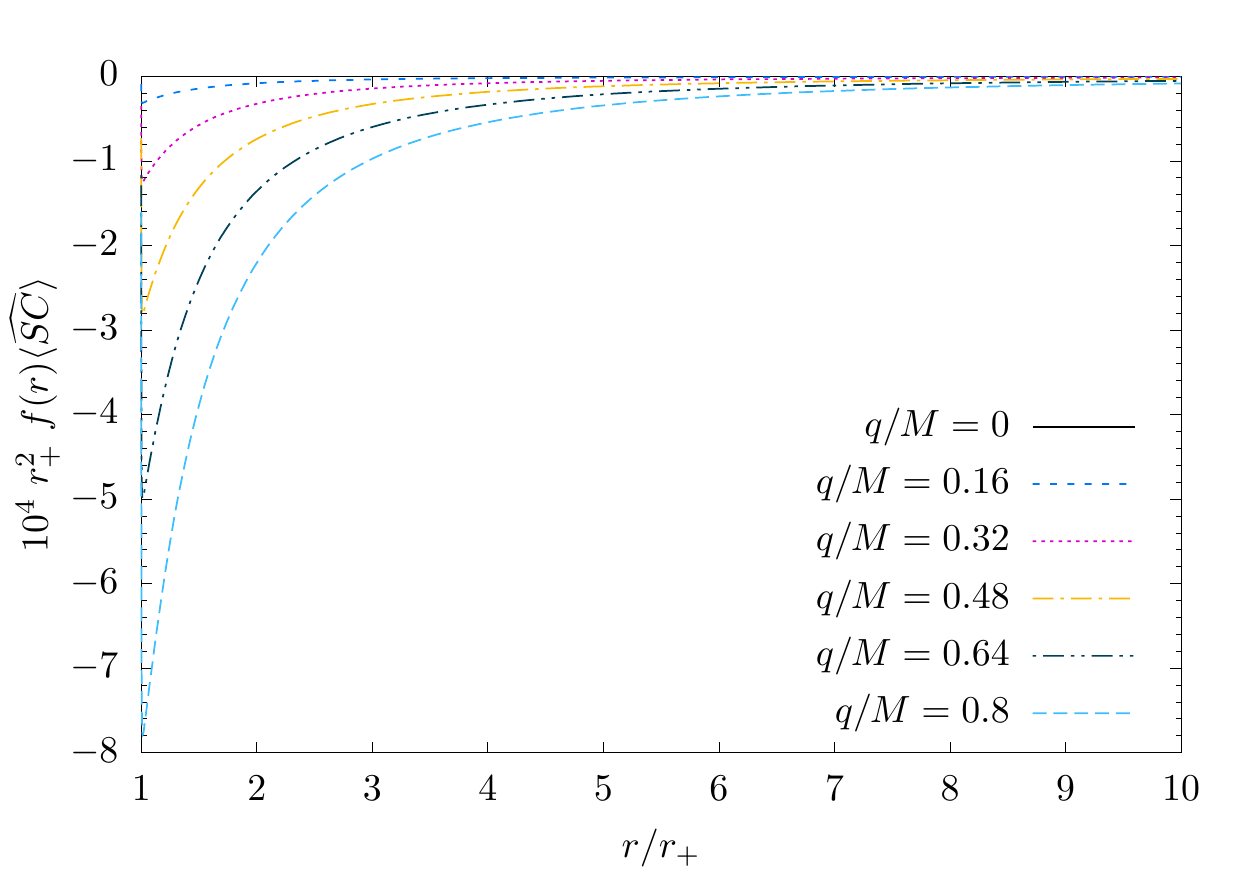}\\ 
	\includegraphics[scale=0.61]{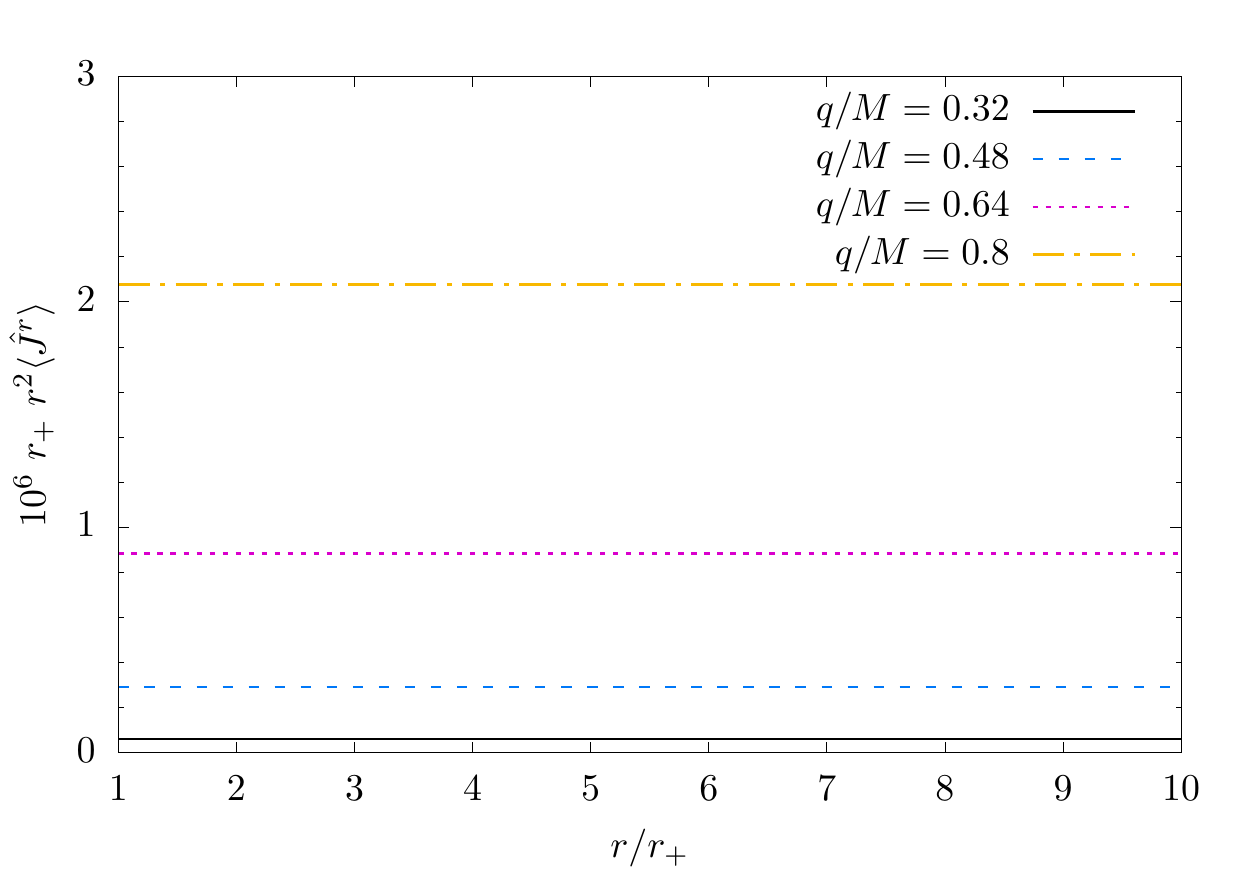}
	\includegraphics[scale=0.61]{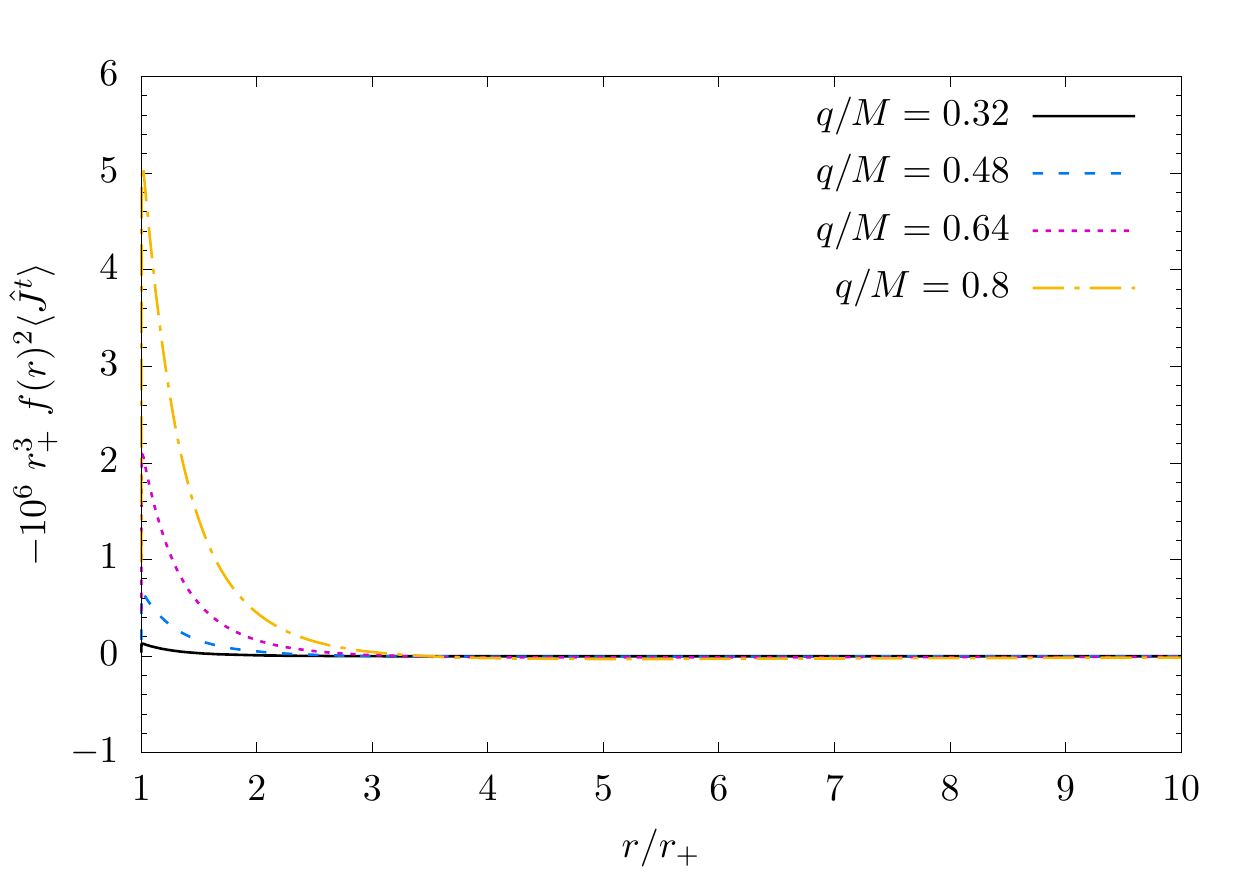}
	\\ \includegraphics[scale=0.61]{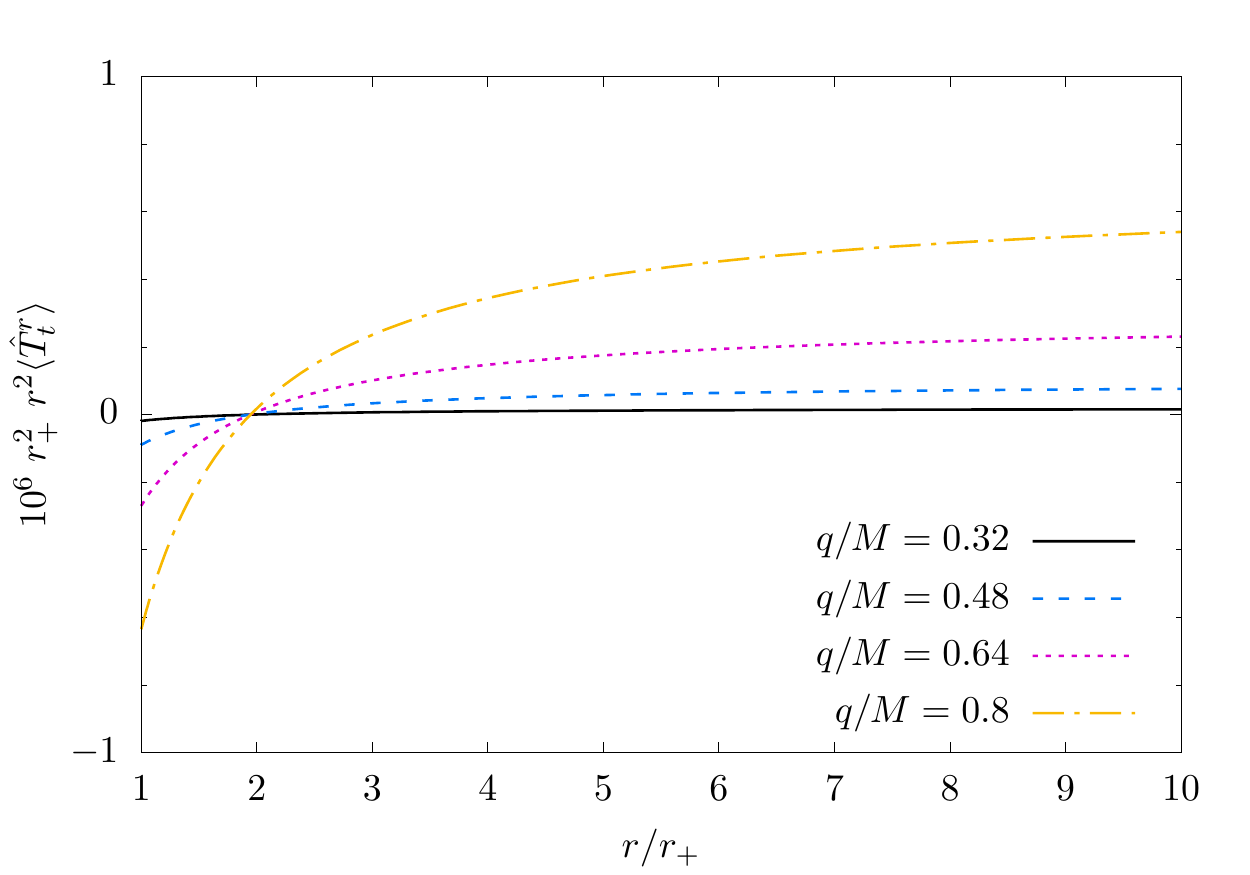}
	\includegraphics[scale=0.61]{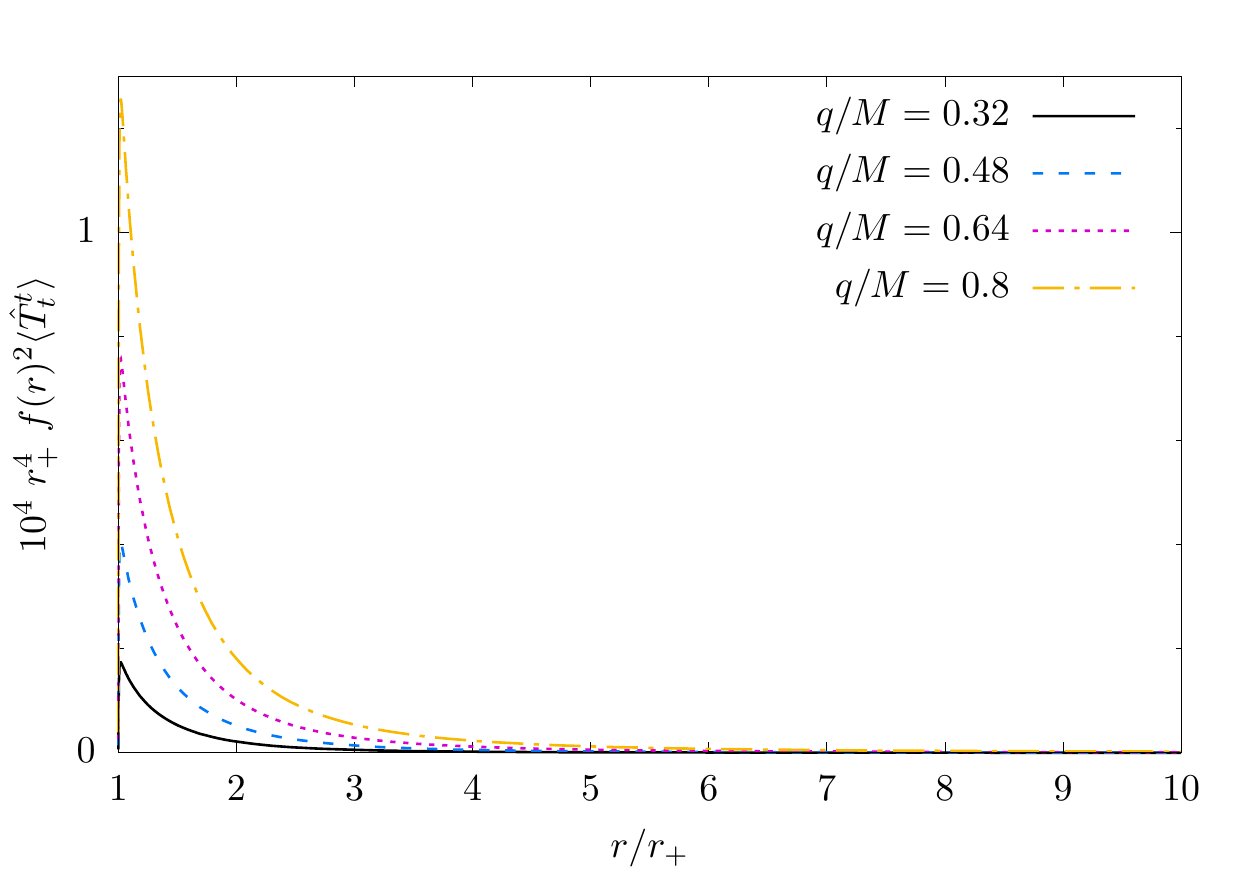} \\
	\includegraphics[scale=0.61]{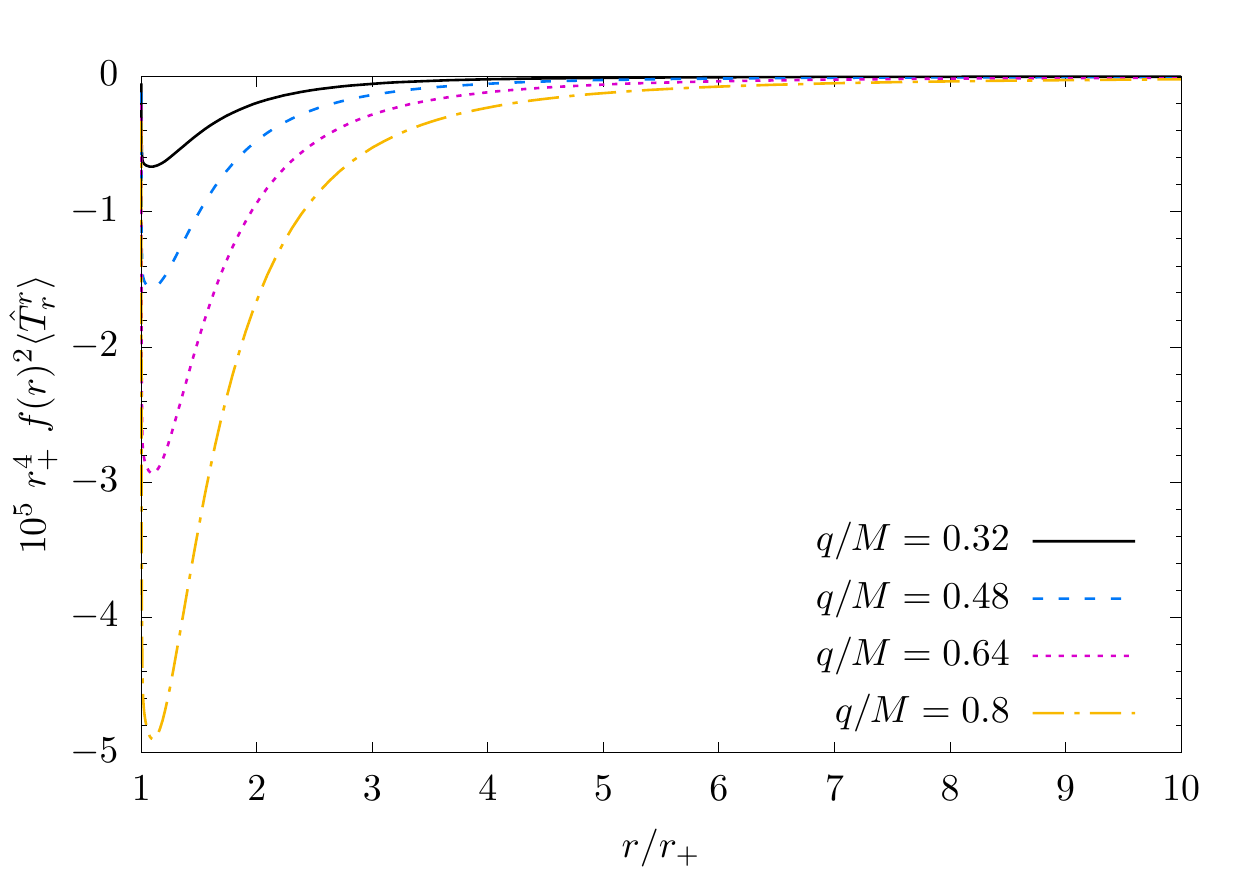}
	\includegraphics[scale=0.61]{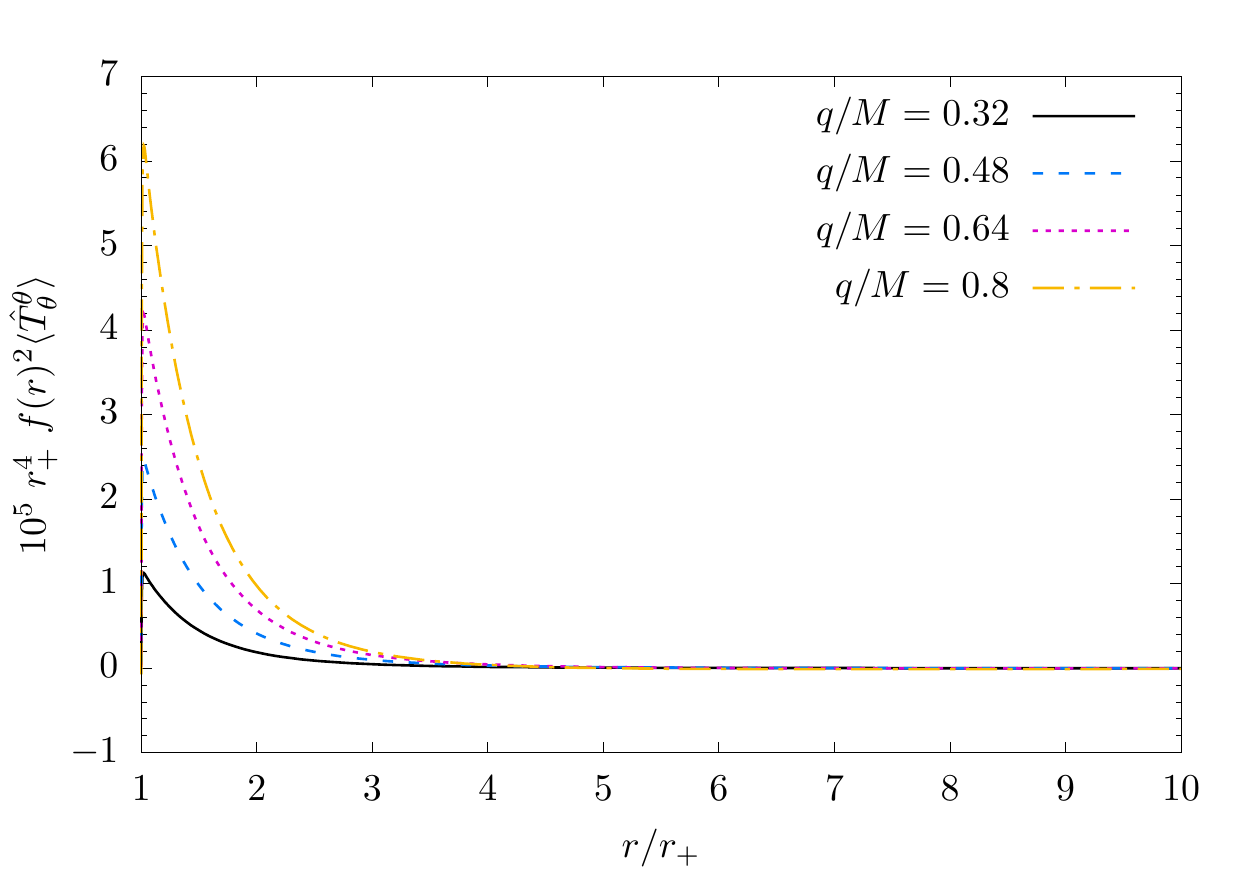} \\
	\vspace{.1cm}
	\caption{Difference in expectation values for the scalar condensate operator and components of the current and stress-energy tensor operators, between the tentative ``Boulware''-like state, $| {\rm {B}} \rangle$, and the ``past'' Boulware state, $|{\rm {B}}^{-} \rangle$, in the spacetime of a RN black hole with $Q=0.8M$. All quantities are multiplied by powers of $f(r)$ so that the resulting quantities are regular at $r=r_{+}$.}
	\label{fig:B-Bminus}
\end{figure}

\begin{figure}
	\includegraphics[scale=0.61]{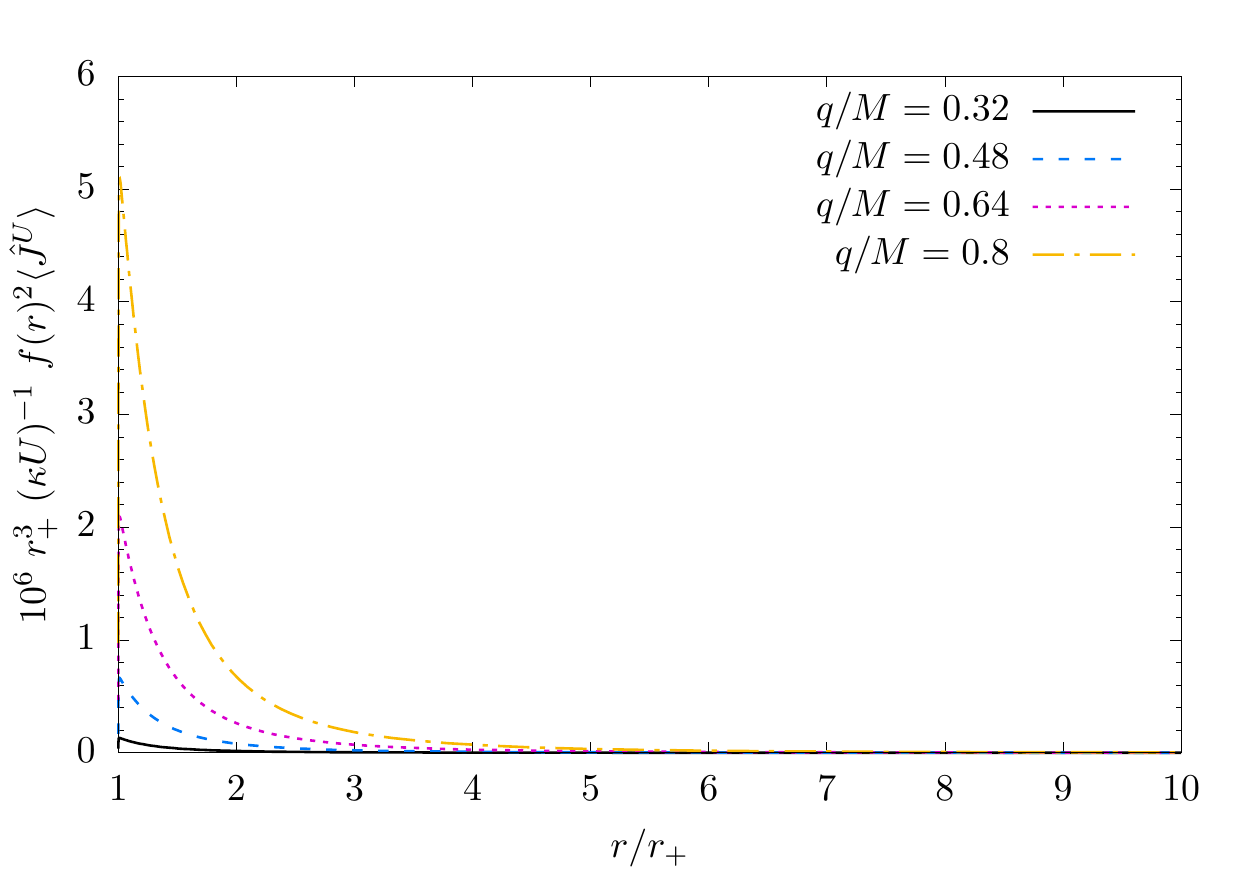}
	\includegraphics[scale=0.61]{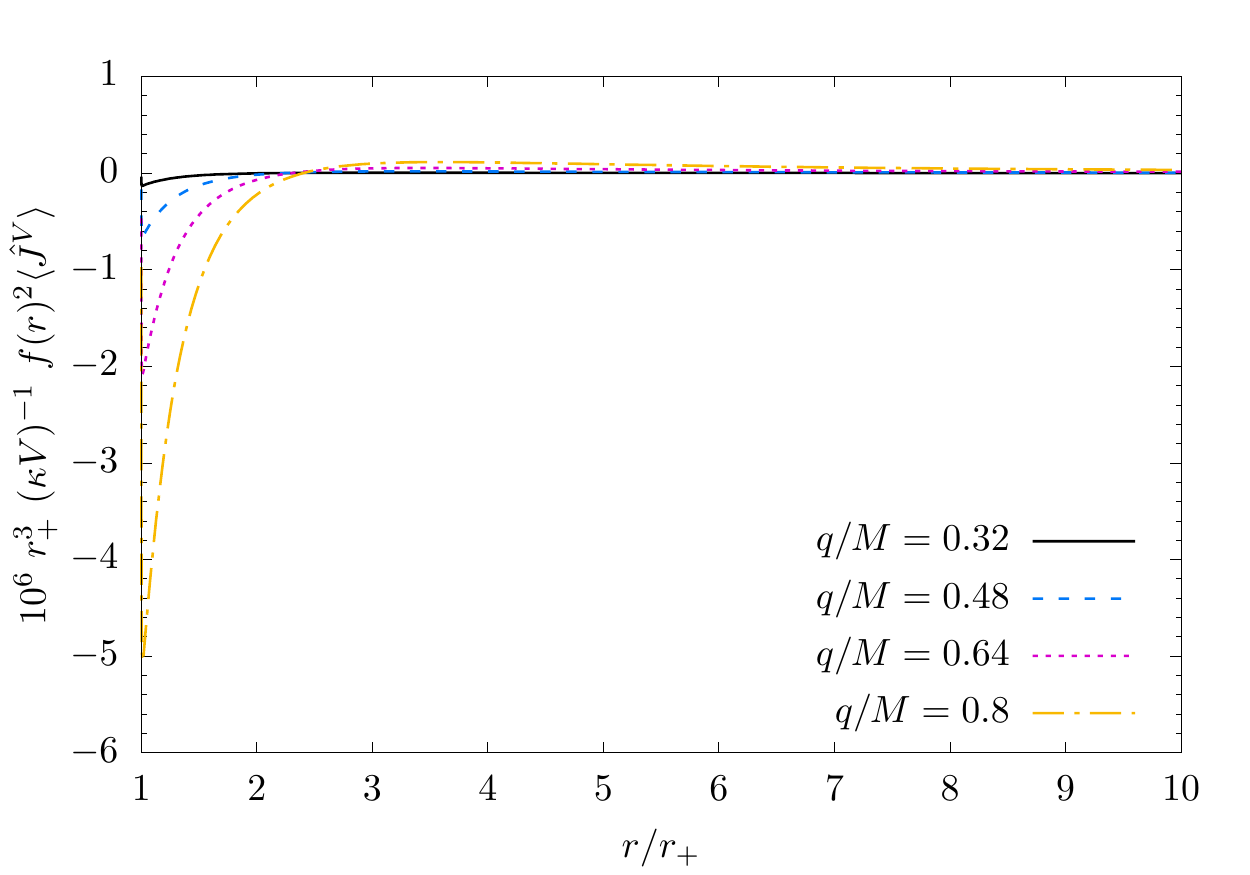}
	\\
	\includegraphics[scale=0.61]{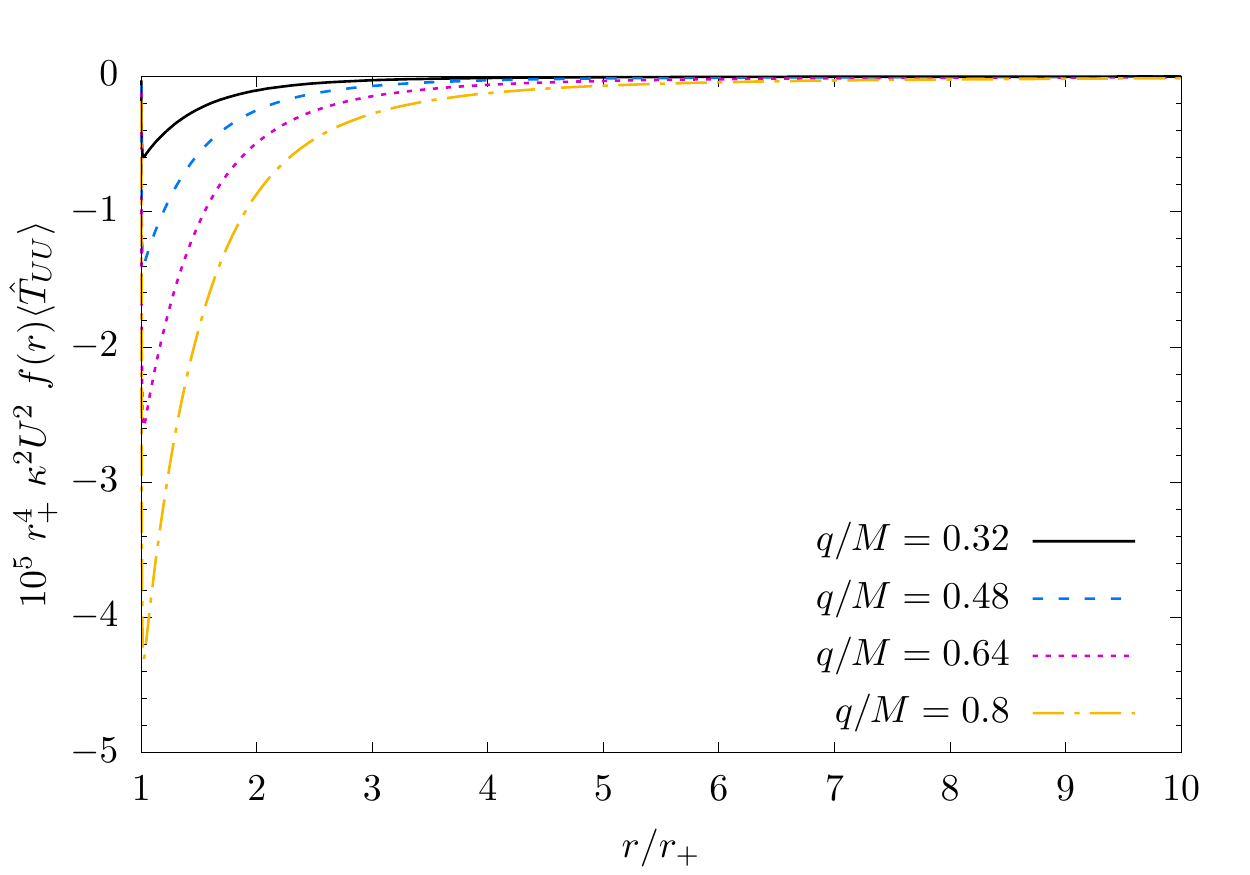}
	\includegraphics[scale=0.61]{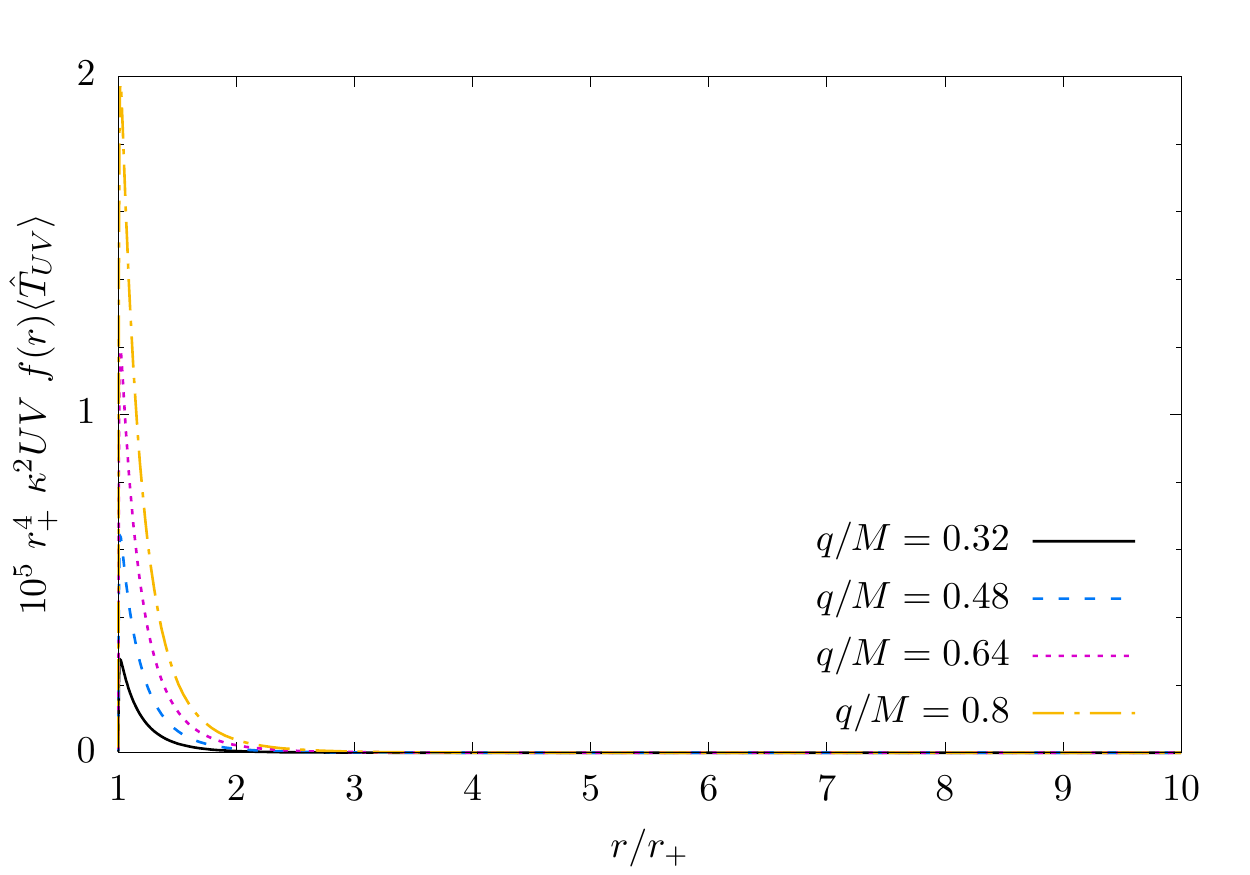}
	\\
	\includegraphics[scale=0.61]{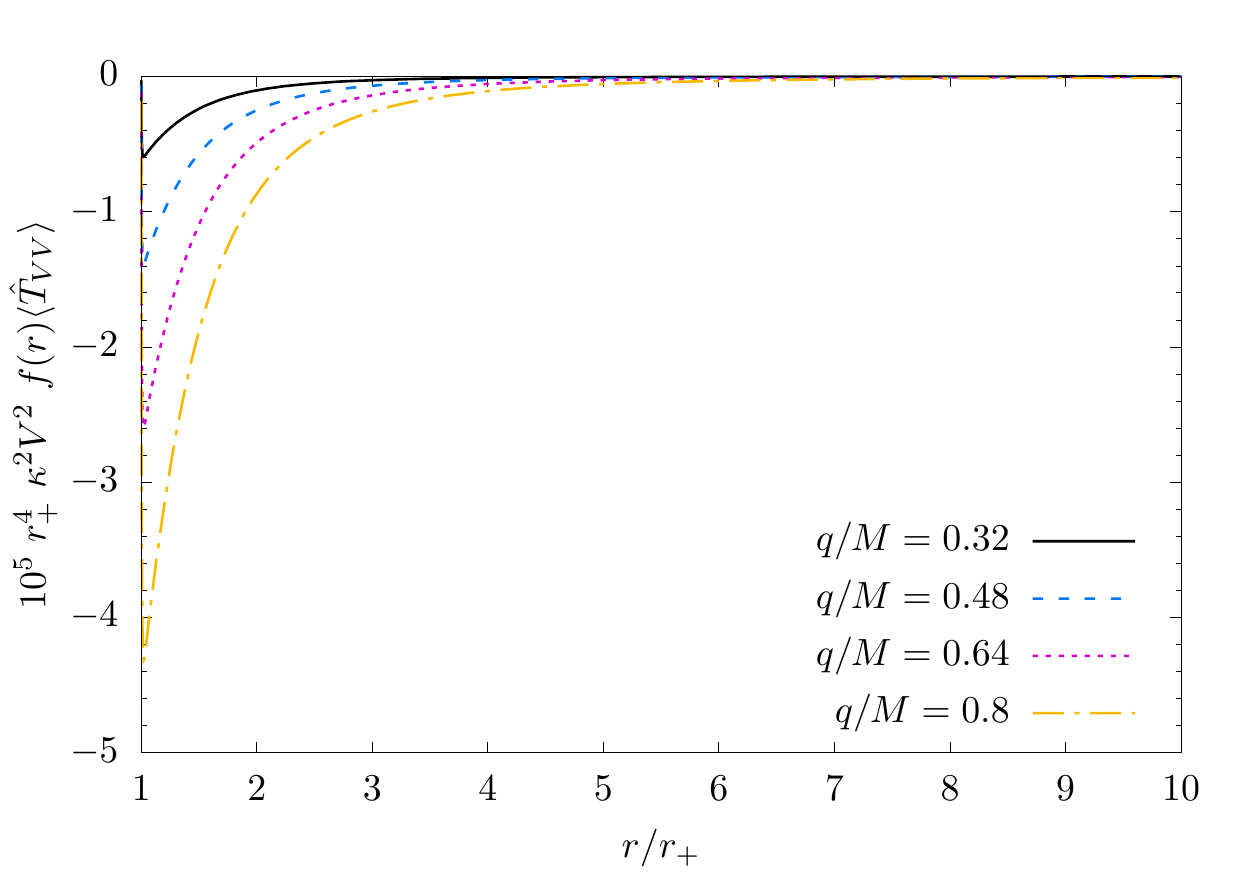}
	\vspace{.1cm}
	\caption{Difference in expectation values for the  components of the current and stress-energy tensor operators in Kruskal coordinates (\ref{eq:Kruskal}), between the tentative ``Boulware''-like state $| {\rm {B}}\rangle$ and the ``past'' Boulware state, $| {\rm {B}}^{-} \rangle$, in the spacetime of a RN black hole with $Q=0.8M$. All expectation values are multiplied by powers of $f(r)$ so that the resulting quantities are regular at $r=r_{+}$.}
	\label{fig:B-Bminus_Kruskal}
\end{figure}

The fluxes of charge and energy across past null infinity are given, as $r\rightarrow \infty $, by the components $J^{U}$, $T_{UU}$ and across future null infinity by $J^{V}$, $T_{VV}$.
These fluxes are found from the components (\ref{eq:BminusBpm}) with respect to Schwarzschild coordinates using the formulae (\ref{eq:JKruskal}, \ref{eq:TKruskal}).
At past null infinity, the differences in expectation values $\langle {\rm {B}} | {\hat {{J}}}^{U} | {\rm {B}} \rangle  
- \langle {\rm {B}}^{-} | {\hat {{J}}}^{U} | {\rm {B}}^{-} \rangle $
and  
$\langle {\rm {B}} | {\hat {{T}}}_{UU} | {\rm {B}} \rangle  
- \langle {\rm {B}}^{-} | {\hat {{T}}}_{UU} | {\rm {B}}^{-} \rangle $
vanish.
By construction, the ``past'' Boulware state has no incoming flux of particles at ${\mathscr{I}}^{-}$, and hence we deduce that the same is true for the Boulware state $|B\rangle$.
Similarly, at future null infinity, the differences in expectation values
$\langle {\rm {B}} | {\hat {{J}}}^{V} | {\rm {B}} \rangle  
- \langle {\rm {B}}^{+} | {\hat {{J}}}^{V} | {\rm {B}}^{+} \rangle $ 
and
$\langle {\rm {B}} | {\hat {{T}}}_{VV} | {\rm {B}} \rangle  
- \langle {\rm {B}}^{+} | {\hat {{T}}}_{VV} | {\rm {B}}^{+} \rangle $
also vanish, and we deduce that $|{\rm{B}}\rangle $ also has no outgoing flux of particles at ${\mathscr{I}}^{+}$.
It thus appears that we have succeeded in defining a state which is as empty as possible at both future and past null infinity.

We can also examine the expectation value of the scalar field condensate. As $r\rightarrow \infty $, this takes the form
\begin{equation}
	\langle {\rm {B}}| {\widehat {{SC}}} | {\rm {B}} \rangle  
	- \langle {\rm {B}}^{-} | {\widehat {{SC}}} | {\rm {B}}^{-} \rangle
	\sim  - \frac{1}{16 \pi^{2} r^{2}}  
	\sum _{\ell =0}^{\infty }
	\int _{\min \left\{\frac{qQ}{r_{+}},0\right\} } ^{\max\left\{\frac{qQ}{r_{+}},0\right\} } d\omega  \, \frac{1}{\left| \widetilde{\omega} \right|} \left( 2\ell + 1 \right) \left| B^{\rm {up}}_{\omega \ell } \right| ^{2} .
\end{equation}
From the Wronskian relations (\ref{eq:wronskians}), the integrand is finite when ${\widetilde {\omega }}=0$. 

Fig.~\ref{fig:B-Bminus} shows the differences in expectation values between the ``Boulware''-like state  $| {\rm {B}} \rangle$ and the ``past'' Boulware state $|{\rm {B}}^{-} \rangle$. 
When the scalar field charge $q=0$, these differences in expectation values vanish, as the two states are identical in this case. 
Unlike the two differences of ``past'' states considered in Sec.~\ref{sec:pastfuture}, the scalar field condensate for this difference of states is negative, decreasing with increasing scalar field charge. The scalar condensate tends to zero far from the black hole, but diverges as the horizon is approached.

The difference in expectation values of the radial component of the current  between the ``Boulware''-like state $| {\rm {B}} \rangle$ and the ``past'' Boulware state $| {\rm {B}} ^{-}\rangle$ is positive everywhere (compare with the previous two differences in states in Sec.~\ref{sec:pastfuture}, for which this quantity was negative). Since ${\mathcal {K}}_{\rm {B}}=0$ (\ref{eq:fluxesB}), only the ``past'' Boulware state contributes to this component of the current. From \cite{Balakumar:2020gli}, the ``past'' Boulware state has a positive flux of charge (we are considering only the case where both the black hole and the scalar field have positive charge), yielding a negative flux of charge for this difference between the ``Boulware''-like state $| {\rm {B}} \rangle$ and the ``past'' Boulware state $|{\rm {B}}^{-} \rangle$.

The charge density $\langle {{\hat {{J}}}}^{t}\rangle$ is negative near the horizon, but becomes small and positive further away, tending rapidly to zero far from the black hole. 
The results for $\langle {{\hat {{J}}}}^{r}\rangle$ presented in Fig.~\ref{fig:B-Bminus} are the negative of those presented in Ref.~\cite{Balakumar:2020gli} for the state $|{\rm {B}}^{-} \rangle$, as anticipated since ${\mathcal {K}}_{B}=0$.
While the ``Boulware''-like state $| {\rm {B}} \rangle$ is time-reversal invariant, the ``past'' Boulware state $|{\rm {B}}^{-} \rangle$ is not, and this is reflected in the  components $\langle {{\hat {{J}}}}^{U}\rangle$ and $\langle {{\hat {{J}}}}^{V}\rangle$ of the current, which can be seen in Fig.~\ref{fig:B-Bminus_Kruskal}. The component $U^{-1}\langle {{\hat {{J}}}}^{U}\rangle $ rapidly decreases to zero as $r\rightarrow \infty $, as expected (since $\langle {{\hat {{J}}}}^{U}\rangle$ vanishes at past infinity where $U\rightarrow \infty $). The component $V^{-1}\langle {{\hat {{J}}}}^{V}\rangle$ also tends to zero at infinity, but not as rapidly. Both $U^{-1}\langle {{\hat {{J}}}}^{U}\rangle$ and $V^{-1}\langle {{\hat {{J}}}}^{V}\rangle$ diverge at the horizon. 

The difference in expectation values of the component $\langle {\hat {{T}}}_t^r \rangle $ of the stress-energy tensor between the ``Boulware''-like and ``past'' Boulware states is simply minus that found in \cite{Balakumar:2020gli} for the ``past'' Boulware state, as expected since ${\mathcal {L}}_{\rm {B}}=0$ (\ref{eq:fluxesB}). The differences in expectation values of the  diagonal components of the SET between the states $| {\rm {B}} \rangle$ and $|{\rm {B}}^{-} \rangle$  all rapidly tend to zero far from the black hole. 
From Fig.~\ref{fig:B-Bminus_Kruskal}, the components of the difference in the current and SET expectation values in Kruskal coordinates all diverge on the event horizon.
Our intuitive expectation is that both the states $| {\rm {B}} \rangle$ and $|{\rm {B}}^{-} \rangle$ will diverge at the horizon. This means that either the ``past'' Boulware state $|{\rm {B}}^{-} \rangle$ diverges more rapidly than our tentative ``Boulware''-like state $| {\rm {B}} \rangle$ as the horizon is approached or that these two states diverge at the same rate, but with different coefficients. We suspect that the latter is more likely, although a computation of renormalized expectation values would be required to settle this question definitively. 

Our results indicate that our proposed ``Boulware''-like state $|{\mathrm {B}}\rangle $ is regular everywhere outside the event horizon, is an equilibrium state, and has no fluxes of charge or energy.
However, from the construction in Sec.~\ref{sec:Boulware}, it is not a conventional vacuum state, since its derivation involved operators satisfying nonstandard commutation relations (\ref{eq:Bccrs}). 
This result is analogous to that on Kerr space-time \cite{Ottewill:2000qh}, where it is shown that there is no vacuum state which is as empty as possible at both future and past null infinity. 
While our results for the state $|{\mathrm {B}}\rangle $ are intriguing, it remains to be seen whether this state can be constructed rigorously or whether the state is pathological in a manner not revealed by our computations.

\subsection{``Hartle-Hawking''-like states}
\label{sec:HHexp}

We now turn to the ``Hartle-Hawking''-like states constructed in Sec.~\ref{sec:HH}. 

\subsubsection{$| {\rm {FT}} \rangle $ state}

We first consider the state $| {\rm {FT}} \rangle $, and in particular the differences in expectation values
\begin{equation}
\langle {\rm {FT}} | {\hat {{O}}} | {\rm {FT}} \rangle  
- \langle {\rm {U}}^{-} | {\hat {{O}}} | {\rm {U}}^{-} \rangle
=
\sum _{\ell =0}^{\infty }\sum _{m=-\ell}^{\ell }
\int _{-\infty }^{\infty }d\omega  \,
\frac{1}{\exp \left| \frac{2\pi {\widetilde {\omega }}}{\kappa } \right| -1 } 
o^{\rm {in}}_{\omega \ell m}. 
\end{equation}
As $r\rightarrow r_{+}$, we find the expectation values of the current and stress-energy tensor take the asymptotic forms:
\begin{subequations}
	\label{eq:FTUminus}
	\begin{align}
		\langle {\rm {FT}} | {\hat {{J}}}^{\mu } | {\rm {FT}} \rangle  
		- \langle {\rm {U}}^{-} | {\hat {{J}}}^{\mu } | {\rm {U}}^{-} \rangle
		\sim & \frac{q}{64 \pi^3 r^2} \sum_{\ell = 0}^\infty  \int_{- \infty}^\infty d \omega \frac{{\widetilde {\omega }}\left( 2\ell + 1 \right)}{\left| \omega \right| \left( \exp \left| \frac{2 \pi \widetilde{\omega}}{\kappa} \right| - 1 \right)} {\left| B^{\mathrm{in}}_{\omega \ell} \right|}^2  {\left( - f(r)^{-1}, 1, 0, 0 \right)}^\intercal ,
		\\
		\langle {\rm {FT}}| {\hat {{T}}}^{\mu }_{\nu } | {\rm {FT}} \rangle  
		- \langle {\rm {U}}^{-} | {\hat {{T}}}^{\mu }_{\nu } | {\rm {U}}^{-} \rangle
		\sim & \frac{1}{16 \pi ^{2} r^2} \sum_{\ell = 0}^\infty  \int_{- \infty}^\infty d \omega \frac{{\widetilde{\omega}}^{2}\left( 2\ell + 1 \right)}{ \left| \omega \right| \left( \exp \left| \frac{2 \pi \widetilde{\omega}}{\kappa} \right| - 1 \right)} { \left| B^{\mathrm{in}}_{\omega \ell} \right| }^2  
		\nonumber \\ 
		& \qquad \qquad  \times \begin{pmatrix}
  - f(r)^{-1} & - f(r)^{-2} & 0 & 0 \\
  1 & f(r)^{-1} & 0 & 0 \\
  0 & 0 & \mathcal{O} \left( 1 \right) & 0 \\
  0 & 0 & 0 & \mathcal{O} \left( 1 \right)
\end{pmatrix} .
	\end{align}
\end{subequations}
The integrands in (\ref{eq:FTUminus}) are regular at both $\omega =0$ and ${\widetilde {\omega }}=0$ from the Wronskian relations (\ref{eq:wronskians}). 
Furthermore, by considering the components of the current and stress-energy tensor in Kruskal coordinates, we find that the leading order divergences in the expectation values (\ref{eq:FTUminus}) cancel on the future horizon but not on the past horizon.
Given that the ``past'' Unruh state is anticipated to be regular on the future horizon but not the past horizon, this implies that the expectation values of the current in the state $|{\mathrm {FT}}\rangle $ will be regular at the future horizon, but that there may be a mild divergence in the SET. 

Combining the relevant components of (\ref{eq:FTUminus}) with (\ref{eq:fluxesUminus}), and using the Wronskian relations (\ref{eq:wronskians}), we find the fluxes of charge and energy in the state $|{\mathrm {FT}}\rangle $ to be
\begin{subequations}
	\label{eq:FTfluxes}
	\begin{align}
		{\mathcal {K}}_{\mathrm {FT}} & = 
		\frac{q}{64\pi ^{3}} \sum _{\ell =0}^{\infty }
		\int _{\min \left\{\frac{qQ}{r_{+}},0\right\} } ^{\max\left\{\frac{qQ}{r_{+}},0\right\} } d\omega  \,
		\frac{\omega \left( 2\ell + 1 \right)}{\left| {\widetilde {\omega }} \right| } \coth \left| \frac{\pi {\widetilde {\omega }}}{\kappa }\right| \left| {{B}}_{\omega \ell }^{\mathrm {up}} \right| ^{2} ,
		\\
		{\mathcal {L}}_{\mathrm {FT}} & = 
		\frac{1}{16\pi ^{2}} \sum _{\ell =0}^{\infty }
		\int _{\min \left\{\frac{qQ}{r_{+}},0\right\} } ^{\max\left\{\frac{qQ}{r_{+}},0\right\} } d\omega  \,
		\frac{\omega ^{2} \left( 2\ell + 1 \right)}{\left| {\widetilde {\omega }} \right| } \coth \left| \frac{\pi {\widetilde {\omega }}}{\kappa }\right| \left| {{B}}_{\omega \ell }^{\mathrm {up}} \right| ^{2} .
	\end{align}
	\end{subequations}
Note that neither of these is zero when $q\neq 0$, so the state $|{\mathrm {FT}}\rangle$ is not an equilibrium state and is not time-reversal invariant.
This is in contrast to the situation on a rotating Kerr black hole, where it is argued in \cite{Frolov:1989jh,Ottewill:2000qh} that the Frolov-Thorne state is an equilibrium state.
Both the fluxes (\ref{eq:FTfluxes}) involve contributions from the superradiant modes only (and vanish when the scalar field charge $q=0$).
 The flux of charge has the same sign as the black hole charge $Q$, and hence the black hole is losing charge. 
The flux of energy ${\mathcal{L}}_{\mathrm {FT}}$ is always positive, and therefore the black hole is also losing energy in this state.

\begin{figure}
 \includegraphics[scale=0.61]{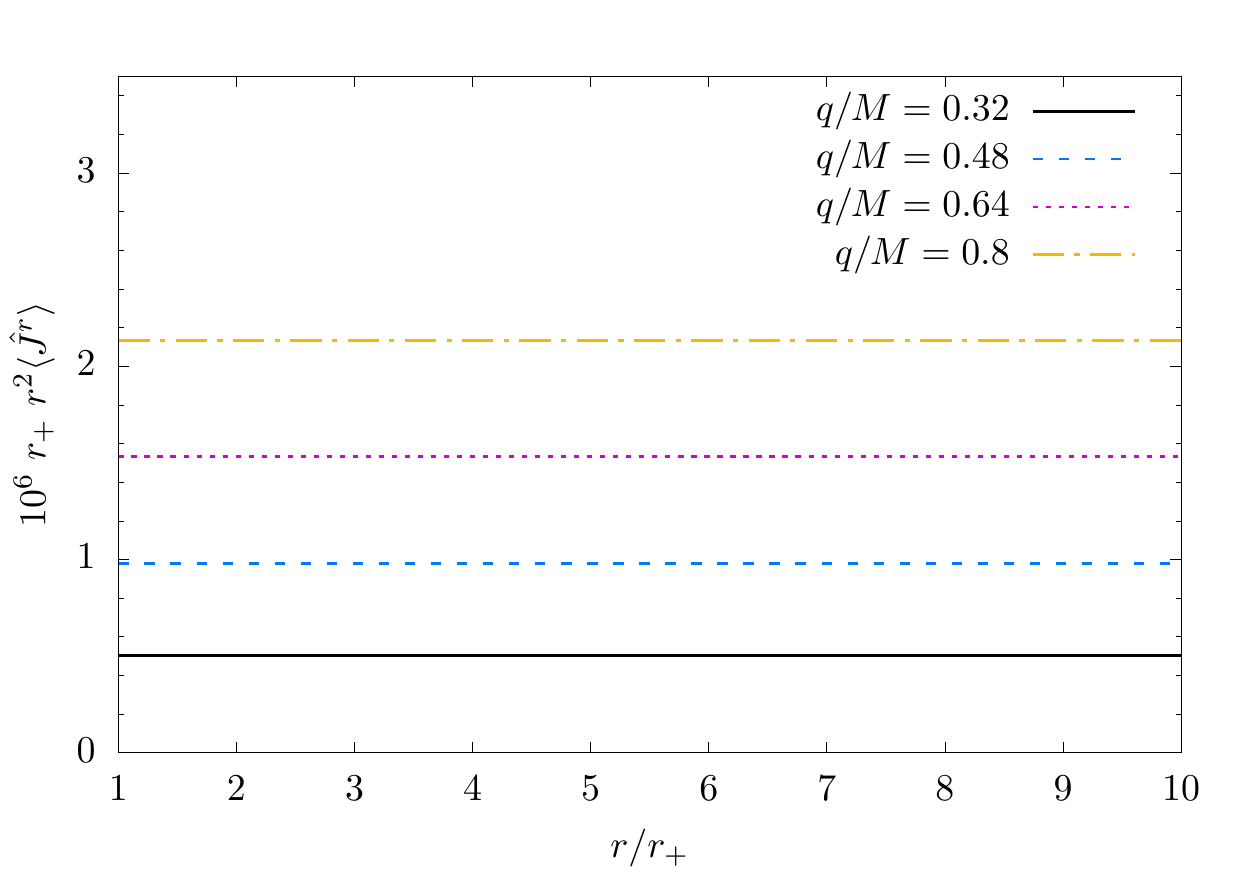}
  \includegraphics[scale=0.61]{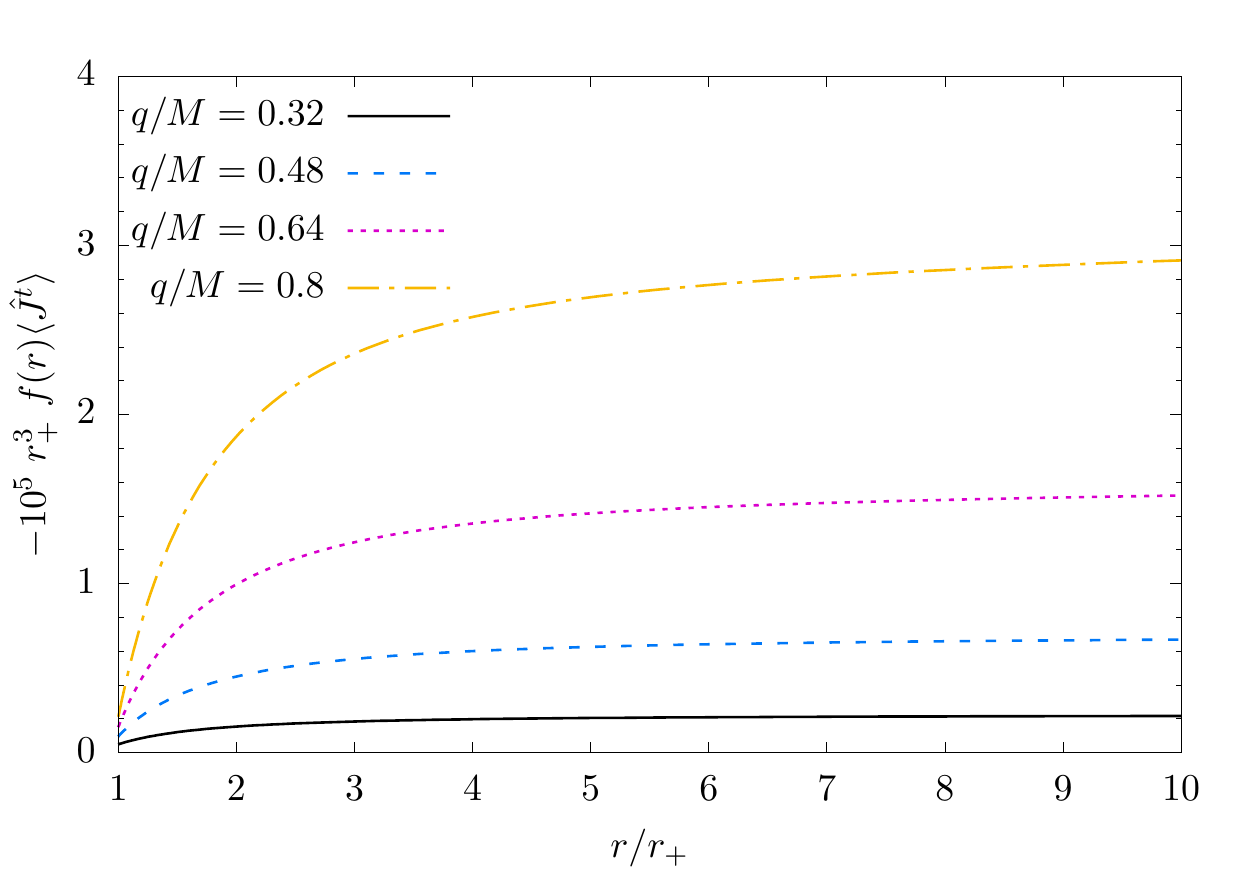}
   \\ \includegraphics[scale=0.61]{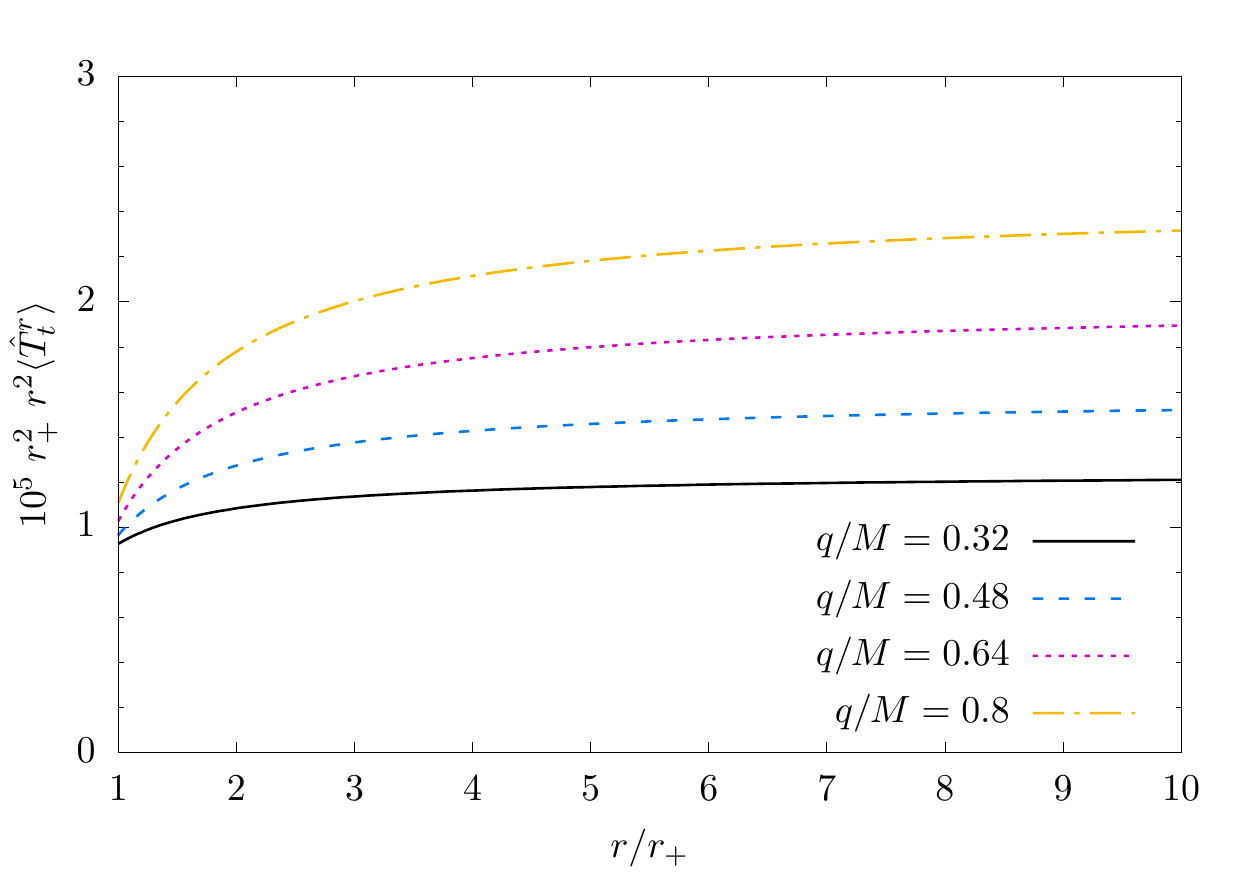}
  \includegraphics[scale=0.61]{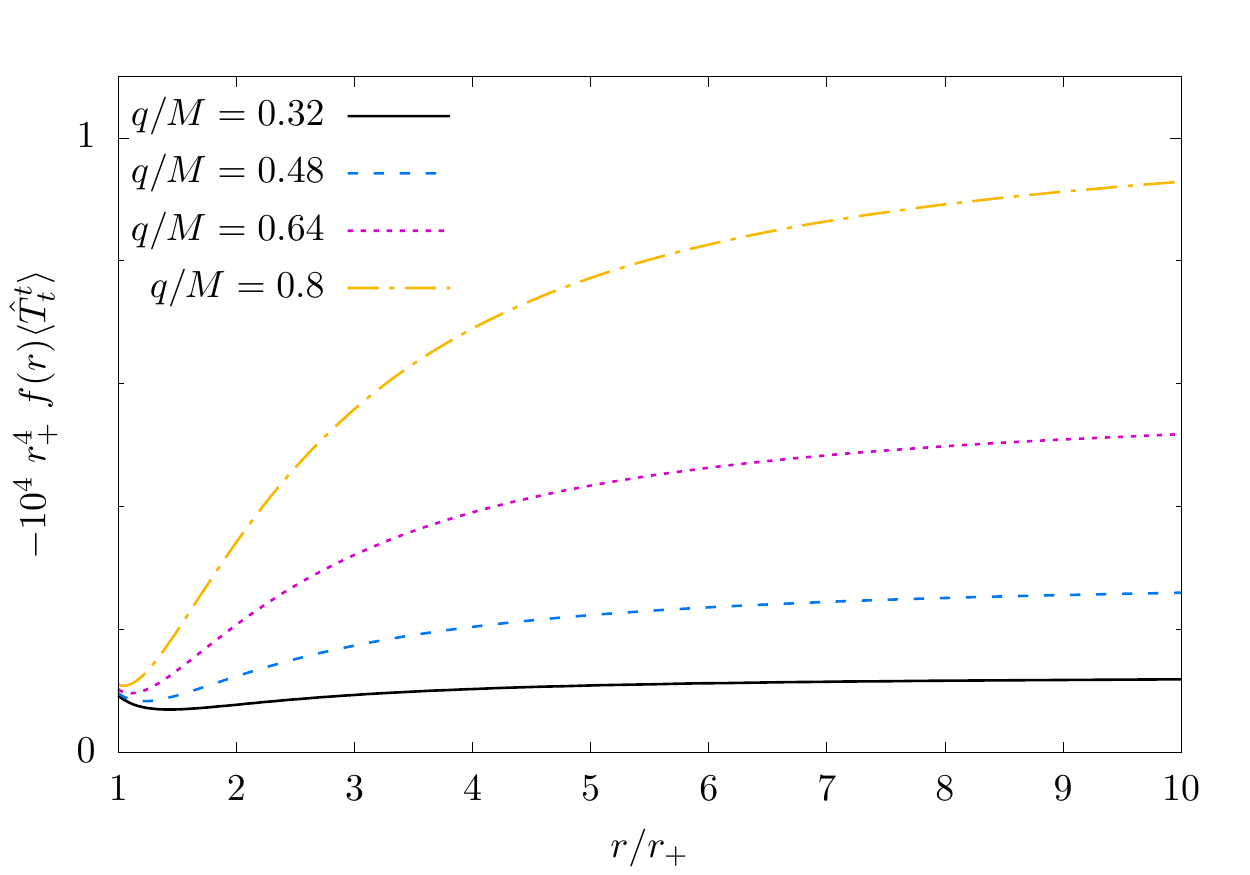} \\
  \includegraphics[scale=0.61]{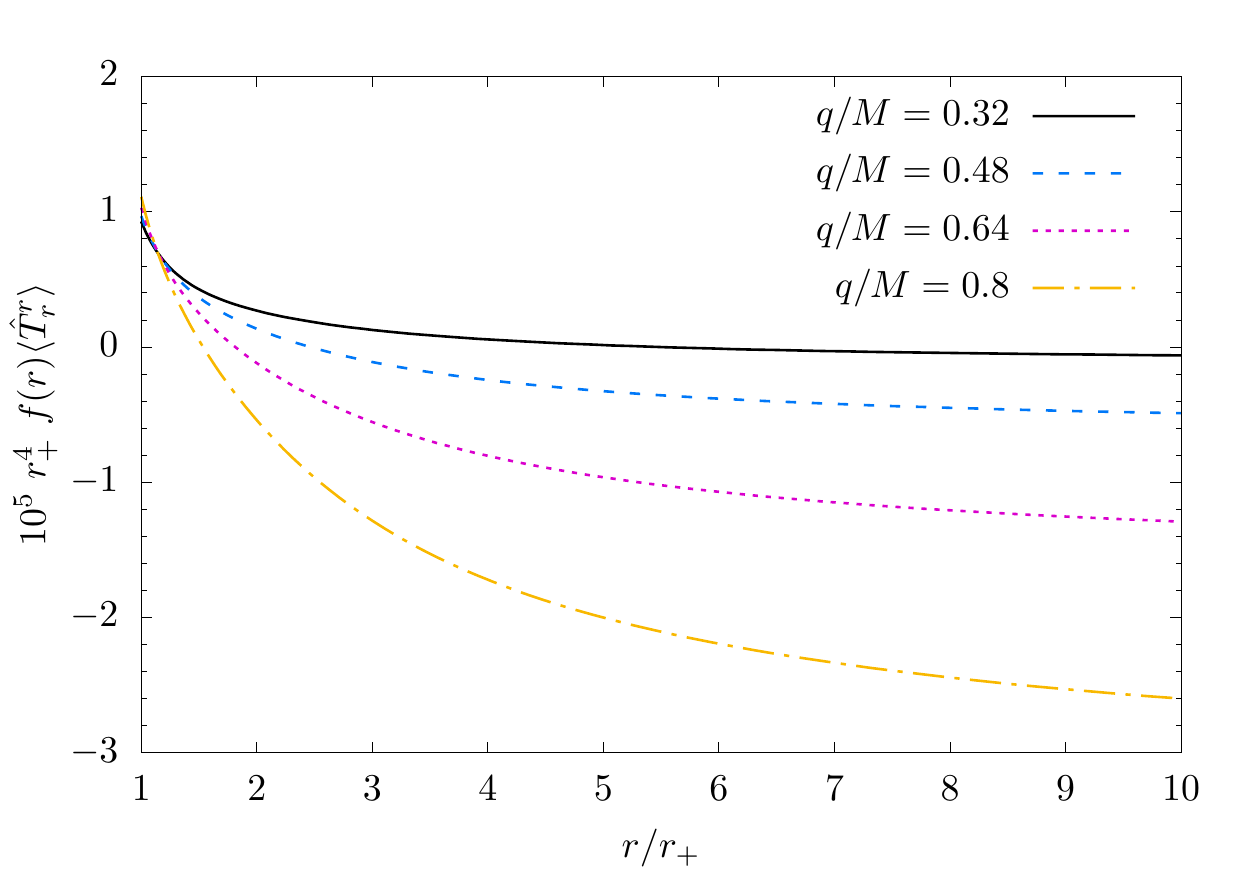}
  \includegraphics[scale=0.61]{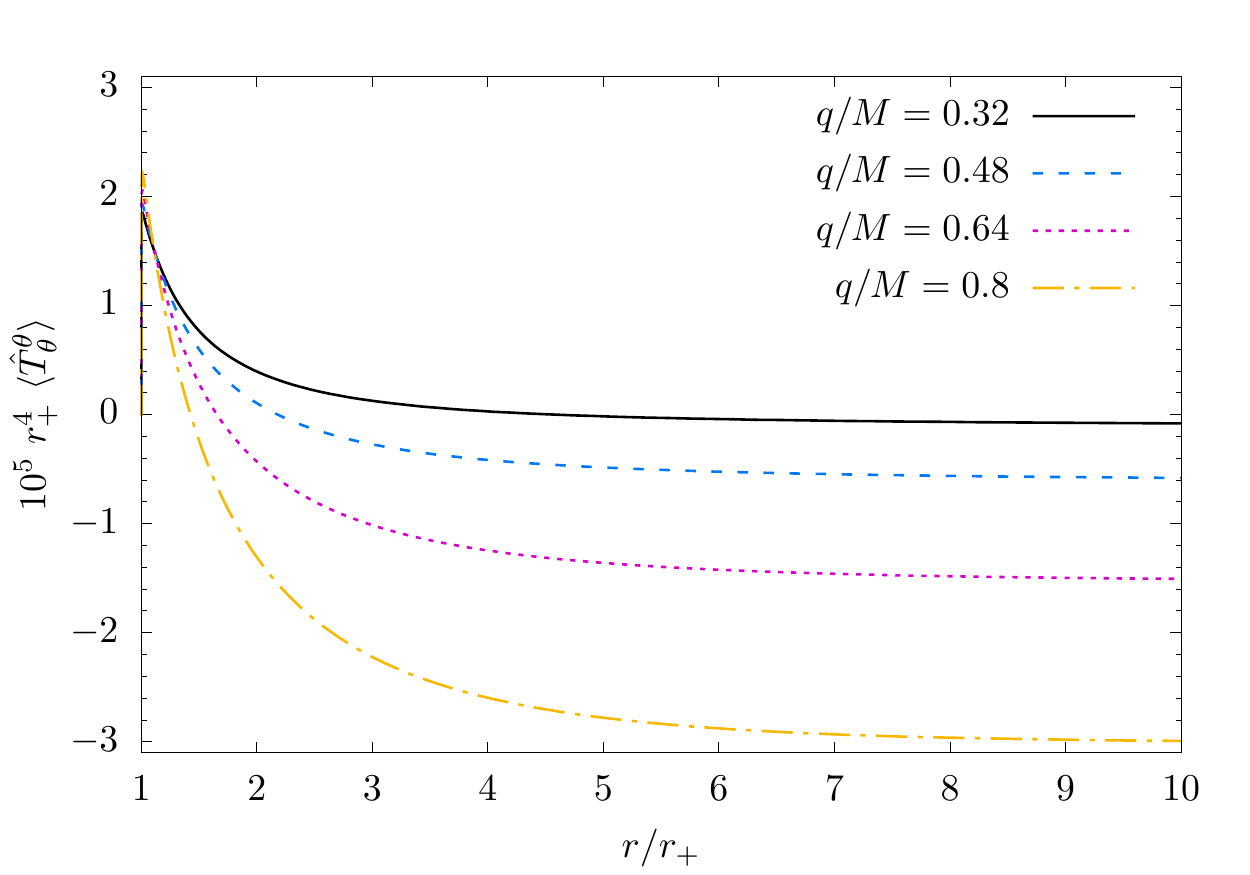} \\
  \vspace{.1cm}
  \caption{Difference in expectation values for the scalar condensate operator and components of the current and stress-energy tensor operators, between the state $| {\rm {FT}} \rangle$ and the ``past'' Unruh state, $|{\rm {U}}^{-} \rangle$, in the spacetime of a RN black hole with $Q=0.8M$. All quantities are multiplied by powers of $f(r)$ so that the resulting quantities are regular at $r=r_{+}$.}
  \label{fig:FT-Uminus}
\end{figure}

\begin{figure}
	\includegraphics[scale=0.61]{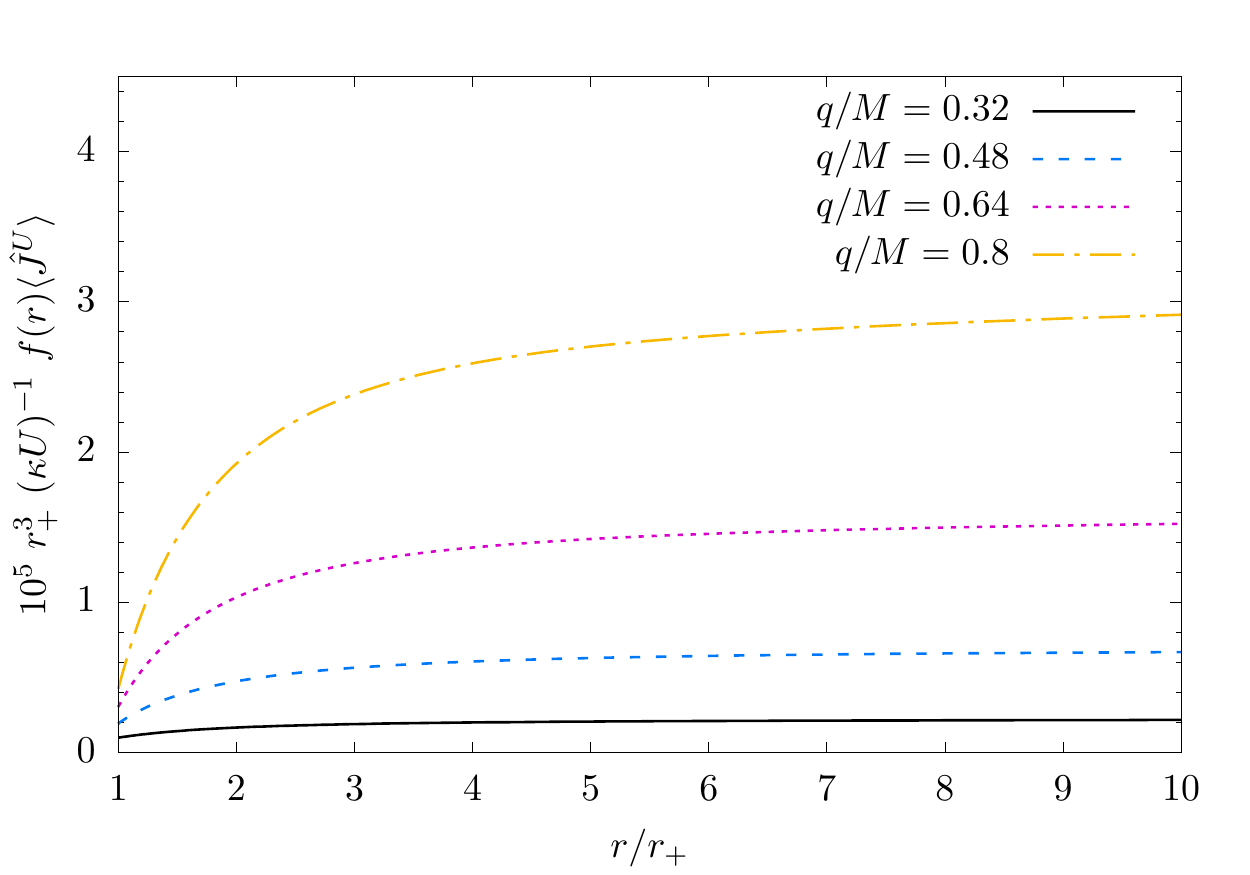}
	\includegraphics[scale=0.61]{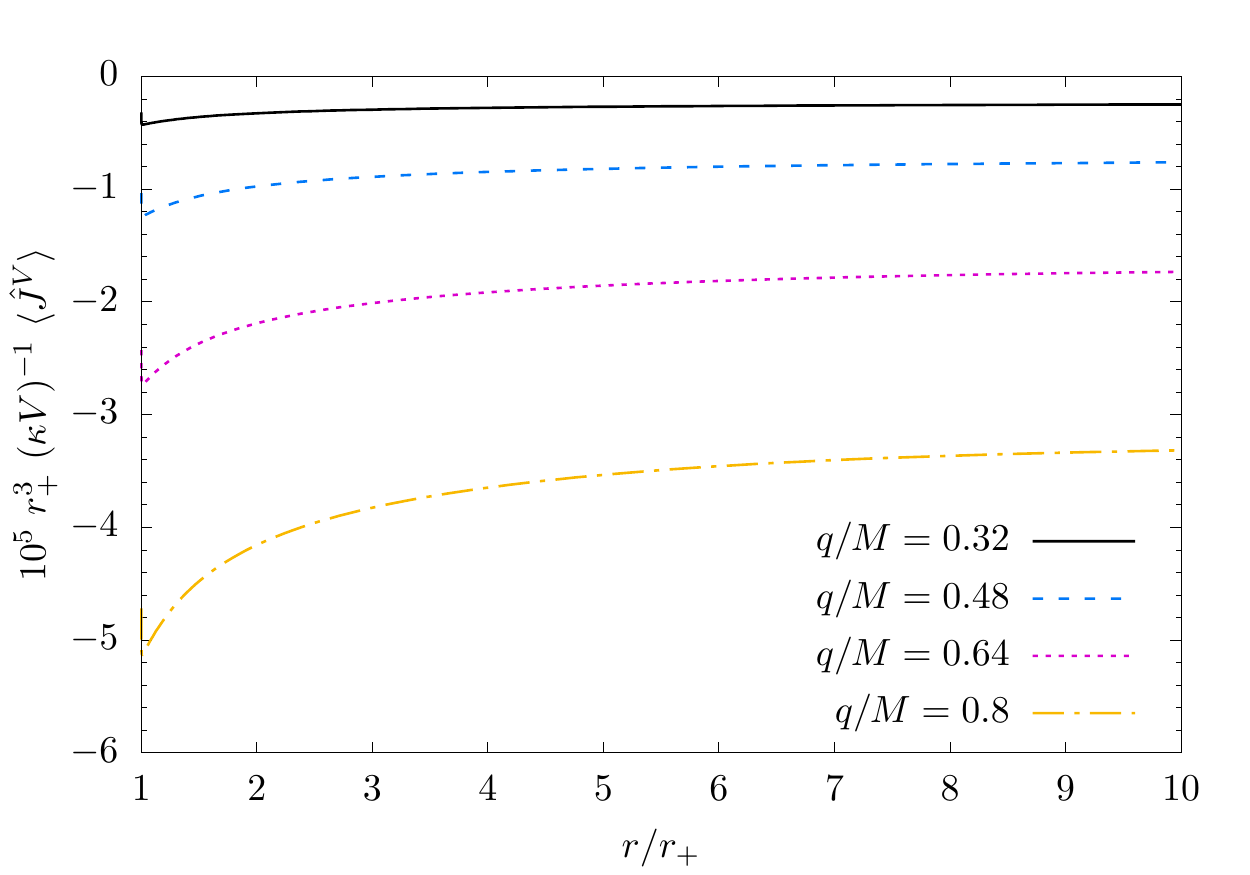}
	\\
	\includegraphics[scale=0.61]{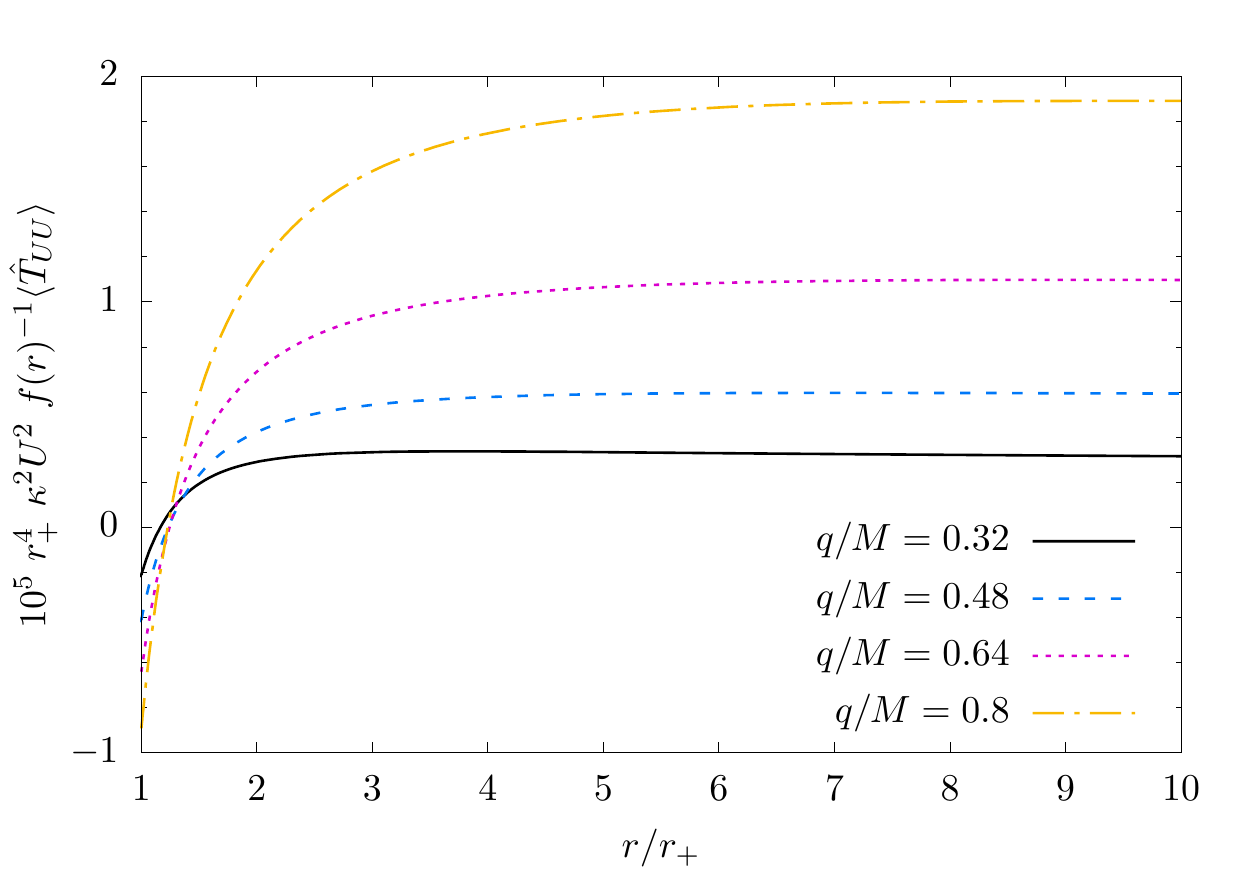}
	\includegraphics[scale=0.61]{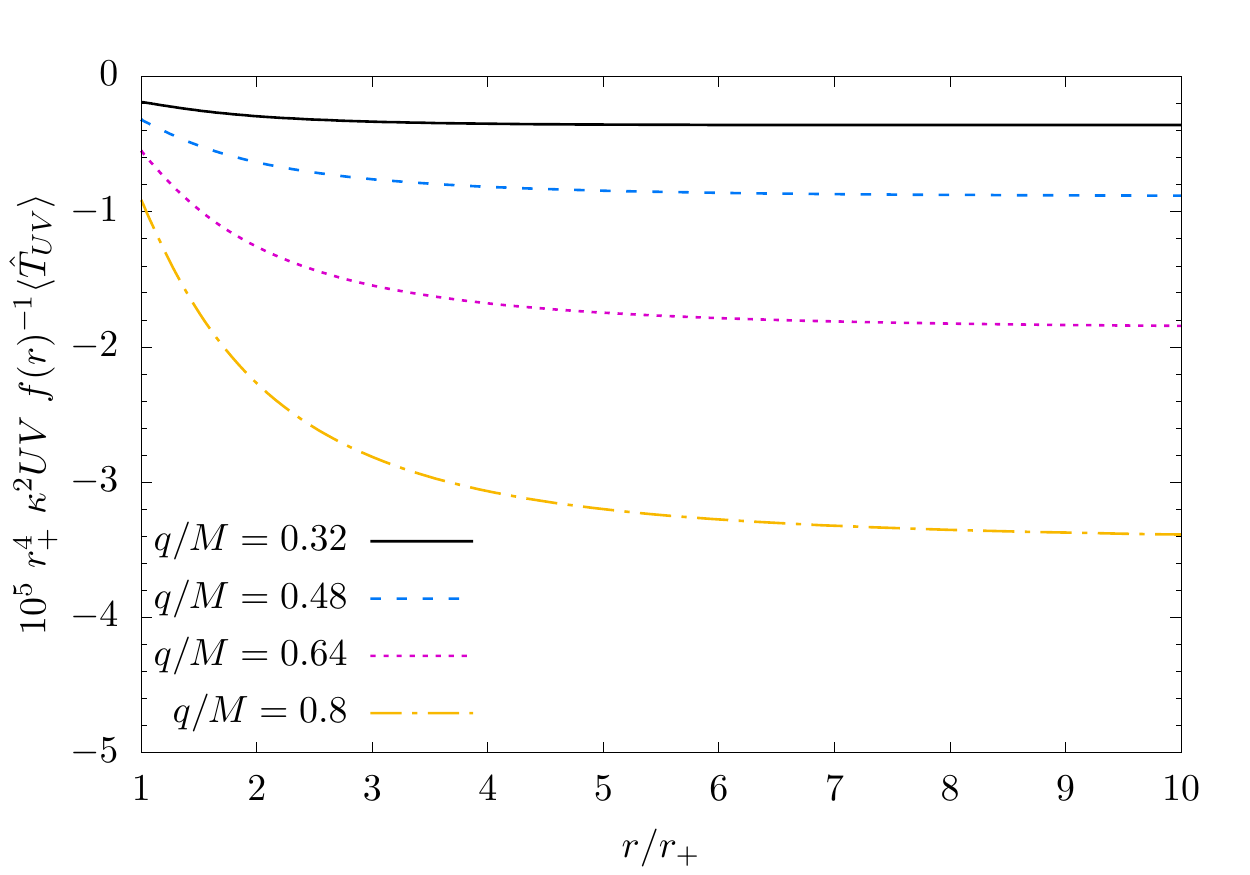}
	\\
	\includegraphics[scale=0.61]{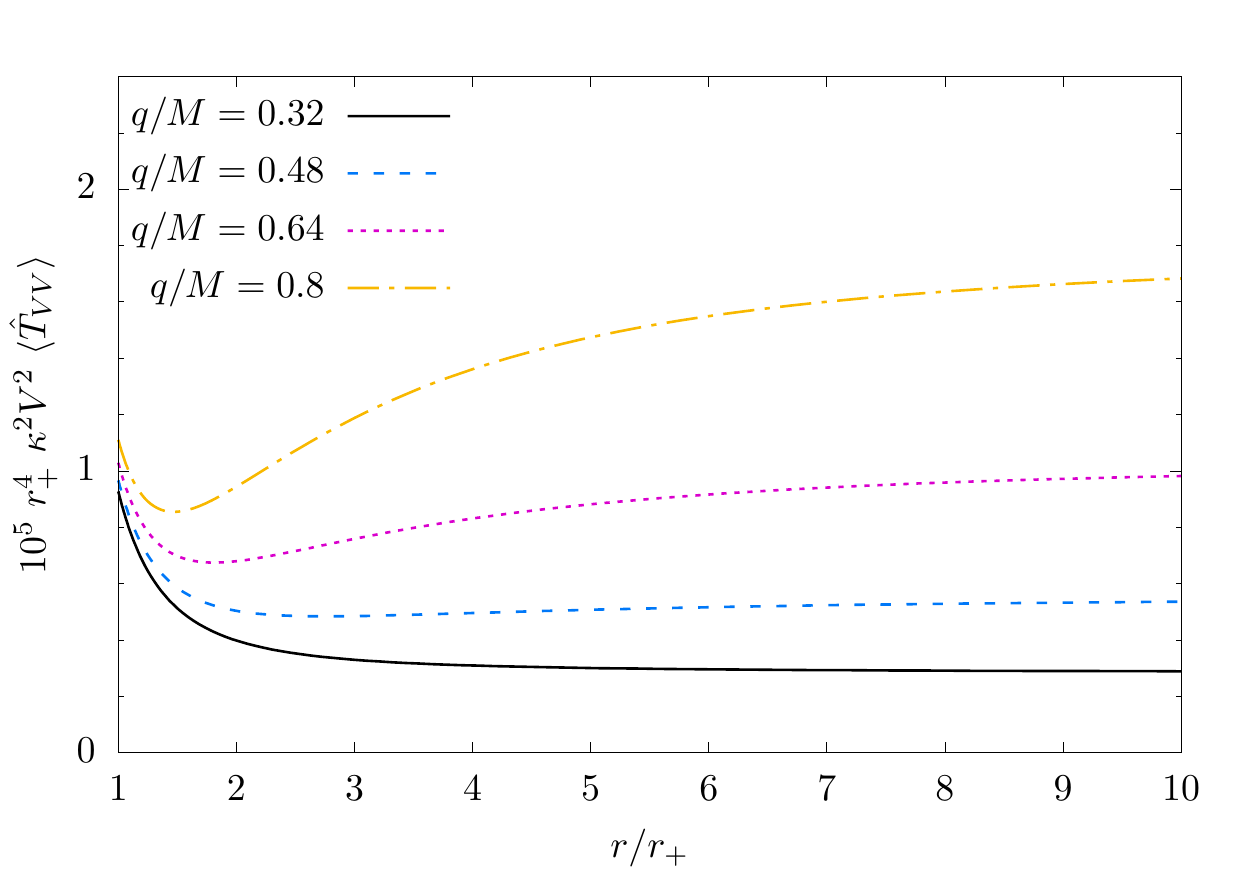}
	\vspace{.1cm}
	\caption{Difference in expectation values for the scalar condensate operator and components of the current and stress-energy tensor operators in Kruskal coordinates (\ref{eq:Kruskal}), between the state $| {\rm {FT}} \rangle$ and the ``past'' Unruh state, $|{\rm {U}}^{-} \rangle$, in the spacetime of a RN black hole with $Q=0.8M$. All quantities are multiplied by powers of $f(r)$ so that the resulting quantities are regular at $r=r_{+}$.}
	\label{fig:FT-Uminus_Kruskal}
\end{figure}

In Fig.~\ref{fig:FT-Uminus} we show the differences in expectation values of the current and SET between the $| {\rm {FT}} \rangle$ state and the ``past'' Unruh state 
$|{\rm {U}}^{-} \rangle$. We will return to the expectation value of the scalar field condensate below. 

From the radial component of the current, we see that the difference in charge flux ${\mathcal  {K}}_{\rm {FT}}-{\mathcal {K}}_{{\rm {U}}^{-}}$ is negative, in contrast to the quantity ${\mathcal  {K}}_{{\rm {CCH}}^{-}}-{\mathcal {K}}_{{\rm {U}}^{-}}$, (see Fig.~\ref{fig:CCHminus-Uminus}) which is positive. Since the charge flux  ${\mathcal  {K}}_{\rm {FT}}$ (\ref{eq:FTfluxes}) contains contributions only from the superradiant modes, this suggests that ${\mathcal  {K}}_{\rm {FT}}$ is small compared to the charge flux in the ``past'' Unruh state ${\mathcal {K}}_{{\rm {U}}^{-}}$ (which contains an outgoing flux of Hawking radiation). 

The difference in expectation values of the charge density $\langle {\hat {{J}}}^{t} \rangle$ between the states $| {\rm {FT}} \rangle$ and  $|{\rm {U}}^{-} \rangle$ also has the opposite sign compared the difference in expectation values between the ``past'' CCH and Unruh states, shown in Fig.~\ref{fig:CCHminus-Uminus}. 
Turning to the components of the current in Kruskal coordinates, shown in Fig.~\ref{fig:FT-Uminus_Kruskal}, we see that $U^{-1}f(r)\langle {\hat {{J}}}^{U} \rangle $ and $V^{-1}\langle {\hat {{J}}}^{V} \rangle $ are regular and nonzero at the horizon. 
 As for the difference between the ``past'' CCH and Unruh states, we deduce that the current is regular across the future horizon where $U=0$ but not the past horizon where $V=0$. 
Assuming that the current in the ``past'' Unruh state is regular across the future horizon, we deduce that the same is true for the current in the $| {\rm {FT}} \rangle$ state.

The difference in expectation values of the component $\langle {\hat {{T}}}_t^r \rangle $ between the state $| {\rm {FT}} \rangle$ and the ``past'' Unruh state is positive, as was found to be the case for the difference in expectation values of this component of the SET between the ``past'' CCH and Unruh states. Therefore the quantity ${\mathcal {L}}_{\rm {FT}}-{\mathcal {L}}_{{\rm {U}}^{-}}$ is negative.  As with the flux of charge discussed above, this result for the energy flux makes physical sense, given that ${\mathcal {L}}_{\rm {FT}}$ arises from a sum over superradiant modes only, while ${\mathcal {L}}_{{\rm {U}}^{-}}$ contains the flux of energy from the Hawking radiation in all field modes. 

The differences in expectation values of the diagonal components of the SET between  the $| {\rm {FT}} \rangle$ and $|{\rm {U}}^{-} \rangle$ states appear to tend to a constant as $r\rightarrow \infty$, at least for small values of the scalar field charge (for larger values of $q$ we would need to consider rather larger values of the radial coordinate $r$ to see this behaviour clearly). 
The difference in expectation values of the component $\langle {\hat {{T}}}_t^t\rangle $  between these two states is negative everywhere outside the horizon, as was the case for the difference between the states $|{\rm {CCH}}^{-} \rangle$ and $|{\rm {U}}^{-} \rangle$. 
	In contrast, far from the black hole, we see that the differences in expectation values of the components $\langle {\hat {{T}}}_r^r\rangle $ and $\langle {\hat {{T}}}_{\theta }^{\theta }\rangle $ between the states   $| {\rm {FT}} \rangle$ and $|{\rm {U}}^{-} \rangle$ have the opposite signs to those observed for the differences between the states $|{\rm {CCH}}^{-} \rangle$ and $|{\rm {U}}^{-} \rangle$. 
			
We now examine the differences in expectation values of the components of the SET between the states $| {\rm {FT}} \rangle$ and $|{\rm {U}}^{-} \rangle$ in Kruskal coordinates, shown in Fig.~\ref{fig:FT-Uminus_Kruskal}. 
First, we see that the difference in expectation values of the quantity $V^{2}\langle {\hat {{T}}}_{VV}\rangle $ is finite and nonzero on the horizon, so  we deduce that the difference in expectation values of the component $\langle {\hat {{T}}}_{VV}\rangle $ diverges on the past horizon where $V=0$.
 In addition, the difference in expectation values of $U^{2}f(r)^{-1}\langle {\hat {{T}}}_{UU}\rangle $ and $UVf(r)^{-1}\langle {\hat {{T}}}_{UV}\rangle $ are finite and nonzero when $r=r_+$.
As for the differences in expectation values between the ``past'' CCH and the ``past'' Unruh states, this means that $\langle {\hat {{T}}}_{UV}\rangle $ is finite on both the future horizon and the past horizon, but that  $\langle {\hat {{T}}}_{UU}\rangle$, while regular on the past horizon, has a mild divergence on the future horizon. 
Since the ``past'' Unruh state $|{\rm {U}}^{-} \rangle$ is expected to be regular on the future (but not the past) horizon, we deduce that the state  $| {\rm {FT}} \rangle$ has a mild divergence on the future horizon. We are unable to make any deductions about its regularity on the past horizon. 
 
From our study of the differences in expectation values between the states 	$| {\rm {FT}} \rangle$ and 	$|{\rm {U}}^{-} \rangle$ of the current and SET, we conclude that the ``Hartle-Hawking''-like state $| {\rm {FT}} \rangle$ is very different physically from the ``past'' CCH state $|{\rm {CCH}}^{-} \rangle$.  The difference in how these states are defined lies in the thermal factor associated with the ``in'' modes. This clearly has a large impact on the expectation values of observables.
However, these two states do share some physical features. For example, at infinity, neither the 	$| {\rm {FT}} \rangle$  state nor the ``past'' CCH state are empty.
			
There is however one more observable that we must consider in our discussion of differences in expectation values between the  $| {\rm {FT}} \rangle$ and $|{\rm {U}}^{-} \rangle$ states, namely the scalar condensate, which is not shown in Fig.~\ref{fig:FT-Uminus}.
Near the horizon $r\rightarrow r_{+}$, we have the asymptotic form:
\begin{equation}
	\langle {\rm {FT}} | {\widehat {{SC}}} | {\rm {FT}} \rangle  
- \langle {\rm {U}}^{-} | {\widehat {{SC}}} | {\rm {U}}^{-} \rangle
\sim  \frac{1}{16 \pi ^{2} r^{2}}  
\sum _{\ell =0}^{\infty }
\int _{-\infty }^{\infty }d\omega \, \frac{ 2\ell + 1 }{\left| \omega \right| \left( \exp \left| \frac{2\pi {\widetilde {\omega }}}{\kappa } \right| -1 \right) }  \left| B^{\rm {in}}_{\omega \ell } \right| ^{2} .
\end{equation}
Using the Wronskian relations (\ref{eq:wronskians}), the integrand is finite at $\omega =0$ but diverges at ${\widetilde {\omega }}=0$. 
We therefore conclude that the expectation value of the scalar field condensate in the state $|{\mathrm {FT}} \rangle $ diverges at the horizon (assuming that the expectation value of the scalar condensate in the ``past'' Unruh state $|{\rm {U}}^{-} \rangle$ is regular there).
Away from the horizon, the expectation value of the scalar field condensate is given by
\begin{equation}
	\langle {\rm {FT}} | {\widehat {{SC}}} | {\rm {FT}} \rangle  
	- \langle {\rm {U}}^{-} | {\widehat {{SC}}} | {\rm {U}}^{-} \rangle
	=
	\sum _{\ell =0}^{\infty }\sum _{m=-\ell}^{\ell }
	\int _{-\infty }^{\infty }d\omega  \,
	\frac{1}{\exp \left| \frac{2\pi {\widetilde {\omega }}}{\kappa } \right| -1 } 
	\left| \phi ^{\rm {in}}_{\omega \ell m} \right|^{2}. 
\end{equation}
The integrand has a pole when ${\widetilde {\omega }}=0$ unless the magnitude of the ``in'' modes vanishes at this frequency.
Numerical investigations reveal that there is at least one ``in'' mode whose magnitude is nonzero when ${\widetilde {\omega }}=0$, and therefore the difference in expectation values of the scalar field condensate between the $|{\mathrm {FT}} \rangle $ and $|{\rm {U}}^{-} \rangle$ states is in fact divergent everywhere outside the event horizon as well.
We therefore deduce that the state $| {\mathrm {FT}} \rangle $ is ill-defined, even though the expectation values of the current and SET appear to be well-behaved in this state.
Similar conclusions were reached on Kerr space-time \cite{Ottewill:2000qh}, namely that the (original) Frolov-Thorne state was ill-defined almost everywhere in the space-time. In particular, the expectation value of the scalar condensate in the Frolov-Thorne state on Kerr is divergent, but there is evidence that the expectation value of the SET (the work \cite{Ottewill:2000qh} considers only a neutral scalar field) is well-behaved, at least close to the horizon. 
In Kerr space-time, on the axis of symmetry the superradiant modes do not contribute \cite{Ottewill:2000qh} and the Frolov-Thorne state reduces to the ``past'' CCH state on this axis. 
In the situation we consider here, namely a charged scalar field on an RN black hole, the superradiant modes contribute everywhere in the space-time exterior to the event horizon and our state $|{\mathrm {FT}} \rangle $ is badly-behaved everywhere outside and on the horizon.

\subsubsection{$| {\rm {H}} \rangle $ state}

Finally, we examine the state $| {\rm {H}}\rangle $. 
As with the other states studied in this paper, we begin with asymptotic expressions.
Comparing (\ref{eq:Hexp}, \ref{eq:FTexp2}), we see that it is convenient to consider
the differences 
\begin{equation}
 \langle {\rm {H}} | {\hat {{O}}} | {\rm {H}} \rangle  
 - \langle {\rm {FT}} | {\hat {{O}}} | {\rm {FT}} \rangle 
=
-
\sum _{\ell =0}^{\infty }\sum _{m=-\ell}^{\ell }
\int _{\min \left\{\frac{qQ}{r_{+}},0\right\} } ^{\max\left\{\frac{qQ}{r_{+}},0\right\} } d\omega  \,
o^{\rm {in}}_{\omega \ell m}
\coth \left| \frac{\pi {\widetilde{\omega }}}{\kappa } \right|  ,
\label{eq:HHminusFT}
\end{equation}
which have contributions only from the superradiant ``in'' modes, as may be expected from the construction of these states in Sec.~\ref{sec:HH}.
In particular, since our analysis above provides evidence that the expectation values of the current and SET are well-defined in the $|{\mathrm {FT}}\rangle $ state, we may consider differences in expectation values of these two quantities between the  $|{\mathrm {H}}\rangle $  and $| {\rm {FT}}\rangle $ states.

 Near the horizon, we find
\begin{subequations}
	\label{eq:HminusFT}
	\begin{align}
		\langle {\rm {H}} | {\hat {{J}}}^{\mu } | {\rm {H}} \rangle  
		- \langle {\rm {FT}}  | {\hat {{J}}}^{\mu } | {\rm {FT}}  \rangle
		\sim & - \frac{q}{64 \pi^3 r^2} \sum_{\ell = 0}^\infty \int_{\mathrm{min} \left\{ \frac{qQ}{r_+}, 0 \right\} }^{\mathrm{max} \left\{ \frac{qQ}{r_+}, 0 \right\} } d \omega \frac{{\widetilde {\omega }}}{\left| \omega \right|} \coth \left| \frac{\pi \widetilde{\omega}}{\kappa} \right| \left( 2\ell + 1 \right) {\left| B^{\mathrm{in}}_{\omega \ell} \right|}^2  {\left( - f \left( r \right)^{-1}, 1, 0, 0 \right)}^\intercal ,
		\\
		\langle {\rm {H}}| {\hat {{T}}}^{\mu }_{\nu } | {\rm {H}}  \rangle  
		- \langle {\rm {FT}} | {\hat {{T}}}^{\mu }_{\nu } | {\rm {FT}}  \rangle
		\sim & - \frac{1}{16 \pi ^{2} r^2} \sum_{\ell = 0}^\infty  \int_{\mathrm{min} \left\{ \frac{qQ}{r_+}, 0 \right\} }^{\mathrm{max} \left\{ \frac{qQ}{r_+}, 0 \right\} } d \omega \, \frac{{\widetilde{\omega}}^{2}}{\left| \omega \right|} \coth \left| \frac{\pi \widetilde{\omega}}{\kappa} \right|  \left( 2\ell + 1 \right){\left| B^{\mathrm{in}}_{\omega \ell} \right|}^2 \nonumber \\ 
  &\times \begin{pmatrix}
   - f \left( r \right) ^{-1} & - f \left( r \right)^{-2} & 0 & 0 \\
   1 &f \left( r \right)^{-1} & 0 & 0 \\
   0 & 0 & \mathcal{O} \left( 1 \right) & 0 \\
   0 & 0 & 0 & \mathcal{O} \left( 1 \right)
  \end{pmatrix} .
	\end{align}
\end{subequations} 
The leading order divergences in the  expectation values (\ref{eq:HminusFT}) cancel on the future horizon but not the past horizon. Combining this result with the fact that the leading order divergences in the expectation values (\ref{eq:FTUminus}) also cancel, and the assumed regularity of the ``past'' Unruh state $| {\rm {U}}^{-} \rangle $ on the future horizon, we may deduce that the expectation value of the current in the state $| {\rm {H}} \rangle $ is regular on the future horizon, and that the expectation value of the SET is at worst divergent as ${\mathcal {O}} (f(r)^{-1})$ as $r\rightarrow r_{+}$. 

Combining the relevant components in (\ref{eq:HminusFT}) and the fluxes (\ref{eq:FTfluxes}), we find that the fluxes of charge and energy in the state $| {\mathrm {H}}\rangle $ vanish:
\begin{equation}
	{\mathcal {K}}_{\mathrm {H}} =0,  \qquad {\mathcal {L}}_{\mathrm {H}} = 0.
	\label{eq:fluxesH}
\end{equation}
Therefore the state $|{\mathrm {H}} \rangle $ is a time-reversal invariant, equilibrium state. 
Hence, if the state $|{\mathrm {H}}\rangle$ is regular on either the past or the future horizon, it is regular on both horizons.

\begin{figure}[p]
	\includegraphics[scale=0.63]{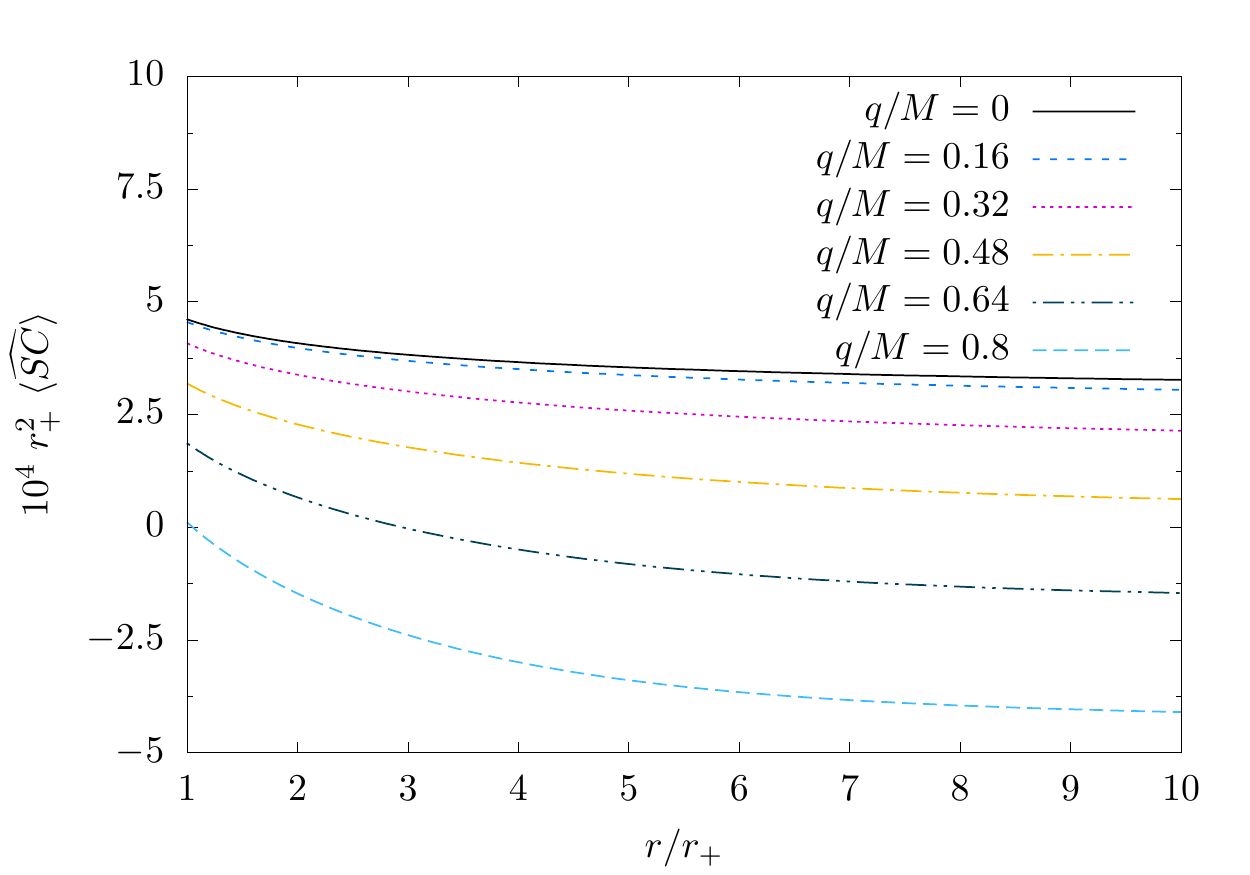}\\ 
	\includegraphics[scale=0.61]{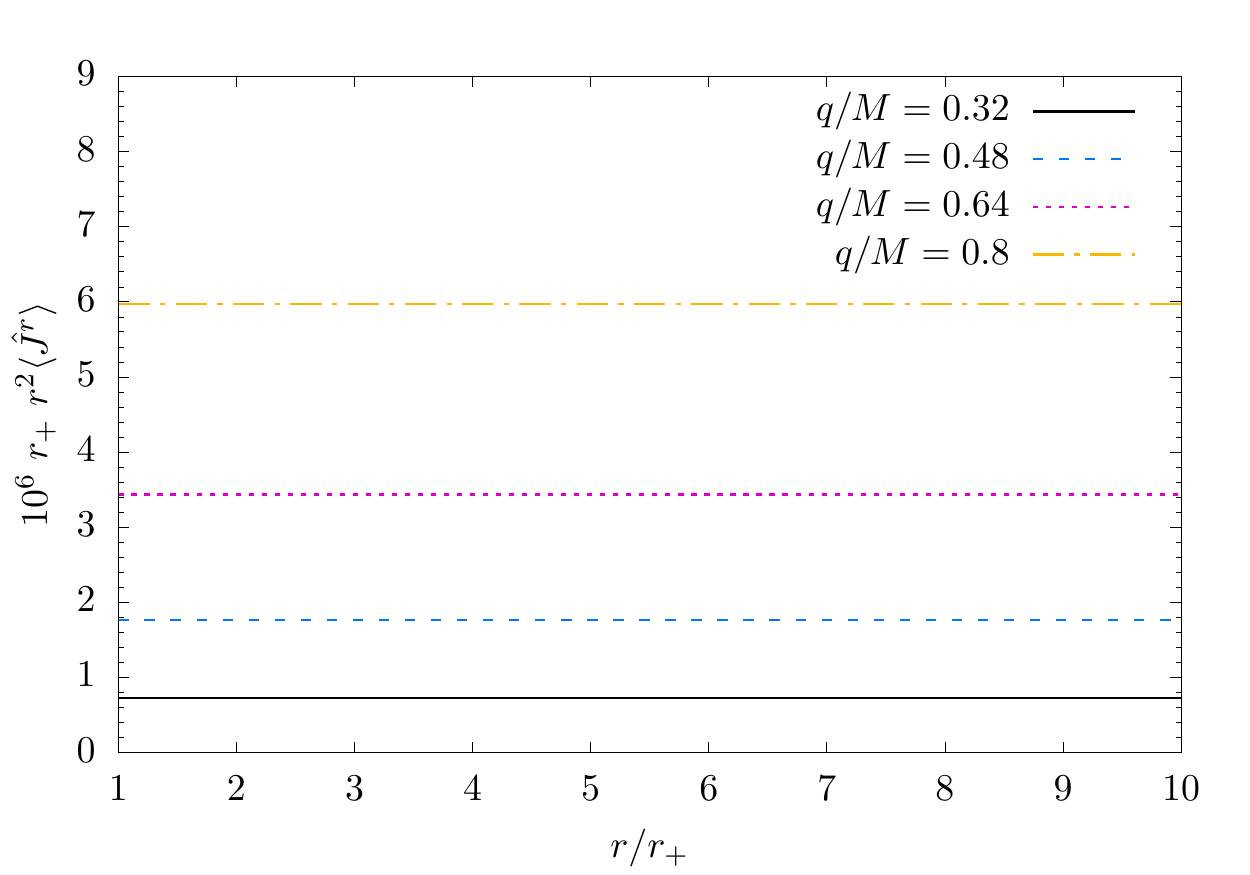}
	\includegraphics[scale=0.61]{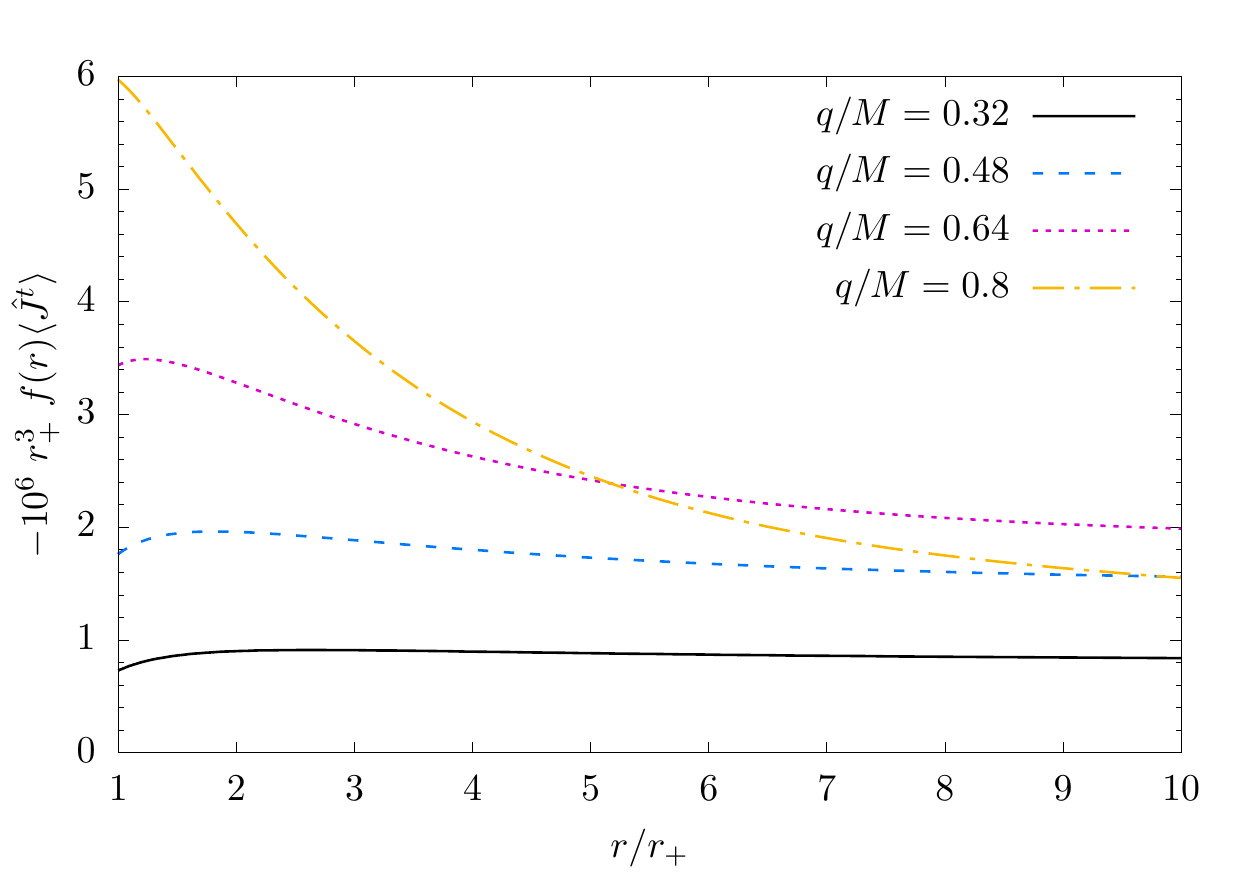}
	\\ \includegraphics[scale=0.61]{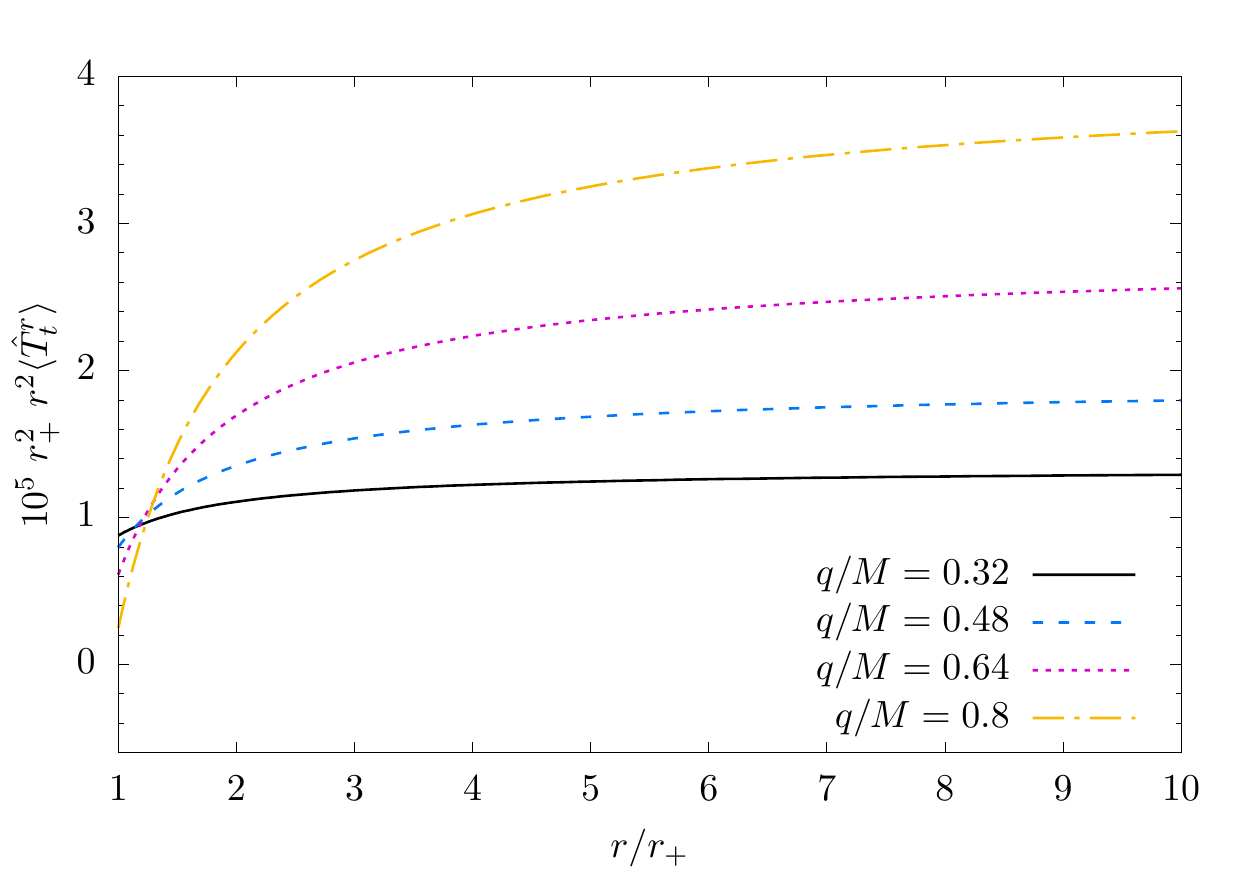}
	\includegraphics[scale=0.61]{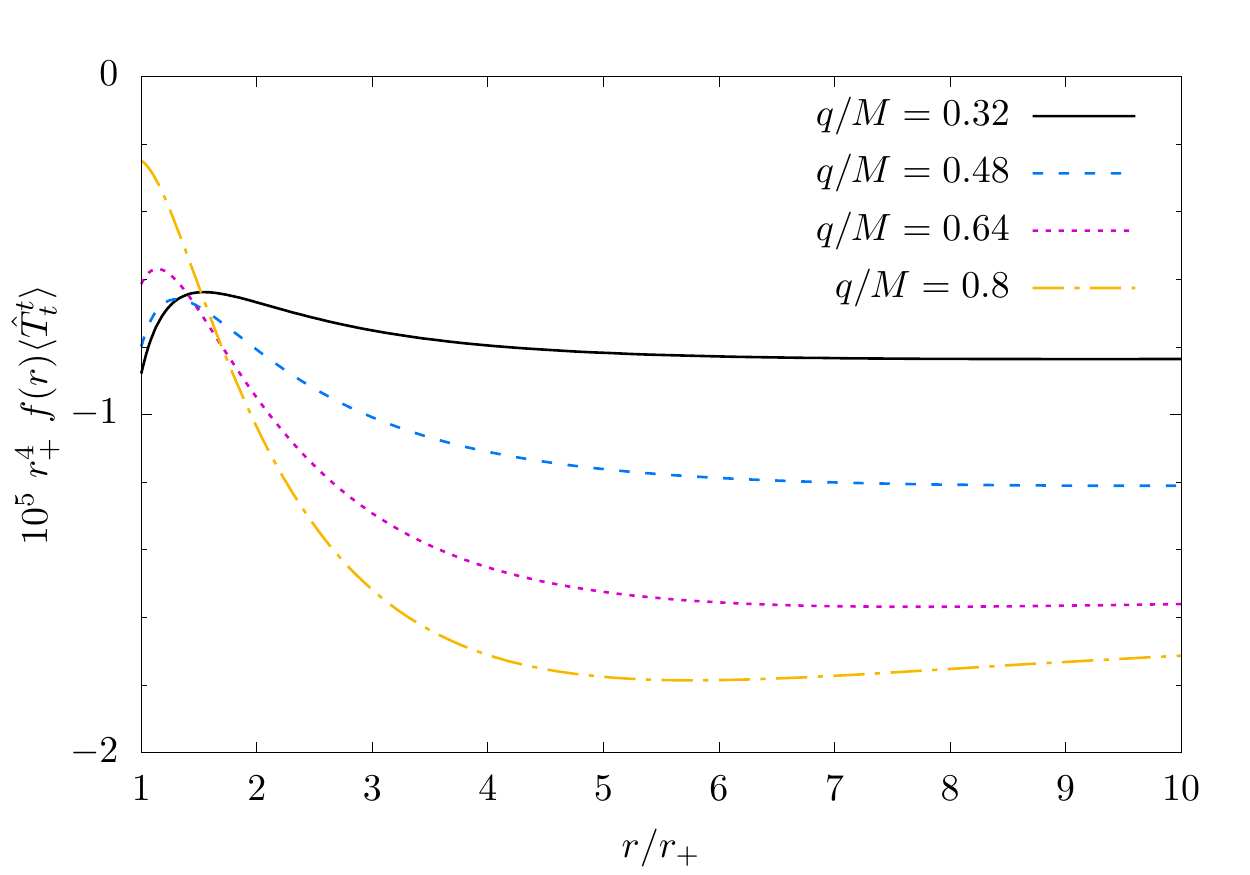} \\
	\includegraphics[scale=0.61]{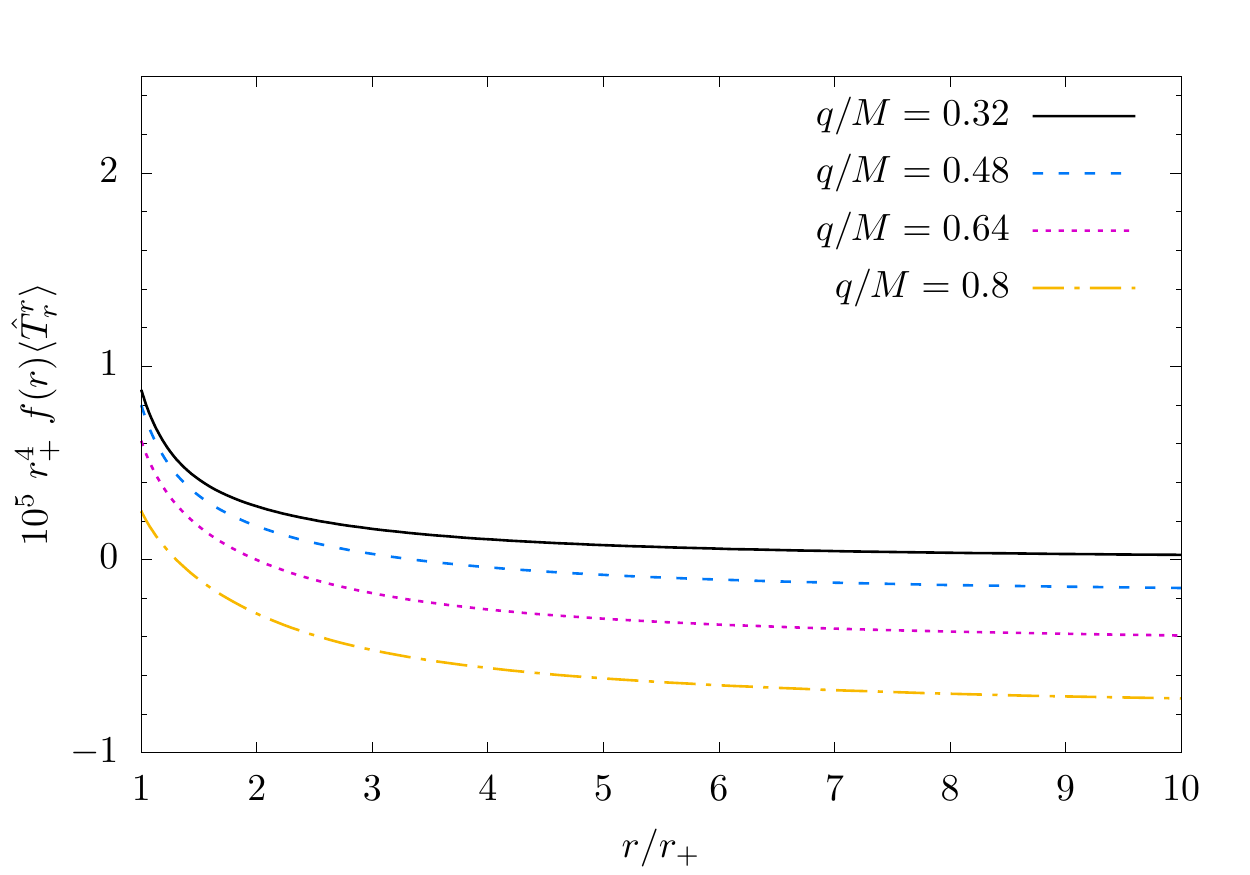}
	\includegraphics[scale=0.61]{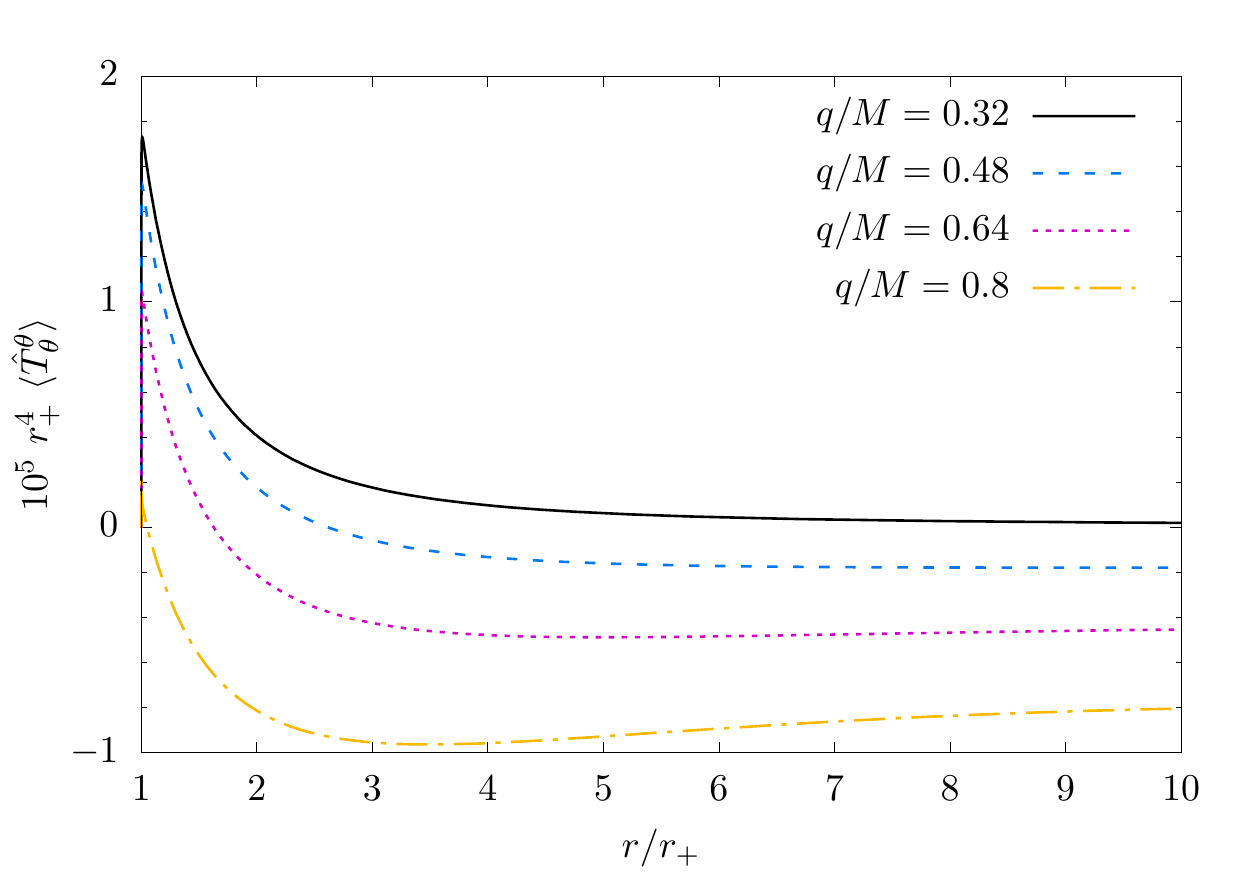} \\
	\vspace{.1cm}
	\caption{Difference in expectation values for the scalar condensate operator and components of the current and stress-energy tensor operators, between the tentative ``Hartle-Hawking''-like state, $| {\rm {H}} \rangle$, and the ``past'' Unruh state, $|{\rm {U}}^{-} \rangle$, in the spacetime of a RN black hole with $Q=0.8M$. All quantities are multiplied by powers of $f(r)$ so that the resulting quantities are regular at $r=r_{+}$.}
	\label{fig:HH-Uminus}
\end{figure}

\begin{figure}
	\includegraphics[scale=0.61]{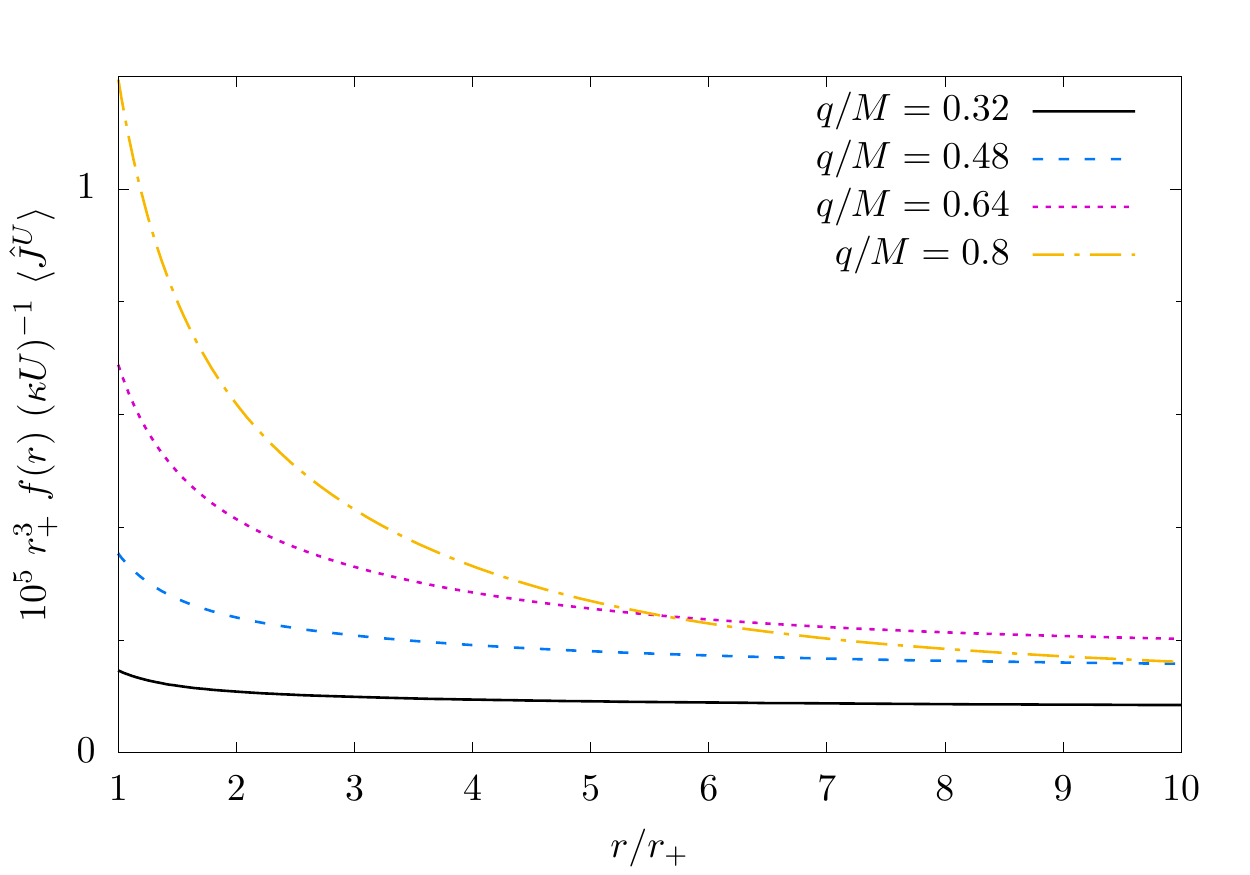}
	\includegraphics[scale=0.61]{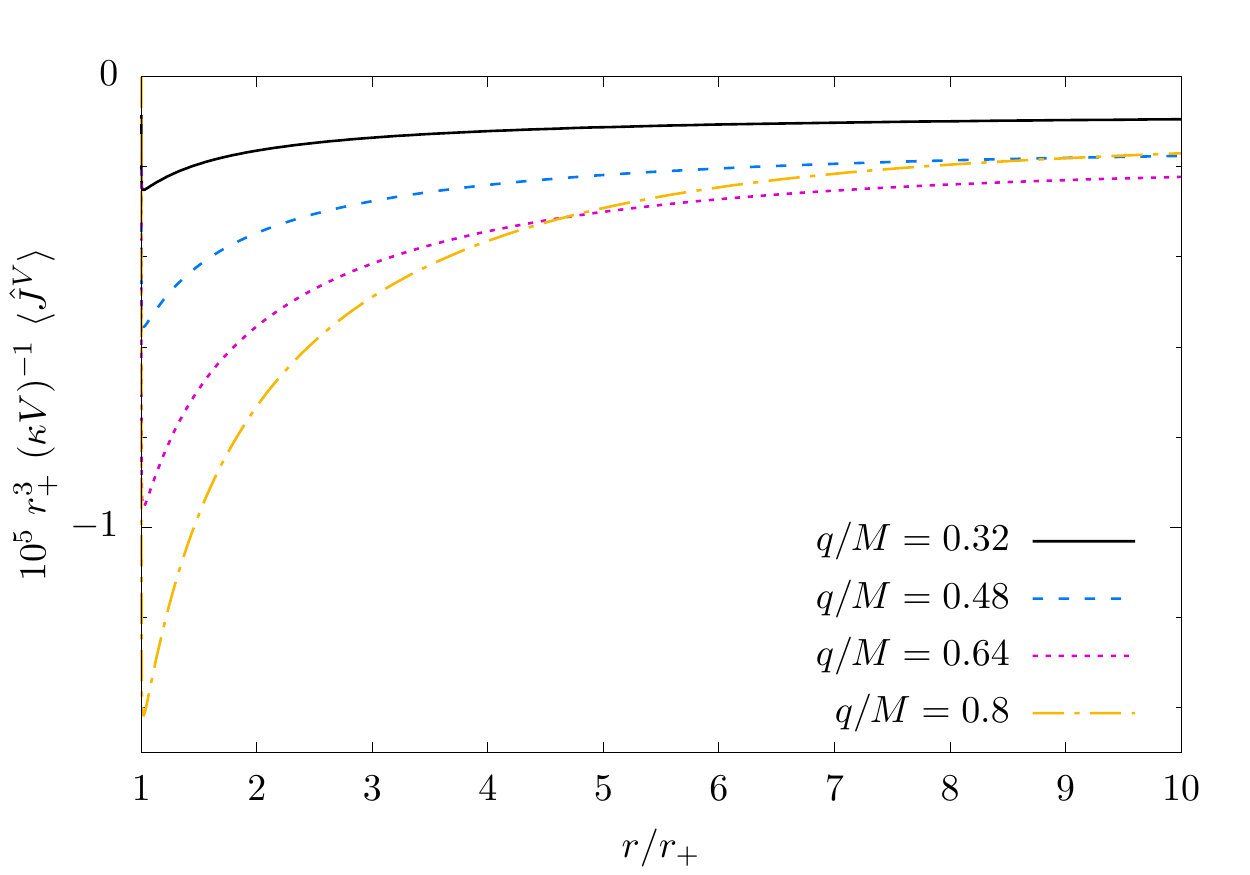}
	\\
	\includegraphics[scale=0.61]{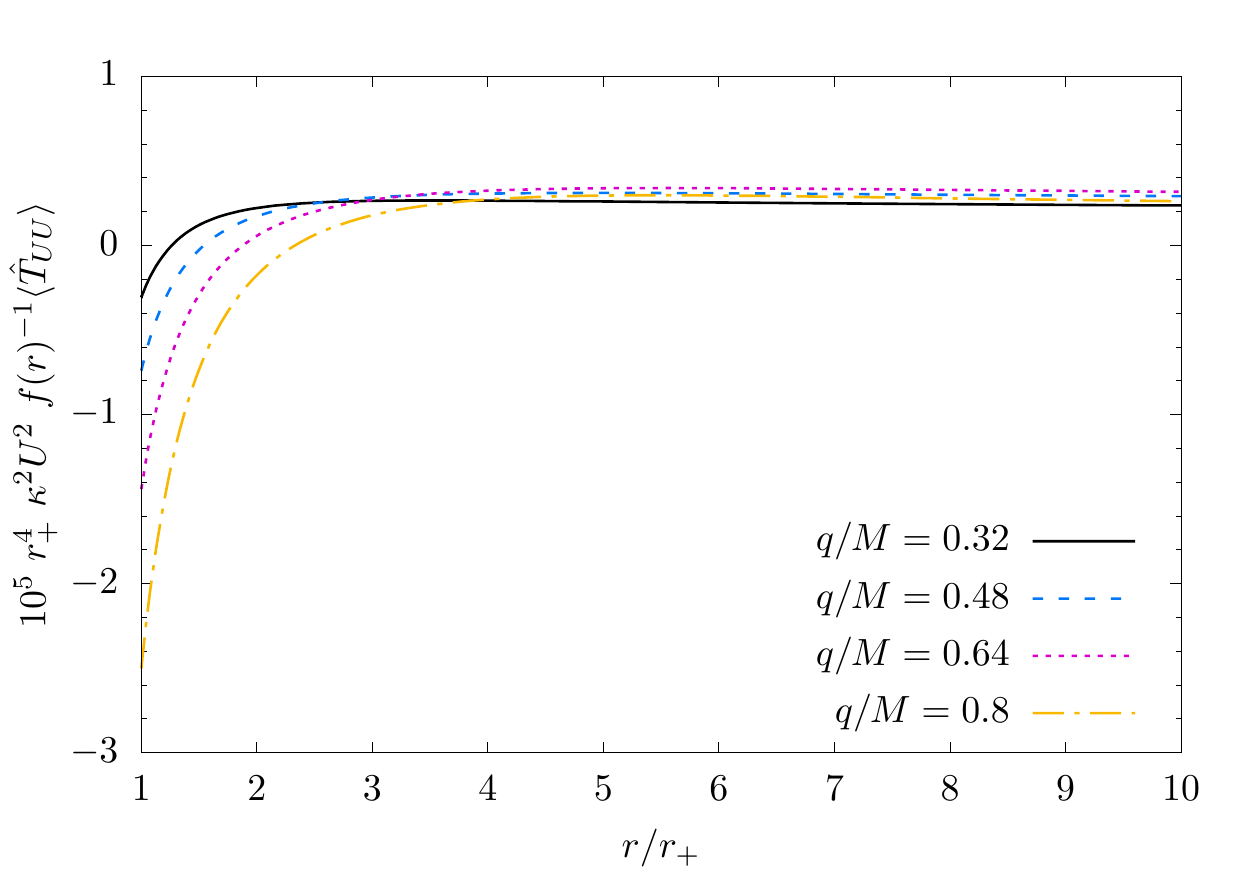}
	\includegraphics[scale=0.61]{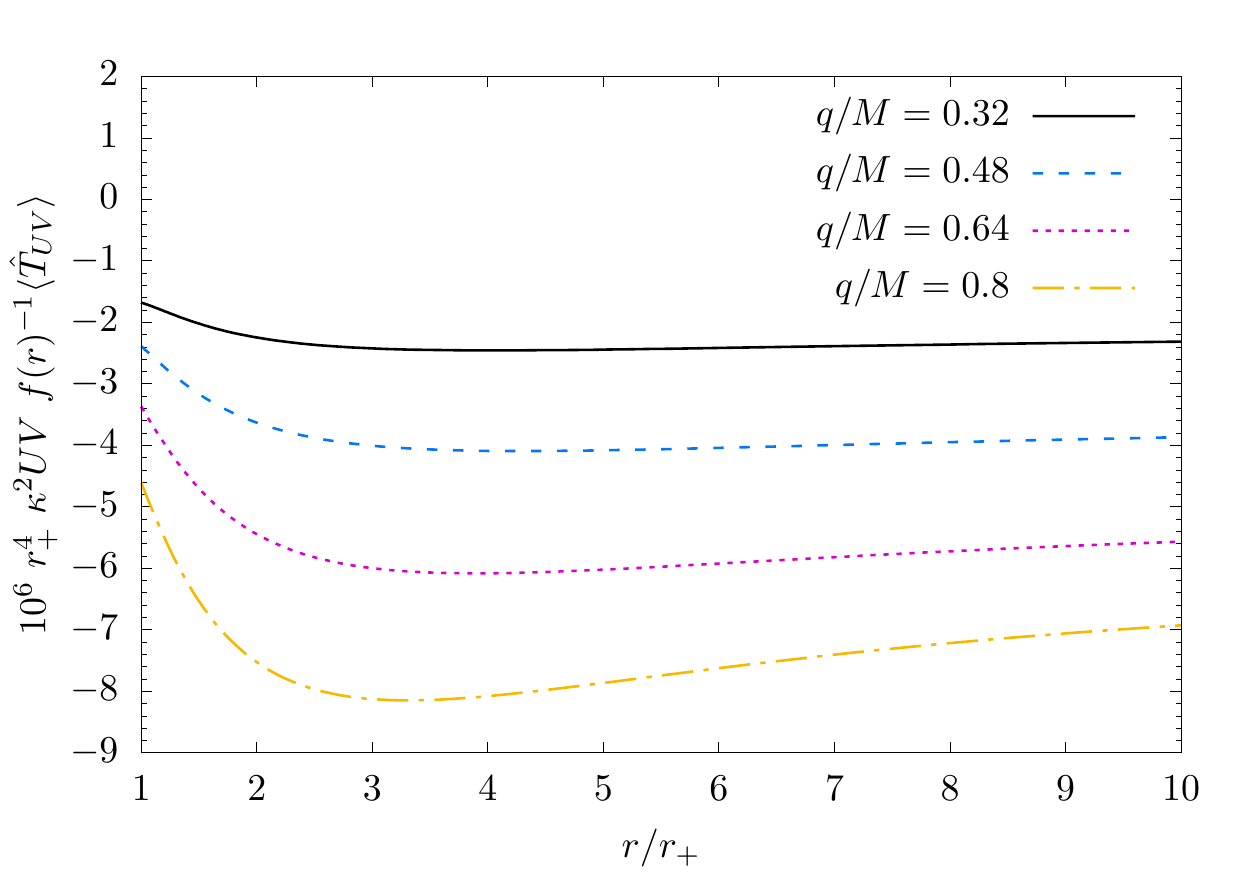}
	\\
	\includegraphics[scale=0.61]{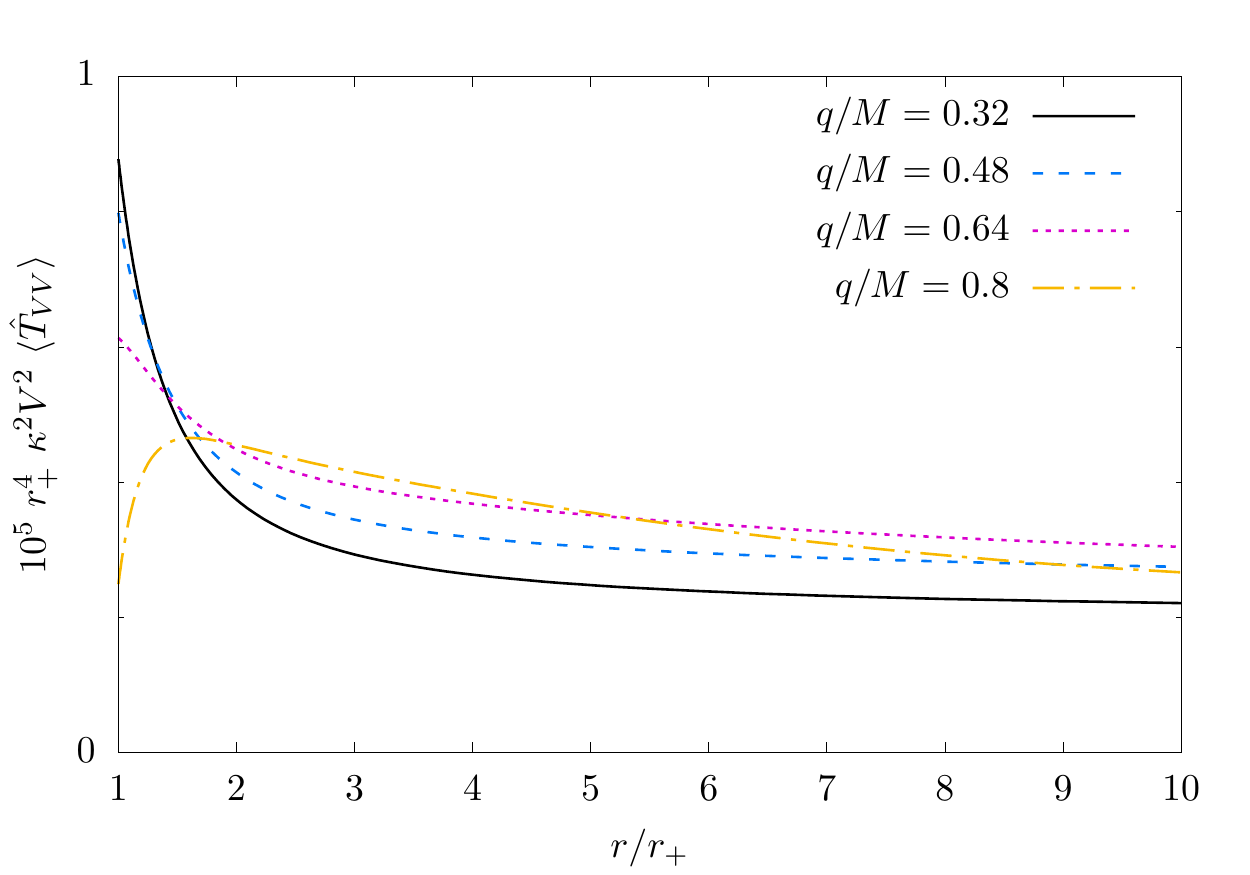}
	\vspace{.1cm}
	\caption{Difference in expectation values for the  components of the current and stress-energy tensor operators in Kruskal coordinates (\ref{eq:Kruskal}), between the tentative ``Hartle-Hawking''-like state $| {\rm {H}}\rangle$ and the ``past'' Unruh state, $| {\rm {U}}^{-} \rangle$, in the spacetime of a RN black hole with $Q=0.8M$. All expectation values are multiplied by powers of $f(r)$ so that the resulting quantities are regular at $r=r_{+}$.}
	\label{fig:HH-Uminus_Kruskal}
\end{figure}

Since the expectation value of the scalar condensate in the state $|{\mathrm {FT}} \rangle$ is divergent, to study the properties of the state $|{\mathrm {H}} \rangle $ in more detail,  we now consider the differences in expectation values between the states $| {\rm {H}} \rangle$ and  $|{\rm {U}}^{-}\rangle$, which take the form
 \begin{equation}
 	\langle {\rm {H}} | {\hat {{O}}} | {\rm {H}} \rangle  
 	- \langle {\rm {U}}^{-} | {\hat {{O}}} | {\rm {U}}^{-} \rangle 
 	=
 	\sum _{\ell =0}^{\infty }\sum _{m=-\ell}^{\ell } 
 	\int _{0}^{\infty }d\omega  \,
 	\left\{ 
 	\frac{1}{\exp \left[ \frac{2\pi {\widetilde {\omega }}}{\kappa } \right] -1 } 
 	o^{\rm {in}}_{\omega \ell m} 
 	+
 		\frac{1}{\exp \left[ \frac{2\pi {\overline {\omega }}}{\kappa } \right] -1 } 
 	o^{\rm {in}}_{-\omega \ell m}
 	\right\}
 	,
 	\label{eq:HHminusU}
 \end{equation}
where ${\overline {\omega }}$ is given by (\ref{eq:baromega}).
In particular, the difference in expectation values of the scalar condensate is
\begin{align}
	\langle {\rm {H}} | {\widehat {{SC}}} | {\rm {H}} \rangle  
	- \langle {\rm {U}}^{-}  | {\widehat {{SC}}} | {\rm {U}}^{-}  \rangle
	& = 
	\sum _{\ell =0}^{\infty } \sum _{m=-\ell }^{\ell }
	\int _{0}^{\infty } d\omega  
	\left\{ 
	\frac{\left| \phi _{\omega \ell m}^{\rm {in}}\right|^{2}}{\exp \left[ \frac{2\pi {\widetilde {\omega }}}{\kappa } \right] -1}
	+ \frac{\left| \phi _{-\omega \ell m}^{\rm {in}}\right|^{2}}{\exp \left[  \frac{2\pi {\overline {\omega }}}{\kappa } \right] -1} 
	\right\} .
	\label{eq:scalarcondensateHminusUminus}
\end{align}
While the two integrands are singular at ${\widetilde {\omega }}=0$ and ${\overline {\omega }}=0$ respectively, the Cauchy principal value of each integral exists. 
Near the horizon, we have
\begin{equation}
	\langle {\rm {H}} | {\widehat {{SC}}} | {\rm {H}} \rangle  
	- \langle {\rm {U}}^{-}  | {\widehat {{SC}}} | {\rm {U}}^{-}  \rangle
	\sim  \frac{1}{16 \pi ^{2} r^{2}}  
	\sum _{\ell =0}^{\infty } 
	\int _{0}^{\infty } d\omega \left( 2\ell + 1 \right) 
	\left\{ 
	\frac{\left| B^{\rm {in}}_{\omega \ell } \right| ^{2} }{\omega \left( \exp \left[ \frac{2\pi {\widetilde {\omega }}}{\kappa } \right] -1 \right) } 
	+\frac{\left| B^{\rm {in}}_{-\omega \ell } \right| ^{2} }{ \omega \left( \exp \left[ \frac{2\pi {\overline {\omega }}}{\kappa } \right] -1 \right) }    
	\right\} 	.
\end{equation}
The quantity (\ref{eq:scalarcondensateHminusUminus}) is shown in the first plot in Fig.~\ref{fig:HH-Uminus}. It can be seen that the difference in expectation values of the scalar condensate in the states $| {\mathrm {H}} \rangle $ and  $| {\mathrm {U}}^{-} \rangle $ is regular everywhere on and outside the event horizon.
Since the ``past'' Unruh state  $| {\mathrm {U}}^{-} \rangle $ is anticipated to be regular on the future horizon, we conclude that the expectation value of the scalar condensate in the state $| {\mathrm {H}}\rangle $, unlike that for the state $|{\mathrm {FT}} \rangle $, is also regular on the event horizon of the black hole.
The scalar condensate (\ref{eq:scalarcondensateHminusUminus}) also does not vanish as $r\rightarrow \infty$, providing evidence that the state $| {\mathrm {H}} \rangle $ is not empty at infinity.
We also see that the scalar condensate varies considerably as the scalar field charge $q$ varies.

In Fig.~\ref{fig:HH-Uminus} we also depict the differences in expectation values of the current and SET between the states $|{\mathrm {H}}\rangle $  and $| {\rm {U}}^{-}\rangle $.
First, looking at the plot of the radial component of the current, we see that the difference in the flux of charge is negative between these two states. Since ${\mathcal {K}}_{\rm {H}}$ is zero (\ref{eq:fluxesH}), we deduce that ${\mathcal {K}}_{{\rm {U}}^{-}}$ (\ref{eq:chargeUminus}) is positive for a black hole and scalar field both having  positive charge. This is as expected: the black hole emits Hawking radiation in such a way as to reduce its charge. 

The difference in expectation values of the charge density between the $|{\mathrm {H}}\rangle $  and $| {\rm {U}}^{-}\rangle $ states is negative and nonzero far from the black hole. The magnitude of the charge density increases significantly as the scalar field charge increases, although we find for large values of the scalar field charge ($q=0.8M$) an interesting effect whereby the magnitude of the charge density near the horizon is large but that at infinity is smaller than for lower values of the scalar field charge. 
To examine whether the difference in expectation values of the current between these two states is regular across the horizon, we turn to  Fig.~\ref{fig:HH-Uminus_Kruskal}.
This shows that, in Kruskal coordinates, the component $V^{-1}\langle {\hat {{J}}}^V\rangle $ is finite as $r\rightarrow r_+$, so this component of the difference in expectation values of the current is regular on both the past and future horizons. In contrast, the component $U^{-1}\langle {\hat {{J}}}^U\rangle $ of the difference in expectation values diverges as $r\rightarrow r_+$. We deduce that the difference in expectation values of the current between the $|{\mathrm {H}}\rangle $  and $| {\rm {U}}^{-}\rangle $  states is regular across the future horizon but not the past horizon.
Since we assume that the ``past'' Unruh state is regular across the future horizon, 
we conclude that the expectation value of the current in the state $|{\mathrm {H}}\rangle $ is also regular across both the future and past horizons. 

We now study the differences in expectation values of the SET between the $| {\rm {H}}\rangle $ and $|{\mathrm {U}}^{-}\rangle $ states. The difference in expectation values of the component $\langle {\hat {{T}}}^r_t\rangle$ is positive far from the black hole. This is to be expected from the fact that ${\mathcal {L}}_{{\rm {H}}}$ (\ref{eq:fluxesH}) vanishes, while ${\mathcal {L}}_{{\rm {U}}^{-}}$ (\ref{eq:energyUminus}) is positive.  
The differences in expectation values of the diagonal components of the SET between these two states appear to approach constant values far from the black hole. The difference in the expectation values of the component  $\langle {\hat {{T}}}_t^t\rangle$ is negative everywhere outside the horizon, and has a magnitude roughly twice that of the difference in the corresponding expectation values between the states $|{\mathrm {FT}}\rangle $ and $|{\mathrm {U}}^{-}\rangle $. 
For all the values of the scalar field charge $q$ studied, the difference in expectation values of the component  $\langle {\hat {{T}}}_r^r\rangle$ is positive close to the horizon, but its sign far from the black hole depends on the magnitude of the scalar field charge.
For smaller values of the scalar field charge, it is positive at infinity, but becomes negative at infinity if the scalar field charge is sufficiently large.  
Similar behaviour is seen in Fig.~\ref{fig:FT-Uminus} for the difference in expectation values of the component $\langle {\hat {{T}}}_r^r\rangle$ between the states $|{\mathrm {FT}}\rangle $ and $|{\mathrm {U}}^{-}\rangle $. 
The difference in the expectation values of the component $\langle {\hat {{T}}}_{\theta }^{\theta }\rangle $  between the states $|{\mathrm {H}}\rangle $ and $|{\mathrm {U}}^{-}\rangle $ also has similar behaviour to that between the states  $|{\mathrm {FT}}\rangle $ and $|{\mathrm {U}}^{-}\rangle $ states. 
For smaller values of the scalar field charge, it is positive everywhere on and outside the horizon, while for intermediate values of the scalar field charge it is positive on the horizon but becomes negative far from the black hole.

The differences in the expectation values of the SET in Kruskal coordinates are shown in Fig.~\ref{fig:HH-Uminus_Kruskal}. 
A similar picture emerges as for the differences in the expectation values between the states $|{\mathrm {CCH}}^{-}\rangle $ and $|{\mathrm {U}}^{-}\rangle $ (Fig.~\ref{fig:CCHminus-Uminus_Kruskal}) as well as for those between the states $|{\mathrm {FT}}\rangle $ and $|{\mathrm {U}}^{-}\rangle $ (Fig.~\ref{fig:FT-Uminus_Kruskal}). 
The component $\langle {\hat {{T}}}_{VV}\rangle $  is divergent on the past horizon where $V\rightarrow 0$, but regular on the future horizon. The component $\langle {\hat {{T}}}_{UV}\rangle $ is regular on both the future and past horizons.
In addition, the component $\langle {\hat {{T}}}_{UU}\rangle $ vanishes on the past horizon where $U$ is finite, but diverges like $f(r)^{-1}$ on the future horizon where $U\rightarrow 0$. 
Since we assume that the state $|{\mathrm {U}}^{-}\rangle $ is regular across the future horizon, we deduce that the state $|{\mathrm {H}}\rangle $ has a mild divergence on the future horizon.
There must also be a mild divergence on the past horizon as the state $|{\mathrm {H}}\rangle $ is time-reversal invariant. 
A full computation of the SET for the state $|{\mathrm {H}}\rangle $ in the vicinity of the horizons would determine whether our deduction is valid. 

The final question we consider in this section is whether the state $|{\mathrm {H}}\rangle $ can be considered as an analogue of the Hartle-Hawking state on Schwarzschild. 
First, the Hartle-Hawking state on Schwarzschild is regular on both the past and future event horizons, and in particular the SET is regular on both horizons. Our numerical results suggest that this is not the case for the state $|{\mathrm {H}}\rangle $.
Second, as discussed in Sec.~\ref{sec:HH}, while the state $|{\mathrm {H}}\rangle $ contains a thermal distribution of particles in the ``up'' modes and nonsuperradiant ``in'' modes, it was constructed using operators satisfying nonstandard commutation relations (\ref{eq:Hccrs}). 
We therefore expect that $|{\mathrm {H}}\rangle $ may not have all the properties required of a ``Hartle-Hawking'' state, although, of all the states constructed in this paper, it is the one which most closely resembles a ``Hartle-Hawking''-like state. 

\end{widetext}

\section{Discussion and conclusions}
\label{sec:conc}

In this paper we have explored the canonical quantization of a charged scalar field on a nonextremal RN black hole background.
Our work was motivated by the aim of disentangling the effects of superradiance and rotation on the construction and properties of quantum states on Kerr black holes,
since in our set-up we have superradiance but no rotation.  
As on Kerr space-time, the presence of superradiant modes complicates the construction of states analogous to the standard Boulware, Unruh and Hartle-Hawking states. 

Nonetheless, in this paper we have constructed a menagerie of states for a charged scalar field on an RN black hole. 
First, we have examined the ``past'' and ``future'' Boulware, Unruh and CCH states, defined here in an analogous manner to the corresponding states on Kerr spacetime \cite{Ottewill:2000qh}. These states are not invariant under time-reversal.
The ``past'' Boulware state is empty far from the black hole except for an outgoing flux of particles in the superradiant modes \cite{Balakumar:2020gli}. 
The ``past'' Unruh state contains an outgoing thermal distribution of particles  with a nonzero chemical potential \cite{Gibbons:1975kk}. 
The ``past'' CCH state is more complicated, as the ``in'' and ``up'' basis modes are thermalized with different thermal factors. 

In addition to these ``past'' and ``future'' states, we have also attempted to construct states analogous to the Boulware and Hartle-Hawking states on Schwarzschild space-time.
We have defined a state $|{\mathrm {B}} \rangle$ which is time-reversal invariant and contains no particles at either future or past null infinity.
However, this is not a vacuum state in the conventional sense, since its construction relies on employing creation and annihilation operators which satisfy modified commutation relations. 

We have also sought to define a thermal equilibrium state. 
Our first attempt, the state $| {\rm {FT}} \rangle$, contains a thermal distribution of particles but is not an equilibrium state. 
It is also ill-defined everywhere on and outside the event horizon.
We have been able to define an equilibrium state $| {\mathrm {H}} \rangle$, which is time-reversal invariant.
However, we have presented some evidence that this state may not be regular at the horizon.
As with the $|{\mathrm {B}} \rangle$ state, the construction of the $| {\mathrm {H}} \rangle$ state relies on having creation and annihilation operators which do not satisfy the usual commutation relations. 
While the Kay-Wald theorem \cite{Kay:1988mu,Kay:1992gr} applies only to a neutral scalar field, one would expect a more general version of the theorem to apply to a charged scalar field. 
We would anticipate that such a theorem would preclude the existence of a thermal equilibrium state for a charged scalar field on an RN black hole. While the state $| {\mathrm {H}} \rangle$ constructed in this paper seems to be a thermal equilibrium state, it is likely to evade a generalized Kay-Wald theorem by failing to satisfy the assumptions of such a theorem.
Specifically, since we have had to introduce nonstandard commutation relations in the construction of $| {\mathrm {H}} \rangle$, we think it likely that this state does not satisfy the usual positivity condition (see the related discussion of the Hartle-Hawking state on Kerr in App.~B of \cite{Frolov:1989jh}).

We are therefore unable to define a conventional vacuum state which is as empty as possible at both future and past null infinity, and our attempts to define a conventional thermal equilibrium state invariant under time-reversal have also been unsuccessful.
Both these results mimic the situation on Kerr space-time, leading us to deduce that it is superradiance which is the dominant effect rather than the rotation, although for Kerr black holes it is the rotation which leads to the superradiance. 

One of our key results is that we have been unable to define an analogue of the Hartle-Hawking state for a charged scalar field.
Of course, this does not prove that no such state exists; but the usual method of canonical quantization, which yields the Hartle-Hawking state on Schwarzschild space-time, fails here, as it does on Kerr black holes.
For a Kerr black hole background, it is possible to define a thermal equilibrium state invariant under time-reversal invariance, if one considers a fermionic rather than a bosonic field \cite{Casals:2012es}.  
It would therefore be interesting to explore the fermionic analogues of our tentative states $| {\rm {B}} \rangle $ and $| {\rm {H}} \rangle $. Since fermionic operators satisfy anticommutation relations rather than commutation relations, one may not need to resort to the unconventional commutation relations we employed in defining these states for charged scalar fields.

The thermal equilibrium state defined for fermions on a Kerr black hole diverges on the speed-of-light surface \cite{Casals:2012es}, the surface on which an observer rigidly rotating with the same angular speed as the black hole event horizon must travel at the speed of light. 
This is similar to the situation in flat space-time, where rigidly rotating thermal states are divergent everywhere for bosonic fields \cite{Duffy:2002ss}, but are regular within the speed-of-light surface for fermionic fields \cite{Ambrus:2014uqa}.
As in flat space-time \cite{Duffy:2002ss}, for a bosonic field on Kerr space-time a thermal equilibrium state can be defined if the black hole is surrounded by a perfectly reflecting mirror located entirely within the speed-of-light surface \cite{Duffy:2005mz}.

A natural question is then whether a Hartle-Hawking-like state can be constructed if the RN black hole is contained within a cavity. 
While there is a generalized concept of an ergosphere for an RN black hole \cite{Denardo:1973pyo,Denardo:1974qis}, there is no surface analogous to the speed-of-light surface in rotating space-times.
While the unbounded RN space-time is stable under charged scalar field perturbations \cite{Hod:2013nn,Hod:2015hza}, 
if the black hole is enclosed by a perfectly reflecting mirror sufficiently far from the event horizon, there is an instability \cite{Herdeiro:2013pia,Degollado:2013bha,Hod:2013fvl,Hod:2016kpm,Dias:2018zjg,DiMenza:2019zli}, leading to a charged analogue of the ``black hole bomb'' \cite{Press:1972zz}, the end-point of which is a stable black hole with charged scalar field hair \cite{Sanchis-Gual:2015lje,Dolan:2015dha,Dias:2018zjg}.
This suggests that, in analogy with the situation on Kerr space-time, it may be possible to define a Hartle-Hawking state for an RN black hole in a cavity if the mirror is sufficiently close to the event horizon.
We plan to return to this question in future work.

In this paper, we have studied the physical properties of the states we have defined by examining  differences in expectation values of observables between two quantum states.
The advantage of studying such differences is that they do not require renormalization.
However, to explore the quantum states in more detail, renormalized expectation values are required.
Renormalized expectation values would also be useful for studying the evolution of an evaporating charged black hole beyond the adiabatic approximation employed in 
\cite{Hiscock:1990ex,Ong:2019rnn,Ong:2019vnv,Xu:2019wak}.
Recently there has been much interest in expectation values of quantum field operators inside the event horizon of a black hole \cite{Lanir:2017oia,Lanir:2018rap,Zilberman:2021vgz,Zilberman:2022iij}, particularly for studying the stability of the inner (Cauchy) horizon of a RN(-de Sitter) or Kerr black hole \cite{Sela:2018xko,Lanir:2018vgb,Zilberman:2019buh,Hollands:2019whz,Hollands:2020qpe,Klein:2021ctt,Zilberman:2022aum}.
Work to date on this question has largely focused on a neutral quantum scalar field (apart from the recent work considering a charged scalar field in \cite{Klein:2021ctt,Klein:2021les}). 
A general formalism for the Hadamard renormalization of  expectation values for a charged quantum scalar field was developed in  \cite{Balakumar:2019djw,Balakumar:2020jhe} (see also \cite{Herman:1995hm,Herman:1998dz} for earlier work based on DeWitt-Schwinger renormalization).   
Using this approach, renormalized expectation values of the current have been computed on an RN-de Sitter black hole \cite{Klein:2021ctt,Klein:2021les} for a charged scalar field in the Unruh state. 
As demonstrated in App.~\ref{sec:nonren}, the $r$-component of the current does not require renormalization, and it is shown in \cite{Klein:2021les} that, with a suitable choice of point-splitting, the $t$-component is renormalized by finite terms, which aids its computation. 
It would be of great interest to extend the work of \cite{Klein:2021ctt,Klein:2021les} to the RN black hole, other quantum states and, ultimately, the expectation value of the stress-energy tensor. 
We leave these questions for future studies.

\appendix

\section{Nonrenormalization of $\langle {\hat{{J}}}^{r} \rangle $ and $\langle {\hat {{T}}}_{tr} \rangle $}
\label{sec:nonren}

Our focus in this Appendix is to show that the components $\langle {\hat{{J}}}^{r} \rangle $ and $\langle {\hat {{T}}}_{tr} \rangle $ of the current and stress-energy tensor respectively do not require renormalization. 
Our method follows that employed in \cite{Frolov:1989jh} to prove the corresponding results for a neutral scalar field on Kerr space-time. 

Let $G_{F}(x,x')$ be the Feynman Green's function for the charged scalar field in a particular, unspecified quantum state.
The renormalized components of the current and stress-energy tensor in this state are given by \cite{Balakumar:2019djw}
\begin{subequations}
\begin{align}
\langle {\hat {{J}}}^{\mu } \rangle = & ~ -\frac{q}{4\pi } \lim _{x'\rightarrow x} \Im \left\{ D^{\mu } \left[- iG_{R}(x,x') \right] \right\} ,
\\
\langle  {\hat {{T}}}_{\mu \nu } \rangle = & ~
\lim _{x'\rightarrow x} \Re \left\{ {\mathscr{T}}_{\mu \nu } \left[ - iG_{R}(x,x') \right] \right\} ,
\end{align}
where ${\mathscr{T}}_{\mu \nu }$ is the second-order differential operator
\begin{equation}
{\mathscr{T}}_{\mu \nu } =  ~
g_{\nu }{}^{\nu '}D_{\mu }D^{*}_{\nu '} 
-\frac{1}{2}g_{\mu \nu }g^{\rho \tau'}D_{\rho }D^{*}_{\tau '} ,
\end{equation}
\end{subequations} 
with $g_{\nu }{}^{\nu '}$ the bivector of parallel transport.
The operator $D_{\mu }$ acts at the space-time point $x$, and $D_{\nu'}$ acts at the space-time point $x'$.
The biscalar $G_{R}(x,x')$ is regular in the coincidence limit $x'\rightarrow x$ and is given by  \cite{Balakumar:2019djw}
\begin{equation}
G_{R}(x,x') = G_{F}(x,x') - G_{S}(x,x') ,
\end{equation}
where $G_{S}(x,x')$ is the singular part of the Hadamard parametrix
\begin{equation}
G_{S}(x,x') =\frac{i}{8\pi ^{2} } \left[ \frac{U(x,x')}{\sigma (x,x')} +  V(x,x')\ln \sigma (x,x') \right]  ,
\end{equation}
with $\sigma (x,x')$ equal to one half of the square of the geodesic distance between the points $x$ and $x'$, assuming that they are connected by a unique geodesic.
The complex biscalars $U(x,x')$ and $V(x,x')$ are regular in the coincidence limit
and can be written as covariant Taylor series expansions. To the order required to perform renormalization in four space-time dimensions, these take the form \cite{Balakumar:2019djw}
\begin{widetext}
\begin{subequations}
	\label{eq:UVexpansions}
	\begin{align}
	U(x,x') =& ~
	U_{00}(x) + U_{01\mu }(x)\sigma ^{;\mu } + U_{02\mu \nu }(x)\sigma ^{;\mu }\sigma ^{;\nu }
	+ U_{03\mu \nu \lambda }(x)\sigma ^{;\mu } \sigma ^{;\nu } \sigma ^{;\lambda }
	+ U_{04 \mu \nu \lambda \tau}(x) \sigma ^{;\mu } \sigma ^{;\nu } \sigma ^{;\lambda }
	\sigma ^{;\tau } + \ldots ,
	\label{eq:HadamardU}
	\\
	V(x,x') = & ~ V_{00}(x) + V_{01\mu }(x)\sigma ^{;\mu } + V_{02\mu \nu }(x)\sigma ^{;\mu }\sigma ^{;\nu } 
	+ V_{10}(x) \sigma + \ldots .
	\label{eq:HadamardV}
	\end{align}
The coefficients in the expansions depend only on the space-time point $x$, and all dependence on $x'$ is contained within $\sigma (x,x')$ and its derivatives.
Since we are considering a massless charged scalar field minimally coupled to the space-time curvature, and the Reissner-Nordstr\"om metric has vanishing Ricci scalar, the coefficients given in \cite{Balakumar:2019djw} simplify to
		\begin{align}
		U_{00} = & ~1, \label{eq:HadamardU1} \\
		U_{01\mu } = & ~ iqA_{\mu }, \\
		U_{02\mu \nu } = & ~
		\frac{1}{12}R_{\mu \nu} - \frac{iq}{2}\nabla _{(\mu } A_{\nu )} -\frac{q^{2}}{2}A_{\mu }A_{\nu },
		\\
		U_{03\mu \nu \lambda } = & ~
		-\frac{1}{24}R_{(\mu \nu ;\lambda )} + \frac{iq}{6}\nabla _{(\mu } \nabla _{\nu }A_{\lambda )} + \frac{q^{2}}{2}A_{(\mu }\nabla _{\nu }A_{\lambda )} - \frac{iq^{3}}{6} A_{\mu }A_{\nu }A_{\lambda }+ \frac{iq}{12}R_{(\mu \nu }A_{\lambda )} ,
		\\
		U_{04\mu \nu  \lambda \tau } = & ~
		\frac{1}{80}R_{(\mu \nu ; \lambda \tau)} + \frac{1}{288}R_{(\mu \nu }R_{\lambda \tau )}
		+ \frac{1}{360} R^{\rho }{}_{(\mu | \psi | \nu } R^{\psi }{}_{\lambda | \rho | \tau) }
		- \frac{iq}{24} \nabla _{(\mu }\nabla _{\nu } \nabla _{\lambda } A_{\tau )}
		- \frac{q^{2}}{6}A_{(\mu }\nabla _{\nu } \nabla _{\lambda } A_{\tau )}
		\nonumber \\ & ~
		- \frac{q^{2}}{8}\left[ \nabla _{(\mu } A_{\nu }\right] \left[ \nabla _{\lambda }A_{\tau )} \right] 
		+ \frac{iq^{3}}{4}A_{(\mu }A_{\nu }\nabla _{\lambda }A_{\tau )}
		+ \frac{q^{4}}{24} A_{\mu }A_{\nu }A_{\lambda } A_{\tau }
		- \frac{iq}{24} A_{(\mu }\nabla _{\nu } R_{\lambda \tau )}
		- \frac{iq}{24}R_{(\mu \nu }\nabla _{\lambda }A_{\tau )}
		\nonumber \\ & ~
		-\frac{q^{2}}{24} R_{(\mu \nu }A_{\lambda }A_{\tau )} ,
		\label{eq:HadamardU2}
		\\
		V_{00} = & ~  0,
		\\
		V_{01\mu } = & ~ -\frac{iq}{12}\nabla ^{\alpha }F_{\alpha \mu },
		\\
		V_{02\mu \nu } = & ~ -\frac{1}{240}\Box R_{\mu \nu } 
		+ \frac{1}{180} R^{\alpha }{}_{\mu }R_{\alpha \nu }
		- \frac{1}{360}R^{\alpha \beta }R_{\alpha \mu \beta \nu }
		- \frac{1}{360} R^{\alpha \beta \gamma }{}_{\mu }R_{\alpha \beta \gamma \nu }
		- \frac{q^{2}}{24}F^{\alpha }{}_{\mu }F_{\nu \alpha }
		- \frac{q^{2}}{12}A_{(\mu }\nabla ^{\alpha }F_{\nu ) \alpha }
		\nonumber \\ & ~
		- \frac{iq}{24} \nabla _{(\mu }\nabla ^{\alpha }F_{\nu ) \alpha },
		\\
		V_{10} = & ~  \frac{1}{720}R^{\alpha \beta \gamma  \delta }R_{\alpha \beta \gamma \delta } - \frac{1}{720} R^{\alpha \beta }R_{\alpha \beta }
		-\frac{q^{2}}{48}F^{\alpha \beta }F_{\alpha \beta },
		\end{align}
	\end{subequations}
\end{widetext}
where brackets round indices denote symmetrization, with vertical lines surrounding those indices not included in the symmetrization.

To show that  $\langle {\hat{{J}}}^{r} \rangle $ and $\langle {\hat {{T}}}_{tr} \rangle $ do not require renormalization, we seek to prove that
\begin{subequations}
	\label{eq:frakF}
\begin{align}
{\mathfrak {F}}_{1}
\equiv & ~
\Im \left\{ D^{r } \left[- iG_{S}(x,x') \right] \right\}  =0,  
\\
{\mathfrak {F}}_{2}  \equiv & ~ \Re \left\{ {\mathscr{T}}_{tr} \left[ - iG_{S}(x,x') \right]  \right\} =0.
\end{align}
\end{subequations}
Since the Reissner-Nordstr\"om metric (\ref{eq:RNmetric}) is static and spherically symmetric, without loss of generality we may consider two space-time points $x$ and $x'$ as follows:
\begin{equation}
x= (0,r,\theta, 0), \qquad x'=(0,r',\theta' ,0) .
\end{equation}
Then the unique geodesic connecting the points $x$ and $x'$ lies in the surface $\Sigma = \{ t=0, \varphi =0\} $.
Using the letter ${\mathcal {X}}$ to denote the indices $t$, $\varphi $, and ${\mathcal {A}}$ to denote $r$, $\theta $, we have \cite{Frolov:1989jh}
\begin{subequations}
	\label{eq:AXcomponents}
	\begin{align}
	\sigma ^{;\mu } = & ~ \delta _{{\mathcal {A}}}^{\mu }\sigma ^{;{\mathcal {A}}} ,
	\\
	g_{\nu }{}^{\nu '} = & ~
	\delta _{{\mathcal {A}}'}^{\nu '}\delta _{\nu }^{{\mathcal {A}}}g_{{\mathcal {A}}}{}^{{\mathcal {A}}'} 
	+ 	\delta _{{\mathcal {X}}'}^{\nu '}\delta _{\nu }^{{\mathcal {X}}}g_{{\mathcal {X}}}{}^{{\mathcal {X}}'} .
	\end{align}
We are considering a purely electric field with gauge potential (\ref{eq:gaugepot}), and hence we can write
\begin{equation}
A_{\mu } = \delta _{\mu }^{{\mathcal {X}}}A_{\mathcal {X}},
\end{equation}
\end{subequations} 
where $A_{\mathcal {X}}$ depends only on ${\mathcal {A}}$ coordinates.
Therefore the quantities (\ref{eq:frakF}) take the form
\begin{subequations}
\begin{align}
{\mathfrak {F}}_{1} = & ~
\Im \left\{ \nabla ^{r} \left[ -iG_{S}(x,x')\right] \right\} ,
\label{eq:F1}
\\
{\mathfrak {F}}_{2} = & ~
\Re \left\{ -i \left[ g_{r}{}^{{\mathcal {A}}'}D_{t}\nabla _{{\mathcal {A}}'}  \right] G_{S}(x,x')\right\} .
\label{eq:F2}
\end{align}
\end{subequations}
The biscalar $\sigma (x,x')$ and its derivatives are real, as are the gauge field potential $A_{\mu }$ and field strength $F_{\mu \nu }$, as well as all curvature tensors and their derivatives. 
From (\ref{eq:AXcomponents}), we have $A_{\mu }\sigma ^{;\mu }=0$, which immediately simplifies the form of $G_{S}(x,x')$.

The symmetries of the metric mean that Christoffel symbols $\Gamma ^{\mu }_{\nu \lambda }$ having an odd number of ${\mathcal {X}}$ indices vanish, while those with an even number of ${\mathcal {X}}$ indices are nonzero. 
Therefore the nonzero components of all covariant derivatives of the gauge potential $A_{\mu }$ contain at least one ${\mathcal {X}}$ index and hence all terms in (\ref{eq:UVexpansions}) containing covariant derivatives of $A_{\mu }$ do not contribute to $U(x,x')$ or $V(x,x')$ when contracted with $\sigma ^{;\mu }$. 
As a result, $U(x,x')$ (\ref{eq:HadamardU}, \ref{eq:HadamardU1}--\ref{eq:HadamardU2}) is real and depends only on curvature tensors; the gauge potential does not contribute.

The gauge field strength has the form
\begin{equation}
F_{\mu \nu } = \left[ \delta _{\mu }^{{\mathcal {A}}}\delta _{\nu }^{{\mathcal {X}}}-
\delta _{\mu }^{{\mathcal {X}}}\delta _{\nu }^{{\mathcal {A}}} \right] F_{{\mathcal {A}}{\mathcal {X}}},
\end{equation} 
where $F_{{\mathcal {A}}{\mathcal {X}}}$ depends only on the ${\mathcal {A}}$ coordinates.
Hence we have
\begin{equation}
\nabla ^{\alpha }F_{\alpha \mu } = \delta _{\mu }^{{\mathcal {X}}}\nabla ^{{\mathcal {A}} }F_{{\mathcal {A}}{{\mathcal {X}}}}. 
\end{equation}
Therefore $V(x,x')$ (\ref{eq:HadamardV}) is also real. 
We deduce that $-iG_{S}(x,x')$ is real and hence ${\mathfrak {F}}_{1}$ (\ref{eq:F1}) is trivally zero, while ${\mathfrak {F}}_{2}$ (\ref{eq:F2}) simplifies to 
\begin{equation}
{\mathfrak {F}}_{2} = g_{r}{}^{{\mathcal {A}}'}\nabla _{t} \nabla _{{\mathcal {A}}'} \left[ -iG_{S}(x,x') \right]. 
\end{equation}
The derivatives in the above expression commute since they are evaluated at different space-time points and $G_{S}(x,x')$ is a biscalar. 
Furthermore, $G_{S}(x,x')$ depends only on the space-time geometry and the background electromagnetic field. 
Therefore $G_{S}(x,x')$ does not depend on $t$ and thus $\nabla _{t}(-iG_{S})$ must be zero. 
We then have ${\mathfrak {F}}_{2}=0$, as required.

In conclusion, the components $\langle {\hat{{J}}}^{r} \rangle $ and $\langle {\hat {{T}}}_{tr} \rangle $ do not require renormalization.

\section{Components of the current and stress-energy tensor}
\label{sec:JT}

In this Appendix we give the explicit formulae for the mode contributions to the current and stress-energy tensor.
The sums over the azimuthal quantum number $m$ are then performed using properties of the spherical harmonics derived in App.~\ref{sec:spherical}.

\begin{widetext}
The classical mode contributions to the current $J^{\mu }$ are
	\begin{subequations}
		\begin{align}
			j^{t}_{\omega \ell m} = & ~ -\frac{q}{4\pi f(r)}\left( \omega - \frac{qQ}{r} \right) \left| \phi _{\omega \ell m}\right| ^{2} ,
			\\
			j^{r}_{\omega \ell m} = & ~ -\frac{qf(r)}{4\pi } \left| {\mathcal {N}}_{\omega } \right| ^{2} \Im \left[ \frac{X^{*}_{\omega \ell }(r)}{r} \frac{d}{dr} \left( \frac{X_{\omega \ell }(r)}{r} \right)  
			\right] \left| Y_{\ell m}(\theta ,\varphi ) \right| ^{2}  ,
			\\ 
			j^{\theta }_{\omega \ell m} = & ~
			-\frac{q}{4\pi r^{4}} \left| {\mathcal {N}}_{\omega } \right| ^{2}
			\left| X_{\omega \ell }(r) \right| ^{2} \Im \left[ 
			Y_{\ell m}^{*}(\theta, \varphi ) \frac{\partial }{\partial \theta } Y_{\ell m}(\theta, \varphi )
			\right] ,
			\\
			j^{\varphi }_{\omega \ell m} = & ~ -\frac{mq}{4\pi r^{2}\sin ^{2}\theta } \left| \phi _{\omega \ell m} \right| ^{2} ,
		\end{align}
	\end{subequations}
\end{widetext}
where $\Im  $ denotes the imaginary part of a complex quantity.
The component $j^{\theta }_{\omega \ell m}$ vanishes identically for all $\ell $ and $m$ using the properties of the spherical harmonics (\ref{eq:harmonics}).
Although the mode contribution to the current component $j^{\varphi }_{\omega \ell m}$ does not vanish, in all our expectation values we will be summing over $m=-\ell , \ldots \ell $.
From the properties of the spherical harmonics (\ref{eq:harmonics}), we have $\left| \phi _{\omega , \ell , -m}\right|^{2} = \left| \phi _{\omega \ell m} \right| ^{2}$ and hence all sums over $m$ in the expectation values of the $\varphi $ component of the current will vanish.
The only nonzero components of the current will therefore be the $t$ and $r$ components.
The sum over $m$ in these components can be performed explicitly using (\ref{eq:Ylmresult4}), giving
\begin{widetext}
	\begin{subequations}
		\begin{align}
			j^{t}_{\omega \ell} = \sum _{m=-\ell }^{\ell }j^{t}_{\omega \ell m} = & ~ 
			-\frac{q\left(2\ell + 1\right)}{16\pi ^{2} r^{2} f(r)}\left|{\mathcal {N}}_{\omega }\right| ^{2} \left|X_{\omega \ell }(r) \right |^{2}\left( \omega - \frac{qQ}{r} \right)  ,
			\\
			j^{r}_{\omega \ell} = \sum _{m=-\ell }^{\ell }j^{r}_{\omega \ell m} = & ~
			-\frac{qf(r)\left(2 \ell + 1 \right)}{16\pi  ^{2}} \left| {\mathcal {N}}_{\omega } \right| ^{2} \Im \left[ \frac{X^{*}_{\omega \ell }(r)}{r} \frac{d}{dr} \left( \frac{X_{\omega \ell }(r)}{r} \right)  
			\right] .
		\end{align}
	\end{subequations}
	
	The components of the mode contributions to the stress-energy tensor are
	\begin{subequations}
		\begin{align}
			t_{tt,\omega \ell m } = & ~
			\frac{1}{2} 
			\left[  \left(  \omega - \frac{qQ}{r} \right) ^{2} + \frac{m^{2}f(r)}{r^{2}\sin ^{2}\theta }  \right] \left| \phi _{\omega \ell m} \right| ^{2} + \frac{f(r)^{2}}{2} 
			\left| {\mathcal {N}}_{\omega } \right| ^{2} \left| \frac{d}{dr} \left( \frac{X_{\omega \ell } (r)}{r}\right) \right| ^{2} \left| Y_{\ell m} (\theta, \varphi )\right| ^{2} 
			\nonumber \\ & 
			+ \frac{f(r)}{2r^{4}} \left| {\mathcal {N}}_{\omega }\right| ^{2} \left| X_{\omega \ell }(r) \right| ^{2} \left| \frac{\partial }{\partial \theta } Y_{\ell m}(\theta, \varphi ) \right| ^{2}  ,
			\\
			t_{tr,\omega \ell m} = & ~
			- \left( \omega - \frac{qQ}{r}\right)  \left| {\mathcal {N}}_{\omega } \right| ^{2}  \Im \left[ 
			\frac {X_{\omega \ell }^{*}(r)}{r} \frac{d}{dr}\left( \frac{X_{\omega \ell}(r)}{r} \right) 
			\right] \left| Y_{\ell m}(\theta , \varphi ) \right| ^{2}  , 
			\\
			t_{t\theta  ,\omega \ell m } = & ~
			-\frac{1}{r^{2}}\left( \omega - \frac{qQ}{r}\right) \left|  {\mathcal {N}}_{\omega }\right| ^{2}  
			\left| X_{\omega \ell}(r) \right| ^{2} \Im \left[ Y_{\ell m}(\theta, \varphi )
			\frac{\partial }{\partial \theta } Y_{\ell m}^{*}(\theta ,\varphi )
			\right] ,
			\\
			t_{t\varphi ,\omega \ell m } = & ~
			-m\left( \omega - \frac{qQ}{r} \right) \left| \phi _{\omega \ell m}\right| ^{2} ,
			\\
			t_{rr,\omega \ell m} = & ~
			\frac{1}{2} \left| {\mathcal {N}}_{\omega } \right| ^{2} \left| \frac{d}{dr} \left( \frac{X_{\omega \ell}(r)}{r} \right) \right| ^{2} \left| Y_{\ell m}(\theta, \varphi ) \right| ^{2}
			+\frac{1}{2f(r)^{2}} \left[ \left( \omega - \frac{qQ}{r} \right) ^{2} - \frac{m^{2}f(r)}{r^{2}\sin ^{2}\theta }
			\right] \left| \phi _{\omega \ell m}\right| ^{2}
			\nonumber \\ & 
			- \frac{1}{2r^{4}f(r)} \left| {\mathcal {N}}_{\omega }\right| ^{2} \left| X_{\omega \ell }(r) \right| ^{2} \left| \frac{\partial }{\partial \theta } Y_{\ell m}(\theta, \varphi ) \right| ^{2} ,
			\\
			t_{r\theta ,\omega \ell m} = & ~
			\left| {\mathcal {N}}_{\omega }\right| ^{2} \Re \left[  \frac {X_{\omega \ell}(r)}{r} \frac{d}{dr} \left( \frac{X_{\omega \ell}^{*}(r)}{r} \right) Y_{\ell m}^{*}(\theta, \varphi ) \frac{\partial }{\partial \theta }Y_{\ell m}(\theta, \varphi )
			\right] ,
			\\
			t_{r\varphi ,\omega \ell m} = & ~
			-m \left| {\mathcal {N}}_{\omega }\right| ^{2} 
			\Im \left[ \frac{X_{\omega \ell }(r)}{r} \frac{d}{dr} \left( \frac{X_{\omega \ell}^{*}(r)}{r}  \right) \right] 
			\left| Y_{\ell m}(\theta, \varphi ) \right| ^{2} ,
			\\
			t_{\theta \theta ,\omega \ell m} = & ~
			\frac{1}{2r^{2}}\left| {\mathcal {N}}_{\omega } \right| ^{2} \left| X_{\omega \ell}(r) \right| ^{2} \left| \frac{\partial }{\partial \theta } Y_{\ell m}(\theta, \varphi ) \right| ^{2}  
			+ \frac{1}{2} \left[ \frac{r^{2}}{f(r)} \left( \omega - \frac{qQ}{r} \right) ^{2} - \frac{m^{2}}{\sin ^{2}\theta }   \right] \left| \phi _{\omega \ell m}\right| ^{2}
			\nonumber \\ & 
			- \frac{f(r)r^{2}}{2} 
			\left| {\mathcal {N}}_{\omega } \right| ^{2} \left| \frac{d}{dr} \left( \frac{X_{\omega \ell } (r)}{r}\right) \right| ^{2} \left| Y_{\ell m} (\theta, \varphi )\right| ^{2} ,
			\\
			t_{\theta \varphi ,\omega \ell m} = & ~
			-\frac{m}{r^{2}} \left|  {\mathcal {N}}_{\omega }\right| ^{2}  
			\left| X_{\omega \ell}(r) \right| ^{2} \Im \left[ Y_{\ell m}(\theta, \varphi )
			\frac{\partial }{\partial \theta } Y_{\ell m}^{*}(\theta ,\varphi )
			\right] ,
			\\
			t_{\varphi \varphi ,\omega \ell m} = & ~
			\frac{1}{2}\left[ m^{2} + \frac{r^{2}\sin ^{2}\theta }{f(r)} \left( \omega - \frac{qQ}{r}\right) ^{2} \right] \left| \phi _{\omega \ell m}\right| ^{2} 
			- \frac{1}{2}f(r) r^{2}\sin ^{2}\theta  
			\left| {\mathcal {N}}_{\omega } \right| ^{2} \left| \frac{d}{dr} \left( \frac{X_{\omega \ell } (r)}{r}\right) \right| ^{2} \left| Y_{\ell m} (\theta, \varphi )\right| ^{2} 
			\nonumber \\ & 
			-\frac{\sin ^{2}\theta }{2r^{2}}\left| {\mathcal {N}}_{\omega } \right| ^{2} \left| X_{\omega \ell}(r) \right| ^{2} \left| \frac{\partial }{\partial \theta } Y_{\ell m}(\theta, \varphi ) \right| ^{2} ,
		\end{align}
	\end{subequations}
	where $\Re $ denotes the real part of a complex quantity.
	Using (\ref{eq:harmonics}), we immediately have that $t_{t\theta ,\omega \ell m}$ and $t_{\theta \varphi , \omega \ell m}$ vanish identically for all $\ell $ and $m$.
	As for the $\varphi $ component of the current, although the mode contributions to the stress-energy tensor components $t_{t\varphi ,\omega \ell m}$ and $t_{r\varphi ,\omega \ell m}$ are nonzero, when summed over $m$ they vanish.
	Using the identity (\ref{eq:Ylmresult1}),
	it is also the case that $t_{r\theta ,\omega \ell m}$ vanishes when summed over $m$.
	The remaining components can be summed over $m$ and simplified using the results (\ref{eq:Ylmresult4}, \ref{eq:Ylmresult2}, \ref{eq:Ylmresult3}).
	Defining 
	\begin{equation}
		t_{\mu \nu ,\omega \ell } = \sum _{m=-\ell}^{\ell } t_{\mu \nu , \omega \ell m},
	\end{equation}
	we find
	\begin{subequations}
		\begin{align}
			t_{tt, \omega \ell} = & ~
			\frac{2\ell + 1}{8\pi }\left| {\mathcal {N}}_{\omega }\right|^{2}
			\left\{ 
			\left[ \frac{1}{r^{2}} \left(  \omega - \frac{qQ}{r} \right) ^{2} + \frac{\ell \left(\ell + 1 \right) f(r)}{r^{4}}  \right]  \left| X_{\omega \ell }(r) \right|^{2}
			+ f(r)^{2}
			\left| \frac{d}{dr} \left( \frac{X_{\omega \ell } (r)}{r}\right) \right| ^{2} 
			\right\} ,
			\\
			t_{tr, \omega \ell  } = & ~
			- \frac{2\ell + 1}{4\pi }\left( \omega - \frac{qQ}{r}\right)  \left| {\mathcal {N}}_{\omega } \right| ^{2}  \Im \left[ 
			\frac {X_{\omega \ell }^{*}(r)}{r} \frac{d}{dr}\left( \frac{X_{\omega \ell}(r)}{r} \right) 
			\right] ,
			\\
			t_{rr, \omega \ell } = & ~
			\frac{2\ell + 1}{8\pi }\left| {\mathcal {N}}_{\omega }\right|^{2}
			\left\{ 
			\left[ \frac{1}{f(r)^{2}r^{2}} \left(  \omega - \frac{qQ}{r} \right) ^{2} - \frac{\ell \left(\ell + 1 \right) }{r^{4}f(r)}  \right]  \left| X_{\omega \ell }(r) \right|^{2}
			+ 
			\left| \frac{d}{dr} \left( \frac{X_{\omega \ell } (r)}{r}\right) \right| ^{2} 
			\right\} ,
			\\
			t_{\theta \theta , \omega \ell } = & ~
			\frac{2\ell + 1}{8\pi }\left| {\mathcal {N}}_{\omega }\right|^{2}
			\left\{ 
			\frac{1}{f(r)} \left(  \omega - \frac{qQ}{r} \right) ^{2} \left| X_{\omega \ell }(r) \right|^{2}
			- f(r)r^{2} 
			\left| \frac{d}{dr} \left( \frac{X_{\omega \ell } (r)}{r}\right) \right| ^{2} 
			\right\} ,
			\\
			t_{\varphi \varphi ,\omega \ell } = & ~ t_{\theta \theta , \omega \ell }\sin ^{2} \theta  .
		\end{align}
	\end{subequations}
	From these results the mode contribution to the trace of the stress-energy tensor is
	\begin{equation}
		t^{\mu }_{\mu , \omega \ell } =
		\frac{2\ell + 1}{4\pi }\left| {\mathcal {N}}_{\omega }\right|^{2}
		\left\{ 
		\left[ \frac{1}{f(r)r^{2}} \left(  \omega - \frac{qQ}{r} \right) ^{2} - \frac{\ell \left(\ell + 1 \right) }{r^{4}}  \right]  \left| X_{\omega \ell }(r) \right|^{2}
		- f(r)
		\left| \frac{d}{dr} \left( \frac{X_{\omega \ell } (r)}{r}\right) \right| ^{2} 
		\right\} .
	\end{equation}
\end{widetext}
Using the radial equation (\ref{eq:radial}), this simplifies to 
\begin{equation}
	t^{\mu }_{\mu , \omega \ell } = -\frac{2\ell + 1}{8\pi }\left|{\mathcal {N}}_{\omega }\right| ^{2} \Box  \left( \frac{\left| X_{\omega \ell }\right|^{2}}{r^{2}} \right) .
	\label{eq:trace1}
\end{equation}
From (\ref{eq:scmode}), the mode contribution to the scalar condensate is
\begin{equation}
	sc_{\omega \ell m} = \frac{1}{r^{2}}\left|{\mathcal {N}}_{\omega  \ell}\right|  ^{2}  \left| X_{\omega \ell }(r)\right| ^{2} \left|Y_{\ell m} (\theta, \varphi ) \right|^{2},
\end{equation}
and hence, using (\ref{eq:Ylmresult4}),
\begin{equation}
	sc_{\omega \ell} = \sum _{m=-\ell }^{\ell }sc_{\omega \ell m}  = \frac{2\ell +1}{4\pi r^{2}} \left|{\mathcal {N}}_{\omega  \ell}\right|  ^{2}  \left| X_{\omega \ell }(r)\right| ^{2} . 
	\label{eq:trace2}
\end{equation}
Comparing (\ref{eq:trace1}, \ref{eq:trace2}), we see that
\begin{equation}
	t^{\mu }_{\mu , \omega \ell } = - \frac{1}{2}\Box sc_{\omega \ell} .
\end{equation}
This is to be expected from (\ref{eq:trace}), since the curvature terms in that equation result from the renormalization process \cite{Balakumar:2019djw}.

\section{Some properties of spherical harmonics}
\label{sec:spherical}

In this final Appendix we collect some results for the spherical harmonics $Y_{\ell m}(\theta , \varphi )$ which are employed in App.~\ref{sec:JT} for simplifying the components of the current and stress-energy tensor.

We begin with the standard addition theorem for spherical harmonics 
\begin{equation}
P_{\ell }(\cos \gamma ) = \frac{4\pi }{2\ell + 1}\sum _{m=-\ell }^{\ell }Y_{\ell m}(\theta ,\varphi ) Y_{\ell m}^{*}(\theta' , \varphi '),
\label{eq:additiontheorem}
\end{equation}
where
\begin{equation}
\cos \gamma = \cos \theta \cos \theta ' + \sin \theta \sin \theta' \cos \left(\varphi - \varphi ' \right) .
\end{equation}
Taking the coincidence limit  $\theta ' = \theta $, $\varphi' = \varphi $ in (\ref{eq:additiontheorem}) yields the well-known addition formula
\begin{equation}
\sum _{m=-\ell }^{\ell } \left| Y_{\ell m}(\theta ,\varphi ) \right|^{2}
= \frac{2\ell +1}{4\pi } ,
\label{eq:Ylmresult4}
\end{equation}
since $P_{\ell }(1)=1$.

Differentiating both sides of (\ref{eq:additiontheorem}) with respect to $\theta $ gives
\begin{equation}
\frac{4\pi }{2\ell + 1}\sum _{m=-\ell }^{\ell }\frac{\partial Y_{\ell m}(\theta ,\varphi )}{\partial \theta } Y_{\ell m}^{*}(\theta' , \varphi ') 
= \frac{\partial (\cos \gamma)}{\partial \theta } P_{\ell }'(\cos \gamma ) ,
\label{eq:additiontheoremDtheta}
\end{equation}
with
\begin{equation}
\frac{\partial (\cos \gamma)}{\partial \theta } = -\sin \theta \cos \theta ' + \cos \theta \sin \theta ' \cos \left(\varphi - \varphi ' \right) .
\end{equation}
Taking the coincidence limit, we have 
\begin{equation}
\sum _{m=-\ell }^{\ell }\frac{\partial Y_{\ell m}}{\partial \theta } Y_{\ell m}^{*} =0.
\label{eq:Ylmresult1}
\end{equation}
We now differentiate (\ref{eq:additiontheoremDtheta}) with respect to $\theta '$ to obtain
\begin{multline}
\frac{4\pi}{2\ell +1}\sum_{m=-\ell}^{\ell}
\frac{\partial Y_{\ell m}(\theta,\varphi)}{\partial \theta}\frac{\partial Y^{*}_{\ell m}(\theta',\varphi')}{\partial \theta'}
\\
=\frac{\partial^2 (\cos \gamma)}{\partial \theta' \partial \theta}P'_{\ell}(\cos \gamma)+\frac{\partial (\cos \gamma )}{\partial \theta}\frac{\partial (\cos \gamma )}{\partial \theta'}P''_{\ell}(\cos \gamma),
\end{multline}
where
\begin{align}
\frac{\partial (\cos \gamma )}{\partial \theta'}= & ~
-\sin \theta' \cos \theta+\cos \theta' \sin \theta \cos(\varphi-\varphi') ,
\nonumber \\
\frac{\partial^2 (\cos \gamma)}{\partial \theta' \partial \theta}= & 
\sin\theta \sin\theta'+\cos \theta \cos \theta' \cos(\varphi-\varphi').
\end{align}
Taking the coincidence limit yields
\begin{equation}
\sum_{m=-\ell}^{\ell}\left|\frac{\partial Y_{lm}}{\partial \theta}\right|^2=\frac{2\ell+1}{4 \pi}P'_{\ell}(1)=
\frac{1}{8\pi } \ell \left( \ell + 1 \right) \left(
2\ell + 1\right) ,
\label{eq:Ylmresult2}
\end{equation}
since
\begin{equation}
P'_{\ell}(1)=\frac{\ell(\ell+1)}{2}.
\label{eq:Pl1}
\end{equation}
Our final identity is derived by differentiating the addition theorem (\ref{eq:additiontheorem}) with respect to $\varphi $ and then $\varphi' $, which gives
\begin{multline}
\frac{4\pi}{2\ell +1}\sum_{m=-\ell}^{\ell}
\frac{\partial Y_{\ell m}(\theta,\varphi)}{\partial \varphi}
\frac{\partial Y^{*}_{\ell m}(\theta',\varphi')}{\partial \varphi'}
\\ 
=\frac{\partial^2 (\cos \gamma)}{\partial \varphi' \partial \varphi}P'_{\ell}(\cos \gamma)+\frac{\partial (\cos \gamma )}{\partial \varphi}
\frac{\partial (\cos \gamma )}{\partial \varphi'}P''_{\ell}(\cos \gamma),
\end{multline}
with
\begin{align}
\frac{\partial (\cos \gamma )}{\partial \varphi}= & ~ 
-\frac{\partial (\cos \gamma )}{\partial \varphi'}=-\sin \theta \sin \theta' \sin(\varphi-\varphi') ,
\nonumber \\ 
\frac{\partial^2 (\cos \gamma)}{\partial \varphi' \partial \varphi}= & ~
\sin \theta \sin \theta' \cos(\varphi-\varphi') .
\end{align}
Taking the coincidence limit, and using (\ref{eq:harmonics}, \ref{eq:Pl1}), we find
\begin{align}
\sum_{m=-\ell}^{\ell}\left|\frac{\partial Y_{lm}}{\partial \varphi}\right|^2 & =\sum_{m=-\ell}^{\ell}m^2\left|Y_{lm}\right|^2
\nonumber \\
& =
\frac{1}{8\pi } \ell \left( \ell + 1 \right) \left(
2\ell + 1\right) \sin ^{2}\theta .
\label{eq:Ylmresult3}
\end{align}
\bigskip

\begin{acknowledgments}
	We thank Lu\'is C.~B.~Crispino for numerous invaluable discussions on this project. 
	We thank the anonymous referee for their careful reading of the paper and helpful comments which have improved the presentation.
	V.B.~thanks STFC for the provision of a studentship supporting this work,
	and the Faculdade de F\'isica, Universidade Federal do Par\'a, for hospitality while this work was in progress.
	The work of V.B. is also supported by an EPSRC Mathematical Sciences Research Associateship and a University of Sheffield Postgraduate
	Research Student Publication Scholarship.
	The work of R.P.B.~is financed in part by Coordena\c{c}\~ao de Aperfei\c{c}oamento de Pessoal de N\'ivel Superior (CAPES, Brazil) - Finance Code 001 and by 
	Conselho Nacional de Desenvolvimento Cient\'ifico e Tecnol\'ogico (CNPq, Brazil).
	The work of E.W.~is supported by the Lancaster-Manchester-Sheffield Consortium for Fundamental Physics under STFC grant ST/T001038/1.
	For the purpose of open access, the authors have applied a Creative Commons Attribution (CC-BY) licence to any Author Accepted Manuscript version arising.
	No data were created or analysed in this study.
	E.W.~thanks Bernard Kay, Jorma Louko and David Bruschi for insightful discussions.
	This research has also received funding from the European Union's Horizon 2020 research and innovation program under the H2020-MSCA-RISE-2017 Grant No.~FunFiCO-777740.  
\end{acknowledgments}

\bibliography{charge}

\begin{thebibliography}{87}%
\makeatletter
\providecommand \@ifxundefined [1]{%
 \@ifx{#1\undefined}
}%
\providecommand \@ifnum [1]{%
 \ifnum #1\expandafter \@firstoftwo
 \else \expandafter \@secondoftwo
 \fi
}%
\providecommand \@ifx [1]{%
 \ifx #1\expandafter \@firstoftwo
 \else \expandafter \@secondoftwo
 \fi
}%
\providecommand \natexlab [1]{#1}%
\providecommand \enquote  [1]{``#1''}%
\providecommand \bibnamefont  [1]{#1}%
\providecommand \bibfnamefont [1]{#1}%
\providecommand \citenamefont [1]{#1}%
\providecommand \href@noop [0]{\@secondoftwo}%
\providecommand \href [0]{\begingroup \@sanitize@url \@href}%
\providecommand \@href[1]{\@@startlink{#1}\@@href}%
\providecommand \@@href[1]{\endgroup#1\@@endlink}%
\providecommand \@sanitize@url [0]{\catcode `\\12\catcode `\$12\catcode
  `\&12\catcode `\#12\catcode `\^12\catcode `\_12\catcode `\%12\relax}%
\providecommand \@@startlink[1]{}%
\providecommand \@@endlink[0]{}%
\providecommand \url  [0]{\begingroup\@sanitize@url \@url }%
\providecommand \@url [1]{\endgroup\@href {#1}{\urlprefix }}%
\providecommand \urlprefix  [0]{URL }%
\providecommand \Eprint [0]{\href }%
\providecommand \doibase [0]{https://doi.org/}%
\providecommand \selectlanguage [0]{\@gobble}%
\providecommand \bibinfo  [0]{\@secondoftwo}%
\providecommand \bibfield  [0]{\@secondoftwo}%
\providecommand \translation [1]{[#1]}%
\providecommand \BibitemOpen [0]{}%
\providecommand \bibitemStop [0]{}%
\providecommand \bibitemNoStop [0]{.\EOS\space}%
\providecommand \EOS [0]{\spacefactor3000\relax}%
\providecommand \BibitemShut  [1]{\csname bibitem#1\endcsname}%
\let\auto@bib@innerbib\@empty
\bibitem [{\citenamefont {Hawking}(1974)}]{Hawking:1974rv}%
  \BibitemOpen
  \bibfield  {author} {\bibinfo {author} {\bibfnamefont {S.~W.}\ \bibnamefont
  {Hawking}},\ }\bibfield  {title} {\bibinfo {title} {{Black hole
  explosions}},\ }\href {https://doi.org/10.1038/248030a0} {\bibfield
  {journal} {\bibinfo  {journal} {Nature}\ }\textbf {\bibinfo {volume} {248}},\
  \bibinfo {pages} {30} (\bibinfo {year} {1974})}\BibitemShut {NoStop}%
\bibitem [{\citenamefont {Hawking}(1975)}]{Hawking:1974sw}%
  \BibitemOpen
  \bibfield  {author} {\bibinfo {author} {\bibfnamefont {S.~W.}\ \bibnamefont
  {Hawking}},\ }\bibfield  {title} {\bibinfo {title} {{Particle creation by
  black holes}},\ }\href {https://doi.org/10.1007/BF02345020} {\bibfield
  {journal} {\bibinfo  {journal} {Commun.\ Math.\ Phys.}\ }\textbf {\bibinfo
  {volume} {43}},\ \bibinfo {pages} {199} (\bibinfo {year} {1975})},\ \bibinfo
  {note} {[Erratum: Commun.~Math.~Phys. {\bf {46}}, 206 (1976)]}\BibitemShut
  {NoStop}%
\bibitem [{\citenamefont {Candelas}(1980)}]{Candelas:1980zt}%
  \BibitemOpen
  \bibfield  {author} {\bibinfo {author} {\bibfnamefont {P.}~\bibnamefont
  {Candelas}},\ }\bibfield  {title} {\bibinfo {title} {{Vacuum polarization in
  Schwarzschild space-time}},\ }\href
  {https://doi.org/10.1103/PhysRevD.21.2185} {\bibfield  {journal} {\bibinfo
  {journal} {Phys.\ Rev.\ D}\ }\textbf {\bibinfo {volume} {21}},\ \bibinfo
  {pages} {2185} (\bibinfo {year} {1980})}\BibitemShut {NoStop}%
\bibitem [{\citenamefont {Boulware}(1975)}]{Boulware:1974dm}%
  \BibitemOpen
  \bibfield  {author} {\bibinfo {author} {\bibfnamefont {D.~G.}\ \bibnamefont
  {Boulware}},\ }\bibfield  {title} {\bibinfo {title} {{Quantum field theory in
  Schwarzschild and Rindler spaces}},\ }\href
  {https://doi.org/10.1103/PhysRevD.11.1404} {\bibfield  {journal} {\bibinfo
  {journal} {Phys. Rev. D}\ }\textbf {\bibinfo {volume} {11}},\ \bibinfo
  {pages} {1404} (\bibinfo {year} {1975})}\BibitemShut {NoStop}%
\bibitem [{\citenamefont {Unruh}(1976)}]{Unruh:1976db}%
  \BibitemOpen
  \bibfield  {author} {\bibinfo {author} {\bibfnamefont {W.~G.}\ \bibnamefont
  {Unruh}},\ }\bibfield  {title} {\bibinfo {title} {{Notes on black hole
  evaporation}},\ }\href {https://doi.org/10.1103/PhysRevD.14.870} {\bibfield
  {journal} {\bibinfo  {journal} {Phys.\ Rev.\ D}\ }\textbf {\bibinfo {volume}
  {14}},\ \bibinfo {pages} {870} (\bibinfo {year} {1976})}\BibitemShut
  {NoStop}%
\bibitem [{\citenamefont {Hartle}\ and\ \citenamefont
  {Hawking}(1976)}]{Hartle:1976tp}%
  \BibitemOpen
  \bibfield  {author} {\bibinfo {author} {\bibfnamefont {J.~B.}\ \bibnamefont
  {Hartle}}\ and\ \bibinfo {author} {\bibfnamefont {S.~W.}\ \bibnamefont
  {Hawking}},\ }\bibfield  {title} {\bibinfo {title} {{Path integral derivation
  of black hole radiance}},\ }\href {https://doi.org/10.1103/PhysRevD.13.2188}
  {\bibfield  {journal} {\bibinfo  {journal} {Phys. Rev. D}\ }\textbf {\bibinfo
  {volume} {13}},\ \bibinfo {pages} {2188} (\bibinfo {year}
  {1976})}\BibitemShut {NoStop}%
\bibitem [{\citenamefont {Israel}(1976)}]{Israel:1976ur}%
  \BibitemOpen
  \bibfield  {author} {\bibinfo {author} {\bibfnamefont {W.}~\bibnamefont
  {Israel}},\ }\bibfield  {title} {\bibinfo {title} {{Thermo-field dynamics of
  black holes}},\ }\href {https://doi.org/10.1016/0375-9601(76)90178-X}
  {\bibfield  {journal} {\bibinfo  {journal} {Phys. Lett. A}\ }\textbf
  {\bibinfo {volume} {57}},\ \bibinfo {pages} {107} (\bibinfo {year}
  {1976})}\BibitemShut {NoStop}%
\bibitem [{\citenamefont {Anderson}(1989)}]{Anderson:1989vg}%
  \BibitemOpen
  \bibfield  {author} {\bibinfo {author} {\bibfnamefont {P.~R.}\ \bibnamefont
  {Anderson}},\ }\bibfield  {title} {\bibinfo {title} {{$\left<\phi^2\right>$
  for massive fields in Schwarzschild space-time}},\ }\href
  {https://doi.org/10.1103/PhysRevD.39.3785} {\bibfield  {journal} {\bibinfo
  {journal} {Phys.\ Rev.\ D}\ }\textbf {\bibinfo {volume} {39}},\ \bibinfo
  {pages} {3785} (\bibinfo {year} {1989})}\BibitemShut {NoStop}%
\bibitem [{\citenamefont {Anderson}\ \emph {et~al.}(1993)\citenamefont
  {Anderson}, \citenamefont {Hiscock},\ and\ \citenamefont
  {Samuel}}]{Anderson:1993if}%
  \BibitemOpen
  \bibfield  {author} {\bibinfo {author} {\bibfnamefont {P.~R.}\ \bibnamefont
  {Anderson}}, \bibinfo {author} {\bibfnamefont {W.~A.}\ \bibnamefont
  {Hiscock}},\ and\ \bibinfo {author} {\bibfnamefont {D.~A.}\ \bibnamefont
  {Samuel}},\ }\bibfield  {title} {\bibinfo {title} {{Stress energy tensor of
  quantized scalar fields in static black hole space-times}},\ }\href
  {https://doi.org/10.1103/PhysRevLett.70.1739} {\bibfield  {journal} {\bibinfo
   {journal} {Phys.\ Rev.\ Lett.}\ }\textbf {\bibinfo {volume} {70}},\ \bibinfo
  {pages} {1739} (\bibinfo {year} {1993})}\BibitemShut {NoStop}%
\bibitem [{\citenamefont {Anderson}\ \emph {et~al.}(1995)\citenamefont
  {Anderson}, \citenamefont {Hiscock},\ and\ \citenamefont
  {Samuel}}]{Anderson:1994hg}%
  \BibitemOpen
  \bibfield  {author} {\bibinfo {author} {\bibfnamefont {P.~R.}\ \bibnamefont
  {Anderson}}, \bibinfo {author} {\bibfnamefont {W.~A.}\ \bibnamefont
  {Hiscock}},\ and\ \bibinfo {author} {\bibfnamefont {D.~A.}\ \bibnamefont
  {Samuel}},\ }\bibfield  {title} {\bibinfo {title} {{Stress-energy tensor of
  quantized scalar fields in static spherically symmetric space-times}},\
  }\href {https://doi.org/10.1103/PhysRevD.51.4337} {\bibfield  {journal}
  {\bibinfo  {journal} {Phys.\ Rev.\ D}\ }\textbf {\bibinfo {volume} {51}},\
  \bibinfo {pages} {4337} (\bibinfo {year} {1995})}\BibitemShut {NoStop}%
\bibitem [{\citenamefont {Candelas}\ and\ \citenamefont
  {Howard}(1984)}]{Candelas:1984pg}%
  \BibitemOpen
  \bibfield  {author} {\bibinfo {author} {\bibfnamefont {P.}~\bibnamefont
  {Candelas}}\ and\ \bibinfo {author} {\bibfnamefont {K.~W.}\ \bibnamefont
  {Howard}},\ }\bibfield  {title} {\bibinfo {title} {{Vacuum $\langle \Phi
  ^{2}\rangle $ in Schwarzschild space-time}},\ }\href
  {https://doi.org/10.1103/PhysRevD.29.1618} {\bibfield  {journal} {\bibinfo
  {journal} {Phys.\ Rev.\ D}\ }\textbf {\bibinfo {volume} {29}},\ \bibinfo
  {pages} {1618} (\bibinfo {year} {1984})}\BibitemShut {NoStop}%
\bibitem [{\citenamefont {Carlson}\ \emph {et~al.}(2003)\citenamefont
  {Carlson}, \citenamefont {Hirsch}, \citenamefont {Obermayer}, \citenamefont
  {Anderson},\ and\ \citenamefont {Groves}}]{Carlson:2003ub}%
  \BibitemOpen
  \bibfield  {author} {\bibinfo {author} {\bibfnamefont {E.~D.}\ \bibnamefont
  {Carlson}}, \bibinfo {author} {\bibfnamefont {W.~H.}\ \bibnamefont {Hirsch}},
  \bibinfo {author} {\bibfnamefont {B.}~\bibnamefont {Obermayer}}, \bibinfo
  {author} {\bibfnamefont {P.~R.}\ \bibnamefont {Anderson}},\ and\ \bibinfo
  {author} {\bibfnamefont {P.~B.}\ \bibnamefont {Groves}},\ }\bibfield  {title}
  {\bibinfo {title} {{Stress energy tensor for the massless spin 1/2 field in
  static black hole space-times}},\ }\href
  {https://doi.org/10.1103/PhysRevLett.91.051301} {\bibfield  {journal}
  {\bibinfo  {journal} {Phys.\ Rev.\ Lett.}\ }\textbf {\bibinfo {volume}
  {91}},\ \bibinfo {pages} {051301} (\bibinfo {year} {2003})},\ \Eprint
  {https://arxiv.org/abs/gr-qc/0305045} {arXiv:gr-qc/0305045} \BibitemShut
  {NoStop}%
\bibitem [{\citenamefont {Elster}(1984)}]{Elster:1984hu}%
  \BibitemOpen
  \bibfield  {author} {\bibinfo {author} {\bibfnamefont {T.}~\bibnamefont
  {Elster}},\ }\bibfield  {title} {\bibinfo {title} {{Quantum vacuum energy
  near a black hole: the Maxwell field}},\ }\href
  {https://doi.org/10.1088/0264-9381/1/1/007} {\bibfield  {journal} {\bibinfo
  {journal} {Class.\ Quant.\ Grav.}\ }\textbf {\bibinfo {volume} {1}},\
  \bibinfo {pages} {43} (\bibinfo {year} {1984})}\BibitemShut {NoStop}%
\bibitem [{\citenamefont {Fawcett}(1983)}]{Fawcett:1983dk}%
  \BibitemOpen
  \bibfield  {author} {\bibinfo {author} {\bibfnamefont {M.~S.}\ \bibnamefont
  {Fawcett}},\ }\bibfield  {title} {\bibinfo {title} {{The energy momentum
  tensor near a black hole}},\ }\href {https://doi.org/10.1007/BF01219528}
  {\bibfield  {journal} {\bibinfo  {journal} {Commun.\ Math.\ Phys.}\ }\textbf
  {\bibinfo {volume} {89}},\ \bibinfo {pages} {103} (\bibinfo {year}
  {1983})}\BibitemShut {NoStop}%
\bibitem [{\citenamefont {Howard}\ and\ \citenamefont
  {Candelas}(1984)}]{Howard:1984qp}%
  \BibitemOpen
  \bibfield  {author} {\bibinfo {author} {\bibfnamefont {K.~W.}\ \bibnamefont
  {Howard}}\ and\ \bibinfo {author} {\bibfnamefont {P.}~\bibnamefont
  {Candelas}},\ }\bibfield  {title} {\bibinfo {title} {{Quantum stress tensor
  in Schwarzschild space-time}},\ }\href
  {https://doi.org/10.1103/PhysRevLett.53.403} {\bibfield  {journal} {\bibinfo
  {journal} {Phys.\ Rev.\ Lett.}\ }\textbf {\bibinfo {volume} {53}},\ \bibinfo
  {pages} {403} (\bibinfo {year} {1984})}\BibitemShut {NoStop}%
\bibitem [{\citenamefont {Howard}(1984)}]{Howard:1985yg}%
  \BibitemOpen
  \bibfield  {author} {\bibinfo {author} {\bibfnamefont {K.~W.}\ \bibnamefont
  {Howard}},\ }\bibfield  {title} {\bibinfo {title} {{Vacuum $\langle T_{\mu
  \nu } \rangle $ in Schwarzschild space-time}},\ }\href
  {https://doi.org/10.1103/PhysRevD.30.2532} {\bibfield  {journal} {\bibinfo
  {journal} {Phys.\ Rev.\ D}\ }\textbf {\bibinfo {volume} {30}},\ \bibinfo
  {pages} {2532} (\bibinfo {year} {1984})}\BibitemShut {NoStop}%
\bibitem [{\citenamefont {Jensen}\ and\ \citenamefont
  {Ottewill}(1989)}]{Jensen:1988rh}%
  \BibitemOpen
  \bibfield  {author} {\bibinfo {author} {\bibfnamefont {B.}~\bibnamefont
  {Jensen}}\ and\ \bibinfo {author} {\bibfnamefont {A.~C.}\ \bibnamefont
  {Ottewill}},\ }\bibfield  {title} {\bibinfo {title} {{Renormalized
  electromagnetic stress tensor in Schwarzschild space-time}},\ }\href
  {https://doi.org/10.1103/PhysRevD.39.1130} {\bibfield  {journal} {\bibinfo
  {journal} {Phys.\ Rev.\ D}\ }\textbf {\bibinfo {volume} {39}},\ \bibinfo
  {pages} {1130} (\bibinfo {year} {1989})}\BibitemShut {NoStop}%
\bibitem [{\citenamefont {Jensen}\ \emph {et~al.}(1992)\citenamefont {Jensen},
  \citenamefont {McLaughlin},\ and\ \citenamefont {Ottewill}}]{Jensen:1992mv}%
  \BibitemOpen
  \bibfield  {author} {\bibinfo {author} {\bibfnamefont {B.~P.}\ \bibnamefont
  {Jensen}}, \bibinfo {author} {\bibfnamefont {J.~G.}\ \bibnamefont
  {McLaughlin}},\ and\ \bibinfo {author} {\bibfnamefont {A.~C.}\ \bibnamefont
  {Ottewill}},\ }\bibfield  {title} {\bibinfo {title} {{Anisotropy of the
  quantum thermal state in Schwarzschild space-time}},\ }\href
  {https://doi.org/10.1103/PhysRevD.45.3002} {\bibfield  {journal} {\bibinfo
  {journal} {Phys.\ Rev.\ D}\ }\textbf {\bibinfo {volume} {45}},\ \bibinfo
  {pages} {3002} (\bibinfo {year} {1992})}\BibitemShut {NoStop}%
\bibitem [{\citenamefont {Jensen}\ \emph {et~al.}(1995)\citenamefont {Jensen},
  \citenamefont {McLaughlin},\ and\ \citenamefont {Ottewill}}]{Jensen:1995qv}%
  \BibitemOpen
  \bibfield  {author} {\bibinfo {author} {\bibfnamefont {B.~P.}\ \bibnamefont
  {Jensen}}, \bibinfo {author} {\bibfnamefont {J.~G.}\ \bibnamefont
  {McLaughlin}},\ and\ \bibinfo {author} {\bibfnamefont {A.~C.}\ \bibnamefont
  {Ottewill}},\ }\bibfield  {title} {\bibinfo {title} {{One loop quantum
  gravity in Schwarzschild space-time}},\ }\href
  {https://doi.org/10.1103/PhysRevD.51.5676} {\bibfield  {journal} {\bibinfo
  {journal} {Phys.\ Rev.\ D}\ }\textbf {\bibinfo {volume} {51}},\ \bibinfo
  {pages} {5676} (\bibinfo {year} {1995})},\ \Eprint
  {https://arxiv.org/abs/gr-qc/9412075} {arXiv:gr-qc/9412075} \BibitemShut
  {NoStop}%
\bibitem [{\citenamefont {Levi}\ and\ \citenamefont
  {Ori}(2015)}]{Levi:2015eea}%
  \BibitemOpen
  \bibfield  {author} {\bibinfo {author} {\bibfnamefont {A.}~\bibnamefont
  {Levi}}\ and\ \bibinfo {author} {\bibfnamefont {A.}~\bibnamefont {Ori}},\
  }\bibfield  {title} {\bibinfo {title} {{Pragmatic mode-sum regularization
  method for semiclassical black-hole spacetimes}},\ }\href
  {https://doi.org/10.1103/PhysRevD.91.104028} {\bibfield  {journal} {\bibinfo
  {journal} {Phys.\ Rev.\ D}\ }\textbf {\bibinfo {volume} {91}},\ \bibinfo
  {pages} {104028} (\bibinfo {year} {2015})},\ \Eprint
  {https://arxiv.org/abs/1503.02810} {arXiv:1503.02810 [gr-qc]} \BibitemShut
  {NoStop}%
\bibitem [{\citenamefont {Levi}\ and\ \citenamefont
  {Ori}(2016{\natexlab{a}})}]{Levi:2016quh}%
  \BibitemOpen
  \bibfield  {author} {\bibinfo {author} {\bibfnamefont {A.}~\bibnamefont
  {Levi}}\ and\ \bibinfo {author} {\bibfnamefont {A.}~\bibnamefont {Ori}},\
  }\bibfield  {title} {\bibinfo {title} {{Versatile method for renormalized
  stress-energy computation in black-hole spacetimes}},\ }\href
  {https://doi.org/10.1103/PhysRevLett.117.231101} {\bibfield  {journal}
  {\bibinfo  {journal} {Phys.\ Rev.\ Lett.}\ }\textbf {\bibinfo {volume}
  {117}},\ \bibinfo {pages} {231101} (\bibinfo {year} {2016}{\natexlab{a}})},\
  \Eprint {https://arxiv.org/abs/1608.03806} {arXiv:1608.03806 [gr-qc]}
  \BibitemShut {NoStop}%
\bibitem [{\citenamefont {Levi}\ and\ \citenamefont
  {Ori}(2016{\natexlab{b}})}]{Levi:2016esr}%
  \BibitemOpen
  \bibfield  {author} {\bibinfo {author} {\bibfnamefont {A.}~\bibnamefont
  {Levi}}\ and\ \bibinfo {author} {\bibfnamefont {A.}~\bibnamefont {Ori}},\
  }\bibfield  {title} {\bibinfo {title} {{Mode-sum regularization of
  $\left\langle \phi^{2} \right\rangle$ in the angular-splitting method}},\
  }\href {https://doi.org/10.1103/PhysRevD.94.044054} {\bibfield  {journal}
  {\bibinfo  {journal} {Phys.\ Rev.\ D}\ }\textbf {\bibinfo {volume} {94}},\
  \bibinfo {pages} {044054} (\bibinfo {year} {2016}{\natexlab{b}})},\ \Eprint
  {https://arxiv.org/abs/1606.08451} {arXiv:1606.08451 [gr-qc]} \BibitemShut
  {NoStop}%
\bibitem [{\citenamefont {Levi}(2017)}]{Levi:2016paz}%
  \BibitemOpen
  \bibfield  {author} {\bibinfo {author} {\bibfnamefont {A.}~\bibnamefont
  {Levi}},\ }\bibfield  {title} {\bibinfo {title} {{Renormalized stress-energy
  tensor for stationary black holes}},\ }\href
  {https://doi.org/10.1103/PhysRevD.95.025007} {\bibfield  {journal} {\bibinfo
  {journal} {Phys.\ Rev.\ D}\ }\textbf {\bibinfo {volume} {95}},\ \bibinfo
  {pages} {025007} (\bibinfo {year} {2017})},\ \Eprint
  {https://arxiv.org/abs/1611.05889} {arXiv:1611.05889 [gr-qc]} \BibitemShut
  {NoStop}%
\bibitem [{\citenamefont {Starobinsky}(1973)}]{Starobinsky:1973aij}%
  \BibitemOpen
  \bibfield  {author} {\bibinfo {author} {\bibfnamefont {A.~A.}\ \bibnamefont
  {Starobinsky}},\ }\bibfield  {title} {\bibinfo {title} {{Amplification of
  waves reflected from a rotating ``black hole''}},\ }\href@noop {} {\bibfield
  {journal} {\bibinfo  {journal} {Sov.\ Phys.\ JETP}\ }\textbf {\bibinfo
  {volume} {37}},\ \bibinfo {pages} {28} (\bibinfo {year} {1973})}\BibitemShut
  {NoStop}%
\bibitem [{\citenamefont {Unruh}(1974)}]{Unruh:1974bw}%
  \BibitemOpen
  \bibfield  {author} {\bibinfo {author} {\bibfnamefont {W.~G.}\ \bibnamefont
  {Unruh}},\ }\bibfield  {title} {\bibinfo {title} {{Second quantization in the
  Kerr metric}},\ }\href {https://doi.org/10.1103/PhysRevD.10.3194} {\bibfield
  {journal} {\bibinfo  {journal} {Phys.\ Rev.\ D}\ }\textbf {\bibinfo {volume}
  {10}},\ \bibinfo {pages} {3194} (\bibinfo {year} {1974})}\BibitemShut
  {NoStop}%
\bibitem [{\citenamefont {Chandrasekhar}(1985)}]{Chandrasekhar:1985kt}%
  \BibitemOpen
  \bibfield  {author} {\bibinfo {author} {\bibfnamefont {S.}~\bibnamefont
  {Chandrasekhar}},\ }\href@noop {} { {\bibinfo {title} {{{\it {The mathematical
  theory of black holes}}}}}}\ (\bibinfo  {publisher} {Oxford University Press},\
  \bibinfo {address} {Oxford, United Kingdom},\ \bibinfo {year}
  {1985})\BibitemShut {NoStop}%
\bibitem [{\citenamefont {Brito}\ \emph {et~al.}(2015)\citenamefont {Brito},
  \citenamefont {Cardoso},\ and\ \citenamefont {Pani}}]{Brito:2015oca}%
  \BibitemOpen
  \bibfield  {author} {\bibinfo {author} {\bibfnamefont {R.}~\bibnamefont
  {Brito}}, \bibinfo {author} {\bibfnamefont {V.}~\bibnamefont {Cardoso}},\
  and\ \bibinfo {author} {\bibfnamefont {P.}~\bibnamefont {Pani}},\ }\href
  {https://doi.org/10.1007/978-3-319-19000-6} { {\bibinfo {title}
  {{\it {{Superradiance}: {energy extraction, black-hole bombs and implications for
  astrophysics and particle physics}}}}}},\ Lect.\ Notes Phys.\ \bibinfo {volume} {{\bf {906}}}\
  (\bibinfo  {publisher} {Springer},\ \bibinfo {year} {2015})\ \Eprint
  {https://arxiv.org/abs/1501.06570} {arXiv:1501.06570 [gr-qc]} \BibitemShut
  {NoStop}%
\bibitem [{\citenamefont {Frolov}\ and\ \citenamefont
  {Thorne}(1989)}]{Frolov:1989jh}%
  \BibitemOpen
  \bibfield  {author} {\bibinfo {author} {\bibfnamefont {V.~P.}\ \bibnamefont
  {Frolov}}\ and\ \bibinfo {author} {\bibfnamefont {K.~S.}\ \bibnamefont
  {Thorne}},\ }\bibfield  {title} {\bibinfo {title} {{Renormalized
  stress-energy tensor near the horizon of a slowly evolving, rotating black
  hole}},\ }\href {https://doi.org/10.1103/PhysRevD.39.2125} {\bibfield
  {journal} {\bibinfo  {journal} {Phys. Rev. D}\ }\textbf {\bibinfo {volume}
  {39}},\ \bibinfo {pages} {2125} (\bibinfo {year} {1989})}\BibitemShut
  {NoStop}%
\bibitem [{\citenamefont {Ottewill}\ and\ \citenamefont
  {Winstanley}(2000)}]{Ottewill:2000qh}%
  \BibitemOpen
  \bibfield  {author} {\bibinfo {author} {\bibfnamefont {A.~C.}\ \bibnamefont
  {Ottewill}}\ and\ \bibinfo {author} {\bibfnamefont {E.}~\bibnamefont
  {Winstanley}},\ }\bibfield  {title} {\bibinfo {title} {{The renormalized
  stress tensor in Kerr space-time: general results}},\ }\href
  {https://doi.org/10.1103/PhysRevD.62.084018} {\bibfield  {journal} {\bibinfo
  {journal} {Phys. Rev. D}\ }\textbf {\bibinfo {volume} {62}},\ \bibinfo
  {pages} {084018} (\bibinfo {year} {2000})},\ \Eprint
  {https://arxiv.org/abs/gr-qc/0004022} {arXiv:gr-qc/0004022 [gr-qc]}
  \BibitemShut {NoStop}%
\bibitem [{\citenamefont {Casals}\ \emph {et~al.}(2013)\citenamefont {Casals},
  \citenamefont {Dolan}, \citenamefont {Nolan}, \citenamefont {Ottewill},\ and\
  \citenamefont {Winstanley}}]{Casals:2012es}%
  \BibitemOpen
  \bibfield  {author} {\bibinfo {author} {\bibfnamefont {M.}~\bibnamefont
  {Casals}}, \bibinfo {author} {\bibfnamefont {S.~R.}\ \bibnamefont {Dolan}},
  \bibinfo {author} {\bibfnamefont {B.~C.}\ \bibnamefont {Nolan}}, \bibinfo
  {author} {\bibfnamefont {A.~C.}\ \bibnamefont {Ottewill}},\ and\ \bibinfo
  {author} {\bibfnamefont {E.}~\bibnamefont {Winstanley}},\ }\bibfield  {title}
  {\bibinfo {title} {{Quantization of fermions on Kerr space-time}},\ }\href
  {https://doi.org/10.1103/PhysRevD.87.064027} {\bibfield  {journal} {\bibinfo
  {journal} {Phys.\ Rev.\ D}\ }\textbf {\bibinfo {volume} {87}},\ \bibinfo
  {pages} {064027} (\bibinfo {year} {2013})},\ \Eprint
  {https://arxiv.org/abs/1207.7089} {arXiv:1207.7089 [gr-qc]} \BibitemShut
  {NoStop}%
\bibitem [{\citenamefont {Casals}\ and\ \citenamefont
  {Ottewill}(2005)}]{Casals:2005kr}%
  \BibitemOpen
  \bibfield  {author} {\bibinfo {author} {\bibfnamefont {M.}~\bibnamefont
  {Casals}}\ and\ \bibinfo {author} {\bibfnamefont {A.~C.}\ \bibnamefont
  {Ottewill}},\ }\bibfield  {title} {\bibinfo {title} {{Canonical quantization
  of the electromagnetic field on the Kerr background}},\ }\href
  {https://doi.org/10.1103/PhysRevD.71.124016} {\bibfield  {journal} {\bibinfo
  {journal} {Phys.\ Rev.\ D}\ }\textbf {\bibinfo {volume} {71}},\ \bibinfo
  {pages} {124016} (\bibinfo {year} {2005})},\ \Eprint
  {https://arxiv.org/abs/gr-qc/0501005} {arXiv:gr-qc/0501005} \BibitemShut
  {NoStop}%
\bibitem [{\citenamefont {Duffy}\ and\ \citenamefont
  {Ottewill}(2008)}]{Duffy:2005mz}%
  \BibitemOpen
  \bibfield  {author} {\bibinfo {author} {\bibfnamefont {G.}~\bibnamefont
  {Duffy}}\ and\ \bibinfo {author} {\bibfnamefont {A.~C.}\ \bibnamefont
  {Ottewill}},\ }\bibfield  {title} {\bibinfo {title} {{The renormalized stress
  tensor in Kerr space-time: numerical results for the Hartle-Hawking
  vacuum}},\ }\href {https://doi.org/10.1103/PhysRevD.77.024007} {\bibfield
  {journal} {\bibinfo  {journal} {Phys.\ Rev.\ D}\ }\textbf {\bibinfo {volume}
  {77}},\ \bibinfo {pages} {024007} (\bibinfo {year} {2008})},\ \Eprint
  {https://arxiv.org/abs/gr-qc/0507116} {arXiv:gr-qc/0507116} \BibitemShut
  {NoStop}%
\bibitem [{\citenamefont {Matacz}\ \emph {et~al.}(1993)\citenamefont {Matacz},
  \citenamefont {Davies},\ and\ \citenamefont {Ottewill}}]{Matacz:1993hs}%
  \BibitemOpen
  \bibfield  {author} {\bibinfo {author} {\bibfnamefont {A.~L.}\ \bibnamefont
  {Matacz}}, \bibinfo {author} {\bibfnamefont {P.~C.~W.}\ \bibnamefont
  {Davies}},\ and\ \bibinfo {author} {\bibfnamefont {A.~C.}\ \bibnamefont
  {Ottewill}},\ }\bibfield  {title} {\bibinfo {title} {{Quantum vacuum
  instability near rotating stars}},\ }\href
  {https://doi.org/10.1103/PhysRevD.47.1557} {\bibfield  {journal} {\bibinfo
  {journal} {Phys.\ Rev.\ D}\ }\textbf {\bibinfo {volume} {47}},\ \bibinfo
  {pages} {1557} (\bibinfo {year} {1993})},\ \Eprint
  {https://arxiv.org/abs/gr-qc/9212004} {arXiv:gr-qc/9212004} \BibitemShut
  {NoStop}%
\bibitem [{\citenamefont {Levi}\ \emph {et~al.}(2017)\citenamefont {Levi},
  \citenamefont {Eilon}, \citenamefont {Ori},\ and\ \citenamefont {van~de
  Meent}}]{Levi:2016exv}%
  \BibitemOpen
  \bibfield  {author} {\bibinfo {author} {\bibfnamefont {A.}~\bibnamefont
  {Levi}}, \bibinfo {author} {\bibfnamefont {E.}~\bibnamefont {Eilon}},
  \bibinfo {author} {\bibfnamefont {A.}~\bibnamefont {Ori}},\ and\ \bibinfo
  {author} {\bibfnamefont {M.}~\bibnamefont {van~de Meent}},\ }\bibfield
  {title} {\bibinfo {title} {{Renormalized stress-energy tensor of an
  evaporating spinning black hole}},\ }\href
  {https://doi.org/10.1103/PhysRevLett.118.141102} {\bibfield  {journal}
  {\bibinfo  {journal} {Phys.\ Rev.\ Lett.}\ }\textbf {\bibinfo {volume}
  {118}},\ \bibinfo {pages} {141102} (\bibinfo {year} {2017})},\ \Eprint
  {https://arxiv.org/abs/1610.04848} {arXiv:1610.04848 [gr-qc]} \BibitemShut
  {NoStop}%
\bibitem [{\citenamefont {Kay}\ and\ \citenamefont {Wald}(1991)}]{Kay:1988mu}%
  \BibitemOpen
  \bibfield  {author} {\bibinfo {author} {\bibfnamefont {B.~S.}\ \bibnamefont
  {Kay}}\ and\ \bibinfo {author} {\bibfnamefont {R.~M.}\ \bibnamefont {Wald}},\
  }\bibfield  {title} {\bibinfo {title} {{Theorems on the uniqueness and
  thermal properties of stationary, nonsingular, quasifree states on
  space-times with a bifurcate Killing horizon}},\ }\href
  {https://doi.org/10.1016/0370-1573(91)90015-E} {\bibfield  {journal}
  {\bibinfo  {journal} {Phys.\ Rept.}\ }\textbf {\bibinfo {volume} {207}},\
  \bibinfo {pages} {49} (\bibinfo {year} {1991})}\BibitemShut {NoStop}%
\bibitem [{\citenamefont {Kay}(1993)}]{Kay:1992gr}%
  \BibitemOpen
  \bibfield  {author} {\bibinfo {author} {\bibfnamefont {B.~S.}\ \bibnamefont
  {Kay}},\ }\bibfield  {title} {\bibinfo {title} {{Sufficient conditions for
  quasifree states and an improved uniqueness theorem for quantum fields on
  space-times with horizons}},\ }\href {https://doi.org/10.1063/1.530354}
  {\bibfield  {journal} {\bibinfo  {journal} {J.\ Math.\ Phys.}\ }\textbf
  {\bibinfo {volume} {34}},\ \bibinfo {pages} {4519} (\bibinfo {year}
  {1993})}\BibitemShut {NoStop}%
\bibitem [{\citenamefont {Candelas}\ \emph {et~al.}(1981)\citenamefont
  {Candelas}, \citenamefont {Chrzanowski},\ and\ \citenamefont
  {Howard}}]{Candelas:1981zv}%
  \BibitemOpen
  \bibfield  {author} {\bibinfo {author} {\bibfnamefont {P.}~\bibnamefont
  {Candelas}}, \bibinfo {author} {\bibfnamefont {P.}~\bibnamefont
  {Chrzanowski}},\ and\ \bibinfo {author} {\bibfnamefont {K.~W.}\ \bibnamefont
  {Howard}},\ }\bibfield  {title} {\bibinfo {title} {{Quantization of
  electromagnetic and gravitational perturbations of a Kerr black hole}},\
  }\href {https://doi.org/10.1103/PhysRevD.24.297} {\bibfield  {journal}
  {\bibinfo  {journal} {Phys. Rev. D}\ }\textbf {\bibinfo {volume} {24}},\
  \bibinfo {pages} {297} (\bibinfo {year} {1981})}\BibitemShut {NoStop}%
\bibitem [{\citenamefont {Letaw}\ and\ \citenamefont
  {Pfautsch}(1980)}]{Letaw:1979wy}%
  \BibitemOpen
  \bibfield  {author} {\bibinfo {author} {\bibfnamefont {J.~R.}\ \bibnamefont
  {Letaw}}\ and\ \bibinfo {author} {\bibfnamefont {J.~D.}\ \bibnamefont
  {Pfautsch}},\ }\bibfield  {title} {\bibinfo {title} {{The quantized scalar
  field in rotating coordinates}},\ }\href
  {https://doi.org/10.1103/PhysRevD.22.1345} {\bibfield  {journal} {\bibinfo
  {journal} {Phys.\ Rev.\ D}\ }\textbf {\bibinfo {volume} {22}},\ \bibinfo
  {pages} {1345} (\bibinfo {year} {1980})}\BibitemShut {NoStop}%
\bibitem [{\citenamefont {Vilenkin}(1980)}]{Vilenkin:1980zv}%
  \BibitemOpen
  \bibfield  {author} {\bibinfo {author} {\bibfnamefont {A.}~\bibnamefont
  {Vilenkin}},\ }\bibfield  {title} {\bibinfo {title} {{Quantum field theory at
  finite temperature in a rotating system}},\ }\href
  {https://doi.org/10.1103/PhysRevD.21.2260} {\bibfield  {journal} {\bibinfo
  {journal} {Phys.\ Rev.\ D}\ }\textbf {\bibinfo {volume} {21}},\ \bibinfo
  {pages} {2260} (\bibinfo {year} {1980})}\BibitemShut {NoStop}%
\bibitem [{\citenamefont {Davies}\ \emph {et~al.}(1996)\citenamefont {Davies},
  \citenamefont {Dray},\ and\ \citenamefont {Manogue}}]{Davies:1996ks}%
  \BibitemOpen
  \bibfield  {author} {\bibinfo {author} {\bibfnamefont {P.~C.~W.}\
  \bibnamefont {Davies}}, \bibinfo {author} {\bibfnamefont {T.}~\bibnamefont
  {Dray}},\ and\ \bibinfo {author} {\bibfnamefont {C.~A.}\ \bibnamefont
  {Manogue}},\ }\bibfield  {title} {\bibinfo {title} {{The rotating quantum
  vacuum}},\ }\href {https://doi.org/10.1103/PhysRevD.53.4382} {\bibfield
  {journal} {\bibinfo  {journal} {Phys.\ Rev.\ D}\ }\textbf {\bibinfo {volume}
  {53}},\ \bibinfo {pages} {4382} (\bibinfo {year} {1996})},\ \Eprint
  {https://arxiv.org/abs/gr-qc/9601034} {arXiv:gr-qc/9601034} \BibitemShut
  {NoStop}%
\bibitem [{\citenamefont {Duffy}\ and\ \citenamefont
  {Ottewill}(2003)}]{Duffy:2002ss}%
  \BibitemOpen
  \bibfield  {author} {\bibinfo {author} {\bibfnamefont {G.}~\bibnamefont
  {Duffy}}\ and\ \bibinfo {author} {\bibfnamefont {A.~C.}\ \bibnamefont
  {Ottewill}},\ }\bibfield  {title} {\bibinfo {title} {{The rotating quantum
  thermal distribution}},\ }\href {https://doi.org/10.1103/PhysRevD.67.044002}
  {\bibfield  {journal} {\bibinfo  {journal} {Phys.\ Rev.\ D}\ }\textbf
  {\bibinfo {volume} {67}},\ \bibinfo {pages} {044002} (\bibinfo {year}
  {2003})},\ \Eprint {https://arxiv.org/abs/hep-th/0211096}
  {arXiv:hep-th/0211096} \BibitemShut {NoStop}%
\bibitem [{\citenamefont {Iyer}(1982)}]{Iyer:1982ah}%
  \BibitemOpen
  \bibfield  {author} {\bibinfo {author} {\bibfnamefont {B.~R.}\ \bibnamefont
  {Iyer}},\ }\bibfield  {title} {\bibinfo {title} {{Dirac field theory in
  rotating coordinates}},\ }\href {https://doi.org/10.1103/PhysRevD.26.1900}
  {\bibfield  {journal} {\bibinfo  {journal} {Phys.\ Rev.\ D}\ }\textbf
  {\bibinfo {volume} {26}},\ \bibinfo {pages} {1900} (\bibinfo {year}
  {1982})}\BibitemShut {NoStop}%
\bibitem [{\citenamefont {Ambru\c{s}}\ and\ \citenamefont
  {Winstanley}(2014)}]{Ambrus:2014uqa}%
  \BibitemOpen
  \bibfield  {author} {\bibinfo {author} {\bibfnamefont {V.~E.}\ \bibnamefont
  {Ambru\c{s}}}\ and\ \bibinfo {author} {\bibfnamefont {E.}~\bibnamefont
  {Winstanley}},\ }\bibfield  {title} {\bibinfo {title} {{Rotating quantum
  states}},\ }\href {https://doi.org/10.1016/j.physletb.2014.05.031} {\bibfield
   {journal} {\bibinfo  {journal} {Phys.\ Lett.\ B}\ }\textbf {\bibinfo
  {volume} {734}},\ \bibinfo {pages} {296} (\bibinfo {year} {2014})},\ \Eprint
  {https://arxiv.org/abs/1401.6388} {arXiv:1401.6388 [hep-th]} \BibitemShut
  {NoStop}%
\bibitem [{\citenamefont {Bekenstein}(1973)}]{Bekenstein:1973mi}%
  \BibitemOpen
  \bibfield  {author} {\bibinfo {author} {\bibfnamefont {J.~D.}\ \bibnamefont
  {Bekenstein}},\ }\bibfield  {title} {\bibinfo {title} {{Extraction of energy
  and charge from a black hole}},\ }\href
  {https://doi.org/10.1103/PhysRevD.7.949} {\bibfield  {journal} {\bibinfo
  {journal} {Phys. Rev. D}\ }\textbf {\bibinfo {volume} {7}},\ \bibinfo {pages}
  {949} (\bibinfo {year} {1973})}\BibitemShut {NoStop}%
\bibitem [{\citenamefont {Benone}\ and\ \citenamefont
  {Crispino}(2016)}]{Benone:2015bst}%
  \BibitemOpen
  \bibfield  {author} {\bibinfo {author} {\bibfnamefont {C.~L.}\ \bibnamefont
  {Benone}}\ and\ \bibinfo {author} {\bibfnamefont {L.~C.~B.}\ \bibnamefont
  {Crispino}},\ }\bibfield  {title} {\bibinfo {title} {{Superradiance in static
  black hole spacetimes}},\ }\href {https://doi.org/10.1103/PhysRevD.93.024028}
  {\bibfield  {journal} {\bibinfo  {journal} {Phys. Rev. D}\ }\textbf {\bibinfo
  {volume} {93}},\ \bibinfo {pages} {024028} (\bibinfo {year} {2016})},\
  \Eprint {https://arxiv.org/abs/1511.02634} {arXiv:1511.02634 [gr-qc]}
  \BibitemShut {NoStop}%
\bibitem [{\citenamefont {Di~Menza}\ and\ \citenamefont
  {Nicolas}(2015)}]{DiMenza:2014vpa}%
  \BibitemOpen
  \bibfield  {author} {\bibinfo {author} {\bibfnamefont {L.}~\bibnamefont
  {Di~Menza}}\ and\ \bibinfo {author} {\bibfnamefont {J.-P.}\ \bibnamefont
  {Nicolas}},\ }\bibfield  {title} {\bibinfo {title} {{Superradiance on the
  Reissner--Nordstr\"om metric}},\ }\href
  {https://doi.org/10.1088/0264-9381/32/14/145013} {\bibfield  {journal}
  {\bibinfo  {journal} {Class.\ Quant.\ Grav.}\ }\textbf {\bibinfo {volume}
  {32}},\ \bibinfo {pages} {145013} (\bibinfo {year} {2015})},\ \Eprint
  {https://arxiv.org/abs/1411.3988} {arXiv:1411.3988 [math-ph]} \BibitemShut
  {NoStop}%
\bibitem [{\citenamefont {Gibbons}(1975)}]{Gibbons:1975kk}%
  \BibitemOpen
  \bibfield  {author} {\bibinfo {author} {\bibfnamefont {G.~W.}\ \bibnamefont
  {Gibbons}},\ }\bibfield  {title} {\bibinfo {title} {{Vacuum polarization and
  the spontaneous loss of charge by black holes}},\ }\href
  {https://doi.org/10.1007/BF01609829} {\bibfield  {journal} {\bibinfo
  {journal} {Commun.\ Math.\ Phys.}\ }\textbf {\bibinfo {volume} {44}},\
  \bibinfo {pages} {245} (\bibinfo {year} {1975})}\BibitemShut {NoStop}%
\bibitem [{\citenamefont {Balakumar}\ \emph {et~al.}(2020)\citenamefont
  {Balakumar}, \citenamefont {Winstanley}, \citenamefont {Bernar},\ and\
  \citenamefont {Crispino}}]{Balakumar:2020gli}%
  \BibitemOpen
  \bibfield  {author} {\bibinfo {author} {\bibfnamefont {V.}~\bibnamefont
  {Balakumar}}, \bibinfo {author} {\bibfnamefont {E.}~\bibnamefont
  {Winstanley}}, \bibinfo {author} {\bibfnamefont {R.~P.}\ \bibnamefont
  {Bernar}},\ and\ \bibinfo {author} {\bibfnamefont {L.~C.~B.}\ \bibnamefont
  {Crispino}},\ }\bibfield  {title} {\bibinfo {title} {{Quantum superradiance
  on static black hole space-times}},\ }\href
  {https://doi.org/10.1016/j.physletb.2020.135904} {\bibfield  {journal}
  {\bibinfo  {journal} {Phys. Lett. B}\ }\textbf {\bibinfo {volume} {811}},\
  \bibinfo {pages} {135904} (\bibinfo {year} {2020})},\ \Eprint
  {https://arxiv.org/abs/2010.01630} {arXiv:2010.01630 [gr-qc]} \BibitemShut
  {NoStop}%
\bibitem [{\citenamefont {Carter}(1974)}]{Carter:1974yx}%
  \BibitemOpen
  \bibfield  {author} {\bibinfo {author} {\bibfnamefont {B.}~\bibnamefont
  {Carter}},\ }\bibfield  {title} {\bibinfo {title} {{Charge and particle
  conservation in black hole decay}},\ }\href
  {https://doi.org/10.1103/PhysRevLett.33.558} {\bibfield  {journal} {\bibinfo
  {journal} {Phys.\ Rev.\ Lett.}\ }\textbf {\bibinfo {volume} {33}},\ \bibinfo
  {pages} {558} (\bibinfo {year} {1974})}\BibitemShut {NoStop}%
\bibitem [{\citenamefont {Page}(1977)}]{Page:1977um}%
  \BibitemOpen
  \bibfield  {author} {\bibinfo {author} {\bibfnamefont {D.~N.}\ \bibnamefont
  {Page}},\ }\bibfield  {title} {\bibinfo {title} {{Particle emission rates
  from a black hole. 3. charged leptons from a nonrotating hole}},\ }\href
  {https://doi.org/10.1103/PhysRevD.16.2402} {\bibfield  {journal} {\bibinfo
  {journal} {Phys.\ Rev.\ D}\ }\textbf {\bibinfo {volume} {16}},\ \bibinfo
  {pages} {2402} (\bibinfo {year} {1977})}\BibitemShut {NoStop}%
\bibitem [{\citenamefont {Sampaio}(2010)}]{Sampaio:2009tp}%
  \BibitemOpen
  \bibfield  {author} {\bibinfo {author} {\bibfnamefont {M.~O.~P.}\
  \bibnamefont {Sampaio}},\ }\bibfield  {title} {\bibinfo {title}
  {{Distributions of charged massive scalars and fermions from evaporating
  higher-dimensional black holes}},\ }\href
  {https://doi.org/10.1007/JHEP02(2010)042} {\bibfield  {journal} {\bibinfo
  {journal} {JHEP}\ }\textbf {\bibinfo {volume} {02}},\ \bibinfo {pages}
  {042}},\ \Eprint {https://arxiv.org/abs/0911.0688} {arXiv:0911.0688 [hep-th]}
  \BibitemShut {NoStop}%
\bibitem [{\citenamefont {Sampaio}(2009)}]{Sampaio:2009ra}%
  \BibitemOpen
  \bibfield  {author} {\bibinfo {author} {\bibfnamefont {M.~O.~P.}\
  \bibnamefont {Sampaio}},\ }\bibfield  {title} {\bibinfo {title} {{Charge and
  mass effects on the evaporation of higher-dimensional rotating black
  holes}},\ }\href {https://doi.org/10.1088/1126-6708/2009/10/008} {\bibfield
  {journal} {\bibinfo  {journal} {JHEP}\ }\textbf {\bibinfo {volume} {10}},\
  \bibinfo {pages} {008}},\ \Eprint {https://arxiv.org/abs/0907.5107}
  {arXiv:0907.5107 [hep-th]} \BibitemShut {NoStop}%
\bibitem [{\citenamefont {Hiscock}\ and\ \citenamefont
  {Weems}(1990)}]{Hiscock:1990ex}%
  \BibitemOpen
  \bibfield  {author} {\bibinfo {author} {\bibfnamefont {W.~A.}\ \bibnamefont
  {Hiscock}}\ and\ \bibinfo {author} {\bibfnamefont {L.~D.}\ \bibnamefont
  {Weems}},\ }\bibfield  {title} {\bibinfo {title} {{Evolution of charged
  evaporating black holes}},\ }\href {https://doi.org/10.1103/PhysRevD.41.1142}
  {\bibfield  {journal} {\bibinfo  {journal} {Phys.\ Rev.\ D}\ }\textbf
  {\bibinfo {volume} {41}},\ \bibinfo {pages} {1142} (\bibinfo {year}
  {1990})}\BibitemShut {NoStop}%
\bibitem [{\citenamefont {Ong}(2019)}]{Ong:2019rnn}%
  \BibitemOpen
  \bibfield  {author} {\bibinfo {author} {\bibfnamefont {Y.~C.}\ \bibnamefont
  {Ong}},\ }\bibfield  {title} {\bibinfo {title} {{The charge of electron, weak
  gravity conjecture and black hole evolution}},\ }\href@noop {} {\  (\bibinfo
  {year} {2019})},\ \Eprint {https://arxiv.org/abs/1909.09977}
  {arXiv:1909.09977 [gr-qc]} \BibitemShut {NoStop}%
\bibitem [{\citenamefont {Ong}(2021)}]{Ong:2019vnv}%
  \BibitemOpen
  \bibfield  {author} {\bibinfo {author} {\bibfnamefont {Y.~C.}\ \bibnamefont
  {Ong}},\ }\bibfield  {title} {\bibinfo {title} {{The attractor of evaporating
  Reissner\textendash{}Nordstr\"om black holes}},\ }\href
  {https://doi.org/10.1140/epjp/s13360-020-00995-4} {\bibfield  {journal}
  {\bibinfo  {journal} {Eur. Phys. J. Plus}\ }\textbf {\bibinfo {volume}
  {136}},\ \bibinfo {pages} {61} (\bibinfo {year} {2021})},\ \Eprint
  {https://arxiv.org/abs/1909.09981} {arXiv:1909.09981 [gr-qc]} \BibitemShut
  {NoStop}%
\bibitem [{\citenamefont {Xu}\ \emph {et~al.}(2020)\citenamefont {Xu},
  \citenamefont {Ong},\ and\ \citenamefont {Yung}}]{Xu:2019wak}%
  \BibitemOpen
  \bibfield  {author} {\bibinfo {author} {\bibfnamefont {H.}~\bibnamefont
  {Xu}}, \bibinfo {author} {\bibfnamefont {Y.~C.}\ \bibnamefont {Ong}},\ and\
  \bibinfo {author} {\bibfnamefont {M.-H.}\ \bibnamefont {Yung}},\ }\bibfield
  {title} {\bibinfo {title} {{Cosmic censorship and the evolution of
  d-dimensional charged evaporating black holes}},\ }\href
  {https://doi.org/10.1103/PhysRevD.101.064015} {\bibfield  {journal} {\bibinfo
   {journal} {Phys.\ Rev.\ D}\ }\textbf {\bibinfo {volume} {101}},\ \bibinfo
  {pages} {064015} (\bibinfo {year} {2020})},\ \Eprint
  {https://arxiv.org/abs/1911.11990} {arXiv:1911.11990 [gr-qc]} \BibitemShut
  {NoStop}%
\bibitem [{\citenamefont {Klein}\ \emph {et~al.}(2021)\citenamefont {Klein},
  \citenamefont {Zahn},\ and\ \citenamefont {Hollands}}]{Klein:2021ctt}%
  \BibitemOpen
  \bibfield  {author} {\bibinfo {author} {\bibfnamefont {C.}~\bibnamefont
  {Klein}}, \bibinfo {author} {\bibfnamefont {J.}~\bibnamefont {Zahn}},\ and\
  \bibinfo {author} {\bibfnamefont {S.}~\bibnamefont {Hollands}},\ }\bibfield
  {title} {\bibinfo {title} {{Quantum (dis)charge of black hole interiors}},\
  }\href {https://doi.org/10.1103/PhysRevLett.127.231301} {\bibfield  {journal}
  {\bibinfo  {journal} {Phys. Rev. Lett.}\ }\textbf {\bibinfo {volume} {127}},\
  \bibinfo {pages} {231301} (\bibinfo {year} {2021})},\ \Eprint
  {https://arxiv.org/abs/2103.03714} {arXiv:2103.03714 [gr-qc]} \BibitemShut
  {NoStop}%
\bibitem [{\citenamefont {Klein}\ and\ \citenamefont
  {Zahn}(2021)}]{Klein:2021les}%
  \BibitemOpen
  \bibfield  {author} {\bibinfo {author} {\bibfnamefont {C.}~\bibnamefont
  {Klein}}\ and\ \bibinfo {author} {\bibfnamefont {J.}~\bibnamefont {Zahn}},\
  }\bibfield  {title} {\bibinfo {title} {{Renormalized charged scalar current
  in the Reissner\textendash{}Nordstr\"om\textendash{}de Sitter spacetime}},\
  }\href {https://doi.org/10.1103/PhysRevD.104.025009} {\bibfield  {journal}
  {\bibinfo  {journal} {Phys. Rev. D}\ }\textbf {\bibinfo {volume} {104}},\
  \bibinfo {pages} {025009} (\bibinfo {year} {2021})},\ \Eprint
  {https://arxiv.org/abs/2104.06005} {arXiv:2104.06005 [gr-qc]} \BibitemShut
  {NoStop}%
\bibitem [{\citenamefont {Macedo}\ \emph {et~al.}(2012)\citenamefont {Macedo},
  \citenamefont {Crispino},\ and\ \citenamefont {Cardoso}}]{Macedo:2012zz}%
  \BibitemOpen
  \bibfield  {author} {\bibinfo {author} {\bibfnamefont {C.~F.}\ \bibnamefont
  {Macedo}}, \bibinfo {author} {\bibfnamefont {L.~C.}\ \bibnamefont
  {Crispino}},\ and\ \bibinfo {author} {\bibfnamefont {V.}~\bibnamefont
  {Cardoso}},\ }\bibfield  {title} {\bibinfo {title} {{Semiclassical analysis
  of the scalar geodesic synchrotron radiation in Kerr spacetime}},\ }\href
  {https://doi.org/10.1103/PhysRevD.86.024002} {\bibfield  {journal} {\bibinfo
  {journal} {Phys. Rev. D}\ }\textbf {\bibinfo {volume} {86}},\ \bibinfo
  {pages} {024002} (\bibinfo {year} {2012})}\BibitemShut {NoStop}%
\bibitem [{\citenamefont {Novikov}\ and\ \citenamefont
  {Frolov}(1989)}]{Novikov:1989sz}%
  \BibitemOpen
  \bibfield  {author} {\bibinfo {author} {\bibfnamefont {I.~D.}\ \bibnamefont
  {Novikov}}\ and\ \bibinfo {author} {\bibfnamefont {V.~P.}\ \bibnamefont
  {Frolov}},\ }\href {https://doi.org/10.1007/978-94-017-2651-1} {
  {\bibinfo {title} {{{\it {Physics of black holes}}}}}},\ Vol.~\bibinfo {volume} {27}\
  (\bibinfo  {publisher} {Kluwer Academic},\ \bibinfo {address} {Dordrecht,
  Netherlands},\ \bibinfo {year} {1989})\BibitemShut {NoStop}%
\bibitem [{\citenamefont {Balakumar}\ and\ \citenamefont
  {Winstanley}(2020{\natexlab{a}})}]{Balakumar:2019djw}%
  \BibitemOpen
  \bibfield  {author} {\bibinfo {author} {\bibfnamefont {V.}~\bibnamefont
  {Balakumar}}\ and\ \bibinfo {author} {\bibfnamefont {E.}~\bibnamefont
  {Winstanley}},\ }\bibfield  {title} {\bibinfo {title} {{Hadamard
  renormalization for a charged scalar field}},\ }\href
  {https://doi.org/10.1088/1361-6382/ab6b6e} {\bibfield  {journal} {\bibinfo
  {journal} {Class.\ Quant.\ Grav.}\ }\textbf {\bibinfo {volume} {37}},\
  \bibinfo {pages} {065004} (\bibinfo {year} {2020}{\natexlab{a}})},\ \Eprint
  {https://arxiv.org/abs/1910.03666} {arXiv:1910.03666 [gr-qc]} \BibitemShut
  {NoStop}%
\bibitem [{\citenamefont {Denardo}\ and\ \citenamefont
  {Ruffini}(1973)}]{Denardo:1973pyo}%
  \BibitemOpen
  \bibfield  {author} {\bibinfo {author} {\bibfnamefont {G.}~\bibnamefont
  {Denardo}}\ and\ \bibinfo {author} {\bibfnamefont {R.}~\bibnamefont
  {Ruffini}},\ }\bibfield  {title} {\bibinfo {title} {{On the energetics of
  Reissner Nordstr\"om geometries}},\ }\href
  {https://doi.org/10.1016/0370-2693(73)90198-6} {\bibfield  {journal}
  {\bibinfo  {journal} {Phys.\ Lett.\ B}\ }\textbf {\bibinfo {volume} {45}},\
  \bibinfo {pages} {259} (\bibinfo {year} {1973})}\BibitemShut {NoStop}%
\bibitem [{\citenamefont {Denardo}\ \emph {et~al.}(1974)\citenamefont
  {Denardo}, \citenamefont {Hively},\ and\ \citenamefont
  {Ruffini}}]{Denardo:1974qis}%
  \BibitemOpen
  \bibfield  {author} {\bibinfo {author} {\bibfnamefont {G.}~\bibnamefont
  {Denardo}}, \bibinfo {author} {\bibfnamefont {L.}~\bibnamefont {Hively}},\
  and\ \bibinfo {author} {\bibfnamefont {R.}~\bibnamefont {Ruffini}},\
  }\bibfield  {title} {\bibinfo {title} {{On the generalized ergosphere of the
  Kerr-Newman geometry}},\ }\href
  {https://doi.org/10.1016/0370-2693(74)90557-7} {\bibfield  {journal}
  {\bibinfo  {journal} {Phys.\ Lett.\ B}\ }\textbf {\bibinfo {volume} {50}},\
  \bibinfo {pages} {270} (\bibinfo {year} {1974})}\BibitemShut {NoStop}%
\bibitem [{\citenamefont {Hod}(2013{\natexlab{a}})}]{Hod:2013nn}%
  \BibitemOpen
  \bibfield  {author} {\bibinfo {author} {\bibfnamefont {S.}~\bibnamefont
  {Hod}},\ }\bibfield  {title} {\bibinfo {title} {{No-bomb theorem for charged
  Reissner-Nordstroem black holes}},\ }\href
  {https://doi.org/10.1016/j.physletb.2012.12.013} {\bibfield  {journal}
  {\bibinfo  {journal} {Phys.\ Lett.\ B}\ }\textbf {\bibinfo {volume} {718}},\
  \bibinfo {pages} {1489} (\bibinfo {year} {2013}{\natexlab{a}})}\BibitemShut
  {NoStop}%
\bibitem [{\citenamefont {Hod}(2015)}]{Hod:2015hza}%
  \BibitemOpen
  \bibfield  {author} {\bibinfo {author} {\bibfnamefont {S.}~\bibnamefont
  {Hod}},\ }\bibfield  {title} {\bibinfo {title} {{Stability of highly-charged
  Reissner-Nordstr\"om black holes to charged scalar perturbations}},\ }\href
  {https://doi.org/10.1103/PhysRevD.91.044047} {\bibfield  {journal} {\bibinfo
  {journal} {Phys.\ Rev.\ D}\ }\textbf {\bibinfo {volume} {91}},\ \bibinfo
  {pages} {044047} (\bibinfo {year} {2015})},\ \Eprint
  {https://arxiv.org/abs/1504.00009} {arXiv:1504.00009 [gr-qc]} \BibitemShut
  {NoStop}%
\bibitem [{\citenamefont {Herdeiro}\ \emph {et~al.}(2013)\citenamefont
  {Herdeiro}, \citenamefont {Degollado},\ and\ \citenamefont
  {R\'unarsson}}]{Herdeiro:2013pia}%
  \BibitemOpen
  \bibfield  {author} {\bibinfo {author} {\bibfnamefont {C.~A.~R.}\
  \bibnamefont {Herdeiro}}, \bibinfo {author} {\bibfnamefont {J.~C.}\
  \bibnamefont {Degollado}},\ and\ \bibinfo {author} {\bibfnamefont {H.~F.}\
  \bibnamefont {R\'unarsson}},\ }\bibfield  {title} {\bibinfo {title} {{Rapid
  growth of superradiant instabilities for charged black holes in a cavity}},\
  }\href {https://doi.org/10.1103/PhysRevD.88.063003} {\bibfield  {journal}
  {\bibinfo  {journal} {Phys. Rev.}\ }\textbf {\bibinfo {volume} {D88}},\
  \bibinfo {pages} {063003} (\bibinfo {year} {2013})},\ \Eprint
  {https://arxiv.org/abs/1305.5513} {arXiv:1305.5513 [gr-qc]} \BibitemShut
  {NoStop}%
\bibitem [{\citenamefont {Degollado}\ and\ \citenamefont
  {Herdeiro}(2014)}]{Degollado:2013bha}%
  \BibitemOpen
  \bibfield  {author} {\bibinfo {author} {\bibfnamefont {J.~C.}\ \bibnamefont
  {Degollado}}\ and\ \bibinfo {author} {\bibfnamefont {C.~A.~R.}\ \bibnamefont
  {Herdeiro}},\ }\bibfield  {title} {\bibinfo {title} {{Time evolution of
  superradiant instabilities for charged black holes in a cavity}},\ }\href
  {https://doi.org/10.1103/PhysRevD.89.063005} {\bibfield  {journal} {\bibinfo
  {journal} {Phys. Rev.}\ }\textbf {\bibinfo {volume} {D89}},\ \bibinfo {pages}
  {063005} (\bibinfo {year} {2014})},\ \Eprint
  {https://arxiv.org/abs/1312.4579} {arXiv:1312.4579 [gr-qc]} \BibitemShut
  {NoStop}%
\bibitem [{\citenamefont {Hod}(2013{\natexlab{b}})}]{Hod:2013fvl}%
  \BibitemOpen
  \bibfield  {author} {\bibinfo {author} {\bibfnamefont {S.}~\bibnamefont
  {Hod}},\ }\bibfield  {title} {\bibinfo {title} {{Analytic treatment of the
  charged black-hole-mirror bomb in the highly explosive regime}},\ }\href
  {https://doi.org/10.1103/PhysRevD.88.064055} {\bibfield  {journal} {\bibinfo
  {journal} {Phys. Rev. D}\ }\textbf {\bibinfo {volume} {88}},\ \bibinfo
  {pages} {064055} (\bibinfo {year} {2013}{\natexlab{b}})},\ \Eprint
  {https://arxiv.org/abs/1310.6101} {arXiv:1310.6101 [gr-qc]} \BibitemShut
  {NoStop}%
\bibitem [{\citenamefont {Hod}(2016)}]{Hod:2016kpm}%
  \BibitemOpen
  \bibfield  {author} {\bibinfo {author} {\bibfnamefont {S.}~\bibnamefont
  {Hod}},\ }\bibfield  {title} {\bibinfo {title} {{The charged black-hole bomb:
  a lower bound on the charge-to-mass ratio of the explosive scalar field}},\
  }\href {https://doi.org/10.1016/j.physletb.2016.02.009} {\bibfield  {journal}
  {\bibinfo  {journal} {Phys.\ Lett.\ B}\ }\textbf {\bibinfo {volume} {755}},\
  \bibinfo {pages} {177} (\bibinfo {year} {2016})},\ \Eprint
  {https://arxiv.org/abs/1606.00444} {arXiv:1606.00444 [gr-qc]} \BibitemShut
  {NoStop}%
\bibitem [{\citenamefont {Dias}\ and\ \citenamefont
  {Masachs}(2018)}]{Dias:2018zjg}%
  \BibitemOpen
  \bibfield  {author} {\bibinfo {author} {\bibfnamefont {O.~J.~C.}\
  \bibnamefont {Dias}}\ and\ \bibinfo {author} {\bibfnamefont {R.}~\bibnamefont
  {Masachs}},\ }\bibfield  {title} {\bibinfo {title} {{Charged black hole bombs
  in a Minkowski cavity}},\ }\href {https://doi.org/10.1088/1361-6382/aad70b}
  {\bibfield  {journal} {\bibinfo  {journal} {Class.\ Quant.\ Grav.}\ }\textbf
  {\bibinfo {volume} {35}},\ \bibinfo {pages} {184001} (\bibinfo {year}
  {2018})},\ \Eprint {https://arxiv.org/abs/1801.10176} {arXiv:1801.10176
  [gr-qc]} \BibitemShut {NoStop}%
\bibitem [{\citenamefont {Di~Menza}\ \emph {et~al.}(2020)\citenamefont
  {Di~Menza}, \citenamefont {Nicolas},\ and\ \citenamefont
  {Pellen}}]{DiMenza:2019zli}%
  \BibitemOpen
  \bibfield  {author} {\bibinfo {author} {\bibfnamefont {L.}~\bibnamefont
  {Di~Menza}}, \bibinfo {author} {\bibfnamefont {J.-p.}\ \bibnamefont
  {Nicolas}},\ and\ \bibinfo {author} {\bibfnamefont {M.}~\bibnamefont
  {Pellen}},\ }\bibfield  {title} {\bibinfo {title} {{A new type of charged
  black hole bomb}},\ }\href {https://doi.org/10.1007/s10714-020-2656-5}
  {\bibfield  {journal} {\bibinfo  {journal} {Gen.\ Rel.\ Grav.}\ }\textbf
  {\bibinfo {volume} {52}},\ \bibinfo {pages} {8} (\bibinfo {year} {2020})},\
  \Eprint {https://arxiv.org/abs/1903.02941} {arXiv:1903.02941 [gr-qc]}
  \BibitemShut {NoStop}%
\bibitem [{\citenamefont {Press}\ and\ \citenamefont
  {Teukolsky}(1972)}]{Press:1972zz}%
  \BibitemOpen
  \bibfield  {author} {\bibinfo {author} {\bibfnamefont {W.~H.}\ \bibnamefont
  {Press}}\ and\ \bibinfo {author} {\bibfnamefont {S.~A.}\ \bibnamefont
  {Teukolsky}},\ }\bibfield  {title} {\bibinfo {title} {{Floating orbits,
  superradiant scattering and the black-hole bomb}},\ }\href
  {https://doi.org/10.1038/238211a0} {\bibfield  {journal} {\bibinfo  {journal}
  {Nature}\ }\textbf {\bibinfo {volume} {238}},\ \bibinfo {pages} {211}
  (\bibinfo {year} {1972})}\BibitemShut {NoStop}%
\bibitem [{\citenamefont {Sanchis-Gual}\ \emph {et~al.}(2016)\citenamefont
  {Sanchis-Gual}, \citenamefont {Degollado}, \citenamefont {Montero},
  \citenamefont {Font},\ and\ \citenamefont {Herdeiro}}]{Sanchis-Gual:2015lje}%
  \BibitemOpen
  \bibfield  {author} {\bibinfo {author} {\bibfnamefont {N.}~\bibnamefont
  {Sanchis-Gual}}, \bibinfo {author} {\bibfnamefont {J.~C.}\ \bibnamefont
  {Degollado}}, \bibinfo {author} {\bibfnamefont {P.~J.}\ \bibnamefont
  {Montero}}, \bibinfo {author} {\bibfnamefont {J.~A.}\ \bibnamefont {Font}},\
  and\ \bibinfo {author} {\bibfnamefont {C.}~\bibnamefont {Herdeiro}},\
  }\bibfield  {title} {\bibinfo {title} {{Explosion and final state of an
  unstable Reissner-Nordstr\"om black hole}},\ }\href
  {https://doi.org/10.1103/PhysRevLett.116.141101} {\bibfield  {journal}
  {\bibinfo  {journal} {Phys.\ Rev.\ Lett.}\ }\textbf {\bibinfo {volume}
  {116}},\ \bibinfo {pages} {141101} (\bibinfo {year} {2016})},\ \Eprint
  {https://arxiv.org/abs/1512.05358} {arXiv:1512.05358 [gr-qc]} \BibitemShut
  {NoStop}%
\bibitem [{\citenamefont {Dolan}\ \emph {et~al.}(2015)\citenamefont {Dolan},
  \citenamefont {Ponglertsakul},\ and\ \citenamefont
  {Winstanley}}]{Dolan:2015dha}%
  \BibitemOpen
  \bibfield  {author} {\bibinfo {author} {\bibfnamefont {S.~R.}\ \bibnamefont
  {Dolan}}, \bibinfo {author} {\bibfnamefont {S.}~\bibnamefont
  {Ponglertsakul}},\ and\ \bibinfo {author} {\bibfnamefont {E.}~\bibnamefont
  {Winstanley}},\ }\bibfield  {title} {\bibinfo {title} {{Stability of black
  holes in Einstein-charged scalar field theory in a cavity}},\ }\href
  {https://doi.org/10.1103/PhysRevD.92.124047} {\bibfield  {journal} {\bibinfo
  {journal} {Phys. Rev.}\ }\textbf {\bibinfo {volume} {D92}},\ \bibinfo {pages}
  {124047} (\bibinfo {year} {2015})},\ \Eprint
  {https://arxiv.org/abs/1507.02156} {arXiv:1507.02156 [gr-qc]} \BibitemShut
  {NoStop}%
\bibitem [{\citenamefont {Lanir}\ \emph
  {et~al.}(2018{\natexlab{a}})\citenamefont {Lanir}, \citenamefont {Levi},
  \citenamefont {Ori},\ and\ \citenamefont {Sela}}]{Lanir:2017oia}%
  \BibitemOpen
  \bibfield  {author} {\bibinfo {author} {\bibfnamefont {A.}~\bibnamefont
  {Lanir}}, \bibinfo {author} {\bibfnamefont {A.}~\bibnamefont {Levi}},
  \bibinfo {author} {\bibfnamefont {A.}~\bibnamefont {Ori}},\ and\ \bibinfo
  {author} {\bibfnamefont {O.}~\bibnamefont {Sela}},\ }\bibfield  {title}
  {\bibinfo {title} {{Two-point function of a quantum scalar field in the
  interior region of a Reissner-Nordstr\"om black hole}},\ }\href
  {https://doi.org/10.1103/PhysRevD.97.024033} {\bibfield  {journal} {\bibinfo
  {journal} {Phys.\ Rev.\ D}\ }\textbf {\bibinfo {volume} {97}},\ \bibinfo
  {pages} {024033} (\bibinfo {year} {2018}{\natexlab{a}})},\ \Eprint
  {https://arxiv.org/abs/1710.07267} {arXiv:1710.07267 [gr-qc]} \BibitemShut
  {NoStop}%
\bibitem [{\citenamefont {Lanir}\ \emph
  {et~al.}(2018{\natexlab{b}})\citenamefont {Lanir}, \citenamefont {Levi},\
  and\ \citenamefont {Ori}}]{Lanir:2018rap}%
  \BibitemOpen
  \bibfield  {author} {\bibinfo {author} {\bibfnamefont {A.}~\bibnamefont
  {Lanir}}, \bibinfo {author} {\bibfnamefont {A.}~\bibnamefont {Levi}},\ and\
  \bibinfo {author} {\bibfnamefont {A.}~\bibnamefont {Ori}},\ }\bibfield
  {title} {\bibinfo {title} {{Mode-sum renormalization of
  $\langle\Phi^{2}\rangle$ for a quantum scalar field inside a Schwarzschild
  black hole}},\ }\href {https://doi.org/10.1103/PhysRevD.98.084017} {\bibfield
   {journal} {\bibinfo  {journal} {Phys.\ Rev.\ D}\ }\textbf {\bibinfo {volume}
  {98}},\ \bibinfo {pages} {084017} (\bibinfo {year} {2018}{\natexlab{b}})},\
  \Eprint {https://arxiv.org/abs/1808.06195} {arXiv:1808.06195 [gr-qc]}
  \BibitemShut {NoStop}%
\bibitem [{\citenamefont {Zilberman}\ and\ \citenamefont
  {Ori}(2021)}]{Zilberman:2021vgz}%
  \BibitemOpen
  \bibfield  {author} {\bibinfo {author} {\bibfnamefont {N.}~\bibnamefont
  {Zilberman}}\ and\ \bibinfo {author} {\bibfnamefont {A.}~\bibnamefont
  {Ori}},\ }\bibfield  {title} {\bibinfo {title} {{Quantum fluxes at the inner
  horizon of a near-extremal spherical charged black hole}},\ }\href
  {https://doi.org/10.1103/PhysRevD.104.024066} {\bibfield  {journal} {\bibinfo
   {journal} {Phys. Rev. D}\ }\textbf {\bibinfo {volume} {104}},\ \bibinfo
  {pages} {024066} (\bibinfo {year} {2021})},\ \Eprint
  {https://arxiv.org/abs/2105.06521} {arXiv:2105.06521 [gr-qc]} \BibitemShut
  {NoStop}%
\bibitem [{\citenamefont {Zilberman}\ \emph
  {et~al.}(2022{\natexlab{a}})\citenamefont {Zilberman}, \citenamefont
  {Casals}, \citenamefont {Ori},\ and\ \citenamefont
  {Ottewill}}]{Zilberman:2022iij}%
  \BibitemOpen
  \bibfield  {author} {\bibinfo {author} {\bibfnamefont {N.}~\bibnamefont
  {Zilberman}}, \bibinfo {author} {\bibfnamefont {M.}~\bibnamefont {Casals}},
  \bibinfo {author} {\bibfnamefont {A.}~\bibnamefont {Ori}},\ and\ \bibinfo
  {author} {\bibfnamefont {A.~C.}\ \bibnamefont {Ottewill}},\ }\bibfield
  {title} {\bibinfo {title} {{Two-point function of a quantum scalar field in
  the interior region of a Kerr black hole}},\ }\href@noop {} {\  (\bibinfo
  {year} {2022}{\natexlab{a}})},\ \Eprint {https://arxiv.org/abs/2203.07780}
  {arXiv:2203.07780 [gr-qc]} \BibitemShut {NoStop}%
\bibitem [{\citenamefont {Sela}(2018)}]{Sela:2018xko}%
  \BibitemOpen
  \bibfield  {author} {\bibinfo {author} {\bibfnamefont {O.}~\bibnamefont
  {Sela}},\ }\bibfield  {title} {\bibinfo {title} {{Quantum effects near the
  Cauchy horizon of a Reissner-Nordstr\"om black hole}},\ }\href
  {https://doi.org/10.1103/PhysRevD.98.024025} {\bibfield  {journal} {\bibinfo
  {journal} {Phys.\ Rev.\ D}\ }\textbf {\bibinfo {volume} {98}},\ \bibinfo
  {pages} {024025} (\bibinfo {year} {2018})},\ \Eprint
  {https://arxiv.org/abs/1803.06747} {arXiv:1803.06747 [gr-qc]} \BibitemShut
  {NoStop}%
\bibitem [{\citenamefont {Lanir}\ \emph {et~al.}(2019)\citenamefont {Lanir},
  \citenamefont {Ori}, \citenamefont {Zilberman}, \citenamefont {Sela},
  \citenamefont {Maline},\ and\ \citenamefont {Levi}}]{Lanir:2018vgb}%
  \BibitemOpen
  \bibfield  {author} {\bibinfo {author} {\bibfnamefont {A.}~\bibnamefont
  {Lanir}}, \bibinfo {author} {\bibfnamefont {A.}~\bibnamefont {Ori}}, \bibinfo
  {author} {\bibfnamefont {N.}~\bibnamefont {Zilberman}}, \bibinfo {author}
  {\bibfnamefont {O.}~\bibnamefont {Sela}}, \bibinfo {author} {\bibfnamefont
  {A.}~\bibnamefont {Maline}},\ and\ \bibinfo {author} {\bibfnamefont
  {A.}~\bibnamefont {Levi}},\ }\bibfield  {title} {\bibinfo {title} {{Analysis
  of quantum effects inside spherical charged black holes}},\ }\href
  {https://doi.org/10.1103/PhysRevD.99.061502} {\bibfield  {journal} {\bibinfo
  {journal} {Phys.\ Rev.\ D}\ }\textbf {\bibinfo {volume} {99}},\ \bibinfo
  {pages} {061502} (\bibinfo {year} {2019})},\ \Eprint
  {https://arxiv.org/abs/1811.03672} {arXiv:1811.03672 [gr-qc]} \BibitemShut
  {NoStop}%
\bibitem [{\citenamefont {Zilberman}\ \emph {et~al.}(2020)\citenamefont
  {Zilberman}, \citenamefont {Levi},\ and\ \citenamefont
  {Ori}}]{Zilberman:2019buh}%
  \BibitemOpen
  \bibfield  {author} {\bibinfo {author} {\bibfnamefont {N.}~\bibnamefont
  {Zilberman}}, \bibinfo {author} {\bibfnamefont {A.}~\bibnamefont {Levi}},\
  and\ \bibinfo {author} {\bibfnamefont {A.}~\bibnamefont {Ori}},\ }\bibfield
  {title} {\bibinfo {title} {{Quantum fluxes at the inner horizon of a
  spherical charged black hole}},\ }\href
  {https://doi.org/10.1103/PhysRevLett.124.171302} {\bibfield  {journal}
  {\bibinfo  {journal} {Phys.\ Rev.\ Lett.}\ }\textbf {\bibinfo {volume}
  {124}},\ \bibinfo {pages} {171302} (\bibinfo {year} {2020})},\ \Eprint
  {https://arxiv.org/abs/1906.11303} {arXiv:1906.11303 [gr-qc]} \BibitemShut
  {NoStop}%
\bibitem [{\citenamefont {Hollands}\ \emph
  {et~al.}(2020{\natexlab{a}})\citenamefont {Hollands}, \citenamefont {Wald},\
  and\ \citenamefont {Zahn}}]{Hollands:2019whz}%
  \BibitemOpen
  \bibfield  {author} {\bibinfo {author} {\bibfnamefont {S.}~\bibnamefont
  {Hollands}}, \bibinfo {author} {\bibfnamefont {R.~M.}\ \bibnamefont {Wald}},\
  and\ \bibinfo {author} {\bibfnamefont {J.}~\bibnamefont {Zahn}},\ }\bibfield
  {title} {\bibinfo {title} {{Quantum instability of the Cauchy horizon in
  Reissner\textendash{}Nordstr\"om\textendash{}deSitter spacetime}},\ }\href
  {https://doi.org/10.1088/1361-6382/ab8052} {\bibfield  {journal} {\bibinfo
  {journal} {Class. Quant. Grav.}\ }\textbf {\bibinfo {volume} {37}},\ \bibinfo
  {pages} {115009} (\bibinfo {year} {2020}{\natexlab{a}})},\ \Eprint
  {https://arxiv.org/abs/1912.06047} {arXiv:1912.06047 [gr-qc]} \BibitemShut
  {NoStop}%
\bibitem [{\citenamefont {Hollands}\ \emph
  {et~al.}(2020{\natexlab{b}})\citenamefont {Hollands}, \citenamefont {Klein},\
  and\ \citenamefont {Zahn}}]{Hollands:2020qpe}%
  \BibitemOpen
  \bibfield  {author} {\bibinfo {author} {\bibfnamefont {S.}~\bibnamefont
  {Hollands}}, \bibinfo {author} {\bibfnamefont {C.}~\bibnamefont {Klein}},\
  and\ \bibinfo {author} {\bibfnamefont {J.}~\bibnamefont {Zahn}},\ }\bibfield
  {title} {\bibinfo {title} {{Quantum stress tensor at the Cauchy horizon of
  the Reissner\textendash{}Nordstr\"om\textendash{}de Sitter spacetime}},\
  }\href {https://doi.org/10.1103/PhysRevD.102.085004} {\bibfield  {journal}
  {\bibinfo  {journal} {Phys. Rev. D}\ }\textbf {\bibinfo {volume} {102}},\
  \bibinfo {pages} {085004} (\bibinfo {year} {2020}{\natexlab{b}})},\ \Eprint
  {https://arxiv.org/abs/2006.10991} {arXiv:2006.10991 [gr-qc]} \BibitemShut
  {NoStop}%
\bibitem [{\citenamefont {Zilberman}\ \emph
  {et~al.}(2022{\natexlab{b}})\citenamefont {Zilberman}, \citenamefont
  {Casals}, \citenamefont {Ori},\ and\ \citenamefont
  {Ottewill}}]{Zilberman:2022aum}%
  \BibitemOpen
  \bibfield  {author} {\bibinfo {author} {\bibfnamefont {N.}~\bibnamefont
  {Zilberman}}, \bibinfo {author} {\bibfnamefont {M.}~\bibnamefont {Casals}},
  \bibinfo {author} {\bibfnamefont {A.}~\bibnamefont {Ori}},\ and\ \bibinfo
  {author} {\bibfnamefont {A.~C.}\ \bibnamefont {Ottewill}},\ }\bibfield
  {title} {\bibinfo {title} {{Quantum fluxes at the inner horizon of a spinning
  black hole}},\ }\href@noop {} {\  (\bibinfo {year} {2022}{\natexlab{b}})},\
  \Eprint {https://arxiv.org/abs/2203.08502} {arXiv:2203.08502 [gr-qc]}
  \BibitemShut {NoStop}%
\bibitem [{\citenamefont {Balakumar}\ and\ \citenamefont
  {Winstanley}(2020{\natexlab{b}})}]{Balakumar:2020jhe}%
  \BibitemOpen
  \bibfield  {author} {\bibinfo {author} {\bibfnamefont {V.}~\bibnamefont
  {Balakumar}}\ and\ \bibinfo {author} {\bibfnamefont {E.}~\bibnamefont
  {Winstanley}},\ }\bibfield  {title} {\bibinfo {title} {{Hadamard parametrix
  of the Feynman Green\textquoteright{}s function of a five-dimensional charged
  scalar field}},\ }\href {https://doi.org/10.1142/S0218271820410023}
  {\bibfield  {journal} {\bibinfo  {journal} {Int. J. Mod. Phys. D}\ }\textbf
  {\bibinfo {volume} {29}},\ \bibinfo {pages} {2041002} (\bibinfo {year}
  {2020}{\natexlab{b}})},\ \Eprint {https://arxiv.org/abs/2002.09448}
  {arXiv:2002.09448 [gr-qc]} \BibitemShut {NoStop}%
\bibitem [{\citenamefont {Herman}\ and\ \citenamefont
  {Hiscock}(1996)}]{Herman:1995hm}%
  \BibitemOpen
  \bibfield  {author} {\bibinfo {author} {\bibfnamefont {R.}~\bibnamefont
  {Herman}}\ and\ \bibinfo {author} {\bibfnamefont {W.~A.}\ \bibnamefont
  {Hiscock}},\ }\bibfield  {title} {\bibinfo {title} {{Renormalization of the
  charged scalar field in curved space}},\ }\href
  {https://doi.org/10.1103/PhysRevD.53.3285} {\bibfield  {journal} {\bibinfo
  {journal} {Phys.\ Rev.\ D}\ }\textbf {\bibinfo {volume} {53}},\ \bibinfo
  {pages} {3285} (\bibinfo {year} {1996})},\ \Eprint
  {https://arxiv.org/abs/gr-qc/9509015} {arXiv:gr-qc/9509015} \BibitemShut
  {NoStop}%
\bibitem [{\citenamefont {Herman}(1998)}]{Herman:1998dz}%
  \BibitemOpen
  \bibfield  {author} {\bibinfo {author} {\bibfnamefont {R.}~\bibnamefont
  {Herman}},\ }\bibfield  {title} {\bibinfo {title} {{A method for calculating
  the imaginary part of the Hadamard elementary function $G^{(1)}$ in static,
  spherically symmetric space-times}},\ }\href
  {https://doi.org/10.1103/PhysRevD.58.084028} {\bibfield  {journal} {\bibinfo
  {journal} {Phys.\ Rev.\ D}\ }\textbf {\bibinfo {volume} {58}},\ \bibinfo
  {pages} {084028} (\bibinfo {year} {1998})},\ \Eprint
  {https://arxiv.org/abs/gr-qc/9803064} {arXiv:gr-qc/9803064} \BibitemShut
  {NoStop}%
\end{thebibliography}%

\end{document}